\documentclass{aa}  

\usepackage{graphicx}
\usepackage[svgnames]{xcolor}
\usepackage[pdftex,colorlinks,citecolor=DarkBlue]{hyperref}
\usepackage{txfonts}
\usepackage{mathrsfs}
\bibpunct{(}{)}{;}{a}{}{,}

\newcommand{\Ms}{{M_\odot}}

\begin{document} 

   \title{Protoplanetary disk formation from the collapse of a prestellar core}

   \subtitle{}

   \titlerunning{}

   \author{
          Yueh-Ning Lee \inst{1,2,3,4}
          \and
          S\'ebastien Charnoz \inst{3}
          \and
          Patrick Hennebelle\inst{4,5,6}
          }
   \institute{
                Department of Earth Sciences, National Taiwan Normal University, 11677 Taipei, Taiwan \\
                  \email{ynlee@ntnu.edu.tw}
         \and
                Center of Astronomy and Gravitation, National Taiwan Normal University, 11677 Taipei, Taiwan
         \and
                 Institut de Physique du Globe de Paris, Sorbonne Paris Cit\'e, Universit\'e Paris Diderot, UMR 7154 CNRS, F-75005 Paris, France
         \and
                Universit\'{e} Paris Diderot, AIM, Sorbonne Paris Cit\'{e}, CEA, CNRS, F-91191 Gif-sur-Yvette, France
         \and
                IRFU, CEA, Universit\'{e} Paris-Saclay, F-91191 Gif-sur-Yvette, France
         \and                
             LERMA (UMR CNRS 8112), Ecole Normale Sup\'{e}rieure, 75231 Paris Cedex, France\\
             }

  \date{Accepted 14th Feb. 2020}

 
  \abstract
   {Between the two research communities that study star formation and protoplanetary disk evolution, only a few efforts have been made to understand and bridge the gap between studies of a collapsing prestellar core and a developed disk. While it has generally been accepted for about a decade that the magnetic field and its nonideal effects play important roles during the stellar formation, simple models of pure hydrodynamics and angular momentum conservation are still widely employed in the studies of disk assemblage in the framework of the so-called alpha-disk model because these models are simple.}
   {We revisit the assemblage phase of the protoplanetary disk and employ current knowledge of the prestellar core collapse.}
   {We performed 3D magnetohydrodynamic (MHD) simulations with ambipolar diffusion and full radiative transfer to follow the formation of the protoplanetary disk within a collapsing prestellar core. The global evolution of the disk and its internal properties were analyzed to understand how the infalling envelope regulates the buildup and evolution of the disk. We followed the global evolution of the protoplanetary disk from the prestellar core collapse during 100 kyr with a reasonable resolution of  AU. Two snapshots from this reference run were extracted and rerun with significantly increased resolution to resolve the interior of the disk.}
   {The disk that formed under our simulation setup is more realistic and agrees with recent observations of disks around class 0 young stellar objects. The source function of the mass flux that arrives at the disk and the radial mass accretion rate within the disk are measured and compared to analytical self-similar models based on angular momentum conservation. The source function is very centrally peaked compared to classical hydrodynamical models, implying that most of the mass falling onto the star does not transit through the midplane of the disk. We also found that the disk midplane is almost dead to turbulence, whereas upper layers and the disk outer edge are highly turbulent, and this is where the accretion occurs. The snow line, located at about 5-10 AU during the infall phase, is significantly farther away from the center than in a passive disk. This result might be of numerical origin.}
   {We studied self-consistent protoplanetary disk formation from prestellar core collapse, taking nonideal MHD effects into account. We developed a zoomed rerun technique to quickly obtain a reasonable disk that is highly stratified, weakly magnetized inside, and strongly magnetized outside. During the class 0 phase of protoplanetary disk formation, the interaction between the disk and the infalling envelope is important and ought not be neglected. We measured the complex flow pattern and compared it to the classical models of pure hydrodynamical infall. Accretion onto the star is found to mostly depend on dynamics at large scales, that is, the collapsing envelope, and not on the details of the disk structure. }

   \maketitle

\section{Introduction}

It is commonly accepted that the planets of our Solar System (S.S.) formed within 1 to 100 million years (Myr), but several lines of evidence show that planetary formation processes may have started on a much shorter timescale. 
The calcium-aluminum-rich inclusions (CAIs) that define the time origin of the S. S. most likely have formed in a hot environment close to the Sun. 
To explain their presence in carbonaceous chondrites far from the Sun, 
it has been proposed that they formed in the disk during the collapse stage of the prestellar core, when infall of diffuse medium was significant. They were subsequently transported outward during the viscous expansion of the disk \citep[see e.g.][]{Yang12, Desch18} or were transported  in disk winds or stellar outflow \citep{Shu04, Yang12, Pignatale18}. 
Moreover, simulations of stellar cluster formation within a molecular cloud showed that the typical infall time onto individual star-disk systems is about a fraction of a Myr \citep{Padoan14}. 

The low abundance of water in the inner S. S. \citep{Morbidelli16, Hyodo19} and the identification of several isotopic reservoirs in the S. S. \citep{Kruijer17, VanKooten16} argue for an early formation of a dynamical barrier such as Jupiter at about one Myr, but the building blocks of a barrier like this must have formed much earlier. 
Finally, a recent study showed that disks do not contain enough detectable dust to form the exoplanet populations \citep{Manara18}, 
implying that either
(1) most planets form in a few 100 thousand years (kyr) at  the class 0 or class I phase, or 
(2) a continuous dust replenishment process feeds planets over the lifetime of the disk \citep{Manara18}. 
All these elements raise the fundamental question whether planetary accretion started during the assemblage of the protoplanetary disks or after, as is often assumed. 
If accretion did start in the disk during the prestellar core infall, 
this would be a major change in our understanding of planet formation, 
as most studies still presume initial conditions from the {\it Minimum Mass Solar Nebula} model \citep[MMSN,][]{Hayashi79, Weidenschilling77}, which is pertinent only for an isolated disk. 

\section{Current understandings of the protoplanetary disk and the scope of the present work}
\subsection{Beyond the alpha-disk model}
Protoplanetary disks are classically described using the alpha-disk model \citep{Shakura73},  
which considers an isolated disk. 
While this model is highly  idealized, it is a useful tool for studying planet formation because it is flexible, and it is still the basis of most dust growth studies \citep[see, e.g.,][for a review]{Birnstiel12, Testi14} and planet population synthesis relevant for exoplanets \citep{Alibert13}. 
Accretion and transport processes across the disk are described with 
the effective viscosity $\nu=\alpha c_{\rm s}^2/\Omega$, 
where $\alpha$ is the dimensionless local shear stress coefficient that we define below, $c_{\rm s}$ is the thermal sound speed, and $\Omega$ is the Keplerian frequency.

The accretion rate onto the star in steady state, $\dot{M}=3 \pi \nu \Sigma$, is also controlled by $\alpha$. 
The disk is heated by both the protostellar radiation and the viscous dissipation \citep{Hueso05}. 
The value of $\alpha$ is linked to the intensity and the structure of turbulence in the disk. 
It has been thought during the last 15 years that the source of turbulence is \textit{\textup{magnetorotational instability}} (MRI), giving $\alpha > 10^{-3}$  \citep[see, e.g.,][]{Balbus91, Fromang06b} in ideal magneto-hydrodynamic (MHD) cases. 
However, later it was realized that nonideal MHD effects (ambipolar diffusion, ohmic diffusion) may preclude disk magnetization, 
such that $\alpha$ may in fact be $\ll 10^{-3}$  in the planet-forming region \citep[see, e.g.,][ for a review]{Turner14}. 
A low turbulence state like this may favor dust coagulation and planetesimal formation \citep{Bai10, Youdin11},  
but a low value of $\alpha$ cannot account for observed accretion rates onto the star, 
especially for young disks, with $10^{-8}  \Ms /{\rm yr} < \dot{M} < 10^{-5}  \Ms/ {\rm yr}$ \citep{Hartmann98}.

Studies of disks  with nonideal MHD show that the disk may in fact be vertically stratified, 
with a low-$\alpha$ midplane, high-$\alpha$ upper layers that drive accretion \citep{Turner08, Turner14}, 
and possibly additional MHD structures above a few pressure scale-heights ($H$) such as disk winds \citep[see, e.g.,][]{Bai13, Riols16}. 
These new understandings lead to descriptions that are more complex than the $\alpha$-prescription and might be time  and position dependent \citep{Suzuki16}. 
Nevertheless, all these studies still consider isolated disks.

\subsection{Attempts to study an assembling protoplanetary disk}
While many works have addressed the issue of disk formation, 
only a handful have attempted to study early planetary formation processes  (dust coagulation, transport, and planetesimal formation) during the collapse of the prestellar core that feeds the growing disk. 
The main difficulties are (1) modeling the infall of the envelope and (2) connecting this properly to the circumstellar disk. 
Classically, the self-similar singular isothermal sphere collapse solution has been used \citep[SIS,][]{Shu77, Ulrich76}. 
Angular momentum conservation inside the system is always assumed (a critical approximation that neglects magnetic forces). 
Several  works showed \citep{Nakamoto94, Hueso05} that under this hypothesis, disks may form with characteristics similar to the MMSN \citep{Nakamoto94} or to some disks observed in young stellar clusters \citep{Hueso05}. 
Using the SIS infall solution, it was shown that CAIs could form close to the star from condensing gas and would then be transported away through viscous disk relaxation \citep{Yang12, Pignatale18}.  

\citet{Drazkowska18} showed that planetesimals may form during the gas infall. 
In these models, the infall from the envelope is represented as a source term for the disk surface density, 
and the disk transport processes (controlled by  $\alpha$) are set independently. 
For example, \citet{Hueso05} considered a fully active disk, whereas \citet{Pignatale18} considered a disk with a dead zone. 
However, these two critical aspects are in fact linked. 
For example, as the gas falls onto the disk, it changes the ionization state of the disk and thus $\alpha$. 
Second, the turbulent motions inside the disk may be inherited from the turbulence or from the asymmetry in the infalling envelope.
In addition, MHD effects may cause some outflow, and part of the gas may be driven above the midplane. 
Finally, the accretion shock of the infalling envelope may heat the upper layers of the disk, thus changing the thermal structure of the disk. 
Therefore it is indispensable to properly study the gas infall that is coupled to the transport properties of the disk and the resulting disk structure in terms of the surface density, temperature profile, and $\alpha$.

\subsection{Previous numerical studies of the disk properties that include the envelope effect}

\citet{Vorobyov10,Vorobyov15b} studied 2D thin-disk models with radiative cooling, mechanical heating, and irradiation from the protostar and the background. For assumed frozen-in fields, the magnetic tensor was treated as a dilution factor of gravity, and the magnetic pressure was treated as being a few times the thermal pressure. They showed that the accretion processes within the disk are intrinsically variable, while external infall and radiation affect the fragmentation within the disk. \citet{Vorobyov15a} studied the effect of the environment on an embedded disk. The size of the disk is highly subject to the properties of the infalling material, in particular  the angular momentum. The disk fragmentation is affected as a consequence. The global mass accretion history onto the star is consistent with that onto the disk, while the burst modes are largely regulated by fragmentation within the disk. This 2D model does not describe the vertical stratification and thus does not have a source term for the surface density. Whether mass transits through the entire disk before reaching the star or falls through upper layers cannot be distinguished in such setup. 

\citet{Kuffmeier17} have developed a zoom-in technique to follow the evolution of nine sink particles at 2 AU in a simulation of a giant molecular cloud with a resolution of 126 AU with an ideal MHD prescription. They showed that the sink mass accretion is highly sensitive to the large-scale environment and is not as simple as an SIS collapse. \citet{Kuffmeier18} further increased the resolution of one of the zooms to 0.06 AU and demonstrated that numerical convergence is probably not reached. The short-term variation (bursts) of the mass accretion is highly attenuated with increased resolution. 

These works have demonstrated the challenges in studying protoplanetary disk evolution during the embedded phase. 
Compromises including thin-disk approximation and ideal MHD were made to alleviate computational cost.

\subsection{Purpose and structure of the present work}
The purpose of the present paper is to revisit the classical pure hydrodynamical model of disk assemblage \citep{Nakamoto94, Hueso05} with 3D nonideal MHD simulations during the first 100 kyr of the building phase of the disk. 
We wish to investigate 
(1) the mass flux onto the disk as a function of position and time, 
(2)  the mass accretion rate from the core onto the disk and that from the disk onto the
star, 
(3) whether the turbulent stress inside the disk or the envelope infall rate controls the accretion rate onto the star,
(4) the lost of angular momentum during the infall and the accretion,
(5) the disk size and its surface density, 
(6) the temperature profile within the disk and the location of the snow line, and 
(7) the behavior of $\alpha$ inside the disk.

To avoid confusion, we define "infall" as mass falling from the envelope to the disk, and "accretion" as mass falling from the disk onto the star throughout this manuscript.
The difficulty in this type of study is that long-term evolution and detailed structures need to be followed simultaneously, 
which is a computational challenge. 
The present work is an attempt in this direction: We present a detailed study of the disk and inflow structure of an assembling protoplanetary disk.
The paper is structured as follows: 
the numerical simulations are presented in Section \ref{st_simu}. 
In Section \ref{st_model} we briefly review the classical disk models that have been applied in planetary science. 
Detailed analyses of the simulation results are presented in Section \ref{st_ana}. 
The results and comparison with the $\alpha$-disk model are discussed in Section \ref{st_adisk}. 
Comparison to other numerical works is discussed in Section \ref{st_comp_simu}.
We conclude the paper in Section \ref{st_conc}.

\section{Numerical simulations}\label{st_simu}
To study the formation of the protoplanetary disk self-consistently, 
we simulated the collapse on the scale of a prestellar core. 
The code {\sc RAMSES} \citep{Teyssier02, Fromang06}, which treats MHD with an adaptive mesh refinement (AMR) scheme, was used.  
Because the initial prestellar core is of subparsec size and the disk forms as a high-density region of a few dozen AU, 
this collapse problem implies a wide dynamical range of scales, equivalent to four orders of magnitude.  
This type of simulation has been proven to be computationally difficult because the required resolution is high, which means both high computational power and large memory. 
Even with an AMR scheme, the highly structured mesh configuration makes the parallelization of the computation inefficient. 
For this reason, the simulations in this study were typically performed on about 100 cores at most. 

\subsection{Physical setup}
The initial condition was a prestellar core with a flat density profile, 0.02 pc (3800 AU) radius, and total mass of $1~\Ms$. 
The core initially had a weak solid-body rotation around its center. 
The radial profile of the rotation is not a major concern because the initial amount of rotation does not strongly affect the outcome of the disk \citep[][paper I hereafter]{Hennebelle20}. 
A uniform magnetic field threaded the core and was inclined by $\theta_{\rm mag} = 10^\circ$ with respect to the rotation axis. 
The magnetic field strength is described with the mass-to-flux ratio, $\mu$, which is the normalized value with respect to its critical value. 
The value $\mu=1$ implies magnetic criticality, 
and an increase in mass or decrease in field flux leads to gravitational collapse. 
The choice of $1/\mu = 0.3 $, compatible with observations \citep[e.g.,][]{Crutcher04,Maury18},  gives  a field of 0.2 mG at the center of the core.  
This run had no initial turbulent velocity field, although turbulent motion was generated spontaneously during the collapse as a result of density fluctuations. 

The physical properties were defined with dimensionless numbers. 
The thermal virial parameter $\alpha_{\rm vir}=0.4$ specifies the ratio of thermal to gravitational energy, 
and $\beta_{\rm rot}=0.04$ is the ratio of rotational to gravitational energy. 
A weak $m=2$ density perturbation of $10 \%$ was introduced perpendicular to the rotation axis to break the symmetry during the collapse. 
This prevents artificial symmetric pattern formation due to the grids. 
The effects of these physical parameters are discussed in paper I. 

We solved the complete MHD equations considering the ambipolar diffusion \citep[ion-neutral friction,][]{Masson12, Masson16, Marchand16}. 
The temperature was calculated with full radiative transfer with a flux-limited diffusion scheme \citep[FLD,][]{Commercon11}, and the temperature floor was set at 20 K. 
The thermal irradiation from the protostar was also considered using the prescription of \citet{Kuiper13}. 
The accretion luminosity was not included here because its actual value is highly uncertain
\citep[see the discussion in][]{hennebelle2020b}.

We did not take ohmic dissipation and Hall effect into account when treating nonideal MHD. 
Ohmic dissipation is important at higher densities and induces the formation of a small secondary disk around the protostar \citep{Tomida15}. 
Including this effect might not affect our general conclusions because the central high-density part is highly demagnetized, as we show in Sect. \ref{st_beta}. 
On the other hand, the Hall effect seems to indeed play a role during the formation of the disk \citep{Tsukamoto15, Marchand18, Zhao2020}, 
and this remains to be considered in future studies. 
It is hampered by large uncertainties, however, because knowledge on the grain properties is limited.

\subsection{Numerical setup}
The AMR scheme requires that the local Jeans length is resolved by 20 cells. 
The base grid had $2^6 = 64$ cells in each dimension, and the canonical run had a maximum refinement level $\ell_{\rm max}=14$, equivalent to a resolution $dx = 0.93$ AU. 
With the canonical setup, the interior of the disk is not very well resolved, and the dynamics within the disk is not fully physical. 
Therefore the analyses of this run focused more on the infall of the prestellar core envelope onto the disk and on the formation of the disk. 
This canonical choice for the resolution allowed us to follow the disk evolution during almost 100 kyr (150 kyr after the simulation started).  
Increasing the resolution requires a much longer computational time, 
and we therefore only reran the simulation from some snapshots with increased resolution to examine the consequences and to resolve the structure of the disk. 

At the center of the collapse, where the high density is no longer resolved by the fluid description, 
we inserted a sink particle to trap the collapsed mass \citep{Bleuler14}. 
This was necessary because otherwise, the warm and dense gas accumulated at the center expands in a way that is not physical. 
The sink particle forms when the gas molecular number density exceeds $n_{\rm thres}=3 \times 10^{13} {\rm cm}^{-3}$. 
It accretes from its surrounding within $4dx$. 
A threshold accretion scheme was used such that every cell having density higher than $n_{\rm acc} = n_{\rm thres}/3$ gives a fraction of the excessive mass to the sink particle. 
This fraction, $c_{\rm acc}$, was set to 0.1 in this study.

\subsection{Restart simulations with increased resolution}\label{st_restart}

\begin{table*}[]
\caption{Simulation parameters. {\it Columns from left to right:} Label, initial time, final time, restart snapshot origin, maximum level of refinement, physical resolution, and sink accretion density threshold. The canonical simulation is evolved during roughly 100 kyr after the sink formation. This corresponds to 600 orbits at 1 AU distance and 10 orbits at 20 AU for a star of $0.03~\Ms$. Two high-resolution restarts and one run without stellar irradiation are performed.  With increased resolution, the time steps are significantly reduced and the time span of the restarts is very short. }
\label{table_params}
\centering
\begin{tabular}{l r r c c c c c }
\hline\hline
Label   & $t_{\rm int}$ (kyr)  & $t_{\rm fin}$ (kyr)  &  Origin  & $\ell_{\rm max}$ & Resolution (AU) & $n_{\rm acc}$ (cm$^{-3}$)\\
\hline
R\_$\ell$14                   &  0.           & 151.    &       -         & 14   & 0.93  & $1 \times 10^{13}$ \\       
R\_40ky\_$\ell$18        &  102.532  & 103.144   & R\_$\ell$14  & 18   &  0.06  & $1 \times 10^{15}$ \\   
R\_80ky\_$\ell$18        &  138.203  & 138.519   & R\_$\ell$14      & 18   &  0.06 & $2 \times 10^{14}$  \\   
R\_$\ell$14\_nofeed                   &  119.          & 124.    &      R\_$\ell$14         & 14   & 0.93  & $1 \times 10^{13}$ \\ 

\hline
\end{tabular}
\end{table*}

In the canonical run, the entire disk is refined at the highest resolution ($2^{14}$).
The vertical structure of the disk is poorly resolved, however. 
To study the detailed dynamics of the disk interior, two restarts were performed at about 40 (R\_40ky\_$\ell$18) and 80 (R\_80ky\_$\ell$18) kyr after sink particle formation (at simulation time 61 kyr) with increased resolution. 
The model parameters are summarized in Table \ref{table_params}.

\begin{figure}[]
\centering
\includegraphics[trim=52 20 113 20,clip,height=0.16\textheight]{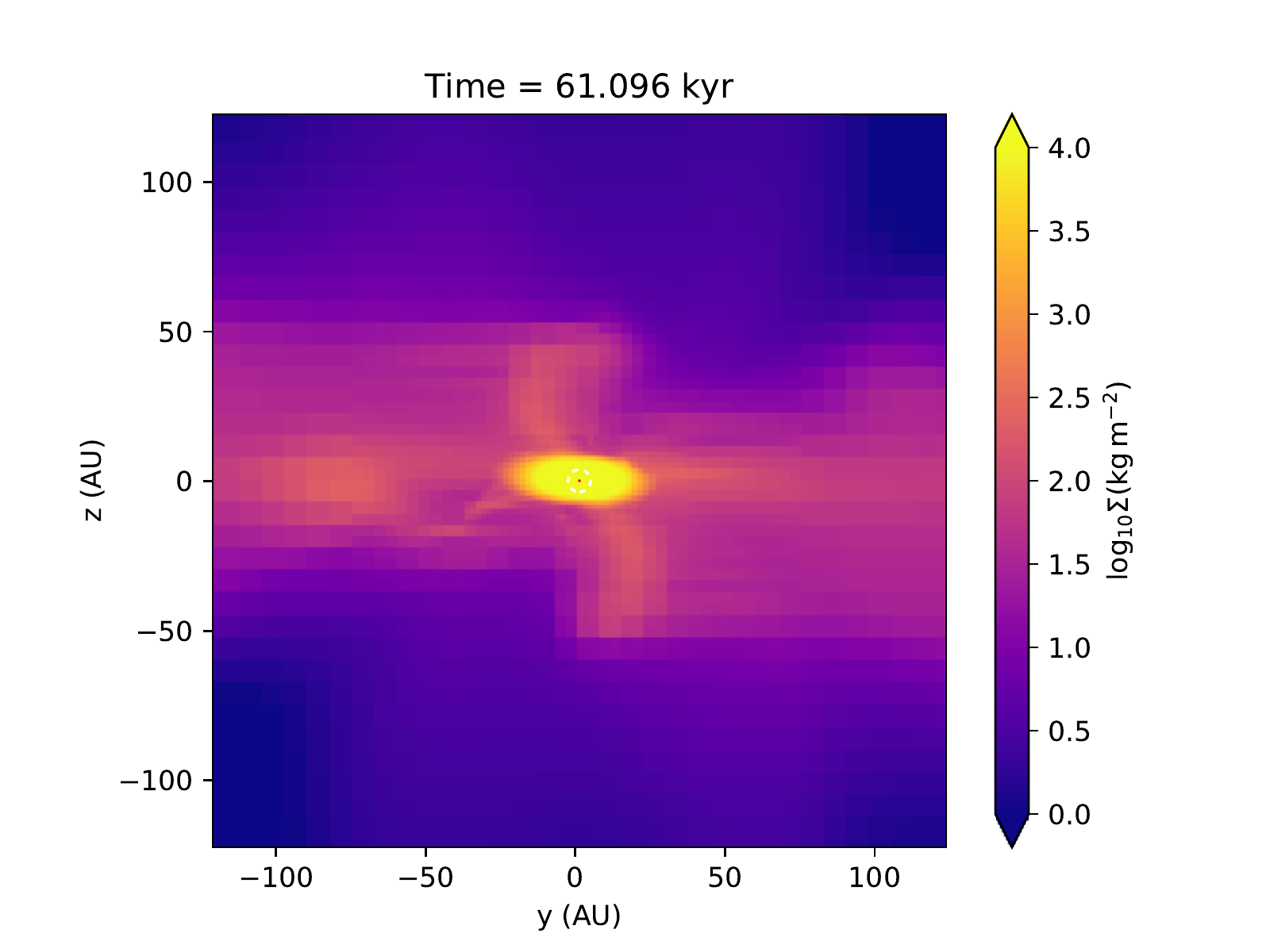}
\includegraphics[trim=52 20 50 20,clip,height=0.16\textheight]{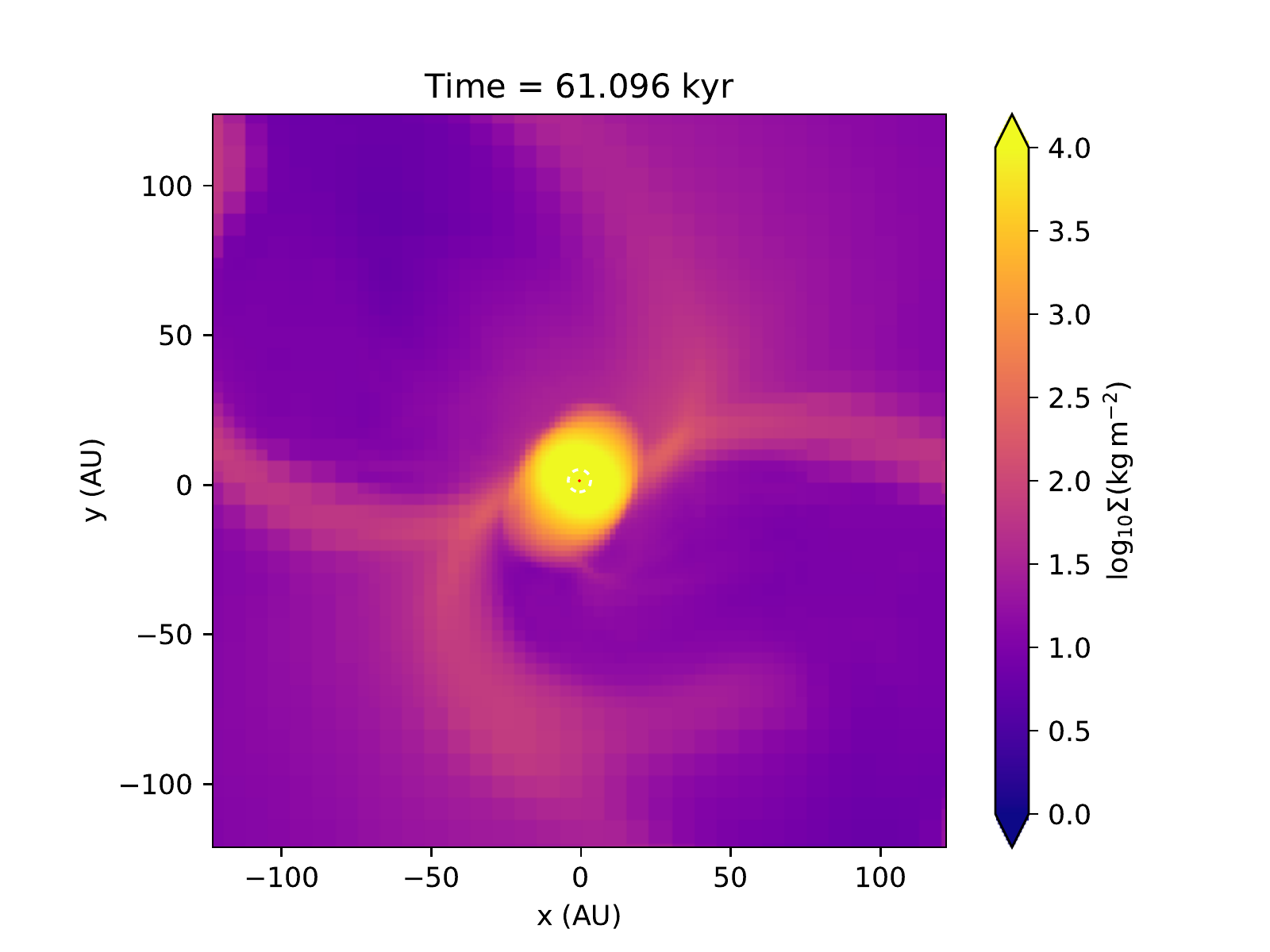} \\
\includegraphics[trim=52 20 113 20,clip,height=0.16\textheight]{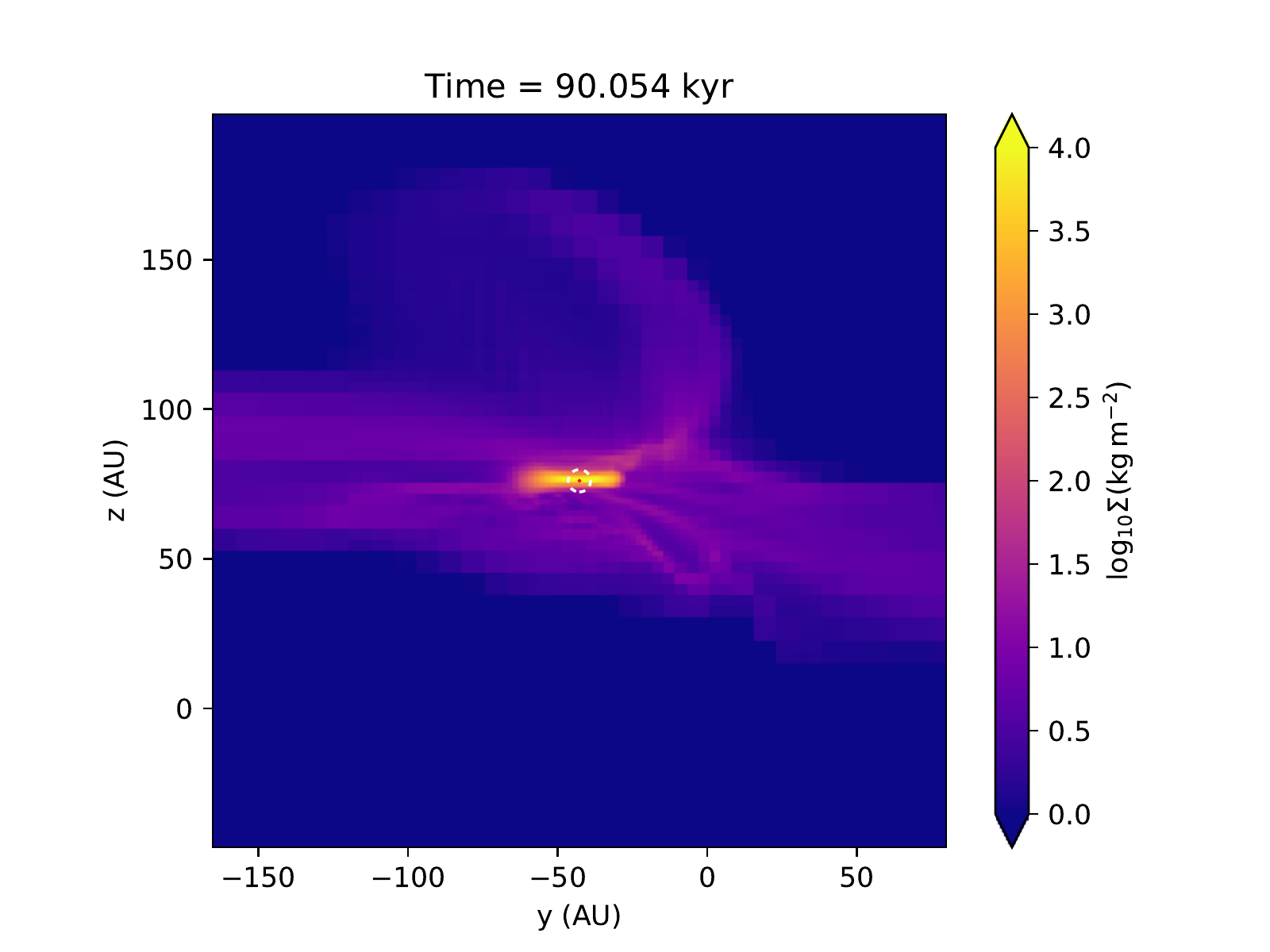} 
\includegraphics[trim=52 20 50 20,clip,height=0.16\textheight]{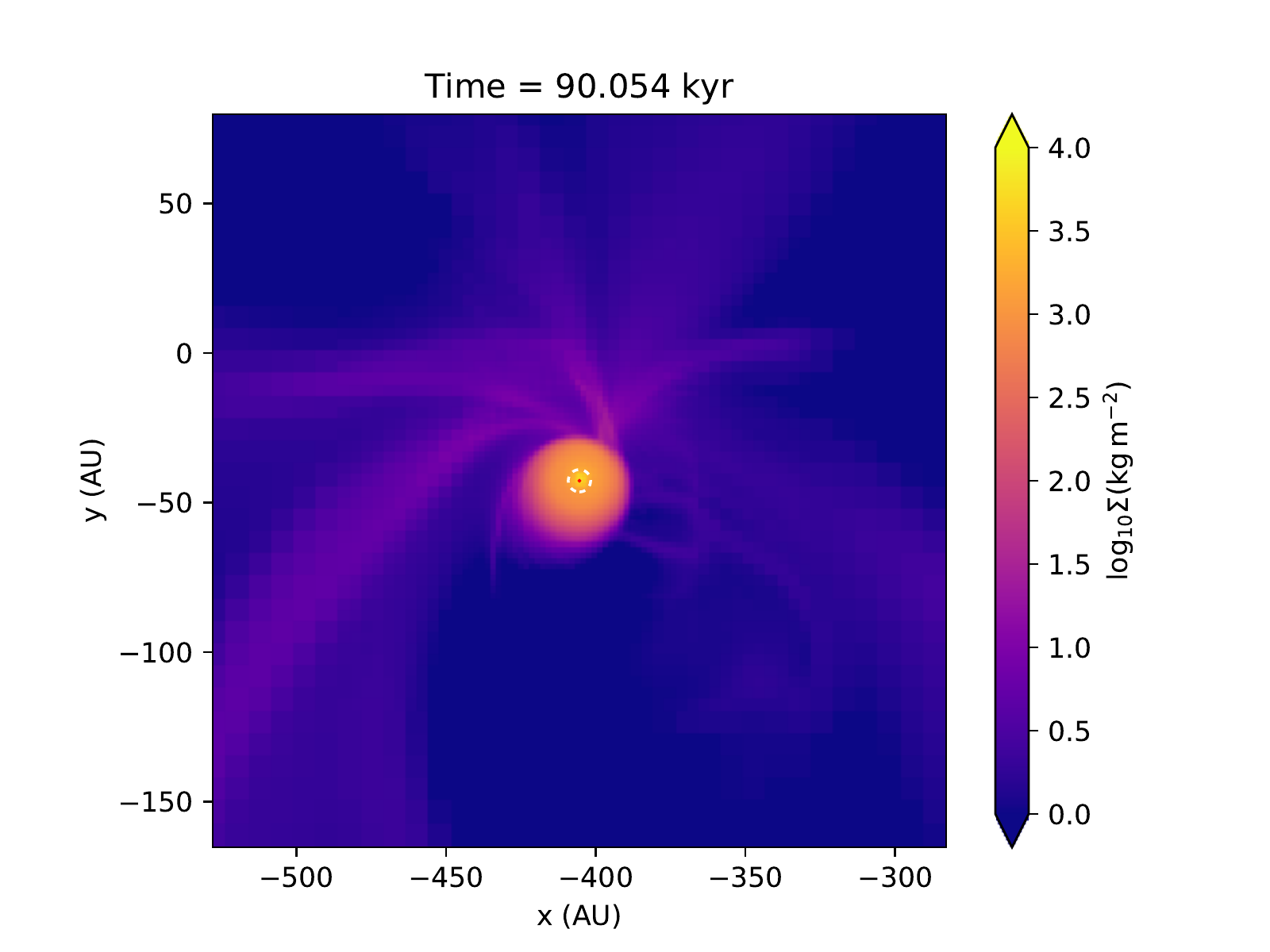} \\
\includegraphics[trim=52 20 113 20,clip,height=0.16\textheight]{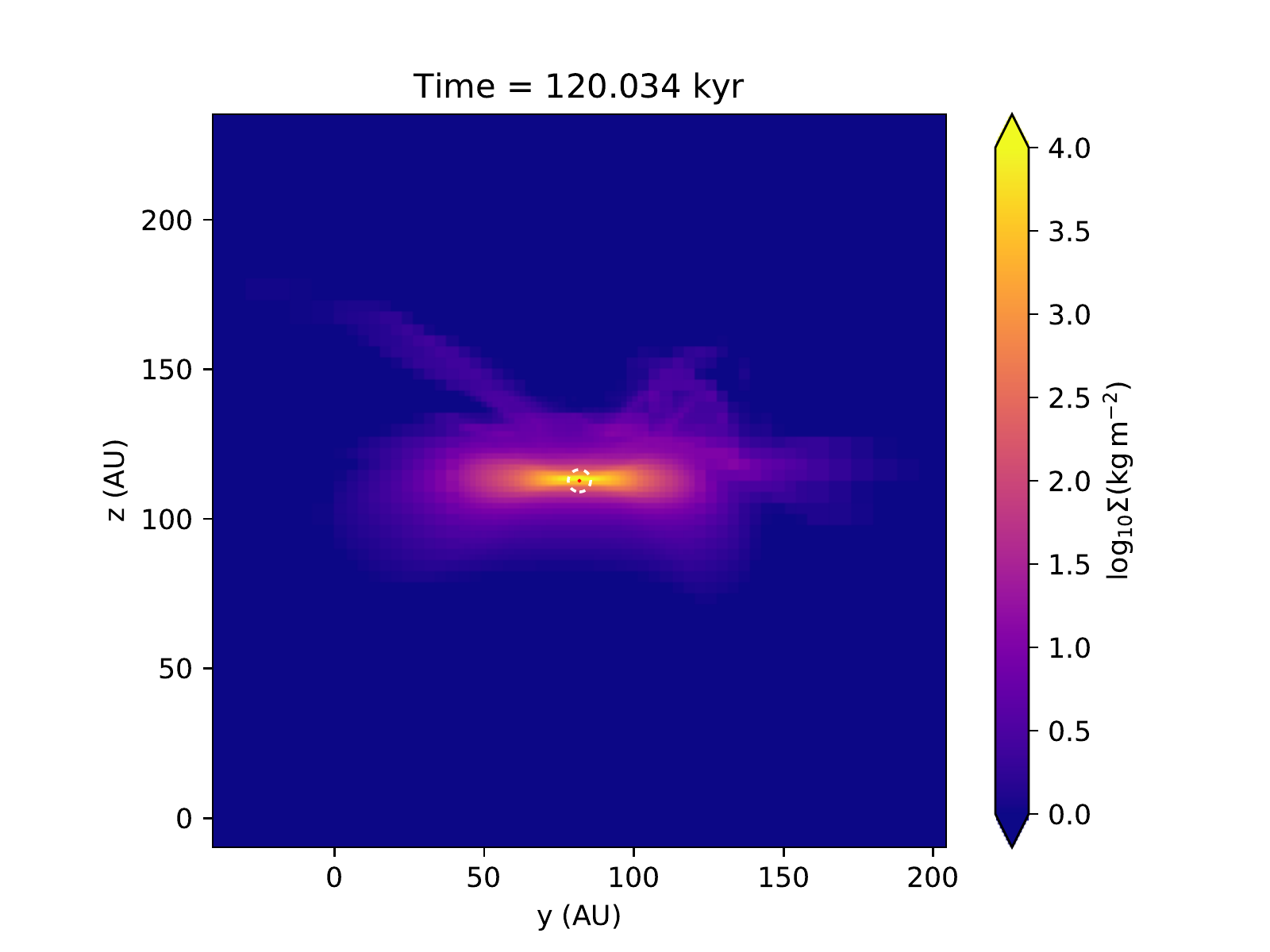}
\includegraphics[trim=52 20 50 20,clip,height=0.16\textheight]{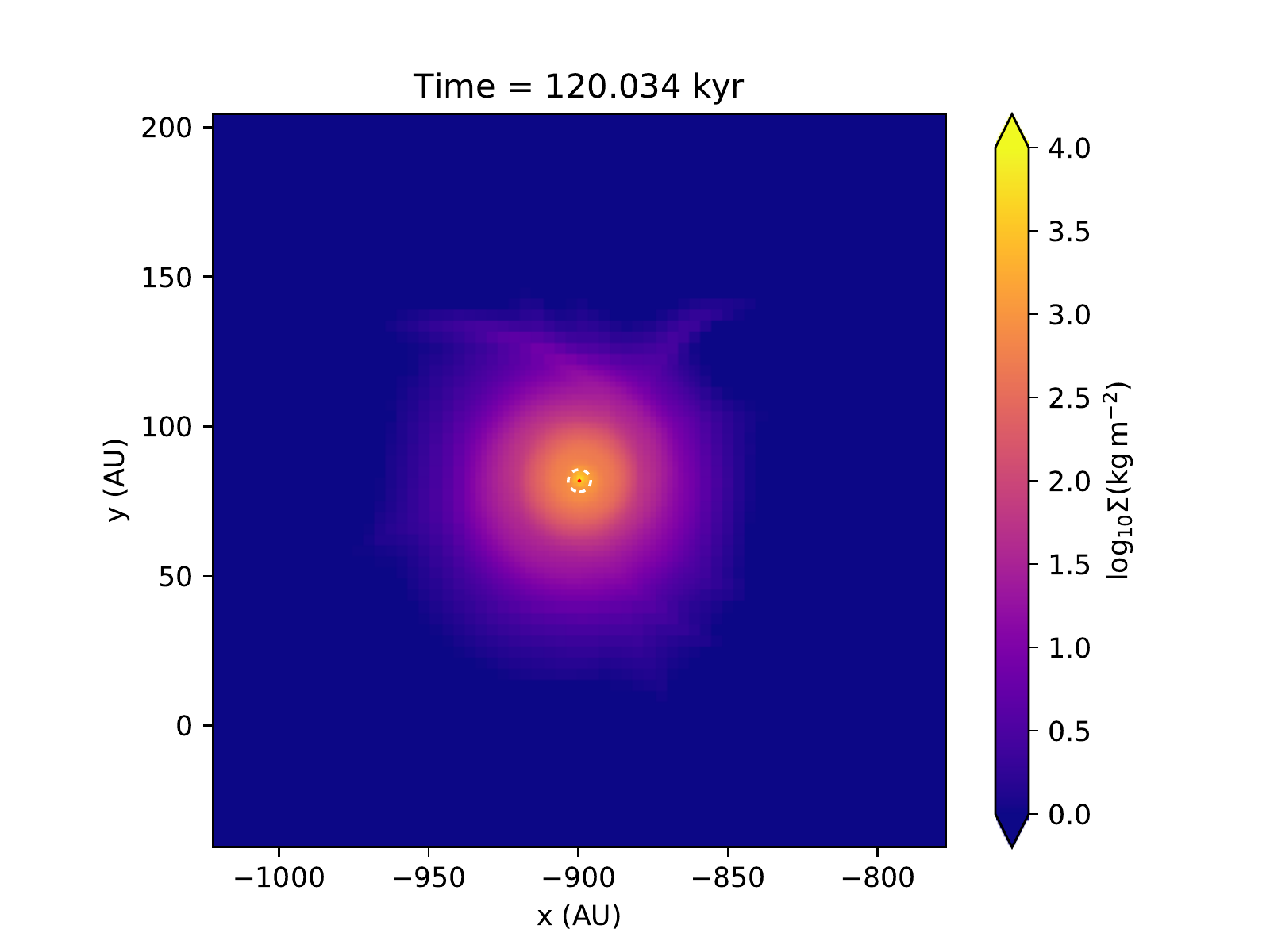}\\
\includegraphics[trim=52 20 113 20,clip,height=0.16\textheight]{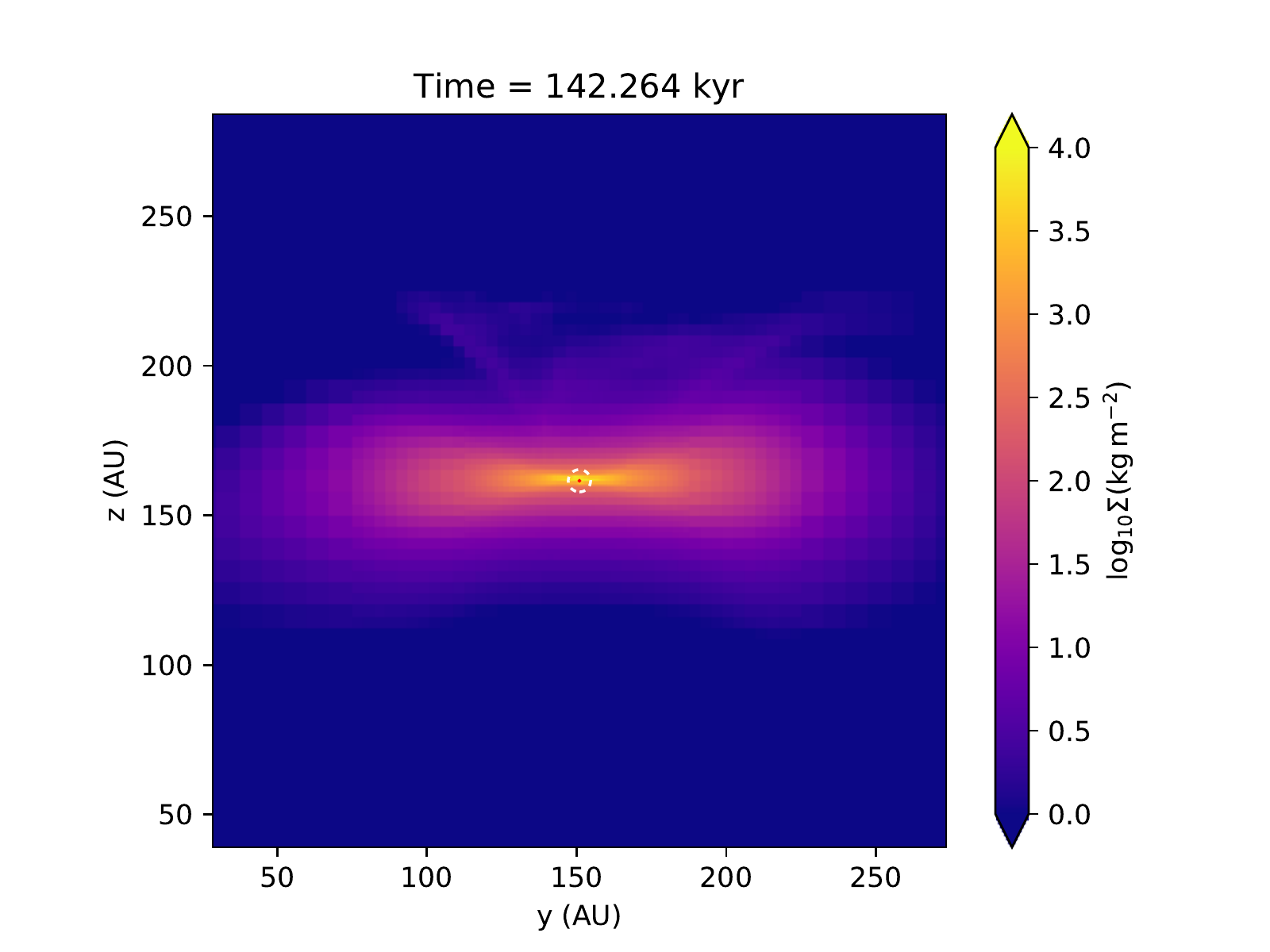} 
\includegraphics[trim=52 20 50 20,clip,height=0.16\textheight]{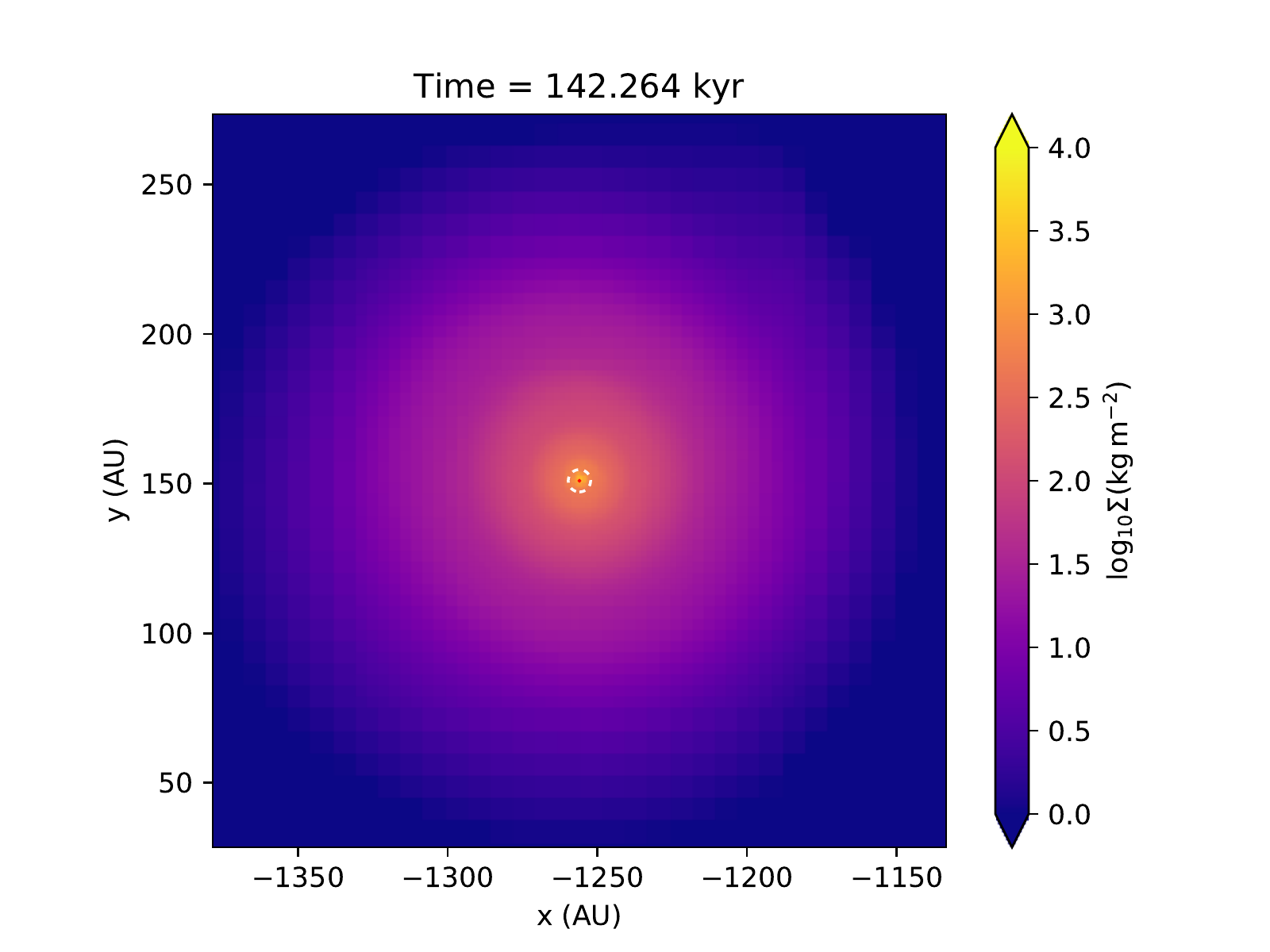} 
\caption{Column density snapshots of the canonical simulation R\_$\ell$14. {\it Left:} Edge-on. {\it Right:} Face-on. The coordinates are in AU, and the origin is at the box center. The images are centered on the sink particle, and the white dashed circle indicates the numerical radius of sink accretion ($4dx$). After the formation of the sink particle, the originally roundish structure of the flattened first Larson-core is quickly accreted and a flat disk forms, with sightly smaller size. The protoplanetary disk does not stay at the center of the computation box and might experience a complex accretion history. The size of the disk grows slowly from its initial radius of about 20 AU and presents fluctuations. At about 120 kyr, the disk grows more rapidly and reaches almost 100 AU. Although only a few snapshots are presented here, the actual picture is highly variant in time.}
\label{fig_l14_evol}
\end{figure}

Figure \ref{fig_l14_evol} shows four column density snapshots of the disk with edge-on and face-on views. 
The first snapshot corresponds to the time immediately after the sink particle formation. 
Shortly before the sink particle forms, the dense core grows and flattens as a result of the infall of the envelope. 
At the same time, the accreted gas with angular momentum leads to the formation of a rotating protoplanetary disk around the central dense core. 
This core-disk structure is initially thick and reaches a mass of almost $0.1~\Ms$. 
After the central star forms, most of the mass of this dense structure is quickly accreted, 
and the disk mass drops and stays at around $2-3\%$ of the stellar mass during the following evolution. 

Despite the inclined initial magnetic field, the disk is almost aligned with the prescribed rotation. 
In the long-term evolution, the disk gradually grows in size from a few dozen to almost 100 AU at simulation time 120 kyr, 
while there are fluctuations due to the turbulent nature of the infalling envelope. 
The disk sometimes decreases in size as a result of sudden strong accretions that are often asymmetric, 
that is, that come from one side of the disk, and as a result of the interchange instability. 
Magnetic field  is also expelled as an outflow in the form of loops or bubbles (see, e.g., the second row of Fig. \ref{fig_l14_evol}), which is highly dynamic.

\begin{figure}[]
\centering
\includegraphics[trim=52 20 113 20,clip,height=0.16\textheight]{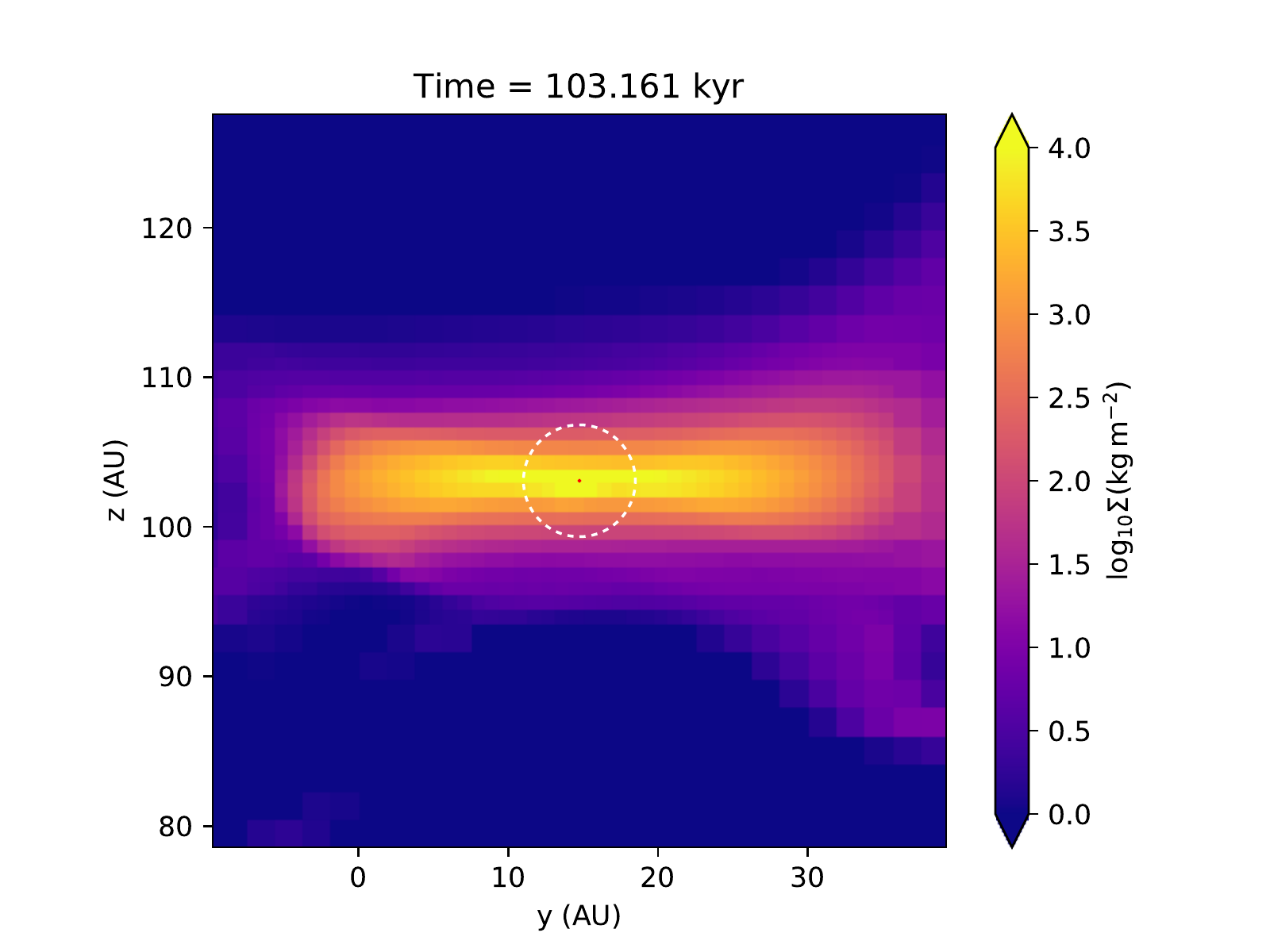}
\includegraphics[trim=52 20 50 20,clip,height=0.16\textheight]{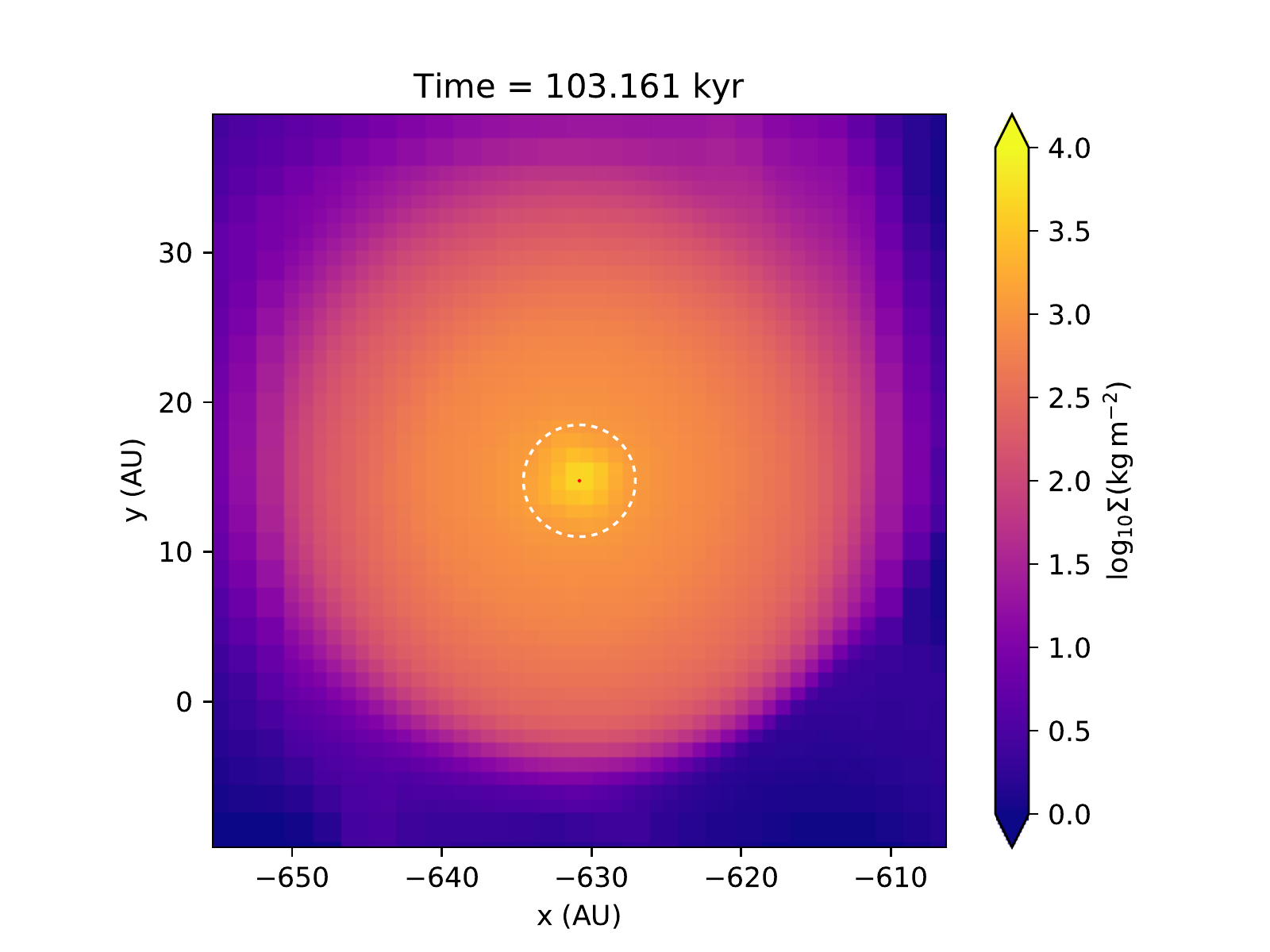} \\
\includegraphics[trim=52 20 113 20,clip,height=0.16\textheight]{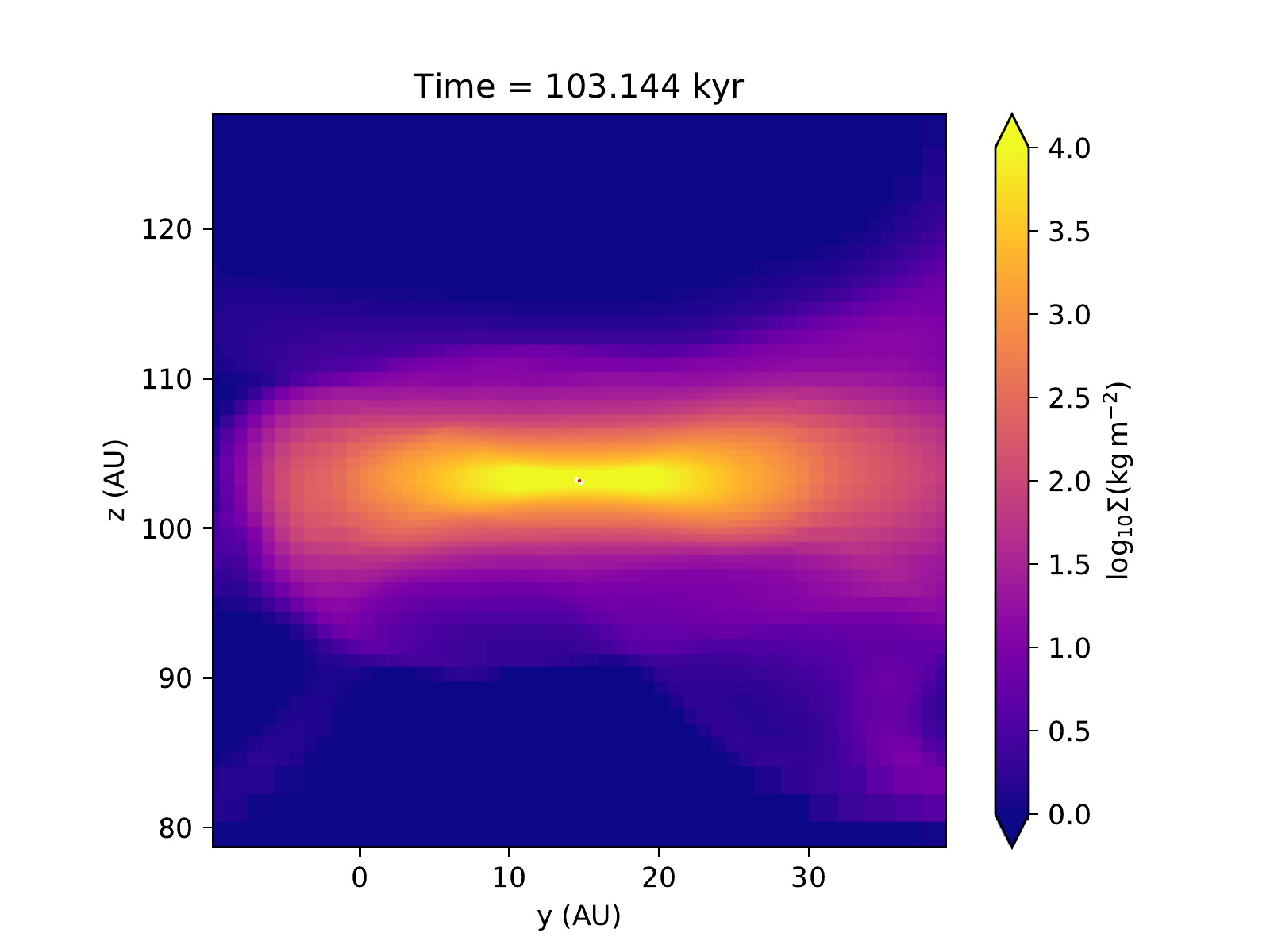}
\includegraphics[trim=52 20 50 20,clip,height=0.16\textheight]{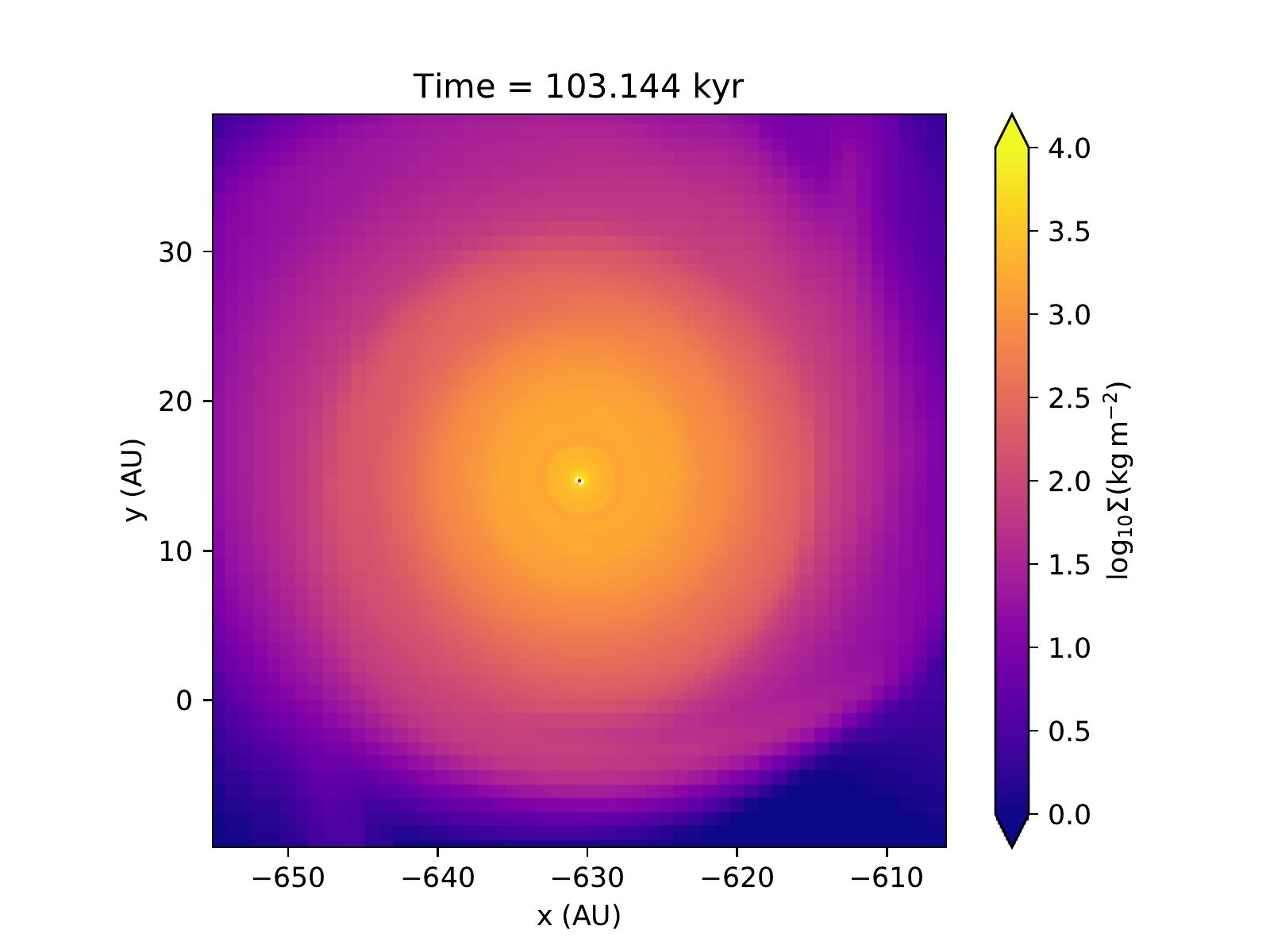} 
\caption{Zoomed views of the disk at age (103.1 kyr in simulation time) from edge-on (left column) and face-on (right column). {\it Top row:} R\_$\ell$14. {\it Bottom row:} R\_40ky\_$\ell$18. The column density is shown as in Fig. \ref{fig_l14_evol}. The disk interior is clearly more resolved in R\_40ky\_$\ell$18 with 0.06 AU resolution, particularly in the vertical direction.} 
\label{fig_40kyr_s}
\end{figure}

Figure \ref{fig_40kyr_s} shows the restart at 40 kyr along with the canonical run. With increased resolution, the disk structure is better resolved, while the global profile stays similar.
The refinement is increased in steps, passing progressively from level 15 to level 18, in order to avoid numerical artifacts from abruptly increased resolution. 
The final resolution at level 18 equals to 0.06 AU. 
When we increased the refinement levels, we had to adjust $n_{\rm acc}$ correspondingly to avoid artifacts. 
As explained in paper I, the disk mass is very sensitive to $n_{\rm acc}$. 
Estimated with a simple analytical model, $n_{\rm acc}$ should scale roughly as $dx^{-2}$.

We caution here that the value of $n_{\rm acc}$ likely sets an inner boundary condition for the disk and determines the density profile throughout the disk, and the temperature is affected in turn \citep[see discussions in][]{Hennebelle20}. 
\citet{Machida14} and \citet{Vorobyov19} have also shown similar results.
When disk simulations are interpreted physically, we therefore need to be cautious. 
As we show below, the effect of the choice of $n_{\rm acc}$ is seen in the disk global density, in the evolution of the snow line, and in the mass accretion rate onto the central star. 

\section{Existing models of disk formation}\label{st_model}
The collapse of the prestellar core is usually treated with axisymmetric and unmagnetized assumptions because they are simple and we lack observational evidence to support the contrary.
Before analyzing our simulations, we briefly review what has been learned from existing models.

\subsection{Collapse of a prestellar core: A purely hydrodynamical scenario}
One classical example of the collapse of a prestellar core is the SIS \citep[][]{Shu77}, 
which describes an inside-out collapse of mass shells that accrete onto a central star with a time-independent mass accretion rate, $\dot{M} = 0.975 c_{\rm s}^3/G$, where 
$G$ the gravitational constant. 
While it should be kept in mind that this solution is just one special case among a whole family of self-similar solutions \citep[e.g.,][]{Whitworth85}, 
it has been applied to many models for its simplicity. 
Furthermore,  in reality, a prestellar core has a finite extent and the mass accretion rate onto the central object decreases during the collapse and eventually stops, which results in an isolated disk. 
Observations have measured mass accretion rates a few times higher than this canonical value ($\sim 2 \times 10^{-6} ~\Ms/{\rm yr}$ for isothermal gas at 10 K) for class 0 objects
that are at the beginning of their protostellar collapse \citep[e.g.,][]{Ohashi99,Hirano02,WardThompson07}. 
Moreover, a declining mass accretion curve has been observed for young stellar objects between class 0 and class I stages, and simple accretion models have been proposed \citep[e.g.,][]{Henriksen97, Whitworth01}.

An easy although not unique \citep[see, e.g.,][]{verliat2020} way to form a disk is to introduce a small amount of rotation into the prestellar core 
under the assumption that the collapse solution remains unaffected at large scales.
Because angular momentum is conserved during the collapse (with at least about 100 times the contrast in size), 
rotation is significantly amplified, and this leads to the formation of a flattened structure in the centermost region around the star. 
While some numerical works exist \citep[e.g.,][]{Li14}, 
very few studies have examined analytically how a prestellar core collapses to form a star that is surrounded by a protoplanetary disk because the breakup of spherical symmetry complicates the problem. 

This simplified picture consists of prescribing each collapsing mass shell with a certain angular velocity and deriving how this mass is distributed radially when it arrives in the equatorial plane.  
The accretion onto the disk is expressed as the source term $S(r,t)$, 
which describes the mass flux as a function of the radial position and time. 
We discuss two widely accepted propositions.
These models provide in a deterministic manner (1) the total mass accretion rate onto the disk as a function of time, (2) the radius of the accreting zone of the disk as a function of time, and (3) the spatial distribution of mass flux onto the disk surface. 

\subsubsection{Free fall along a parabolic trajectory}
As proposed by \citet{Ulrich76}, in spherical coordinates (radius $r$, latitude $\theta$, and longitude $\phi$), all particles passing through radial position $r_0$
are assumed to have angular velocity $\dot{\phi_0}$  
and move in the plane that is instantaneously defined with the vectors $\vec{r}$ and $\vec{\phi}$. 
A parabolic trajectory, with the central star as the focus, is uniquely defined with these two criteria, while the nonzero velocity in the two other directions are not explicitly specified. 
This spherical shell is not a rotating solid body, but this did not concern the author, probably because the introduced amount of rotation is small. 
The particle is considered as accreted onto the disk when its trajectory intercepts the equatorial plane. 
The radial extent, or the centrifugal radius, of this region of accretion is derived as
\begin{align}\label{eq_rd}
r_{\rm c} = {\dot{\phi_0}^2 r_0^4 \over GM}, 
\end{align}
which is the pericenter of the particle that arrives from the equatorial plane, 
with
$M$ being 
the mass of the star-disk system, which is assumed to be dominated by the stellar mass. 

When we assume initial uniform distribution of mass on the shell, the source function can be derived as 
\begin{align}
S_{\rm p}(r,t) = {\dot{M}(t) \over 4 \pi  r_{\rm c}^2(t)} \left({r \over r_{\rm c}(t)}\right)^{-1}  \left(1-{r \over r_{\rm c}(t)}\right)^{-1/2},
\end{align}
where $\dot{M}(t)$ is the mass accretion rate of the mass that falls from the radius $r_0$, 
and is assumed to be a constant of time in most of the existing models, as previously explained. 

\subsubsection{Free fall onto a Keplerian radius}
On the other hand, \citet{Nakamoto94} and \citet{Hueso05} assumed that the particle, conserving its angular momentum, arrives at the radius of Keplerian rotation, regardless of the trajectory. 
With the same setup of initial mass and angular momentum distribution as in the previous model, 
the source function becomes 
\begin{align}
S_{\rm k}(r,t) = {\dot{M}(t) \over 8 \pi r_{\rm c}^2(t)} \left({r \over r_{\rm c}(t)}\right)^{-3/2}\left(1-\sqrt{r \over r_{\rm c}(t)}\right)^{-1/2}.
\end{align}

\subsubsection{Implication of these formalisms}
In the model by \citet{Ulrich76}, the particle has sub-Keplerian rotation when it reaches the disk and will fall inward. 
The exchange in angular momentum caused by friction of particles in the disk is not described, and a disk evolution model is further required to follow the redistribution of matter inside the disk. 
In the model by \citet{Nakamoto94} and \citet{Hueso05}, on the other hand, 
the trajectory is not described and the particle directly reaches its centrifugal radius, 
giving a slightly more centrally concentrated source function. 

\begin{figure}[]
\centering
\includegraphics[trim=0 0 0 0,clip,width=0.5\textwidth]{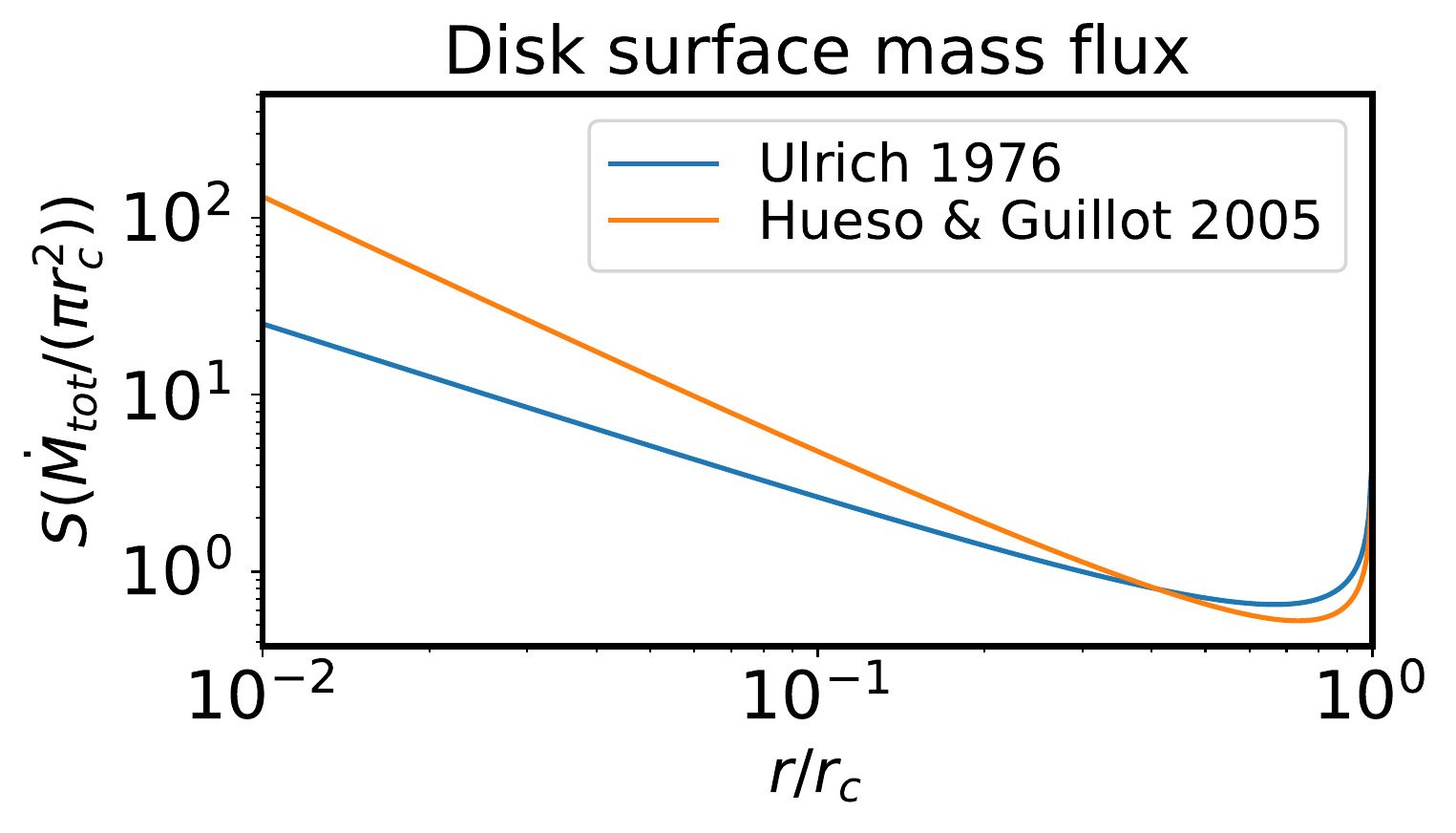}
\includegraphics[trim=0 0 0 0,clip,width=0.5\textwidth]{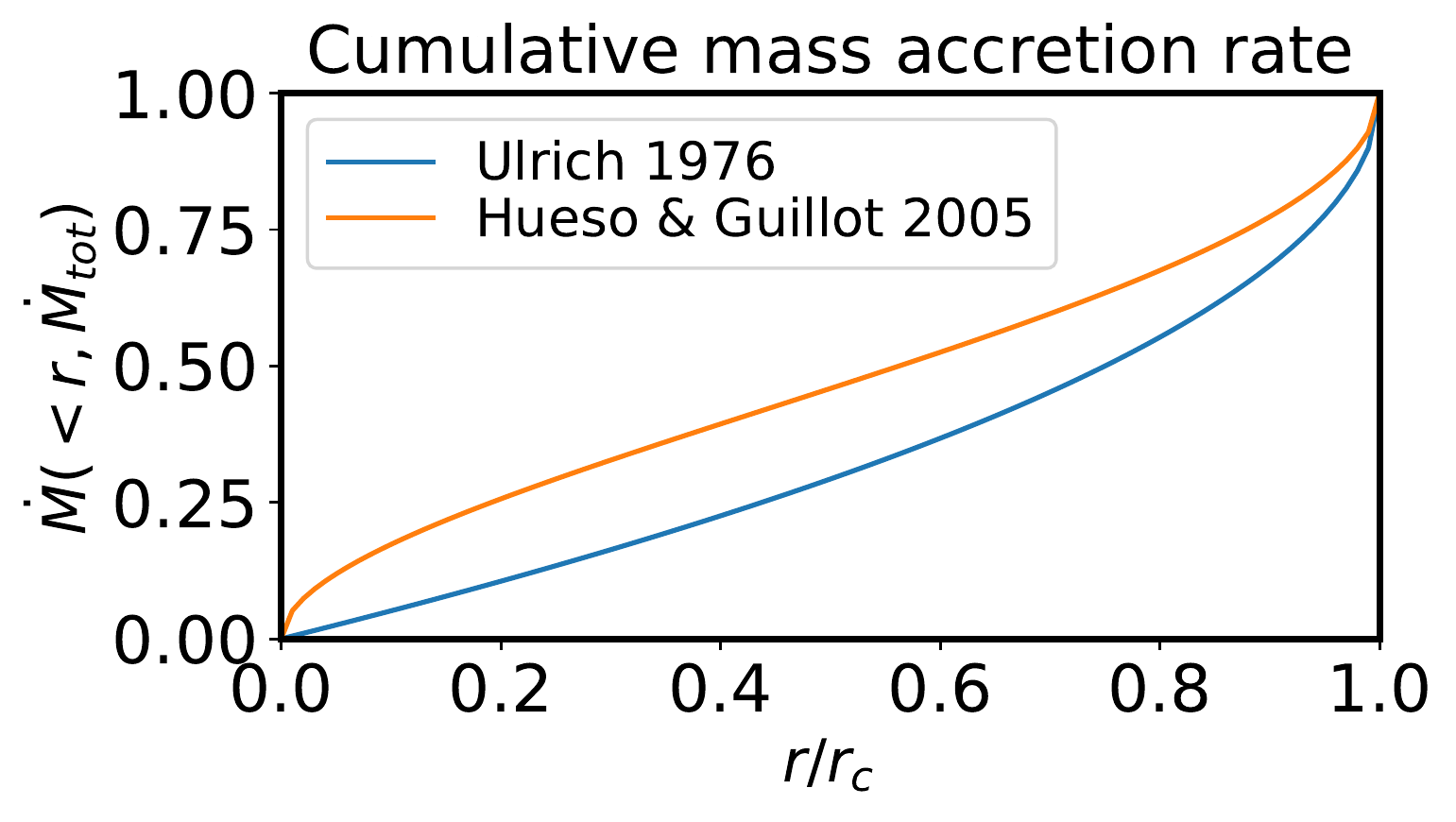}
\caption{{\it Upper panel:} Source function as proposed by \citet{Ulrich76} and \citet{Hueso05}. The radius is normalized to the centrifugal radius $r_{\rm c}$, and the source term is normalized to the mean surface mass flux ($\dot{M}_{\rm tot}/(\pi r_{\rm c}^2)$) such that its surface integral gives unity. {\it Lower panel:} Cumulated mass accretion rate (normalized by the total mass accretion rate), $\dot{M}$, integrated within radius $r$. The values of $\dot{M}_{\rm tot}$ and $r_{\rm c}$ are to be specified as functions of time according to the initial conditions of the prestellar core envelope. }
\label{fig_source_classical}
\end{figure}

The two models are shown in Fig. \ref{fig_source_classical}. 
For a more intuitive understanding, we also show the accumulated mass accretion within radius $r$,
\begin{align}
\dot{M}(<r) = \int_0^r S(r^\prime) 2 \pi r^\prime dr^\prime,
\end{align}
which equals zero at the origin and the total $\dot{M}$ at $r_{\rm c}$.
The difference between the two prescriptions does not appear to strongly affect the evolution of the disk \citep{Hueso05}. 
Both models rely on pure hydrodynamical arguments, and the initial cloud is mapped to the disk deterministically, in a shell-by-shell manner. 
The instantaneous source function entirely depends on the assumption of uniform mass distribution on the shell and constant angular velocity throughout the shell. 
While widely accepted because they are convenient, these models are probably too simplistic. 

The mass accretion rate $\dot{M}(t)$ and centrifugal radius $r_{\rm c}(t)$ are both determined by the initial conditions (density and angular momentum profiles) of the core. 
The inside-out collapse solution of \citet{Shu77} suggests a constant accretion rate $\dot{M}(t)$. 
By assuming a rotation profile of the prestellar core (often solid body, which leads to $r_{\rm c}(t) \propto t^3$), the full accretion history of the disk can be constructed. 
Simple variations of this class of models can be prescribed with different initial density and rotation profiles, 
leading to different histories of $\dot{M}(t)$ and $r_{\rm c}(t)$. 

The source function is of particular interest for the studies of early S.S. evolution because it implies the temperature-pressure history that the matter inside the S.S. can experience. 
In a purely hydrodynamic model, a non-negligible fraction of mass arrives at the inner part of the disk, where the temperature is high due to viscous heating and protostellar irradiation. 
The centrifugal radius grows rapidly in time such that accretion later arrives at the cooler part of the disk. 
This model has been used to explain the composition diversity of chondritic meteorites \citep{Pignatale18}.

\subsection{Effects of magnetic field and nonideal MHD}
Observational evidence increasingly shows that the star-forming gas is likely magnetized, and so is the disk because it is unlikely that the magnetic field at this scale is completely removed. 
In the past decade, numerical simulations started to take nonideal MHD effects in the studies of early disk formation into account. 
The authors found that the size of the disk is regulated by the magnetic field and is smaller than what would have been expected when pure hydrodynamics were considered \citep[see e.g.,][]{Dapp10,Inutsuka12, Li14,Tsukamoto15, Tomida15,Hennebelle16, Machida16, Zhao18, Wurster18, Lam2019}. 
This is also in line with the observations suggesting that most class 0 disks are small \citep[<50 AU,][]{Andrews18, Maury19}.

In this work, we aim to provide a revised model for the assembly of a protoplanetary disk by taking into account the nonideal MHD effects and the finite extent of the prestellar core. 
This can serve as more realistic framework for future studies of the early evolution of protoplanetary disks.

\section{Disk characteristics evaluated from simulations}\label{st_ana}
Before performing the canonical run, we simulated the collapse of a nonmagnetized core. 
The outcome is very similar to what is suggested by classical models: 
A disk forms due to angular momentum conservation, and its size grows quickly in time.
However, this disk is highly unstable and fragments very quickly \citep[see, e.g.,][]{Matsumoto03}. 
Therefore it is unlikely that this scenario can lead to an extended and evolved disk.
Observations \citep[e.g.,][]{ALMA15} also suggest that disks are probably small at early times. 
This motivated us to include the magnetic field along with its nonideal effects. 

In the presence of ambipolar diffusion, the disk size is regulated by the competition between the magnetic braking that decreases the angular momentum of the gas and the ambipolar diffusion that leaks the field lines outward and reduces the braking \citep{Hennebelle16}. 
Given the very different picture with respect to ideal MHD simulations, 
nonideal MHD effects are indeed necessary to understand the evolution of disk structure. 
The magnetic flux is allowed to leak outward, reducing the braking, and a rotation-supported demagnetized disk can form as a consequence.  

The simulation started with a core in free collapse, and thus it takes some time for the mass to accumulate before the protostar and the disk form. 
When the central density increases, 
the gas is heated by the collapse due to the increasing dust opacity, and the pressure locally supports against the self-gravitating collapse. 
A quasi-stationary first hydrostatic dense core \citep[or first Larson core,][]{Larson69,Masunaga98,Vaytet17} forms as a consequence, 
and this core has a slightly flattened shape because of rotation. 
Further mass accumulation leads to the formation of a sink particle that represents the protostar at the center about 61 kyr after the beginning of the simulation. 

We evaluated several disk properties from the simulations. 
All quantities were averaged in the azimuthal direction and between the upper and lower halves of the disk. 

\subsection{Disk geometry and global properties}\label{st_result}
The first step was to clearly identify the disk region and infer its geometrical properties. 
For this purpose, we analyzed the distribution of mass as a function of radius and altitude to derive the scale height, 
surface density, and radial extent of the disk. 
The analyses were made inside a selected cylindrical region with radius $r_{\rm cyl} =100$ AU and half-height $z_{\rm cyl} =100$ AU, 
of which the center and the axis were first evaluated by selecting dense cells over a certain threshold as a first step of disk identification (see details in Appendix \ref{ap_id}). 
At later times when the disk had grown in size, the selected region was increased to $r_{\rm cyl} = z_{\rm cyl}  =1000$ AU. 
As we show below, because the disk is mostly dominated by its central dense part, 
the size of the cylindrical region is not important as long as it is large enough with respect to the disk. 

\subsubsection{Vertical density profiling: Scale height $H$}\label{st_height}

\begin{figure}[]
\centering
\setlength{\unitlength}{0.5\textwidth}
\begin{picture}(1,1.8)
\put(0,1.35){\includegraphics[trim=0 0 0 5,clip,width=0.5\textwidth]{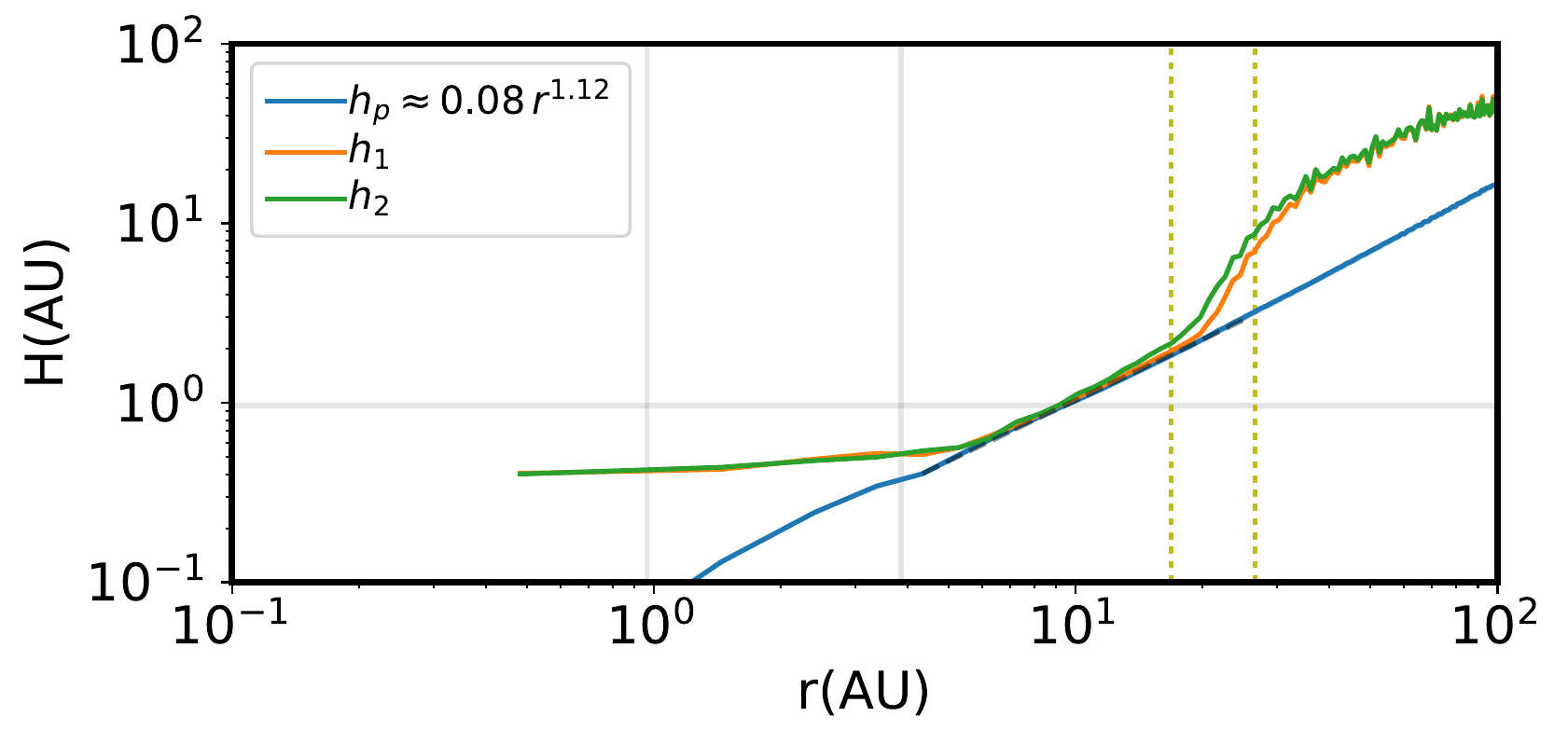}}
\put(0,0.9){\includegraphics[trim=0 0 0 5,clip,width=0.5\textwidth]{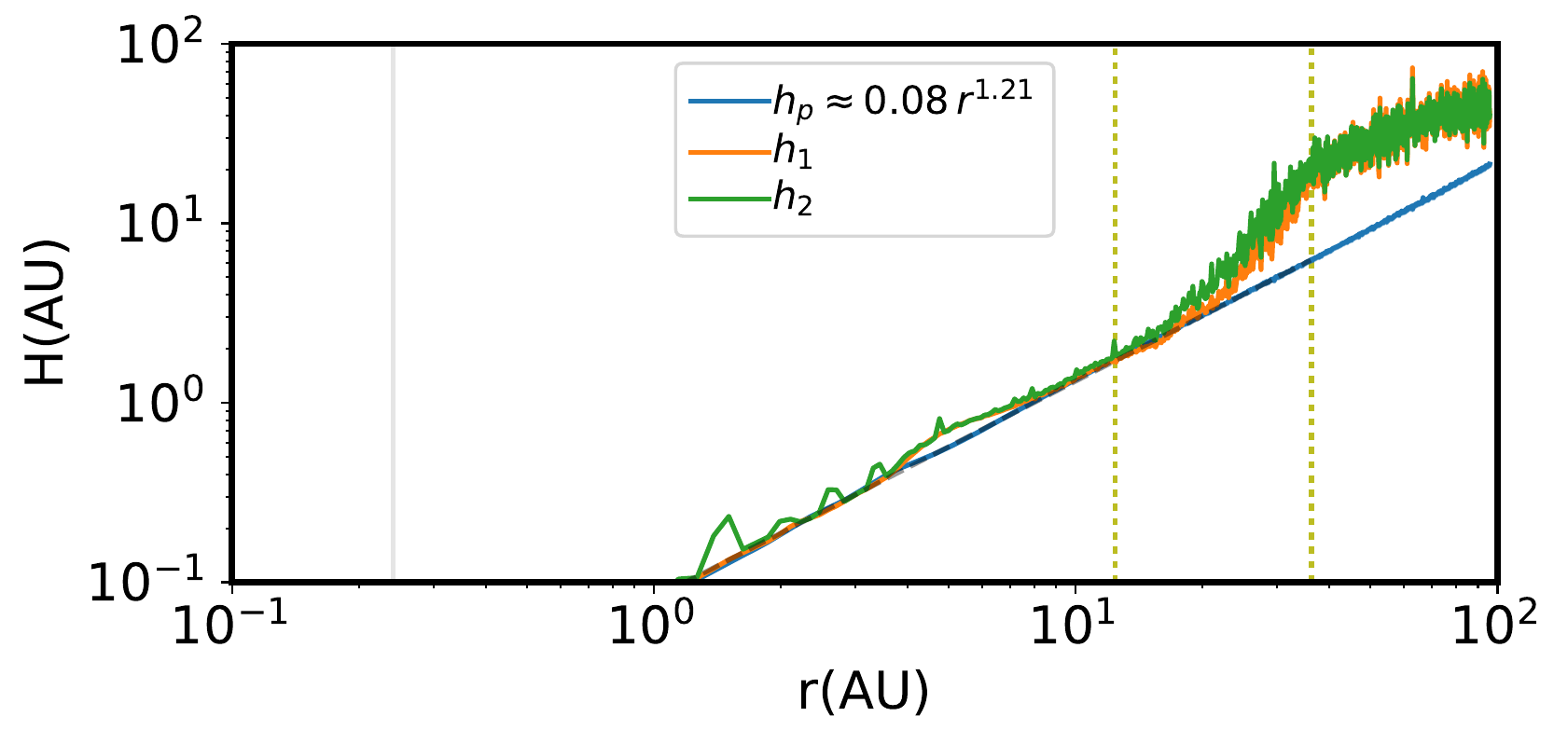}}
\put(0,0.45){\includegraphics[trim=0 0 0 5,clip,width=0.5\textwidth]{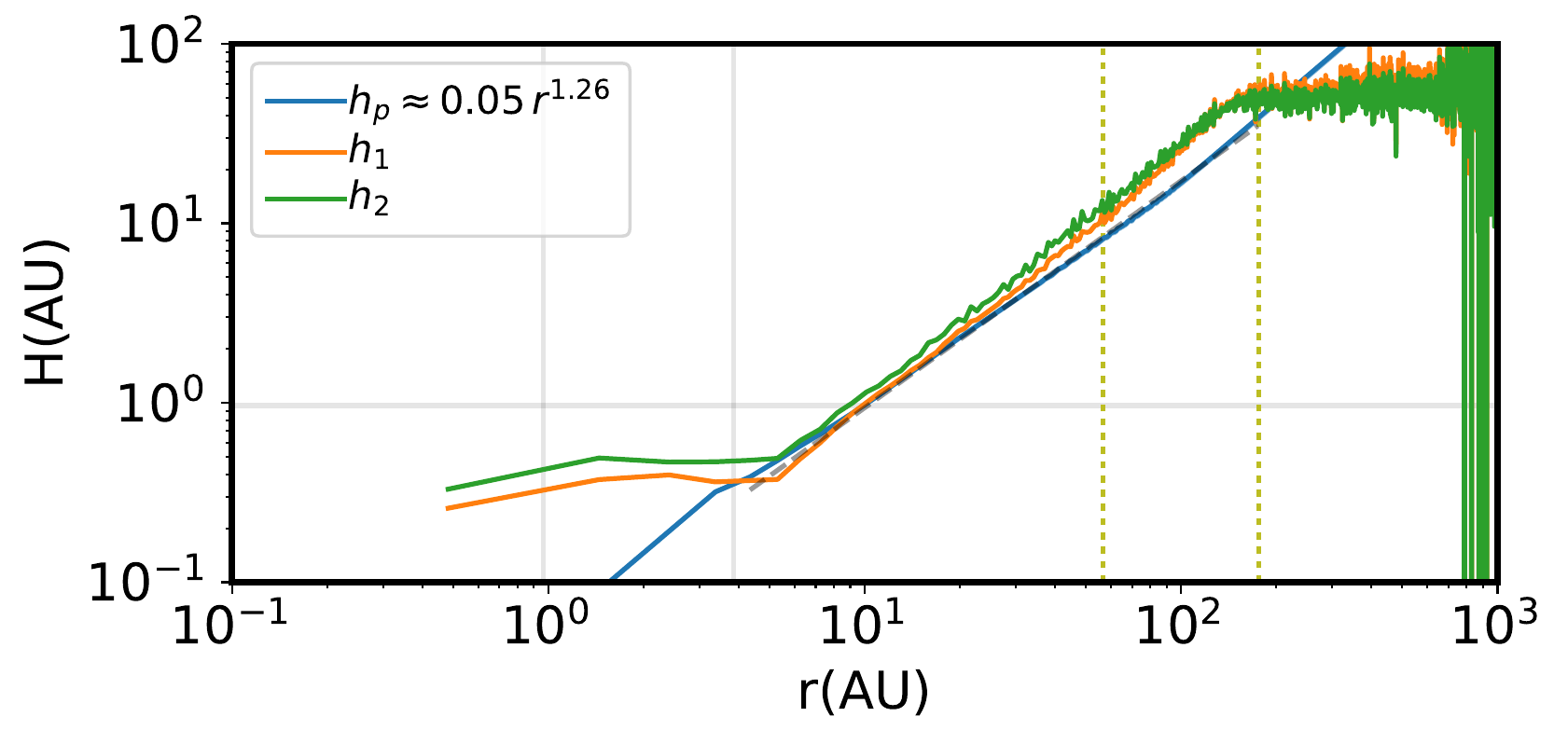}}
\put(0,0){\includegraphics[trim=0 0 0 5,clip,width=0.5\textwidth]{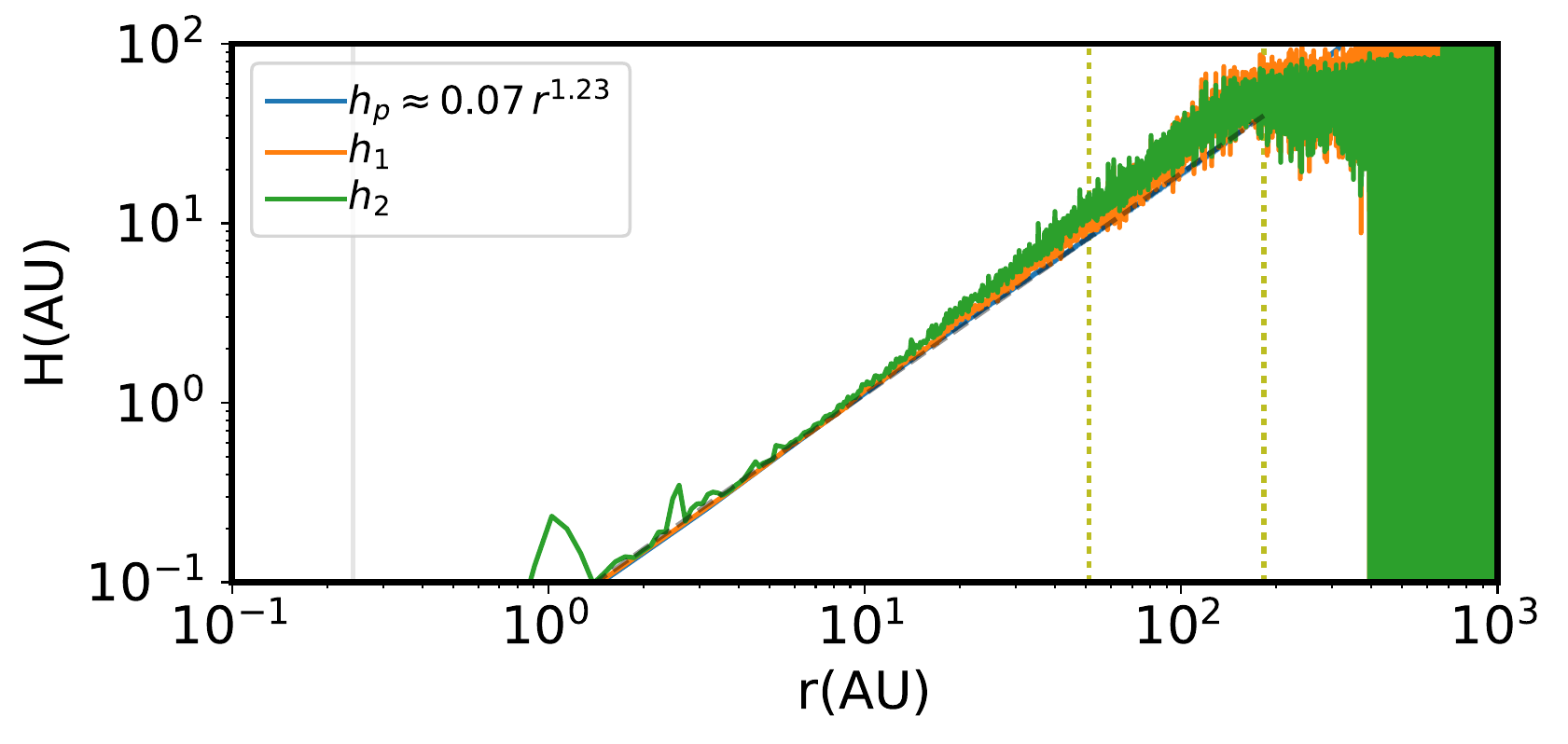}}
\put(0.16,1.47){R\_$\ell$14 (103 kyr)}
\put(0.16,1.12){R\_40ky\_$\ell$18 (103 kyr)}
\put(0.16,0.57){R\_$\ell$14 (138 kyr)}
\put(0.16,0.22){R\_80ky\_$\ell$18 (138 kyr),}
\end{picture}
\caption{Scale heights from vertical hydrostatic equilibrium, $h_{\rm p}$ (blue), from the first ($h_1$, orange) and second ($h_2$, green) moment of mass, plotted against the radius. {\it Top two panels:} R\_$\ell$14 and R\_40ky\_$\ell$18 at 103 kyr. {\it Bottom two panels:} R\_$\ell$14 and R\_80ky\_$\ell$18 at 138 kyr. The horizontal and the left vertical gray lines mark the numerical resolution ($dx$, outside the figure extent, thus not shown for the high-resolution runs). The values are averaged within rings of thickness $dx$ and plotted against the mean radius of the rings. The second vertical gray line marks the radius of the numerical sink accretion scheme ($4dx$). The two vertical dotted lines mark the characteristic radii of the disk ($R_{\rm kep}$ and $R_{\rm mag}$), as we defined in Sect. \ref{st_Sigma}. Within $R_{\rm kep}$ the disk is close to vertical thermal equilibrium and the three definitions of scale height coincide very well. Beyond this radius, magnetic support becomes strong and deviation occurs. The thermal scale height, $h_{\rm p} (r)$, is fitted to a power law, which is shown in the legend (in units of AU).}
\label{fig_H}
\end{figure}
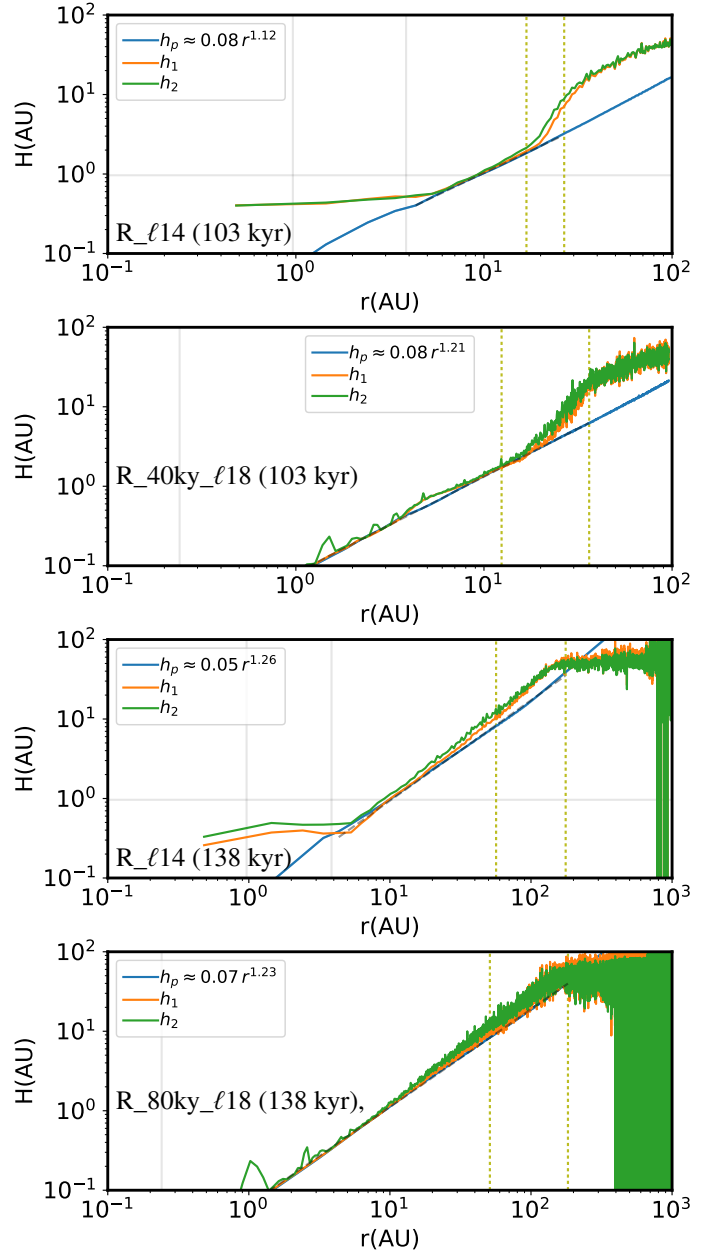

We calculated the azimuthally averaged vertical density profile for all radii. 
The binning size in the radial direction was taken to be $dx$, the smallest cell size (or several times $dx$ if the disk was large with respect to the resolution). 
We evaluated the disk scale height with several approaches, 
which were all expected to give identical results if the disk is vertically isothermal and follows an exact Gaussian density profile in pressure equilibrium. 
The thermal scale height  from vertical hydrostatic equilibrium (assuming a dominating stellar mass) is 
\begin{align}
h_{\rm p} (r) \approx \sqrt{c_{\rm s}^2(r) r^3 \over G[M_\ast+M_{\rm d}(r)]}, 
\end{align}
where $c_{\rm s}(r)$ is the local sound speed, $r$ is the cylindrical radius, $M_\ast$  is the stellar (sink) mass, and $M_{\rm d}(r)$ is the disk mass inside $r$. 
The disk is not a point mass, and the complete gravitational field should be calculated by integrating over the whole disk. 
We used this approximate formula for simplicity because the disk mass is only a few percent of the stellar mass. 
Because the disk is not exactly isothermal, the thermal sound speed was calculated as $c_{\rm s}^2 = \sum{PdV} / \sum{dm}$, 
where $P, dV, dm$ are the pressure, volume, and mass of each cell, and the summation was performed vertically within cylindrical shells. 
Here again, we verified that the vertical upper limit of the sum has no significant effect, and we simply summed over the whole cylinder. 

In the case of a vertical Gaussian profile, 
\begin{align}
\rho(z) = { \Sigma_{\rm tot} \over \sqrt{2\pi} H} \exp{\left[{-z^2 \over 2H^2}\right]}
\end{align}
the scale height can be recovered from the first or second moment of mass: 
\begin{align}
H = \sqrt{\pi\over 2}{1\over \Sigma_{\rm tot} }\int_{-\infty}^{\infty} \rho z dz = \left[{1\over \Sigma_{\rm tot} }\int_{-\infty}^{\infty} \rho z^2 dz\right]^{1/2}.
\end{align}
In practice, we calculated 
\begin{align}
h_1(r) = \sqrt{\pi \over 2} {\sum{zdm} \over  \sum{dm}} ~~~\text{and}~~~
h_2(r) =  \left[{\sum{z^2 dm} \over \sum{dm}}\right]^{1/2},
\end{align}
where the summation was performed vertically within cylindrical shells. 
These quantities were used as estimates of the thermal scale height.  

Figure \ref{fig_H} shows snapshots of the scale height calculated for the two high-resolution restarts compared to the canonical run. 
These results of $h_{\rm p}$, $h_1$, and $h_2$ indeed reasonably coincide with one another inside the disk, 
and the significant incoherence marks the radius at which the disk ends (left dotted vertical line). 
This also confirms that our choice of the cylindrical region size is sufficient to avoid errors because the Gaussian density profile drops very quickly at high altitudes. 

The two vertical dotted lines denote the disk characteristic radii that we define in Sect. \ref{st_Sigma} and further discuss in Sect. \ref{st_radius}. 
The first radius, $R_{\rm kep}$, corresponds to the inner part of the disk that is centrifugally supported in the radial direction and thermally supported in the vertical direction (see Sect. \ref{st_vel}). 
The scale heights determined from three different definitions thus coincide very well with one another. 
The second radius, $R_{\rm mag}$, corresponds to an outer region of the disk where the magnetic support is relatively strong,  
and deviations start to appear. 
Beyond $R_{\rm mag}$, the moments of mass no longer give much information as the infalling envelope does not have a flattened geometry. 
For a vertically uniform density profile (which is almost the case in the outer part of the disk and the envelope), 
$h_1 =\sqrt{\pi/8} z_{\rm cyl} \approx 0.577  z_{\rm cyl}$ and $h_2=\sqrt{1/3} z_{\rm cy} \approx 0.626  z_{\rm cyl}$. 
They are almost indistinguishable either inside or outside the disk, 
thus their discrepancy with $h_{\rm p}$ exhibits the clear change to a nonthermally supported region. 

The overall behavior is similar in the high- and low-resolution runs. 
However, the high-resolution runs provide more precise descriptions of the disk height profile and allowed us to probe the high-density interior of the disk. 
The horizontal gray line in Fig. \ref{fig_H} marks the resolution $dx$. The disk in the canonical run contains only one cell in the vertical direction within a radius of 10 AU,  while the high-resolution restarts resolve the disk down to radius of 1 AU. 
Run R\_40ky\_$\ell$18 shows a larger and better delineated transition zone between the two characteristic radii, 
while in the canonical run this region is barely resolved by a few cells. 
The disk has nearly constant aspect ratio $H/R \approx 0.1$ up to almost 20 AU radius. 
On the other hand, the disk has a larger radius after simulation time 120 kyr, and thus the run R\_80ky\_$\ell$18 resolves the disk center better, but does not give much additional information at large radius because it is already well resolved in the canonical run. 

\subsubsection{Radial density profiling: Surface density $\Sigma$}\label{st_Sigma}

\begin{figure}[]
\centering
\setlength{\unitlength}{0.5\textwidth}
\begin{picture}(1,2.25)
\put(0,1.8){\includegraphics[trim=0 0 0 5,clip,width=0.5\textwidth]{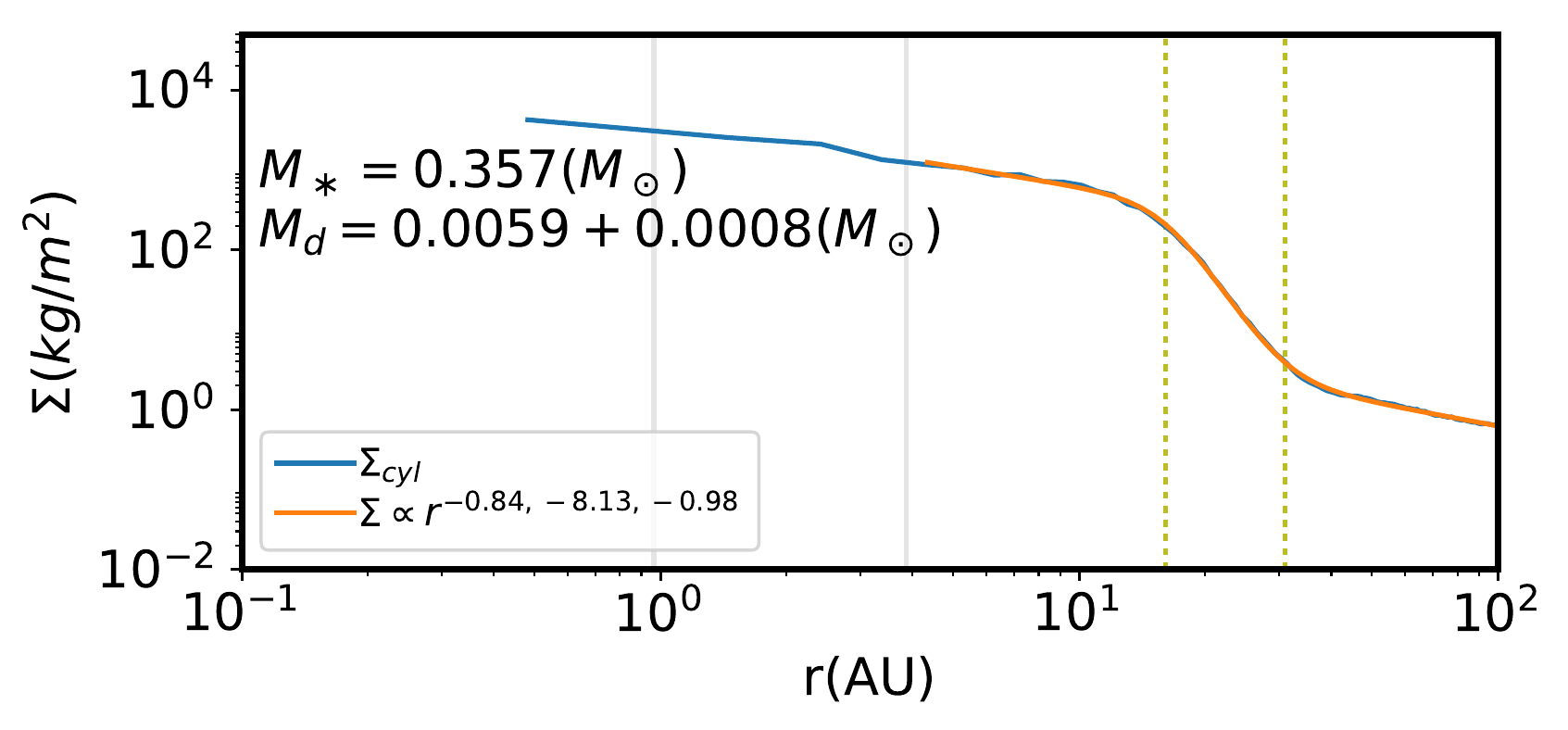}}
\put(0,1.35){\includegraphics[trim=0 0 0 5,clip,width=0.5\textwidth]{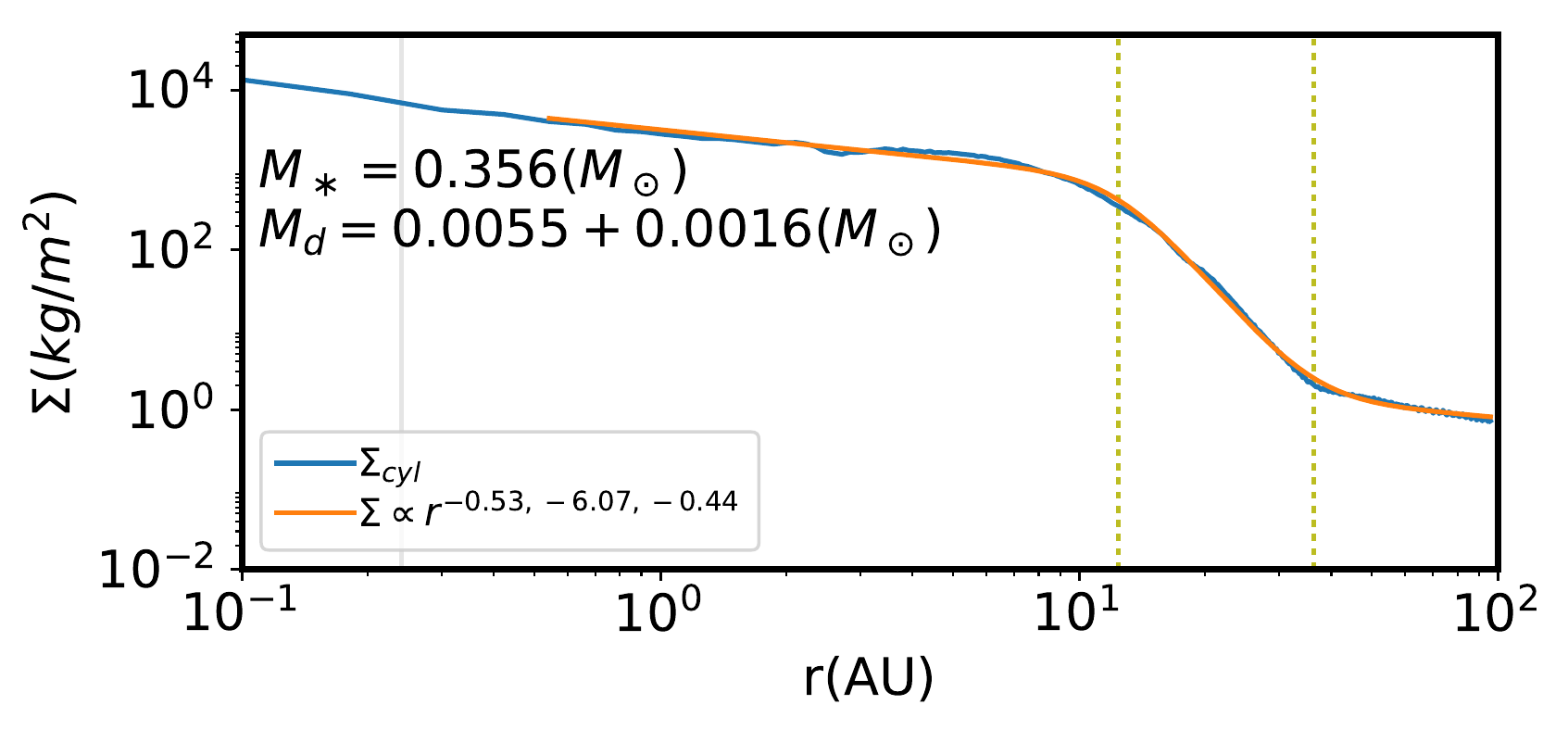}}
\put(0,0.9){\includegraphics[trim=0 0 0 5,clip,width=0.5\textwidth]{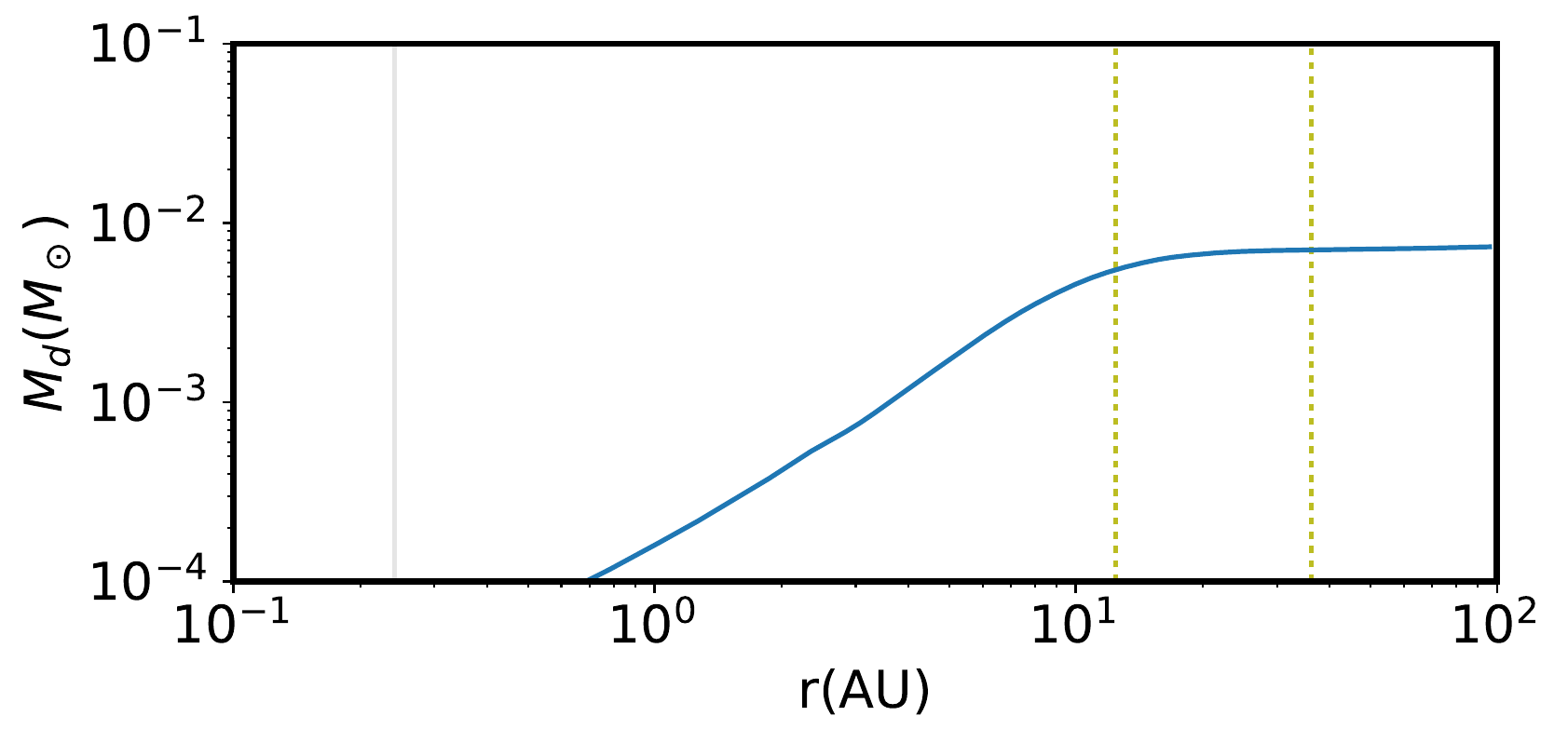}}
\put(0,0.45){\includegraphics[trim=0 0 0 5,clip,width=0.5\textwidth]{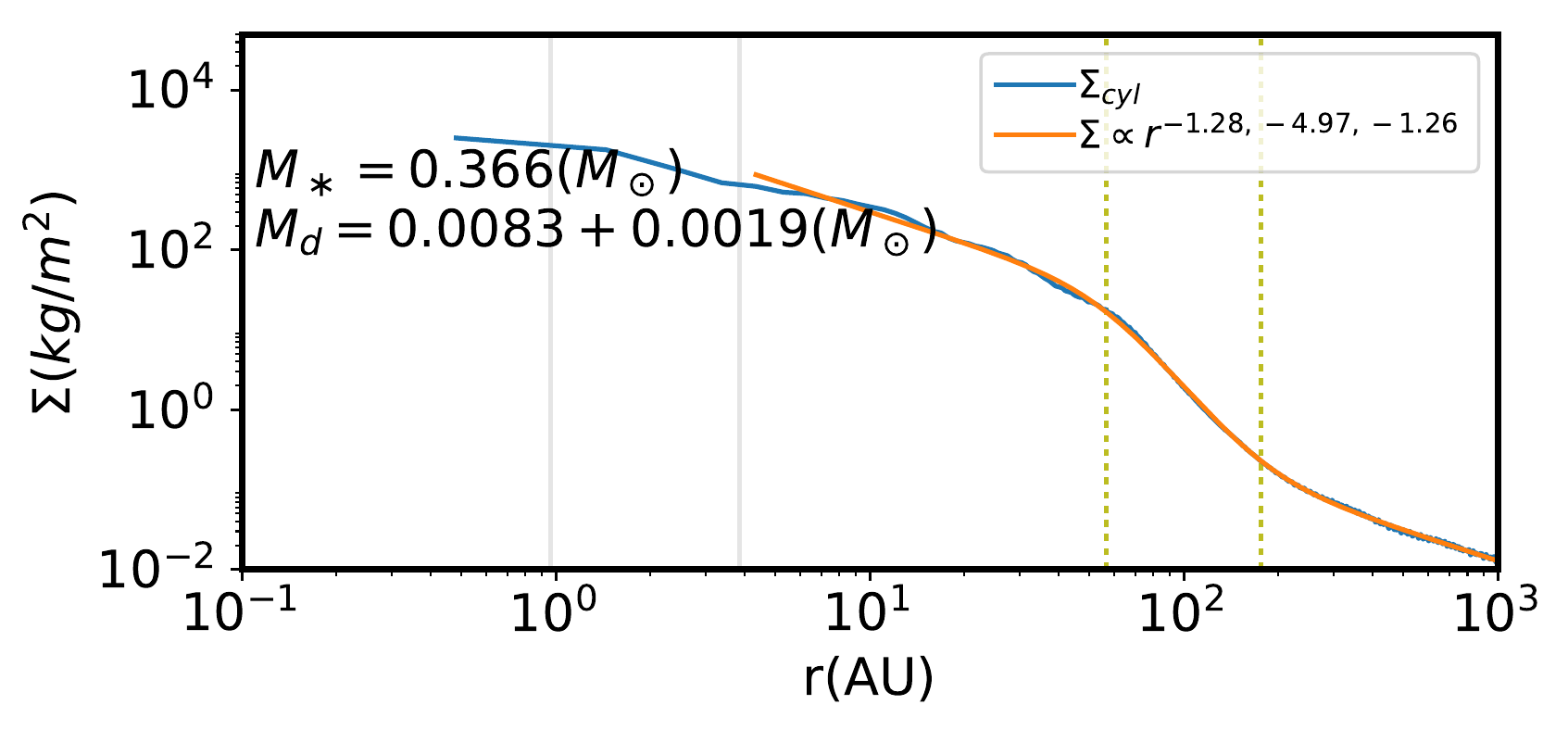}}
\put(0,0){\includegraphics[trim=0 0 0 5,clip,width=0.5\textwidth]{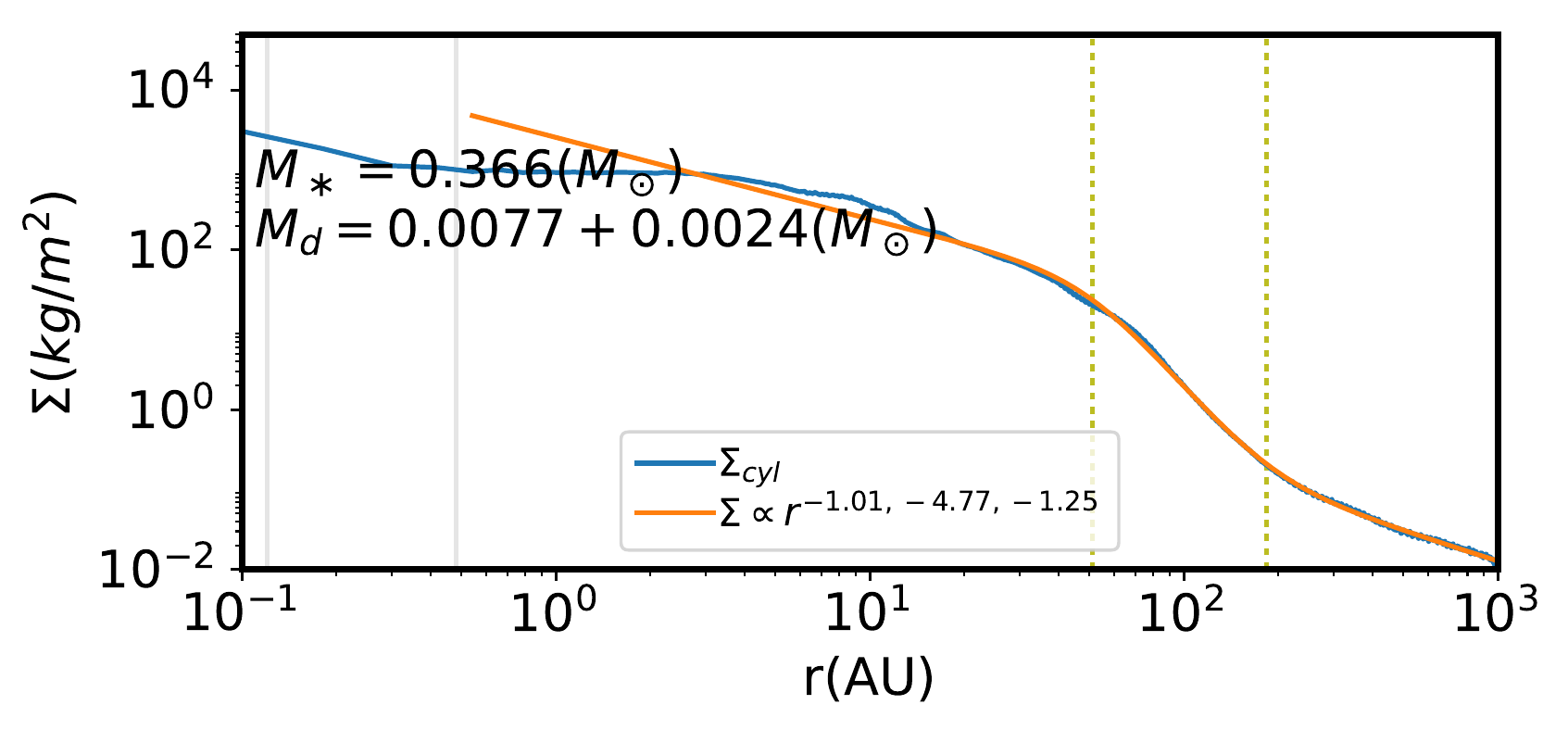}}
\put(0.16,2.21){R\_$\ell$14 (103 kyr)}
\put(0.16,1.76){R\_40ky\_$\ell$18 (103 kyr)}
\put(0.16,1.31){R\_40ky\_$\ell$18 (103 kyr)}
\put(0.16,0.86){R\_$\ell$14 (138 kyr)}
\put(0.16,0.41){R\_80ky\_$\ell$18 (138 kyr)}
\end{picture}
\caption{Disk surface density profile $\Sigma_{\rm cyl}(r)$ measured inside a selected cylindrical region (blue). {\it Top two panels:} R\_$\ell$14 and R\_40ky\_$\ell$18 at 103 kyr. {\it Middle:} Radially accumulated mass of the disk of R\_40ky\_$\ell$18 at 103 kyr. {\it. Bottom two panels:} R\_$\ell$14 and R\_80ky\_$\ell$18 at 138 kyr.  A three-segment power-law fit is overplotted (orange), and the power-law exponents are shown in the legend. Two vertical dotted lines show the characteristic radii $R_{\rm kep}$ and $R_{\rm mag}$ that correspond to the transition of the power-law slope. The masses of the star $M_\ast$ and of the disk $M_{\rm d}$ (inside $R_{\rm kep}$ plus between $R_{\rm kep}$ and  $R_{\rm mag}$) are also displayed. Most of the disk mass is contained within $R_{\rm kep}$. }
\label{fig_Sigma}
\end{figure}
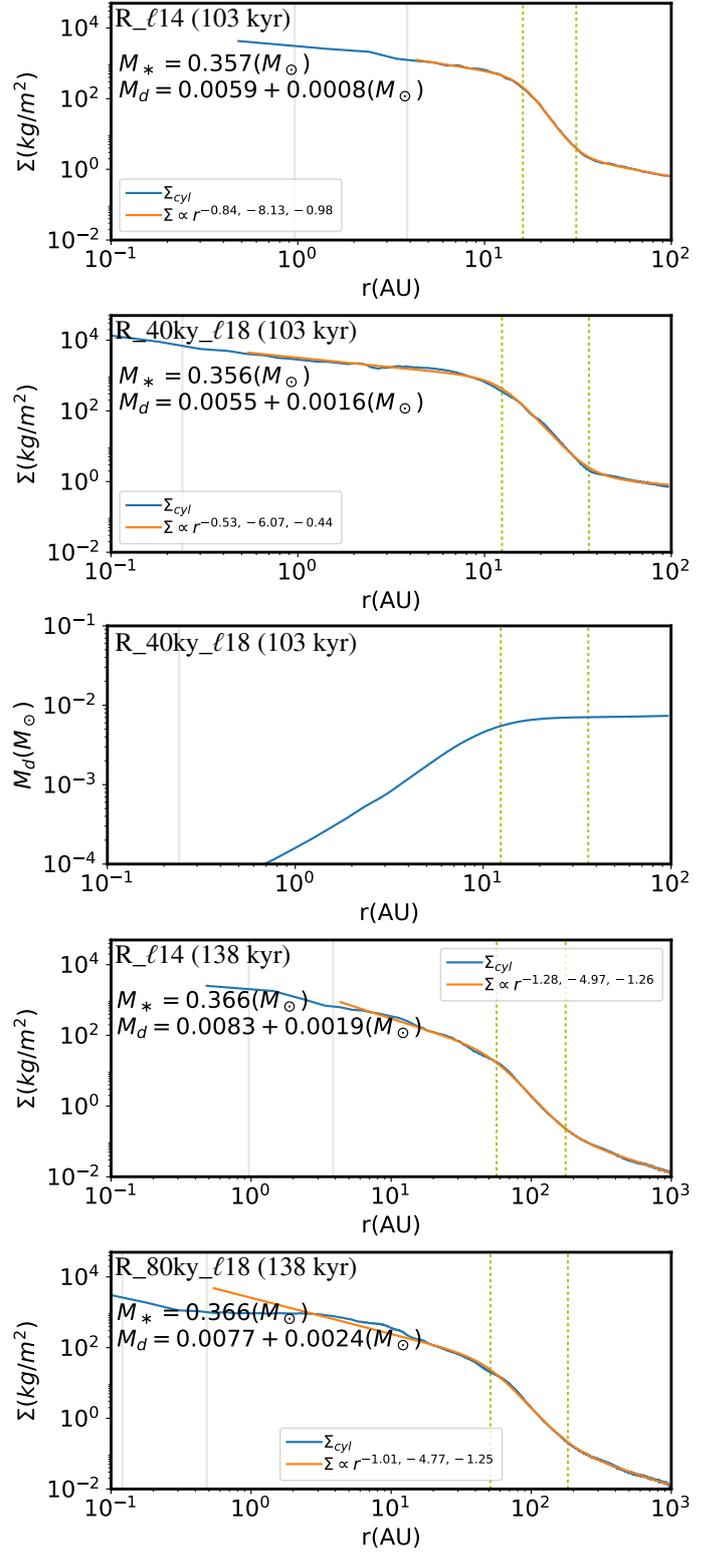

The surface density of the disk is obtained as
\begin{align}
\Sigma_{\rm cyl} (r) = {\sum{dm} \over 2\pi rdr} =  \Sigma_{\rm tot} (r) ~\rm{erf}{\left(z_{\rm cyl}\over \sqrt{2}H\right)} \approx \Sigma_{\rm tot} (r),
\end{align}
where $\Sigma_{\rm cyl}$ is the column density within the cylindrical region, 
and $\Sigma_{\rm tot}$ is the total surface density if the vertical density profile followed an exact Gaussian, with the mass contained within $\pm z_{\rm cyl}$ equaling the numerically measured value.
The error function is a correction for the mass outside the cylinder vertical extent that is not taken into account.  
However, this correction is almost negligible for $H \ll z_{\rm cyl}$. 

Figure \ref{fig_Sigma} shows snapshots of the $\Sigma$ profile. 
At 138 kyr, the surface density profile is close to a power law with a sharp drop near 13 AU, 
and it becomes flat again outside 30 AU.  
The rapidly decreasing part is due to increased magnetic support toward the outer disk (see Sect. \ref{st_beta} for discussions of the magnetic field). 
Most of the disk mass is within the inner region, 
while sometimes the region with sharply decreasing surface density can contain up to a quarter of the total disk mass. 
The outer flat part, on the other hand, is not a real surface density, 
but a truncated value of the diffuse envelope that is more vertically extended. 
We performed a three-segment power-law fit to the measured $\Sigma_{\rm cyl}(r)$, 
and the transition points define the two characteristic radii,  $R_{\rm kep}$ and $R_{\rm mag}$,
which we discussed in the following paragraph.  

\subsubsection{Disk radius $R$}\label{st_radius}
Two characteristic radii, $R_{\rm kep}$ and $R_{\rm mag}$, were defined from a three-segment power-law fit of the surface density profile (Sect. \ref{st_Sigma}). 
This is shown in Fig. \ref{fig_Sigma} with the vertical dotted lines. 
The inner part of the disk has a shallow power-law profile and is almost nonmagnetized. 
The outer part, on the other hand, is supported by the magnetic field, and the surface density drops quickly (see Sect. \ref{st_beta}). 
These two characteristic radii are marked in all the following radial profile figures to show that they indeed correspond to some transitions of the properties inside the disk. We recall that no clear-cut disk radius exists, 
while our choice of characteristic values allows making reasonable evaluations of other disk properties in our analyses. 

\subsection{Global evolution of the disk: Mass and radius}\label{st_global}

\begin{figure}[]
\centering
\includegraphics[trim=0 0 0 0,clip,width=0.5\textwidth]{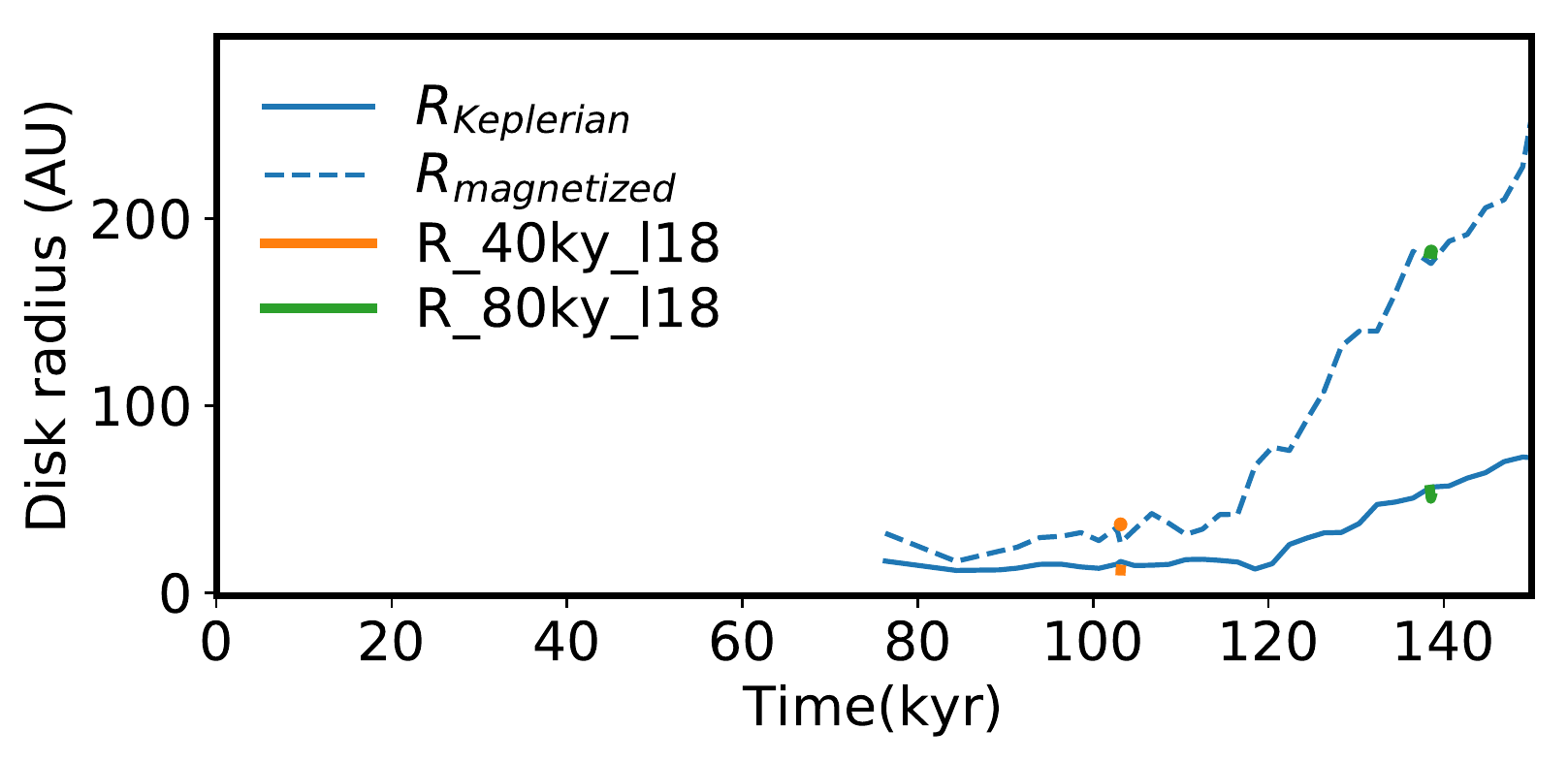}
\caption{Radius evolution of the disk. The radii measured from simulations as described in Sect. \ref{st_radius} are plotted with solid ($R_{\rm kep}$) and dashed ($R_{\rm mag}$) lines for R\_$\ell$14 in blue. The disk initially has a Keplerian radius of about 20 AU for almost 50 kyr and a magnetized envelope that is roughly 1.5 times larger. At later times, the disk grows to nearly 50 AU in radius, and the magnetized region grows even more significantly to more than twice the size of the demagnetized Keplerian disk. Runs R\_40ky\_$\ell$18 and R\_80ky\_$\ell$18 are overplotted (with the same line styles), and the size of the disk is consistent with different resolutions.}
\label{fig_Rdisk}
\end{figure}

With the nonideal MHD effects, the disk becomes self-regulated and its size evolves slowly. 
This is due to the balance between the rotation and the magnetic braking as well as to that between the magnetic field induction through rotation and its loss through diffusion \citep{Hennebelle16}. 
The radius very quickly reaches $\sim 20$ AU and becomes quasi-stationary. 
The disk receives mass from the infalling envelope, and at the same time, some of its mass is lost through accretion onto the central star. 
The mass inside the protoplanetary disk stays at $\sim 0.01 ~\Ms$. 

Figure \ref{fig_Rdisk} shows the disk radius evolution for the canonical simulation at 1AU resolution, 
with $R_{\rm kep}$ plotted as a solid blue line and $R_{\rm mag}$ as a dashed blue line. 
The sink particle forms at simulation time 61 kyr, and we only show measurements from $\sim$ 75 kyr, when most mass of the star-disk system is accreted by the sink particle and a flattened disk is well defined. Fig. \ref{fig_l14_evol} shows that the star-disk system is more like a flattened ellipsoid at 61 kyr when the sink particle has just formed. 
The size, $R_{\rm kep}$ in particular, does not vary strongly within more than 50 kyr. 
At about 120 kyr, the disk starts to grow in size up to $\sim$ 50 AU. 
The magnetized part grows significantly, reaching several times the size of the Keplerian disk. 
The growth of the disk is possibly a consequence of the evolution of the surrounding 
medium properties, and in particular, of the magnetic field. It may also indicate that the coupling with the surrounding 
envelope, with which angular momentum is exchanged, becomes less tight, and thus the efficiency of magnetic braking is reduced. This particular point certainly requires further investigation. 
The high-resolution restarts are overplotted and the disk sizes are consistent. 
With increased resolution, the time steps are significantly reduced such that the evolution is not followed during very long time spans. 

\begin{figure}[]
\centering
\includegraphics[trim=0 0 0 0,clip,width=0.5\textwidth]{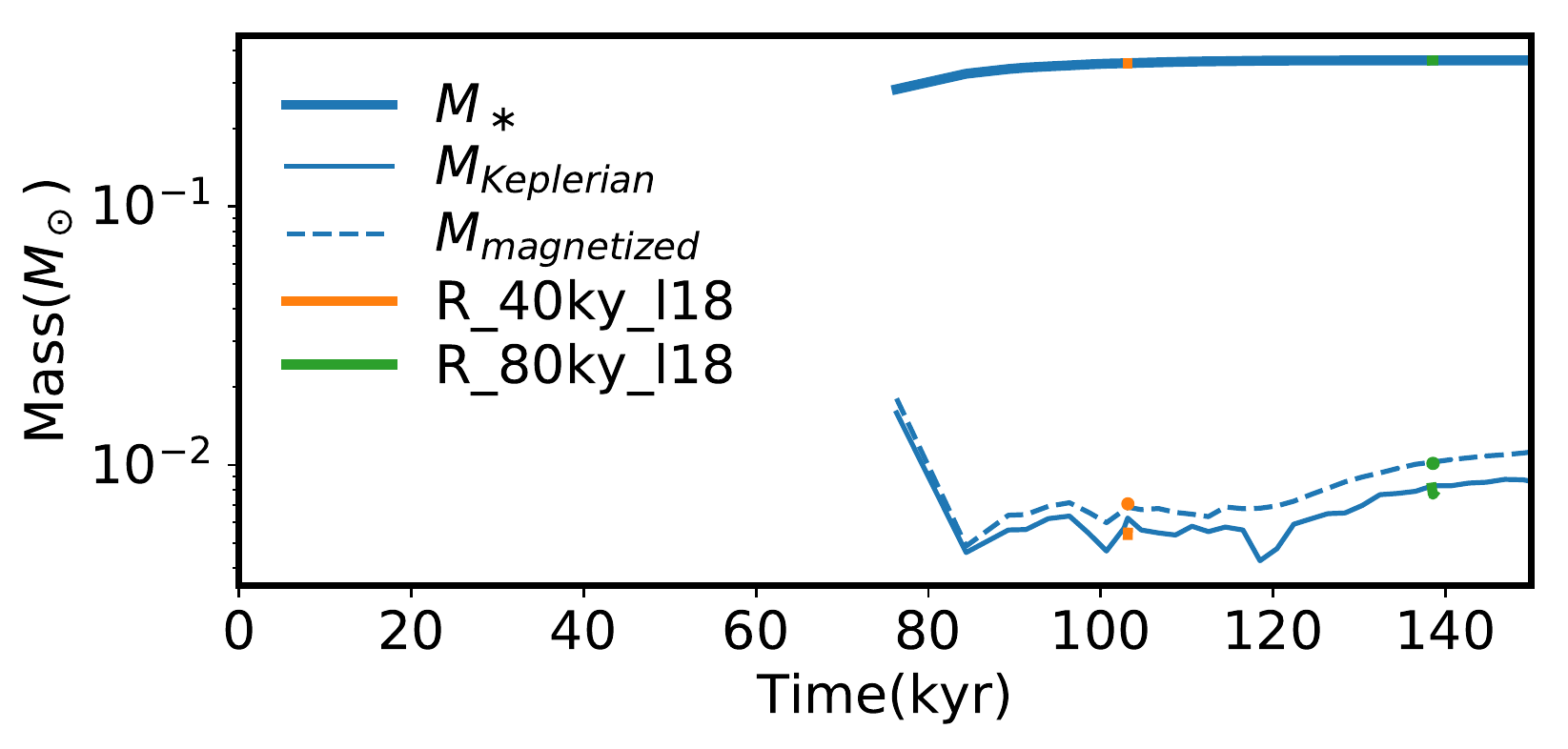}
\caption{Mass evolution of the star-disk system. Run R\_$\ell$14 is plotted as a thick blue line for the sink mass, a thin line for the mass within $R_{\rm kep}$ , and as a dashed line for the mass within $R_{\rm mag}$. A flattened core-disk structure forms at the beginning. After the sink forms, it quickly accretes most of the mass inside the disk, and the disk mass drops to about 1\% of the stellar mass. The magnetized region around the Keplerian disk, though significant in size, contains much less mass.  The high-resolution restarts are overplotted with the same line styles, and there is an overall good agreement. }
\label{fig_Mdisk}
\end{figure}

Figure \ref{fig_Mdisk} shows the mass evolution of the star, the disk mass within $R_{\rm kep}$, and that within $R_{\rm mag}$. 
Before the sink formation, a flattened dense structure is formed. 
Quickly after sink particle formation, most of the mass is accreted by the star and the disk mass drops to stably $\sim 2-3 \%$ of the stellar mass. 
Most of the mass is contained within $R_{\rm kep}$, where the disk is essentially Keplerian. 
When the disk starts to expand at later times, the magnetized region grows significantly, and its mass sometimes reaches one-third of the mass of the Keplerian part. 
The overplotted high-resolution runs are essentially identical.

\begin{figure}[]
\centering
\includegraphics[trim=0 0 0 0,clip,width=0.5\textwidth]{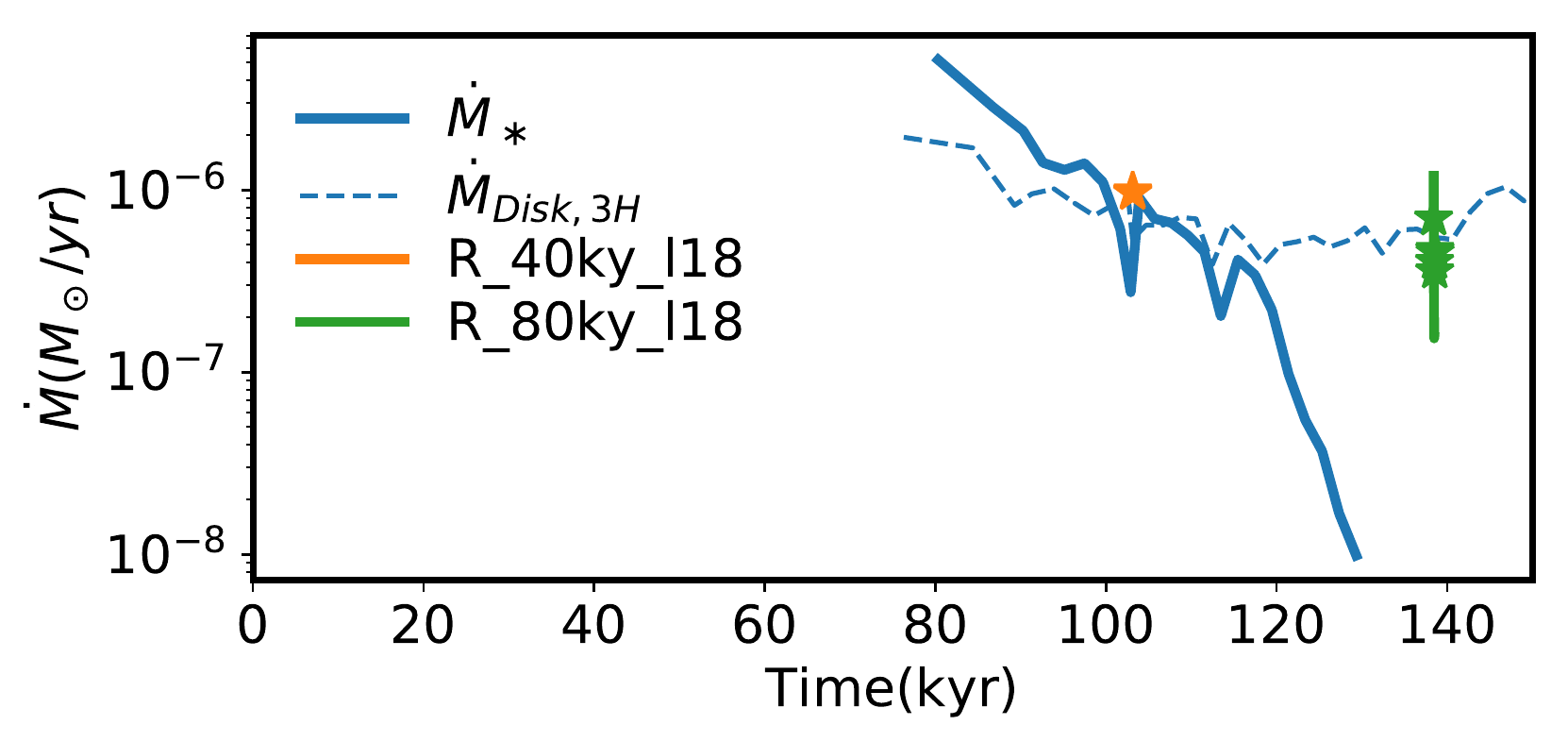}
\caption{Mass accretion rate evolution. The canonical run at 1 AU resolution is plotted in blue. The mass accretion rate of the sink particle (thick line) is derived directly from its mass evolution. The accretion rate onto the disk is the integral of the measured source function (dashed line, as described in Sect. \ref{st_source}). The mass accretion rate is initially high an decreases slowly with time. The mass accretion onto the sink particle, though generally decreasing in time, shows some bursty behavior. This is due to the accumulation at the inner disk border shortly before reaching the accretion threshold. After 130 kyr, the density of the disk drops so much such that the threshold density of the sink accretion algorithm is no longer met and the sink hardly accretes. The high-resolution restarts are overplotted (solid line for sink accretion and stars for disk surface accretion). The mass flux onto the disk surface is unchanged in general, while the sink accretion is sensitive to the inner boundary condition of the disk (i.e., the choice of $n_{\rm acc}$, see Sect. \ref{st_restart}).}
\label{fig_Mdot}
\end{figure}

Figure \ref{fig_Mdot} shows the mass accretion rate of the sink particle and that across the disk surface (at $\min(3h_{\rm p}, 4dx)$).  
The accretion rate of the sink is directly derived from the sink mass history. 
The occasional bursty behavior is probably due to the mass accumulating at the inner disk boundary before the density exceeds $n_{\rm acc}$. 
\citet{Kuffmeier18} also saw a similar behavior in their zoomed-in systems at 2 AU resolution, but this feature diminished with increased resolution. Therefore this phenomenon should be interpreted with care.
The two values are broadly comparable. 
The mass accretion onto the disk surface slowly decreases in time. 
On the other hand, the sink accretion rate experiences a more serious drop. 
This is probably because the accretion becomes numerically more difficult as the disk density slowly decreases while $n_{\rm acc}$ is kept a constant. 
At later times, $\dot{M}_{\rm Disk} > \dot{M}_{\rm sink}$. 
This is also reflected in the growth of mass and size of the disk (Figs. \ref{fig_Rdisk} and \ref{fig_Mdisk}). 

The mass accretion rate across the disk surface in the high-resolution runs is compatible with the canonical run. 
However, R\_40ky\_$\ell$18 presents zero sink accretion. 
This reflects the sensitivity of the sink evolution to the numerical parameters, and $n_{\rm acc}$ was probably not very well adjusted for this case (see Sect. \ref{st_restart}). 
On the other hand, R\_80ky\_$\ell$18 shows a higher sink accretion rate than the canonical run.
This might be due to the better description of the central high-density region, 
or it might suggest that the value of $n_{\rm acc}$ is slightly too low.

\subsection{Temperature inside the disk: Snow line and CAI formation}\label{st_temperature}
One of the important properties of the protoplanetary disk is the temperature distribution because it is important for the condensation of different minerals, and also for water.
Knowing the temperature conditions will allow us to study the formation of precursors of different mineral components of the building blocks of the S.S. 
The water content, on the other hand, plays crucial roles not only in the conditions of chemical reactions, but also but also for the growth of solid particles because water ice represent the most massive reservoir of solid material in the S.S. 
In this section, we discuss the thermal structure of the disk and its evolution. 

\subsubsection{Radial temperature profile}
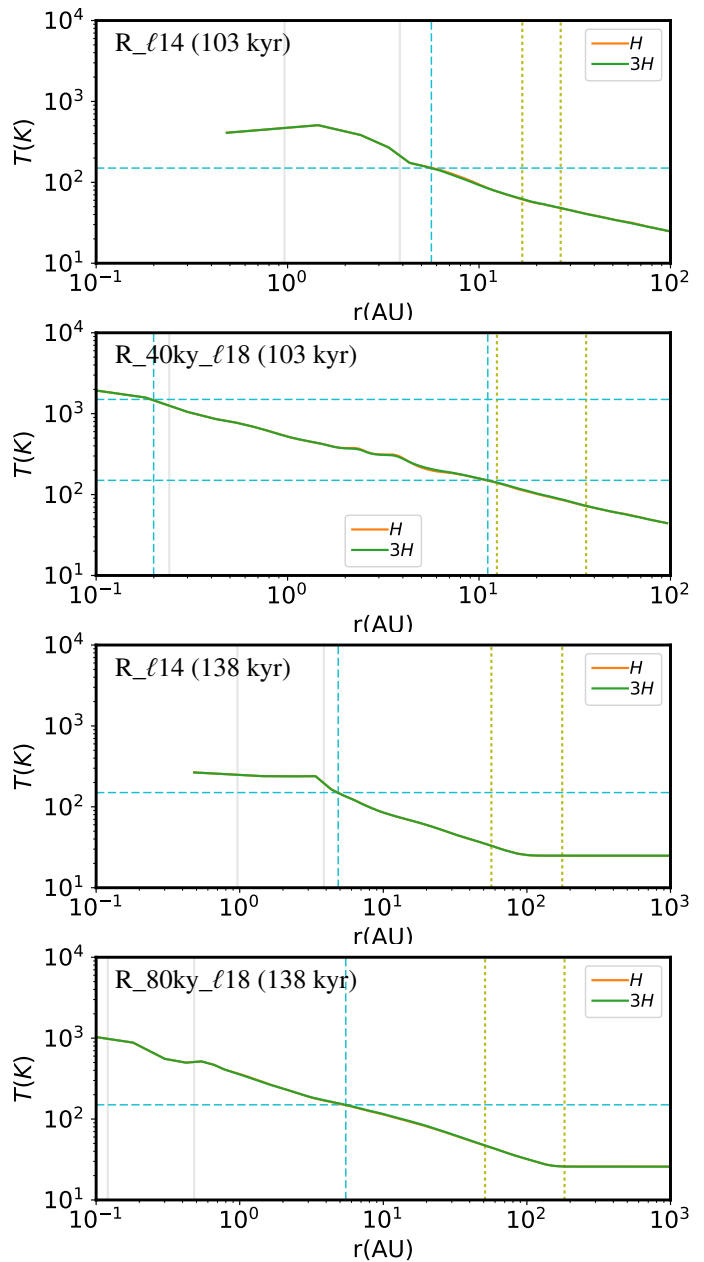
\begin{figure}[]
\centering
\setlength{\unitlength}{0.5\textwidth}
\begin{picture}(1,1.8)
\put(0,1.35){\includegraphics[trim=0 0 0 5,clip,width=0.5\textwidth]{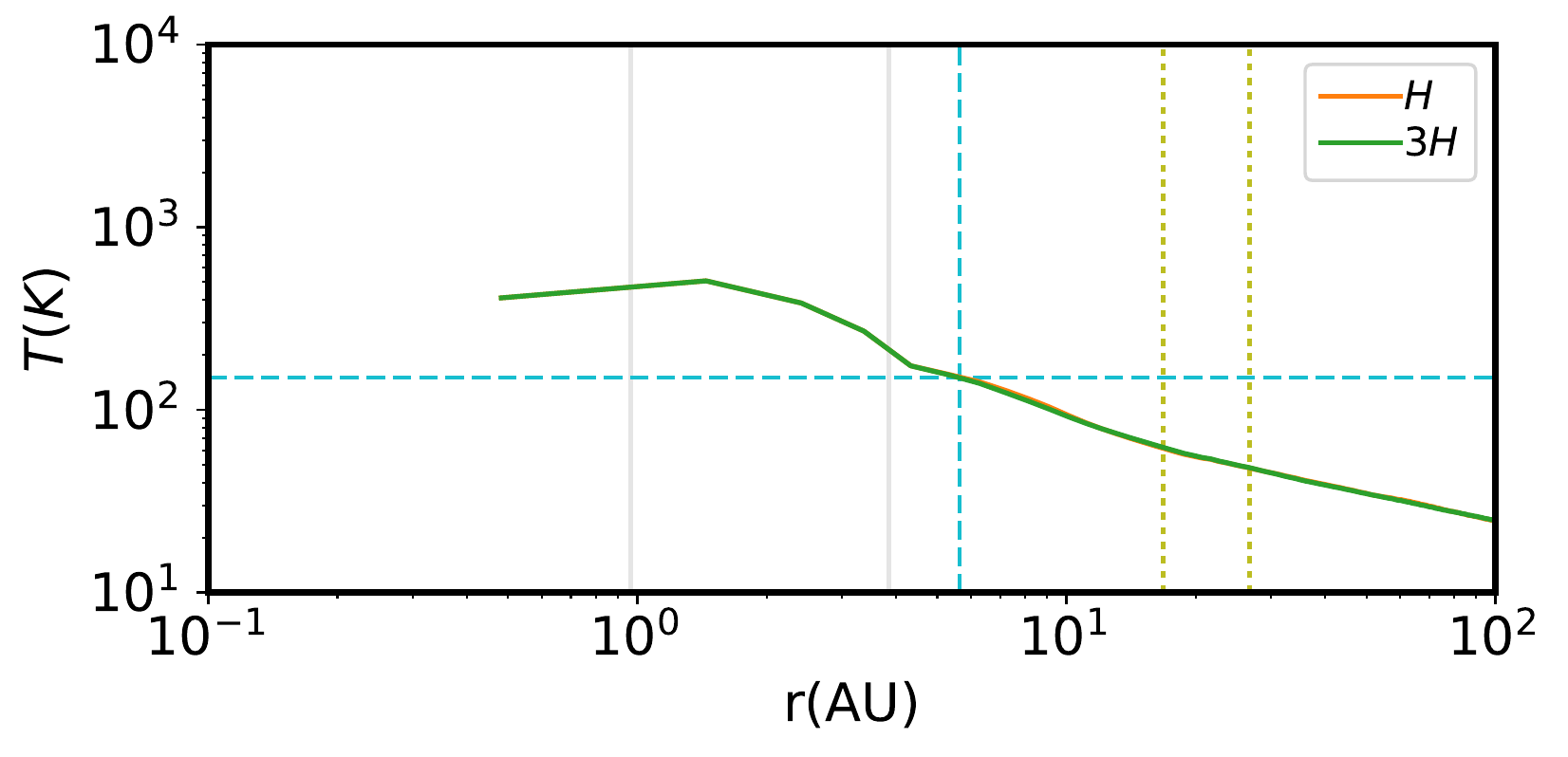}}
\put(0,0.9){\includegraphics[trim=0 0 0 5,clip,width=0.5\textwidth]{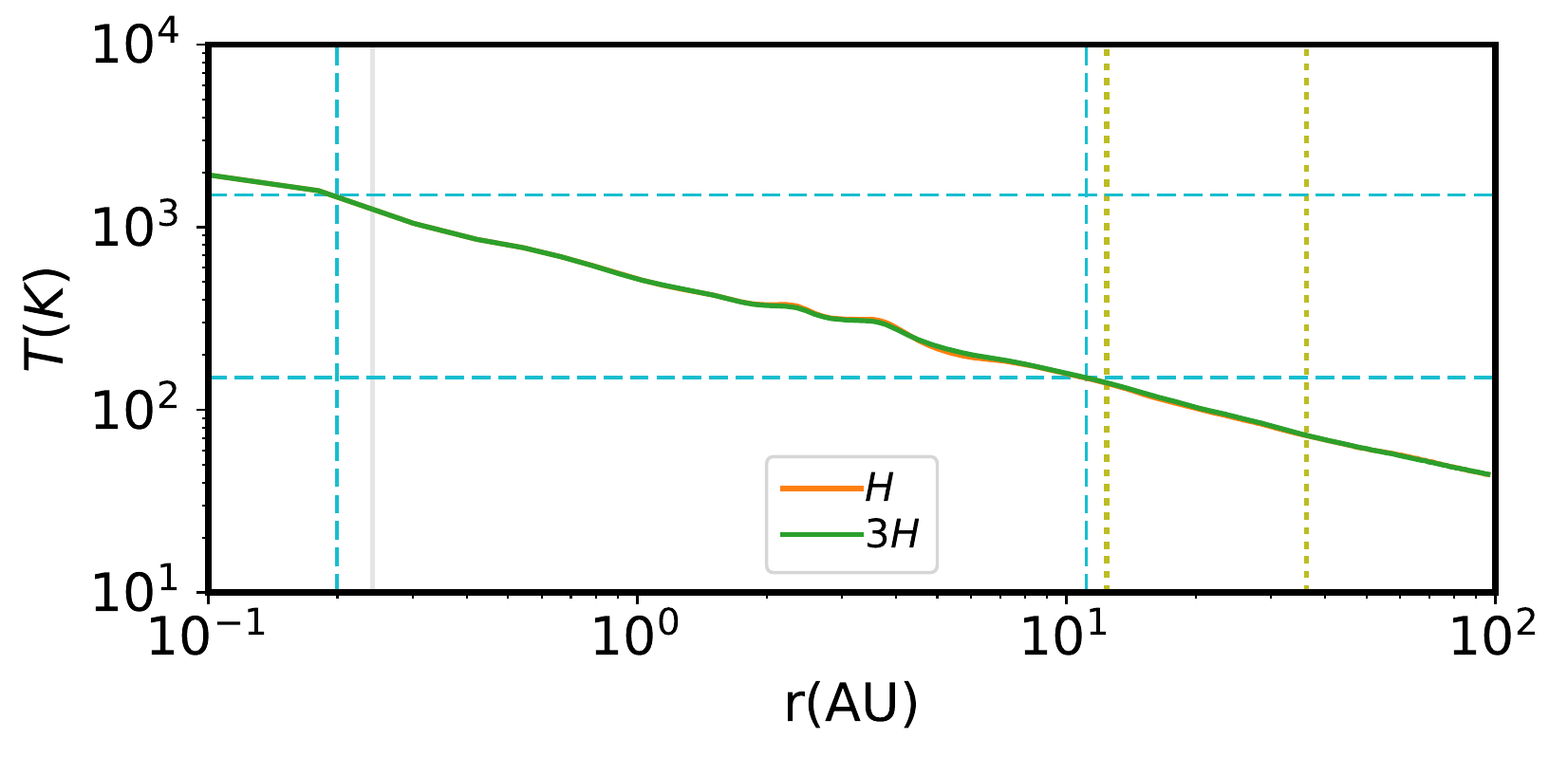}}
\put(0,0.45){\includegraphics[trim=0 0 0 5,clip,width=0.5\textwidth]{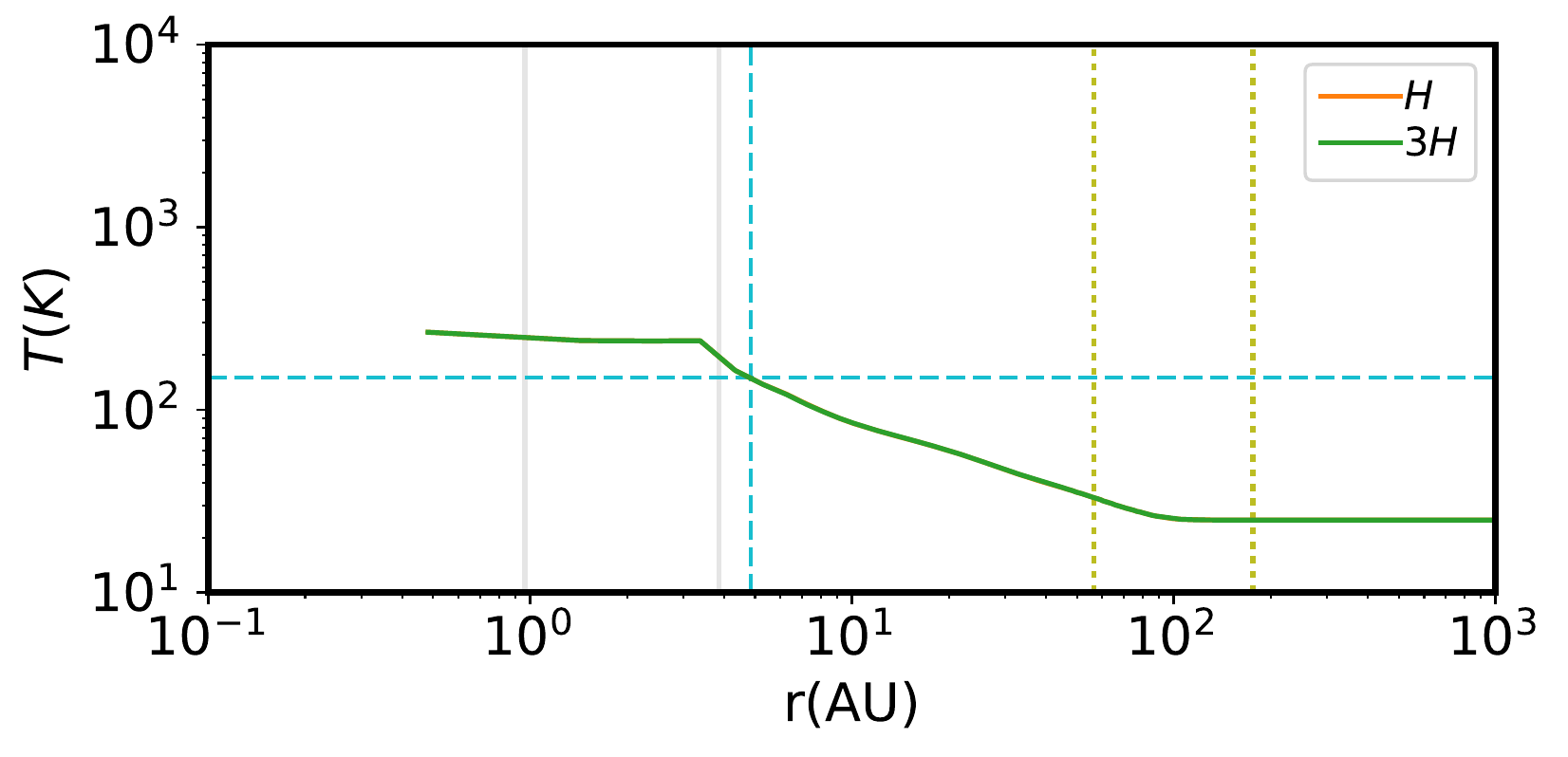}}
\put(0,0){\includegraphics[trim=0 0 0 5,clip,width=0.5\textwidth]{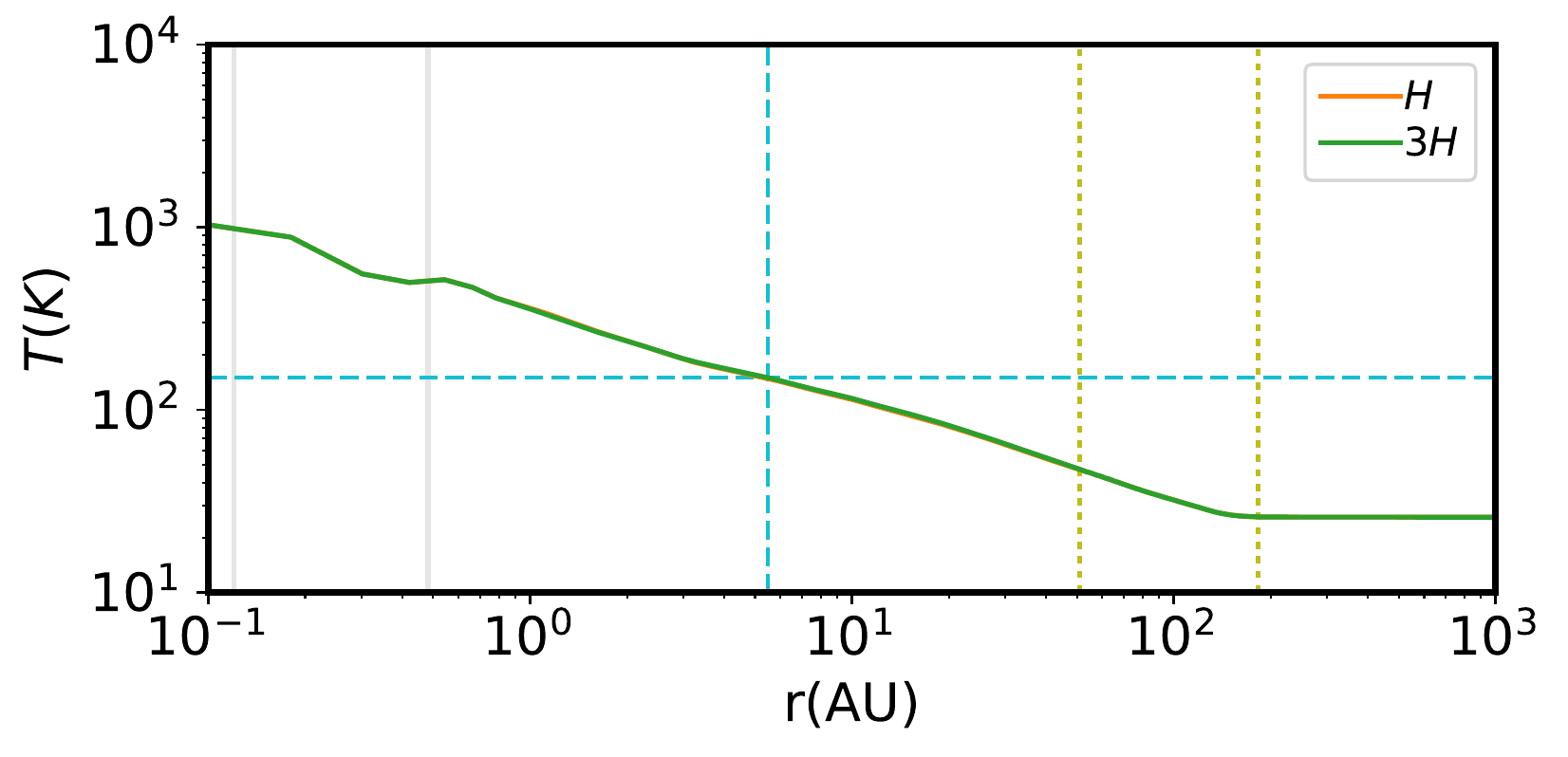}}
\put(0.16,1.76){R\_$\ell$14 (103 kyr)}
\put(0.16,1.31){R\_40ky\_$\ell$18 (103 kyr)}
\put(0.16,0.86){R\_$\ell$14 (138 kyr)}
\put(0.16,0.41){R\_80ky\_$\ell$18 (138 kyr)}
\end{picture}
\caption{Radial temperature profiles of the disk. The same snapshots as in Fig. \ref{fig_H} are shown. The temperature is averaged (mass-weighted) vertically within $H$ (orange) and $3H$ (green). The snow-line ($R_{150}$) and the CAI condensation line ($R_{1500}$, if present) are marked with vertical dashed cyan lines. In the low-resolution run, the snow-line $R_{150}$ is very close to the resolution limit and thus not well resolved. High-resolution restarts are indeed necessary to study the interior structure of the disk correctly. The CAI condensation line $R_{1500}$ appears in R\_40ky\_$\ell$18, while in R\_80ky\_$\ell$18, this line has migrated inward due to the decreased overall density and temperature.}
\label{fig_T}
\end{figure}

Figure \ref{fig_T} shows snapshots of the radial temperature profile. 
We show profiles averaged (mass-weighted) within one and three times the disk scale height $H$ , 
and they are indeed almost identical, confirming the vertical isothermal simplification. 
There is no properly defined midplane temperature because 
we only have access to the temperature at the resolution limit, that is, within one cell height. 
The value of $h_{\rm p}$ is used for $H$, 
thus it is more reliable within $R_{\rm kep}$. 
The gas is relatively diffuse outside and the vertical temperature variation is considerably low. 
As a verification of the thermal sound speed evaluated in Sect. \ref{st_height}, 
we examined the vertical temperature profiles. 
The temperature only deviates significantly  at rather high altitude compared to the disk scale height, and we refer to Appendix \ref{ap_Tz} for more details. 

\begin{figure}[]
\centering
\setlength{\unitlength}{0.5\textwidth}
\begin{picture}(1,0.45)
\put(0,0.){\includegraphics[trim=0 0 0 5,clip,width=0.5\textwidth]{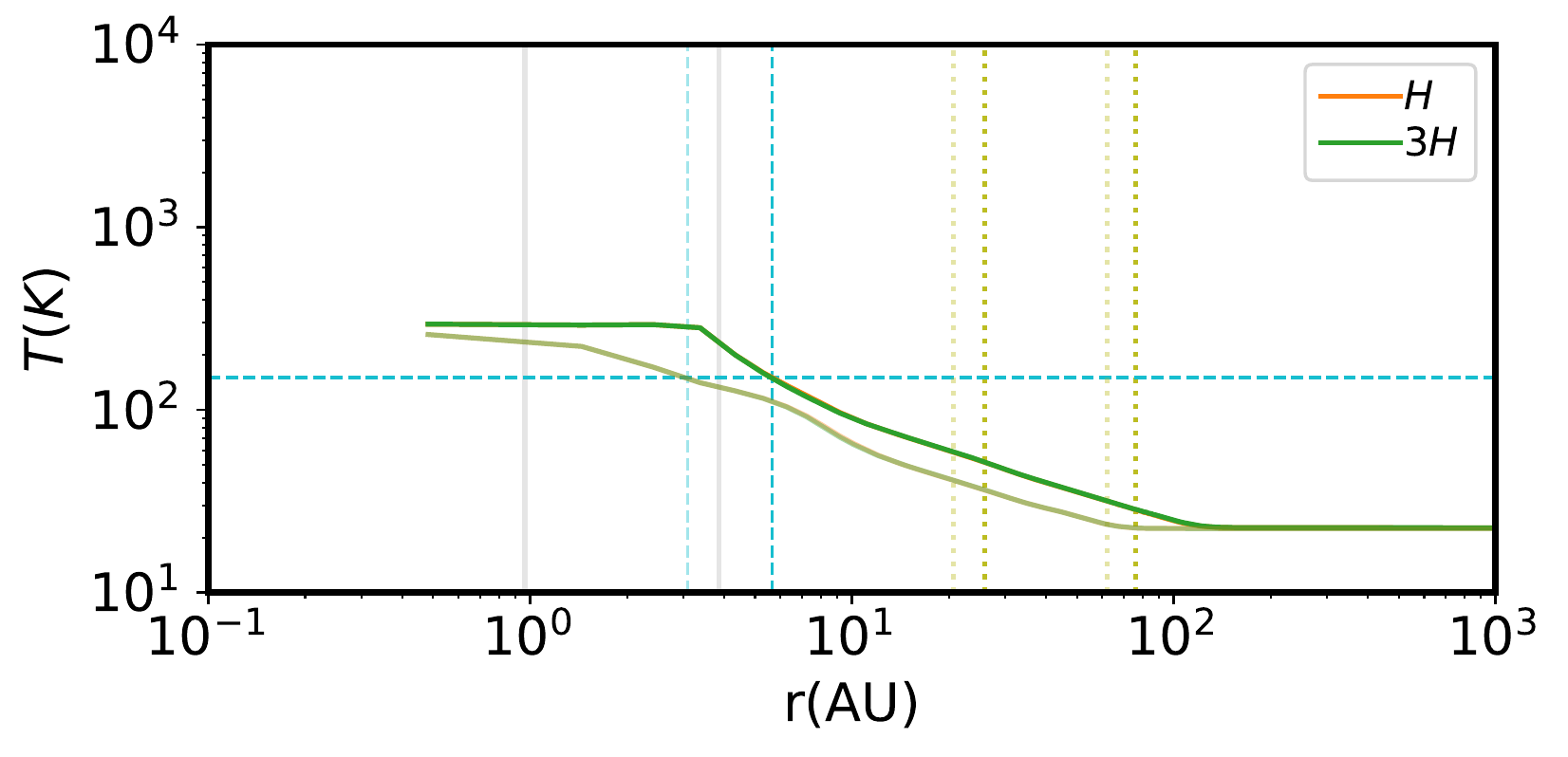}}
\put(0.16,0.41){R\_$\ell$14 vs. R\_$\ell$14\_nofeed (120 kyr),}
\end{picture}
\caption{Radial temperature profiles of the disk with and without stellar irradiation at time 120 kyr. Values of R\_$\ell$14\_nofeed are plotted with lighter colors. The temperature is averaged (mass-weighted) vertically within $H$ (orange) and $3H$ (green). The snow line ($R_{150}$) is marked with a vertical dashed cyan line. The disk temperature is significantly lower when no stellar irradiation is introduced, while the slope of radial profile does not seem to differ significantly. } 
\label{fig_T_nofeed}
\end{figure}

In general, the temperature decreases with increasing distance to the central star. 
This is mostly due to the heating from the central star because the temperature of the disk is significantly lower when the stellar irradiation is not considered (see Figs. \ref{fig_T_nofeed} and \ref{fig_snowline}). 
We caution that the kink seen near 4 AU in the canonical run is not physical. 
It corresponds to the sink accretion radius, where gas behavior is no longer physical. 
The temperature profile is close to what would be expected for a flared disk illuminated by the star, where $T \propto r^{-1/2}$ \citep{Kenyon87}.
This temperature profile is derived by balancing the flux received at the disk surface ($\propto r^{-2}$) with the local black body radiation ($\propto T^4$). 
More specifically, \citet{Chiang97} derived
\begin{align}
T \propto r^{-1/2} \alpha_{\rm g}^{1/4},
\end{align}
where $\alpha_{\rm g} = dh_{\rm p}/dr - h_{\rm p}/r$ is the grazing angle at which the stellar irradiation strikes the disk. 
Combined with our measurements of $h_{\rm p} \propto r^{1.1-1.3}$ (see Fig. \ref{fig_H}), 
we can derive a temperature profile slightly shallower than $r^{-1/2}$.
However, our disk is surrounded by an envelope that is not optically thin, and the geometry of the disk is not very flared (see Fig. \ref{fig_H}). 
The photons emitted from the protostar are probably unable to penetrate the upper layers freely and directly strike the disk surface at large radii. 
It is thus more appropriate to describe the temperature distribution in the diffusion regime, 
where the radial temperature gradient is established through the equilibrium of radiative diffusion, such that \citep[cf. Eq. (5) of][]{hennebelle2020b}
\begin{align}
- {4\pi r^2 \sigma \over 3 \kappa(T) \rho(r) } {\partial T^4 \over \partial r} = L_\ast,
\end{align}
where $L_\ast$ is the luminosity from the protostar, $\kappa$ is the opacity, and $\sigma$ is the Stefan-Boltzmann constant. 
The density profile $\rho(r)$ is slightly shallower than $r^{-2}$ in general \citep[see Fig. 3 of][]{Hennebelle20}.
For roughly constant opacity \citep{Semenov03} and $\rho \sim r^{-2 \sim -1}$, we can derive $T  \propto r^{-3/4 \sim -1/2}$. 
In conclusion, the disk temperature profile is similar to what would be expected for an irradiated disk, while the physical origin of this profile is different. 

For a passive disk that is heated only by viscous dissipation, \citet{Bitsch2013} derived 
$T \propto r^{-3/4 - s/2}$ (constant viscosity) or $T \propto r^{-1/2 -2s/3}$ ($\alpha$ viscosity), where $s$ is the power-law exponent of the surface density profile $\Sigma \propto r^{-s}$, depending on different viscosity models. 
Again, these profiles are very similar to those we derived above.
This shows that the temperature profile is not a very sensitive proxy with which we can distinguish different heating mechanisms of the disk. 

Here we are particularly interested in two temperature values: 
the water snow line at 150 K, and the CAI condensation temperature at 1500 K. 
We used 1500 K as a representative value, while it should be kept in mind that the CAIs are a family of refractory minerals and form in a temperature range around this value. 
We identified the two radii, $R_{150} \sim 7$ AU  and $R_{1500} < 1$ AU. 
Only in the high-resolution runs ($dx = 0.12$ AU) is $R_{1500}$ marginally resolved.

\subsubsection{Evolution of snow lines}
\begin{figure}[]
\centering
\includegraphics[trim=0 0 0 0,clip,width=0.5\textwidth]{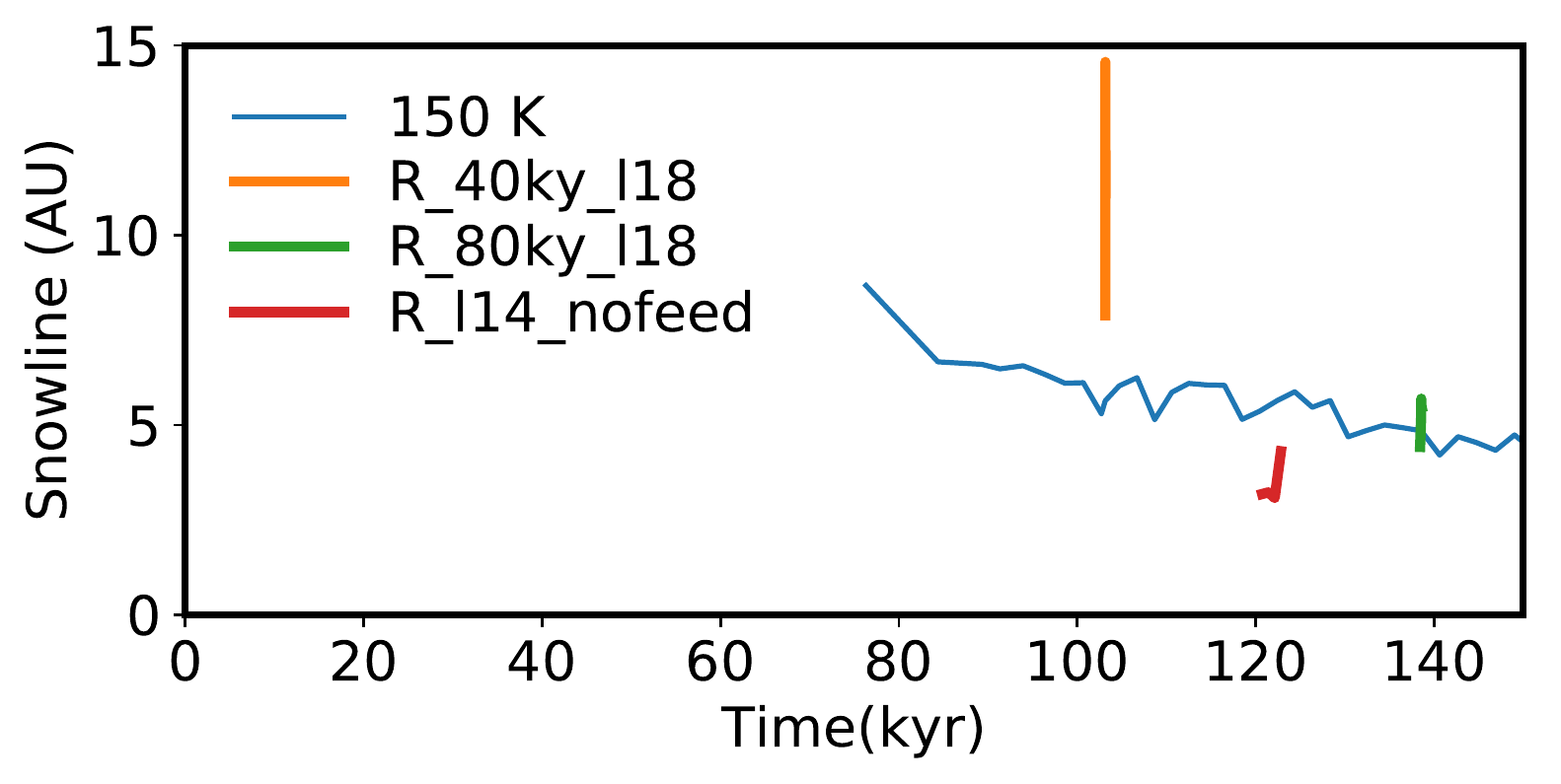}
\caption{Evolution of the water snow line, $R_{150}$. The snow line is initially located at about 10 AU and migrates inward in time as the disk evolves and loses its mass, finally stabilizing at $\sim 5$ AU. 
High-resolution runs are overplotted. R\_80ky\_$\ell$18 has consistent temperature, while in R\_40ky\_$\ell$18 the snow line migrates to a larger radius, possibly due to the mass accumulation in the disk  and the increase in opacity. This reflects the importance of properly choosing $n_{\rm acc}$ with increased resolution (see Sect. \ref{st_restart}). The run with no stellar irradiation is also plotted for comparison, where the temperature is significantly lower. }
\label{fig_snowline}
\end{figure}

Figure \ref{fig_snowline} shows the evolution of $R_{150}$, the water snow line.  
We do not discuss $R_{1500}$ because it is only marginally resolved in the runs with the highest resolution. 

$R_{150}$ starts at about 10 AU and decreases slowly in time, 
as the global density of the disk decreases due to accretion onto the star and redistribution of material in the envelope surrounding the disk.  
After about 80 kyr, the snow line stabilizes at about 5 AU, 
close to the current location of Jupiter, which may have consequences for its formation (see Sect. \ref{st_adisk}). 
$R_{150}$ from high-resolution restarts are overplotted. 
Run R\_80ky\_$\ell$18 shows a consistent snow-line position. 
On the other hand, the R\_40ky\_$\ell$18 is significantly heated,
and the snow line moves very quickly outward due to the high opacity resulting from the accumulated mass. 
This reflects that the choice of $n_{\rm acc}$ is not entirely consistent when we increase the resolution (see Sect. \ref{st_restart}).   
We recall that the sink does not accrete in this high-resolution restart (Sect. \ref{st_global}). 
The slight increase in surface density, which is barely perceptible in the $\Sigma$ profiles, leads to an increased opacity and thus to an increased temperature. 
Due to the same issue, the evolution trend of the snow line we obtained should be regarded as tentative, and its exact position depends on the detailed physics of the stellar accretion.

To evaluate the significance of the stellar irradiation, we performed run R\_$\ell$14\_nofeed, where the intrinsic luminosity from the protostar was turned off at simulation time 119 kyr.
As shown in Fig. \ref{fig_snowline}, the temperature of the disk immediately is significantly lower. 
This implies that in our simulations, the disk temperature is primarily set by the stellar irradiation. 
However, because this run was only performed for a short time span, it is unclear whether the disk will reach a different equilibrium state with slightly higher temperature by increasing the density. 
The effect of irradiation should be investigated in detail, along with the luminosity resulting from mass accretion. We leave this for future studies.

\subsection{Disk dynamical properties}
Some disk properties, such as the growth and radial drift of dust and pebbles, as well as planetesimal formation, are closely linked to its dynamics and evolution. 
Because it is always assumed that the protoplanetary disk has close-to-Keplerian rotation, 
we first confirm the validity of this common assumption. 
We then continue to discuss the magnetization and the transport parameters that are essential in controlling dust-growth and transport within the disk.

The azimuthally averaged quantities are presented, 
and we show both the average within one and three times the disk scale height $H$. 
Because 68.3 \% and 99.7 \% of the mass are contained within these extents for a vertical Gaussian profile, 
these quantities should be representative of the disk. 
All quantities were calculated throughout the selected radius, $r_{\rm cyl}$, 
while the notion of scale height is only valid within the disk radius $R_{\rm kep}$. 
The quantities inferred in the envelope region are therefore only suggestive of the values around midplane. 

\subsubsection{Velocity: Rotation and accretion}\label{st_vel}

\begin{figure}[]
\centering
\setlength{\unitlength}{0.5\textwidth}
\begin{picture}(1,0.9)
\put(0,0.45){\includegraphics[trim=0 0 0 6,clip,width=0.5\textwidth]{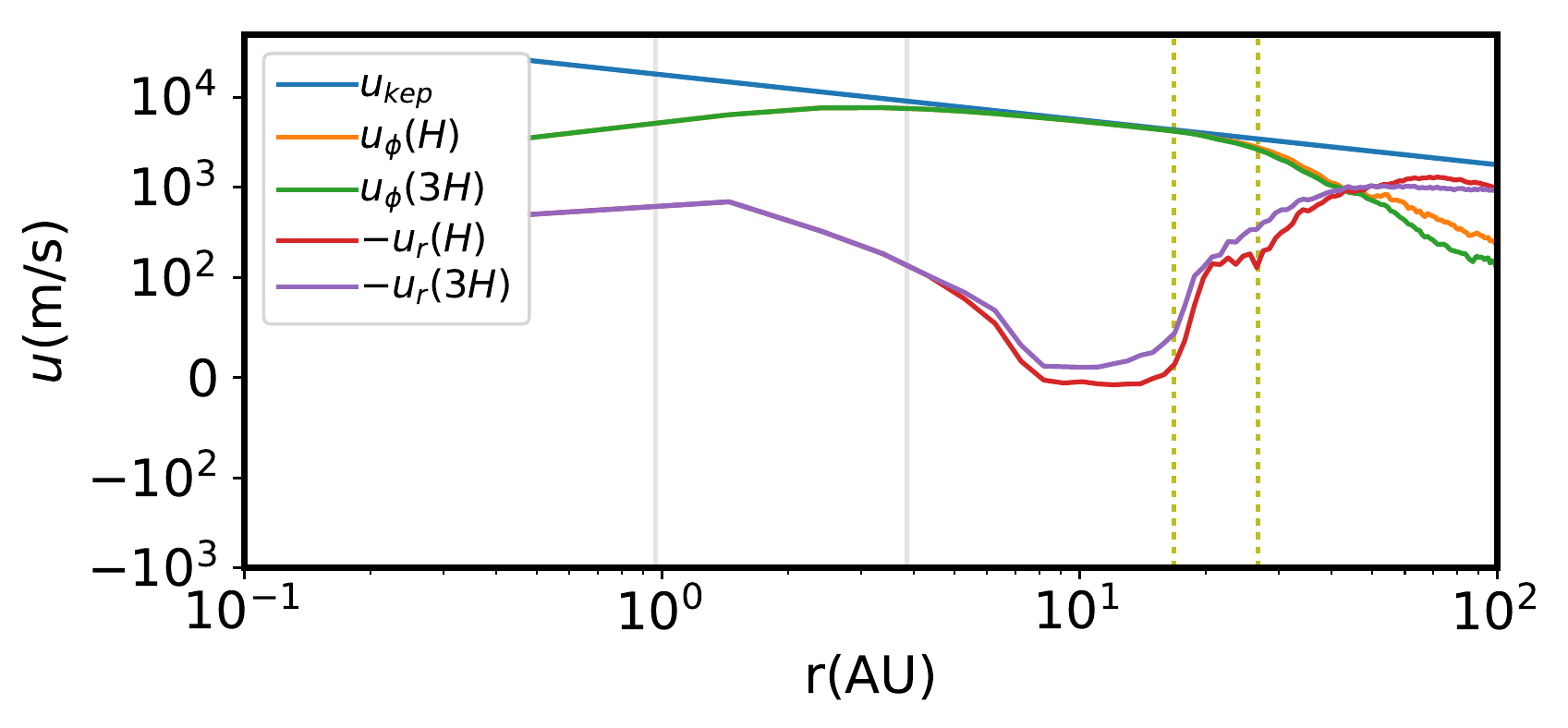}}
\put(0,0){\includegraphics[trim=0 0 0 6,clip,width=0.5\textwidth]{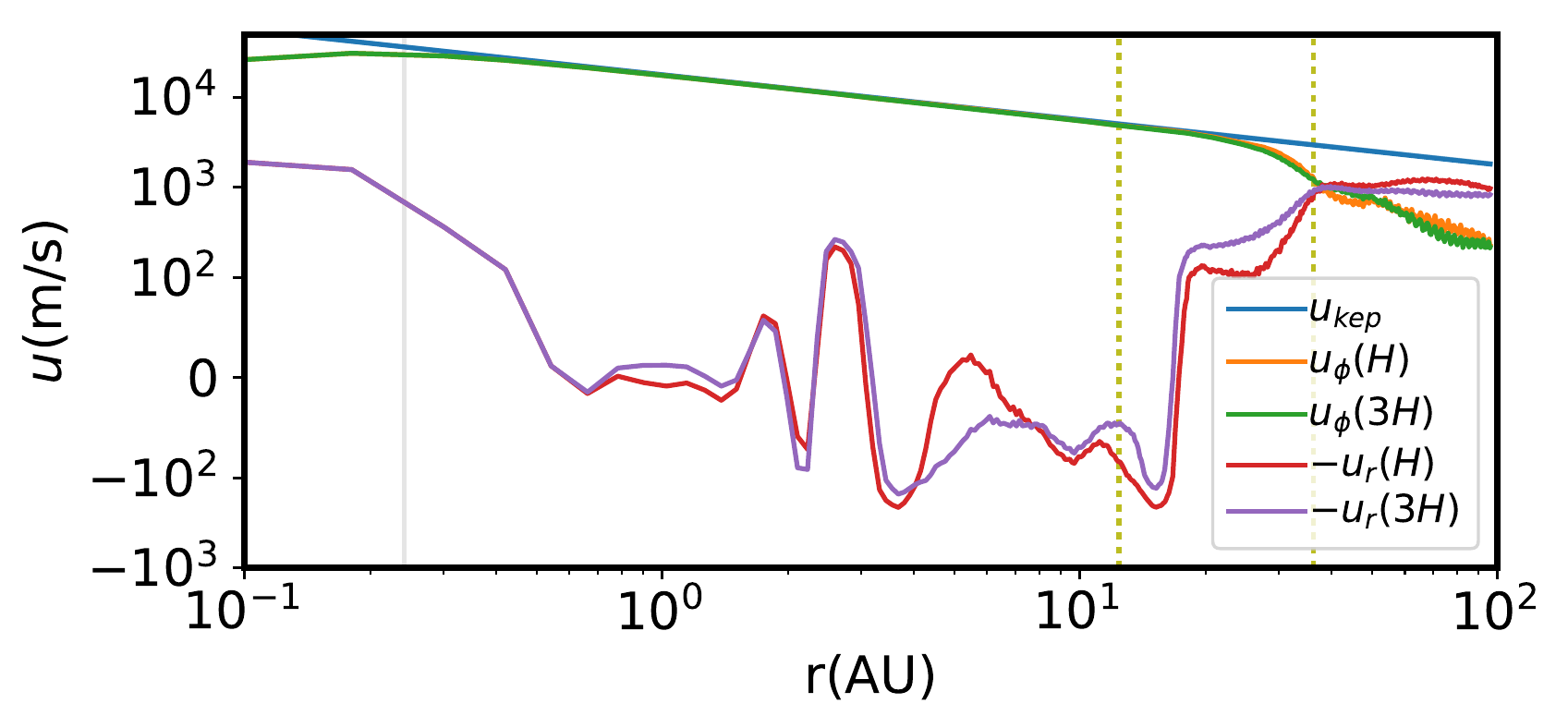}}
\put(0.17,0.57){R\_$\ell$14 (103 kyr)}
\put(0.17,0.12){R\_40ky\_$\ell$18 (103 kyr)}
\end{picture}
\caption{Velocity profiles of the disk at two resolutions: Rotational velocity within $H$ (orange) /$3H$ (green) and infall velocity within $H$ (red)/$3H$ (violet). Both positive and negative values with $|u| > 10^2$ are plotted in logarithmic scale, and this is connected with a linear range in between. The Keplerian rotational velocity is plotted in blue for reference. The inner disk is very close to Keplerian rotation, and the deviation becomes remarkable in the outer magnetized part of the disk. The envelope is far from Keplerian rotation, while the infalling gas has roughly half the free-fall velocity. Below the disk accretion radius (vertical gray line), the velocity is not physically meaningful. The disk characteristic radii $R_{\rm kep}$ and $R_{\rm mag}$ (the dotted vertical lines) mark the transition between different behaviors. }
\label{fig_v}
\end{figure}

We show the rotation velocity, $u_\phi$, and the radial velocity (parallel to the disk plane), $u_r$, in Fig. \ref{fig_v}. 
The Keplerian rotation velocity is plotted for reference: 
\begin{align}
u_{\rm kep}(r) \approx \sqrt{G[M_\ast+M_{\rm d}(r)] \over r}.
\end{align} 
This approximation is valid for a disk mass dominated by the stellar mass. 

The mass-weighted average is calculated within one and three disk scale heights. 
The azimuthal velocity is plotted in orange ($H$) and green ($3H$). 
Different behaviors are clearly seen in regions separated by the disk characteristic radii $R_{\rm kep}$ and $R_{\rm mag}$ (see the definition in Sect. \ref{st_Sigma}, and here we see the reason for giving these names).
The velocity is very close to the Keplerian rotation between $\sim 4$ AU (sink accretion radius) and $\sim 13$ AU ($R_{\rm kep}$). 
Inside the sink accretion radius, the flow property is not physical and should not be considered. 
Beyond $R_{\rm kep}$, the rotation starts to deviate from the Keplerian value because significant support comes from the magnetic field. 
Beyond $R_{\rm mag}$ it is clear that the gas inside the infalling envelope is far from Keplerian rotation. 

On the other hand, the infall velocity (negative radial velocity) is plotted in red ($H$) and purple ($3H$). 
In the envelope, the gas falls radially onto the disk at almost half the free-fall velocity, which has the same expression as the Keplerian velocity. 
In the interior of the disk, the radial velocity is significantly lower than the rotational velocity (by at least two orders of magnitude). 
The radial velocity flips between inward and outward directions with no clear pattern, and it also varies with time. 
The difference between the two runs at different resolutions indicates that the internal dynamics is highly variant and has not yet reached numerical convergence.
Nonetheless, the innermost disk (below 1 AU) always shows accretion toward the central star. 
As we discuss in Sect. \ref{st_transport}, most of the mass accretion onto the star occurs directly through a high-altitude channel and does not pass through the bulk of the disk.
Therefore the disk is not necessarily always in accretion.

\subsubsection{The $\beta$ parameter: Magnetization}\label{st_beta}

\begin{figure}[]
\centering
\setlength{\unitlength}{0.5\textwidth}
\begin{picture}(1,1.8)
\put(0,1.35){\includegraphics[trim=0 0 0 6,clip,width=0.5\textwidth]{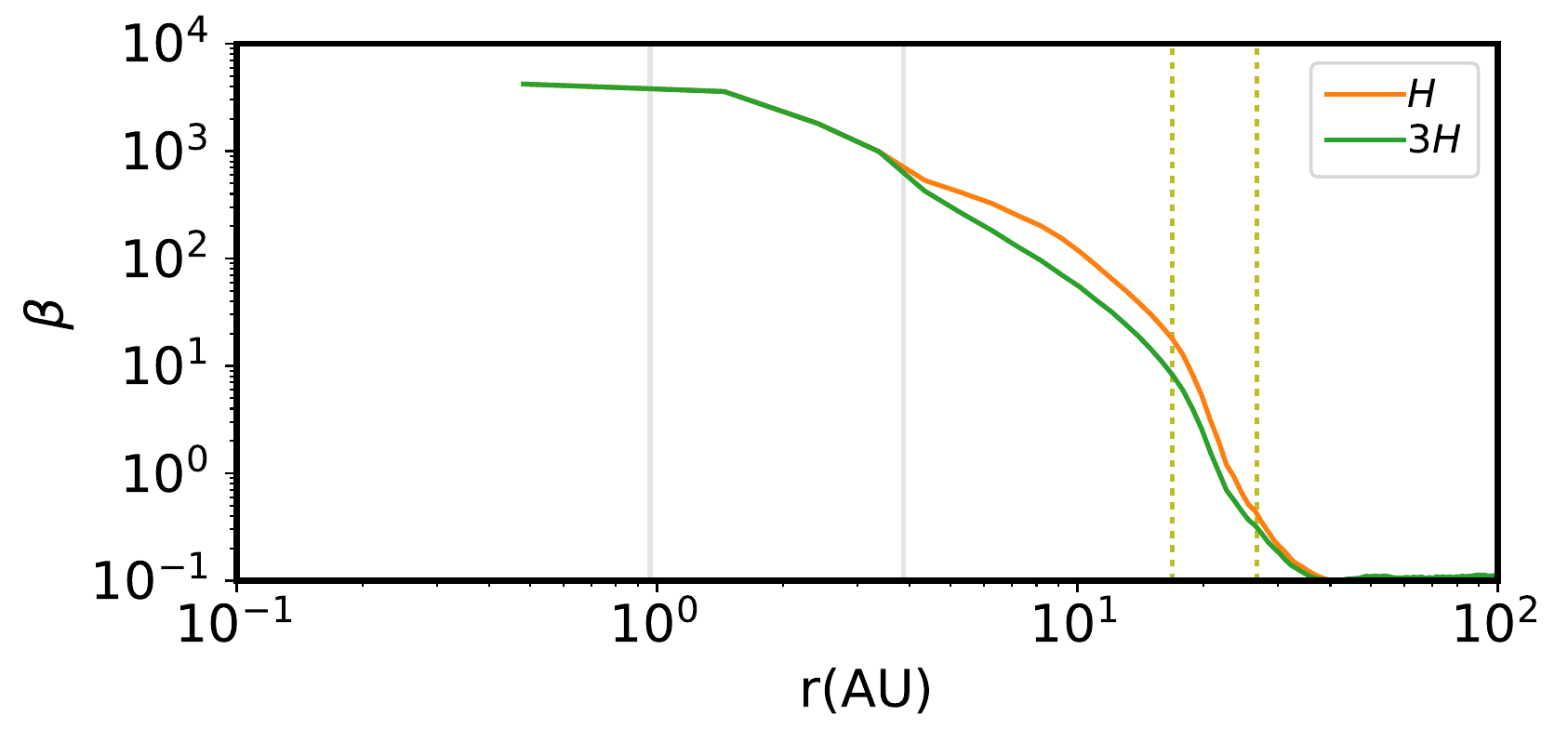}}
\put(0,0.9){\includegraphics[trim=0 0 0 6,clip,width=0.5\textwidth]{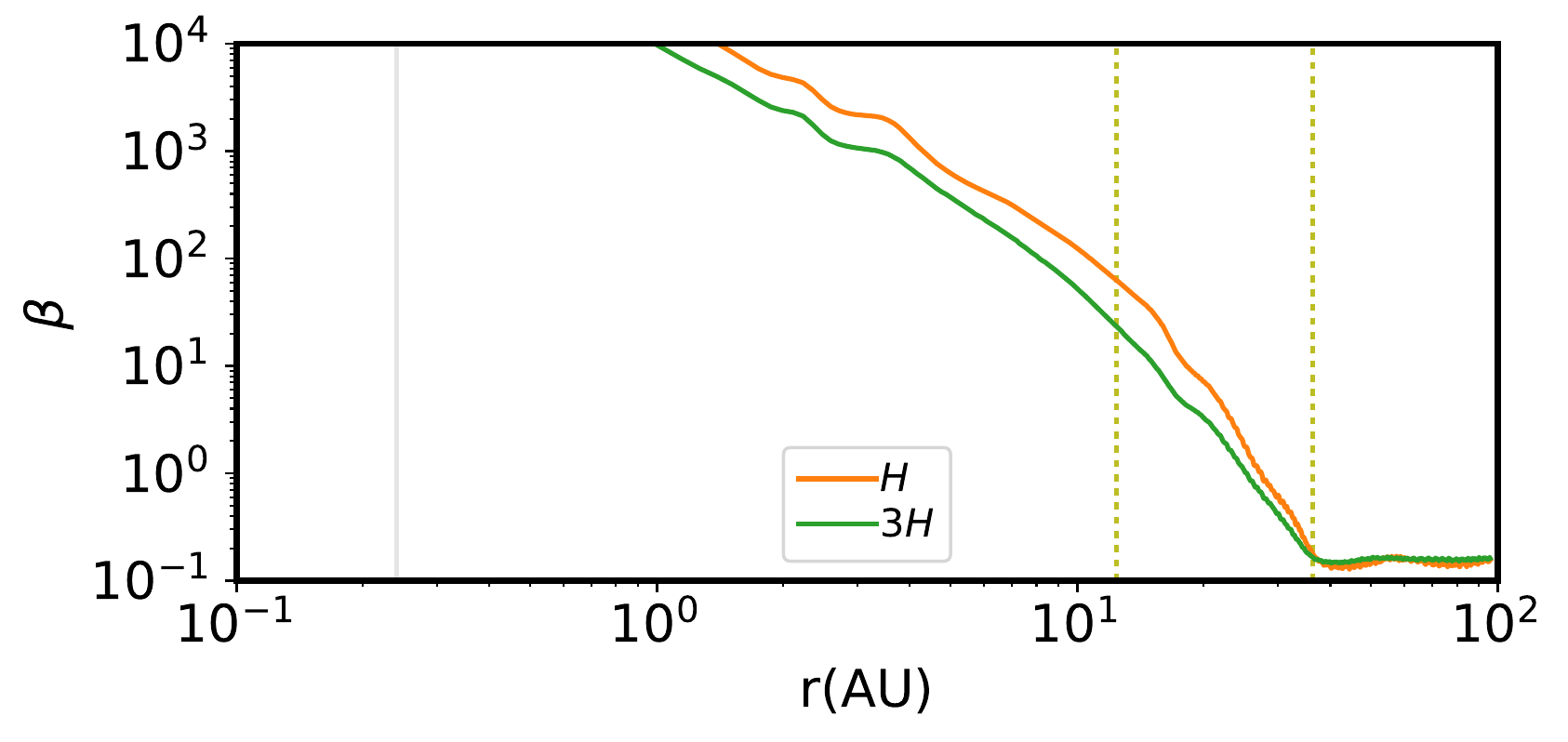}}
\put(0,0.45){\includegraphics[trim=0 0 0 6,clip,width=0.5\textwidth]{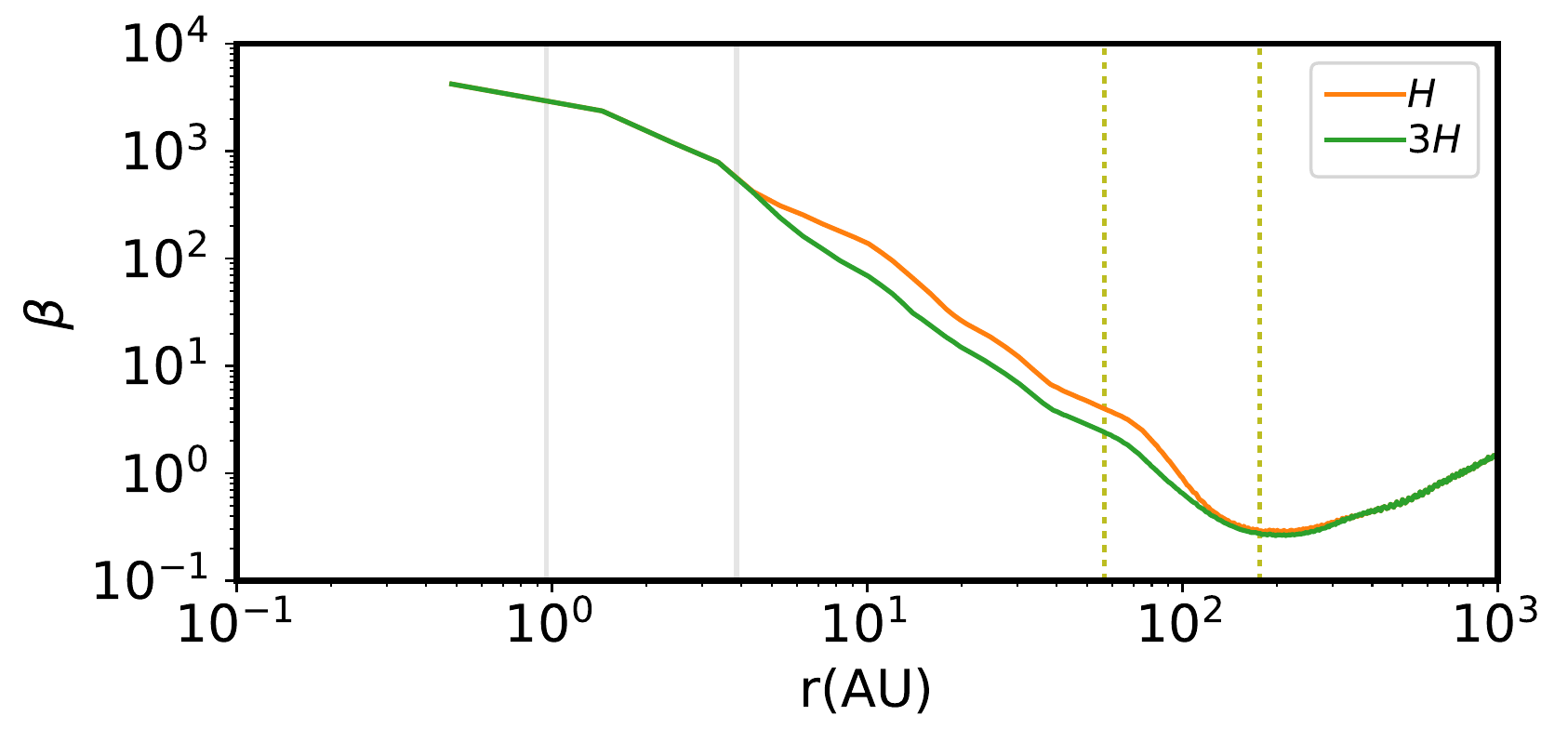}}
\put(0,0){\includegraphics[trim=0 0 0 6,clip,width=0.5\textwidth]{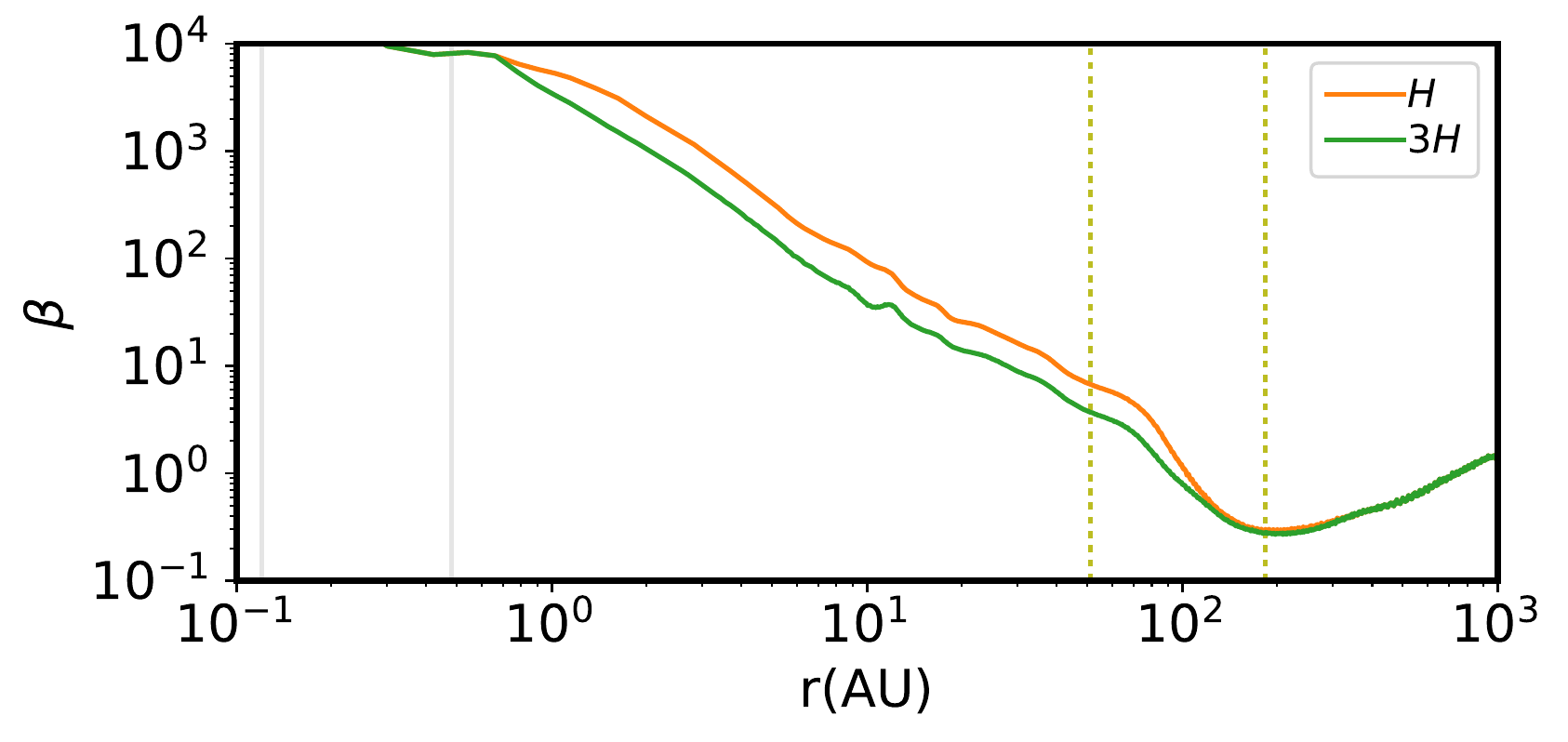}}
\put(0.16,1.47){R\_$\ell$14 (103 kyr)}
\put(0.16,1.02){R\_40ky\_$\ell$18 (138 kyr)}
\put(0.16,0.57){R\_$\ell$14 (103 kyr)}
\put(0.16,0.12){R\_80ky\_$\ell$18 (138 kyr)}
\end{picture}
\caption{Plasma beta, the ratio between thermal and magnetic pressures, $\beta= P_{\rm therm}/P_{\rm mag}$, evaluated within one (orange) and three (green) times the disk scale height. The disk is highly demagnetized because ambipolar diffusion leaves the field lines outside during the collapse. The $\beta$ value passes through unity around the junction between the disk and the envelope. The difference between the two curves comes from the lower magnetization level close to the disk midplane.}
\label{fig_beta}
\end{figure}
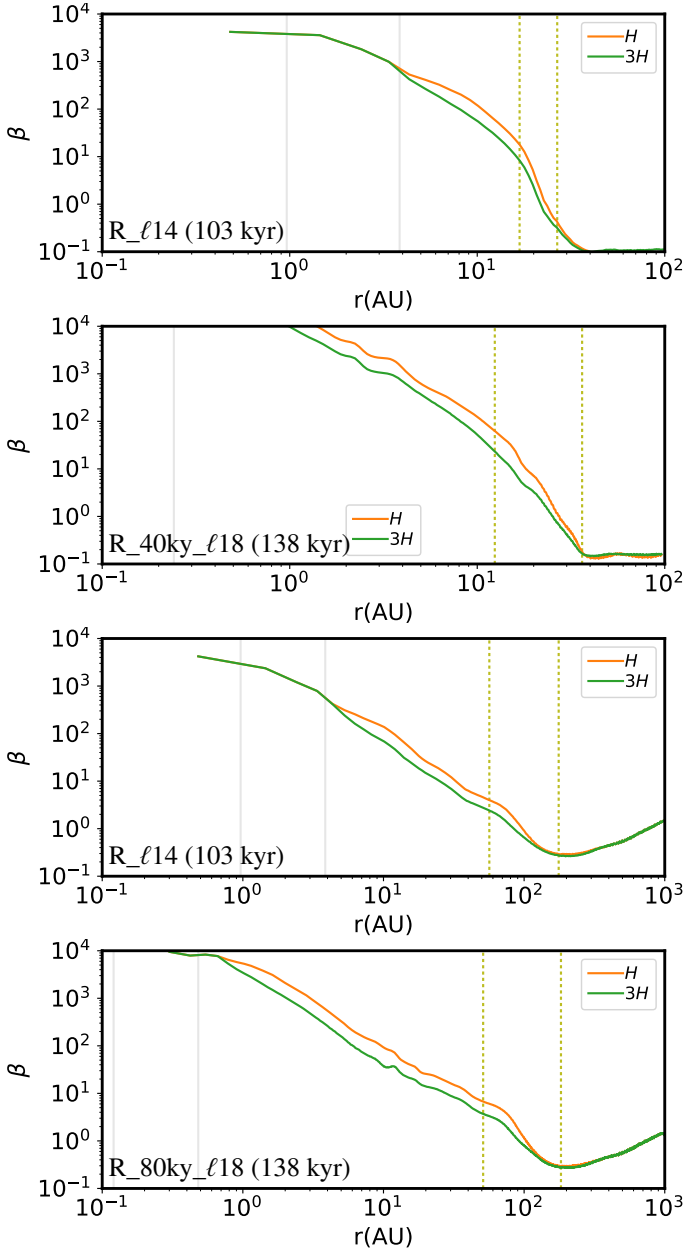

The ratio of the thermal and magnetic pressures, $\beta = P_{\rm therm}/P_{\rm mag} = 8 \pi \rho c_{\rm s}^2 / B^2$, 
indicates which supporting agent is dominant inside and outside the disk region.
This is calculated as
\begin{align}
\beta(r) 
= 8\pi{\sum P dV \over \sum B^2 dV} .
\end{align}

Figure \ref{fig_beta} shows $\beta$ values within $H$ and $3H$. 
Because the temperature inside the disk is high, the thermal pressure largely dominates, with $\beta$ decreasing outward from $10^3$ at 1 AU.
The value of $\beta$ crosses unity inside the magnetized zone of the disk (between $R_
{\rm kep}$ and $R_{\rm mag}$), 
marking a transition to the magnetized infalling envelope. 
As the global collapse proceeds, the central high-density part loses the field lines through ambipolar diffusion, and the magnetic pressure is therefore weak in the inner part of the disk. 
The disk is more magnetized at higher altitudes, giving lower $\beta$ values at $3H$ than at $H$ (because the temperature is almost identical to that in Fig. \ref{fig_T}). 
Although significantly demagnetized, the magnetic field is still stronger than what is typically assumed in studies of disk instabilities. For example, \citet{Bethune2020} used $\beta \in [10^3, 10^8]$.

\subsubsection{The $\alpha$ parameter: Accretion and diffusion}\label{st_alpha}
The $\alpha$ parameter describes the transport in the disk as a result of torques from the turbulence or magnetic field. 
The cross correlation of different components of a vector field generates a tensor term that implies the transportation of mass and angular momentum \citep{Balbus99}.
It has contributions from the turbulent fluctuation (Reynolds tensor), the magnetic field (Maxwell tensor), and the self-gravity of the disk. 
The turbulent diffusion is not very well known within the disk, 
and thus $\nu = \alpha_{\rm Rey} c_{\rm s}^2/\Omega$ is often used as the turbulent viscosity to describe the diffusion within the disk. 
We evaluated radial dependence of these terms inside the disk and discuss their effect on the transport of mass and angular momentum. 

The stress tensor, $T_{\vec{p}\phi}(\vec{p},\phi) \equiv \langle \rho \delta u_{\vec{p}} \delta u_\phi\rangle_\phi$, is an azimuthal average along a circle perpendicular to the axis, 
where the subscript $\vec{p}$ stands for the poloidal direction that has two components $r$ and $z,$ and $\delta$ denotes the deviation from the mean value. 
The dimensionless $\alpha$ expression is defined as $T_{\vec{p}\phi}/\langle \rho c_{\rm s}^2 \rangle_\phi$, 
where the denominator is the azimuthally averaged thermal pressure. 

The radial mass accretion rate of a stationary disk is linked to the stress tensor with the transport equation (see details in Appendix \ref{ap_transport}),
\begin{align}\label{eq_Mdot_alpha}
\dot{M}(r) &= - 2 \pi r \Sigma \langle u_R \rangle_r \\
& = {2 \pi \over (r^2\Omega)^\prime} \left[ { d\over dr} \left( r^2 \int_{-H}^H T_{r\phi} dz \right) + r^2T_{z\phi}|^H_{-H} \right]  \nonumber\\
& =   {2 \pi \over (r^2\Omega)^\prime}  \left[ (r^2\Sigma c_{\rm s}^2 \alpha_r)^\prime +{r^2 \over 2H}~\Sigma c_{\rm s}^2 \alpha_z \right] \nonumber\\
&\approx \sqrt{2 \pi \Sigma r \over G} c_{\rm s}^2 \left( \alpha_r + {r\over 2H} \alpha_z \right), \nonumber
\end{align}
where $^\prime$ signifies the derivative with respect to $r$. 
The approximation is obtained by making several assumptions such that $d/dr \approx 1/r$ and $\Omega^2 \approx G \int 2\pi r \Sigma dr \approx 2 \pi G \Sigma r^2$, which are not too far from reality \citep[see, e.g.,][]{Fromang13}, 
with the purpose of allowing an appreciation of how the two terms in parentheses compare to one another. 
More precisely, $\alpha_r$ is measured in the bulk of the disk, while $\alpha_z$ is measured on the surfaces, and their effects on the mass accretion rate differ by a factor $r/(2H)$. These two values should therefore not be compared directly.

While the stress tensor is defined along a thin circular line, 
 the average is in practice taken within a finite volume. 
The $\alpha$ values are thus calculated as
\begin{align}
\alpha_{{\rm Rey},r} &=\hat{T}_{{\rm Rey},r\phi} / \hat{P}, 
&\alpha_{{\rm Rey},z} &= (\hat{T}_{{\rm Rey},z\phi}^H - \hat{T}_{{\rm Rey},z\phi}^{-H} )/ \hat{P}, \\
\alpha_{{\rm Max},r}&= \hat{T}_{{\rm Max},r\phi} / \hat{P}, 
&\alpha_{{\rm Max},z} &= (\hat{T}_{{\rm Max},z\phi}^H - \hat{T}_{{\rm Max},z\phi}^{-H} )/ \hat{P}, \label{eq_alphaBz}\\
\alpha_{{\rm Grav},r} &= \hat{T}_{{\rm Grav},r\phi} / \hat{P} , 
&\alpha_{{\rm Grav},z} &= (\hat{T}_{{\rm Grav},z\phi}^H - \hat{T}_{{\rm Grav},z\phi}^{-H} )/ \hat{P},
\end{align}
where the hat symbol signifies the volume average. 
The region of the summations is an annulus of radial thickness $dx$, 
while its vertical thickness is to be specified. 

The volume-averaged stress tensors and pressure are defined as
\begin{align}
\hat{T}_{\vec{p}\phi, {\rm Rey}} \sum dV  &= \sum   \delta u_{\vec{p}}  \delta u_\phi \rho dV =\sum   \delta u_{\vec{p}}  \delta u_\phi dm \\
\hat{T}_{\vec{p}\phi, {\rm Max}}  \sum dV  &= {1\over 4 \pi} \sum B_{\vec{p}} B_\phi dV = \sum  u_{A,\vec{p}} u_{A,\phi} dm, \\
\hat{T}_{\vec{p}\phi, {\rm Grav}}  \sum dV &=  {1\over 4 \pi G} \sum g_{\vec{p}} g_\phi dV = \sum u_{G,\vec{p}} u_{G,\phi} dm, \\
\hat{P} \sum dV  & =  \sum P_{\rm therm} dV = \sum c_{\rm s}^2 dm,
\end{align}
where 
\begin{align}
\vec{u}_{A} &= { \vec{B} \over \sqrt{4\pi \rho}},\\
\vec{u}_{G} &= { \nabla{\Phi}  \over \sqrt{4\pi G \rho}} = { -\vec{g}  \over \sqrt{4\pi G \rho}}.
\end{align}
$dV$ and $dm$ represent the volume and mass of each cell.

\begin{figure}[]
\centering
\includegraphics[trim=0 0 0 0,clip,width=0.5\textwidth]{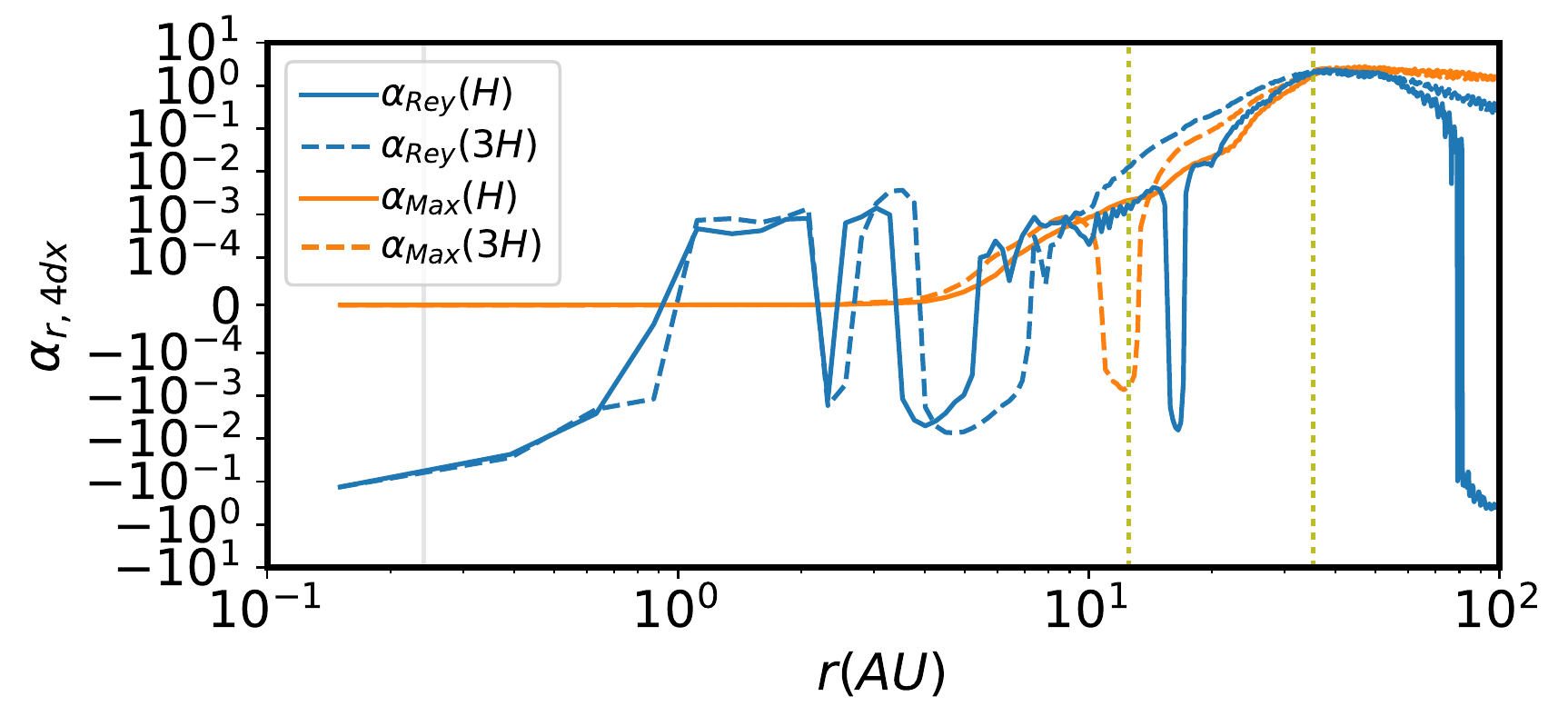}
\includegraphics[trim=0 0 0 0,clip,width=0.5\textwidth]{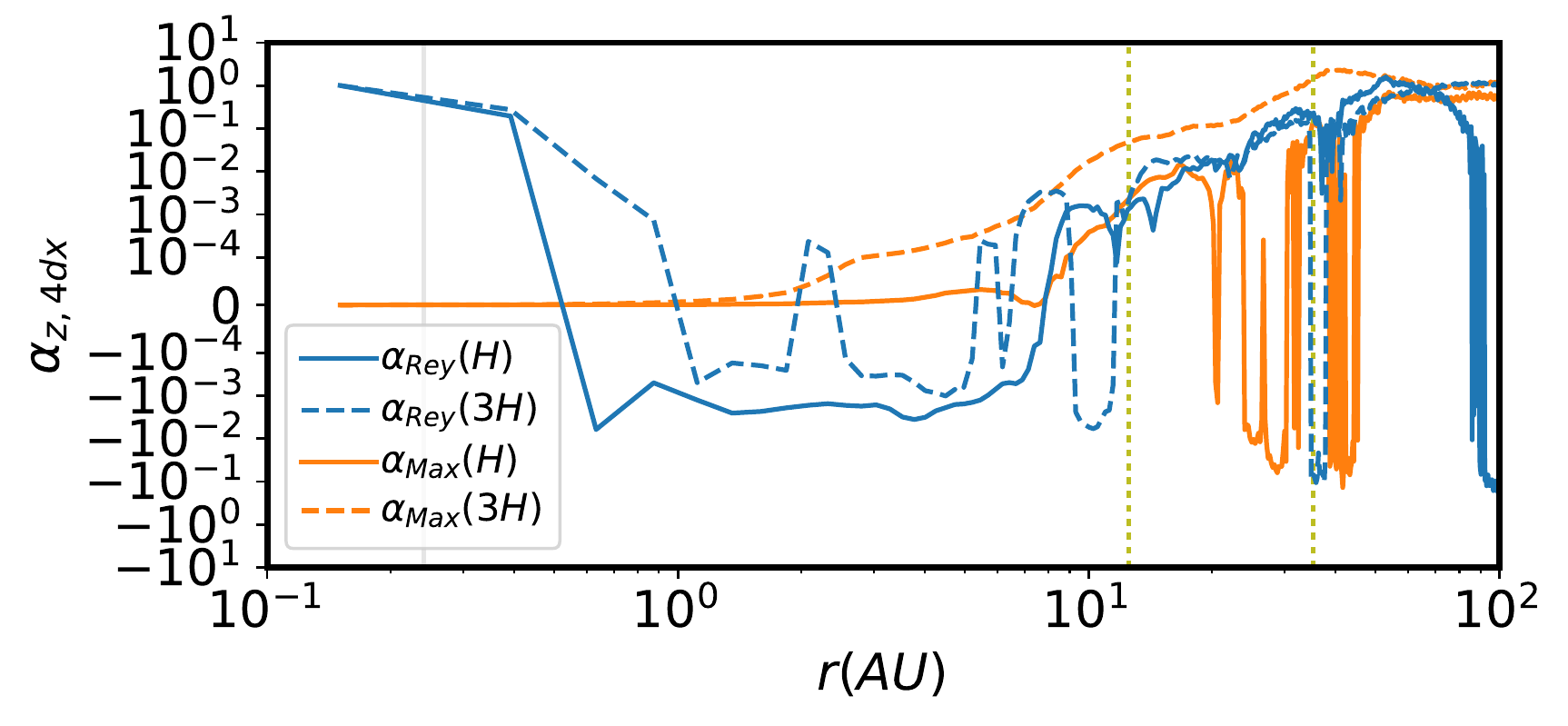}
\caption{Disk $\alpha$ for R\_40ky\_$\ell$18 at 103 kyr. The top panel presents the Reynolds and Maxwell stresses in $r$ direction, and the bottom panel shows the two components in $z$ direction. Both positive and negative values with $|\alpha| > 10^{-4}$ are plotted in logarithmic scale, and this is connected by a linear range in between.}
\label{fig_alpha}
\end{figure}

The three radial terms $\alpha_{{\rm Rey},r}$, $\alpha_{{\rm Max},r}$, and $\alpha_{{\rm Grav},r}$ are related to the radial transport within the disk. 
Their values are therefore calculated as the mean across the disk vertical extent. 
The summations were made within $H$ or $3H$. 
The total $\alpha_r$ is the sum of all the three components. 
On the other hand, $\alpha_{{\rm Rey},z}$, $\alpha_{{\rm Max},z}$, and $\alpha_{{\rm Grav},z}$ are the vertical counterparts that are linked to the accretion or angular momentum transport across the disk surfaces. 
The sum of the three terms,  $\alpha_z$, also leads to radial mass accretion. 
The summation was performed within a vertical thickness of $dx$ around $H$ and $3H$ in this case for the stress tensors, 
while the pressure was still averaged within the disk vertical extent. 
With the definition in Eq. (\ref{eq_Mdot_alpha}), it naturally justifies the choice of normalization by the bulk pressure, whether the stress tensor is averaged in the bulk ($T_{r\phi}$) or only on the surface ($T_{z\phi}$).

Figure \ref{fig_alpha} shows the $r$-(top panel) and $z$-(lower panel) components of $\alpha$.  
The gravitational term is mostly noise, and we do not show it for conciseness (see Appendix \ref{ap_alpha_dx} for more details). 
The solid lines are the values measured at $H$ and the dashed lines are those measured at $3H$.
For the $\alpha_r$ component 
at $H$, the Maxwell term in the inner part of the disk is almost zero because the disk is strongly demagnetized as a consequence of the ambipolar diffusion (see also Sect. \ref{st_beta}).
It grows steadily when the magnetized outer part of the disk is approached and reaches $\sim 10^{-3}$ at $R_{\rm kep}$. 
The Reynolds term fluctuates around zero with a maximum absolute value lower than $10^{-3}$ throughout the disk. 
Both components grow significantly in the outer zone of the disk between $R_{\rm kep}$ and $R_{\rm mag}$. 
The Reynolds component flips between positive and negative values, showing a complex transport pattern in the inner disk, 
which is also seen in the radial velocity (Sect. \ref{st_vel}). 
The values at $H$ and $3H$ are generally comparable, 
implying that the radial stress tensors do not cause significant stratified behavior within the vertical extent of the disk. 
On the other hand, the values of $\alpha_z$ show more significant variation across different altitudes, 
and this is the dominant component that leads to radial accretion within the disk. 
The Maxwell term at $3H$ is higher that at $H$ above $r \sim 5$ AU, 
implying that most of the disk radial accretion occurs in the upper layers of the disk due to magnetic angular momentum transport. 
At $H$, the Reynolds component is significantly negative, which might imply viscous outward transport near the disk midplane. 

\begin{figure*}[]
\centering
\includegraphics[trim=0 0 0 0,clip,width=0.24\textwidth]{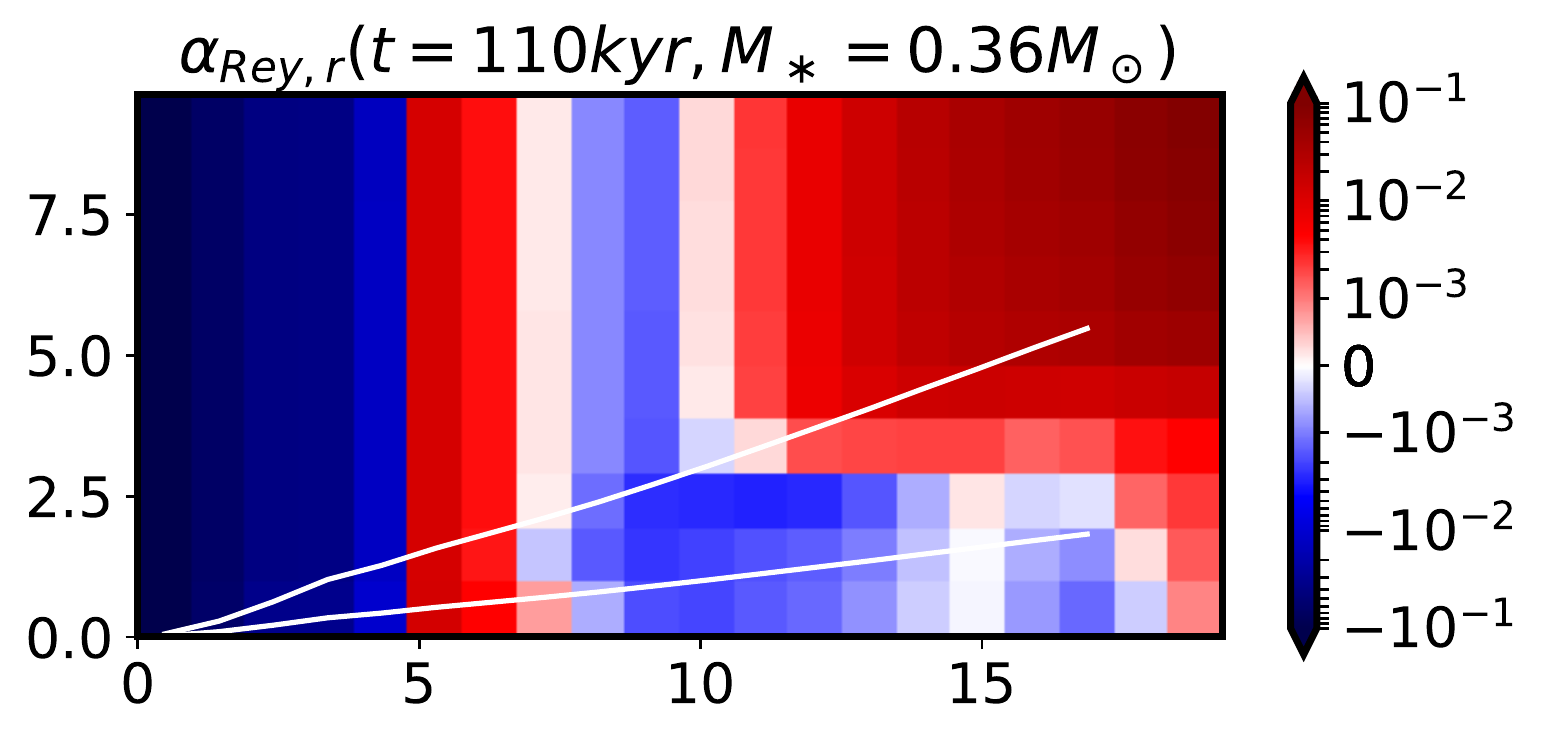}
\includegraphics[trim=0 0 0 0,clip,width=0.24\textwidth]{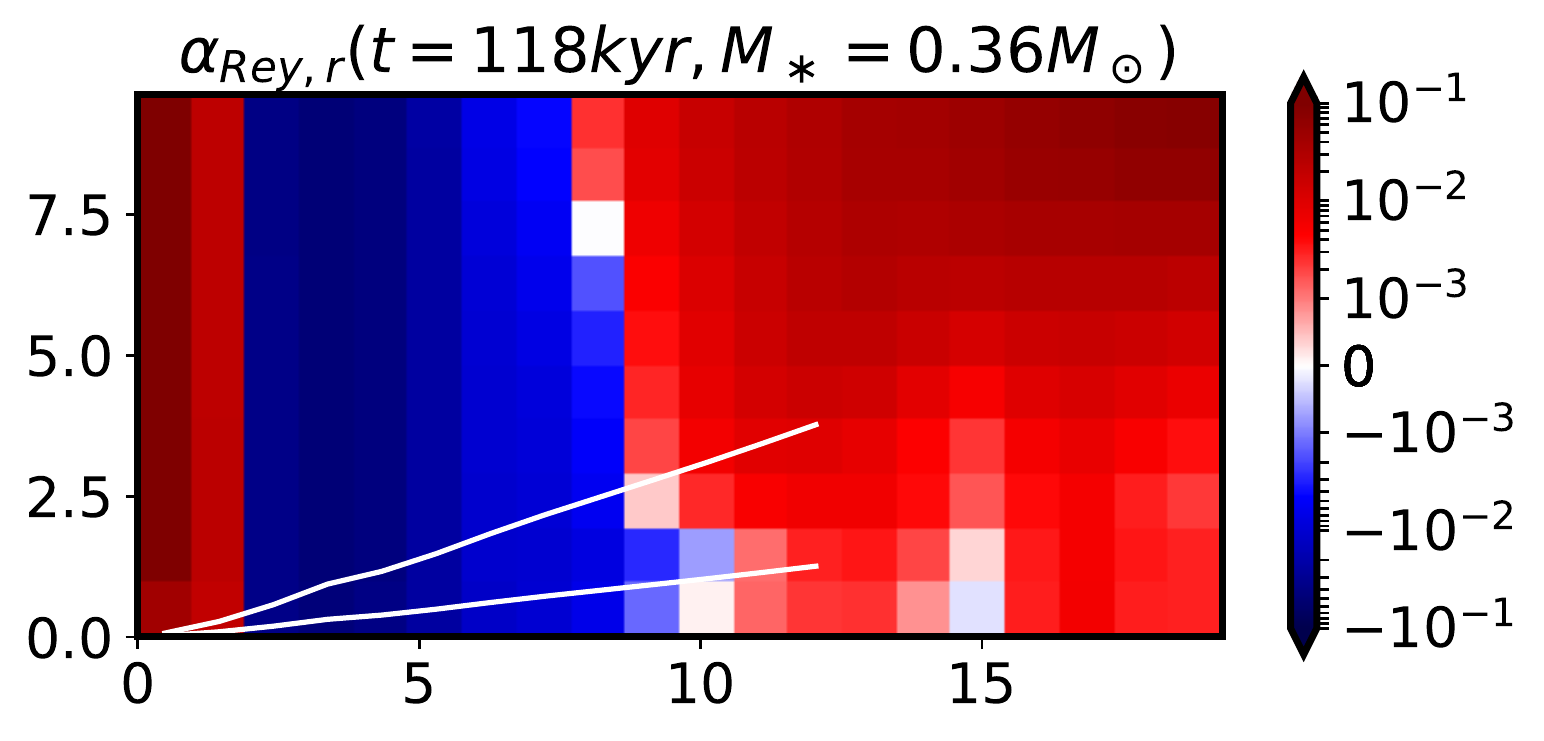}
\includegraphics[trim=0 0 0 0,clip,width=0.24\textwidth]{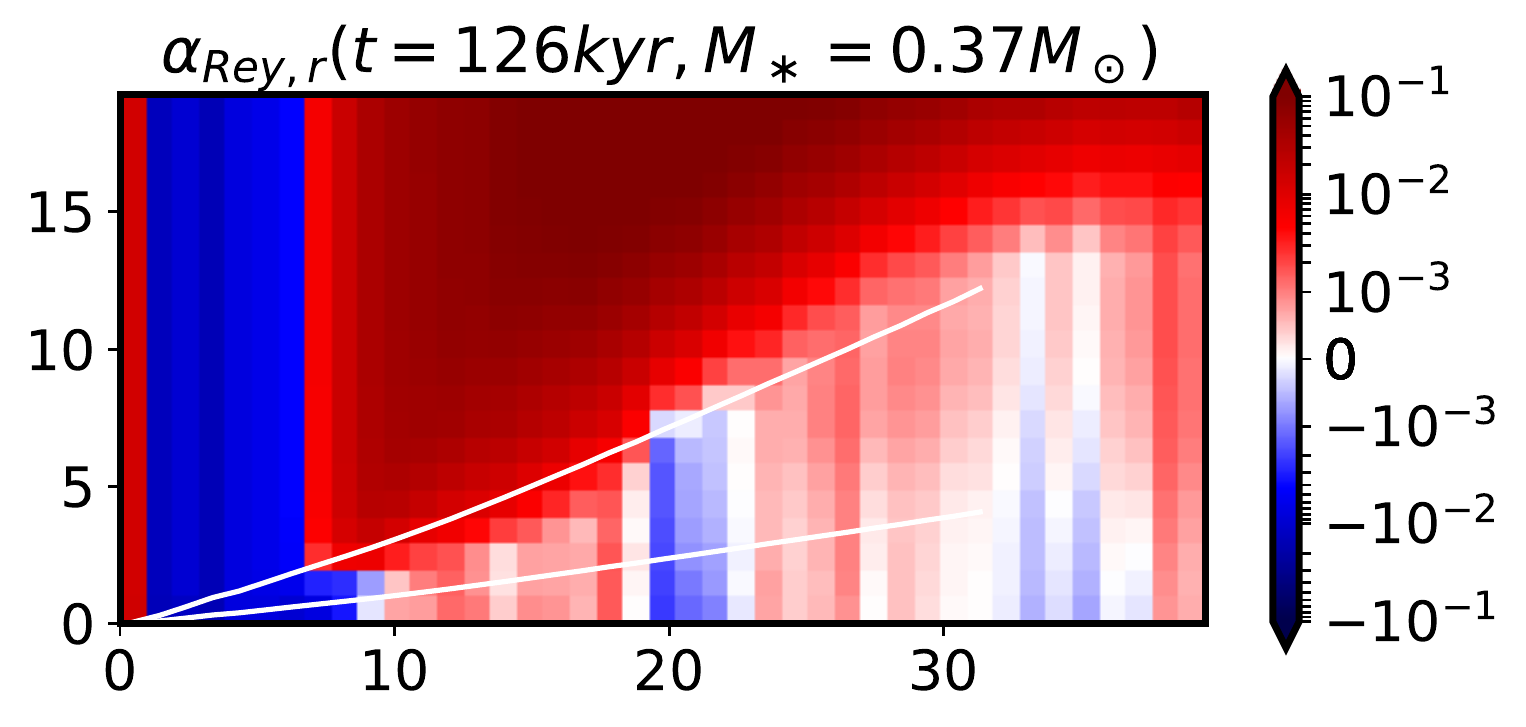}
\includegraphics[trim=0 0 0 0,clip,width=0.24\textwidth]{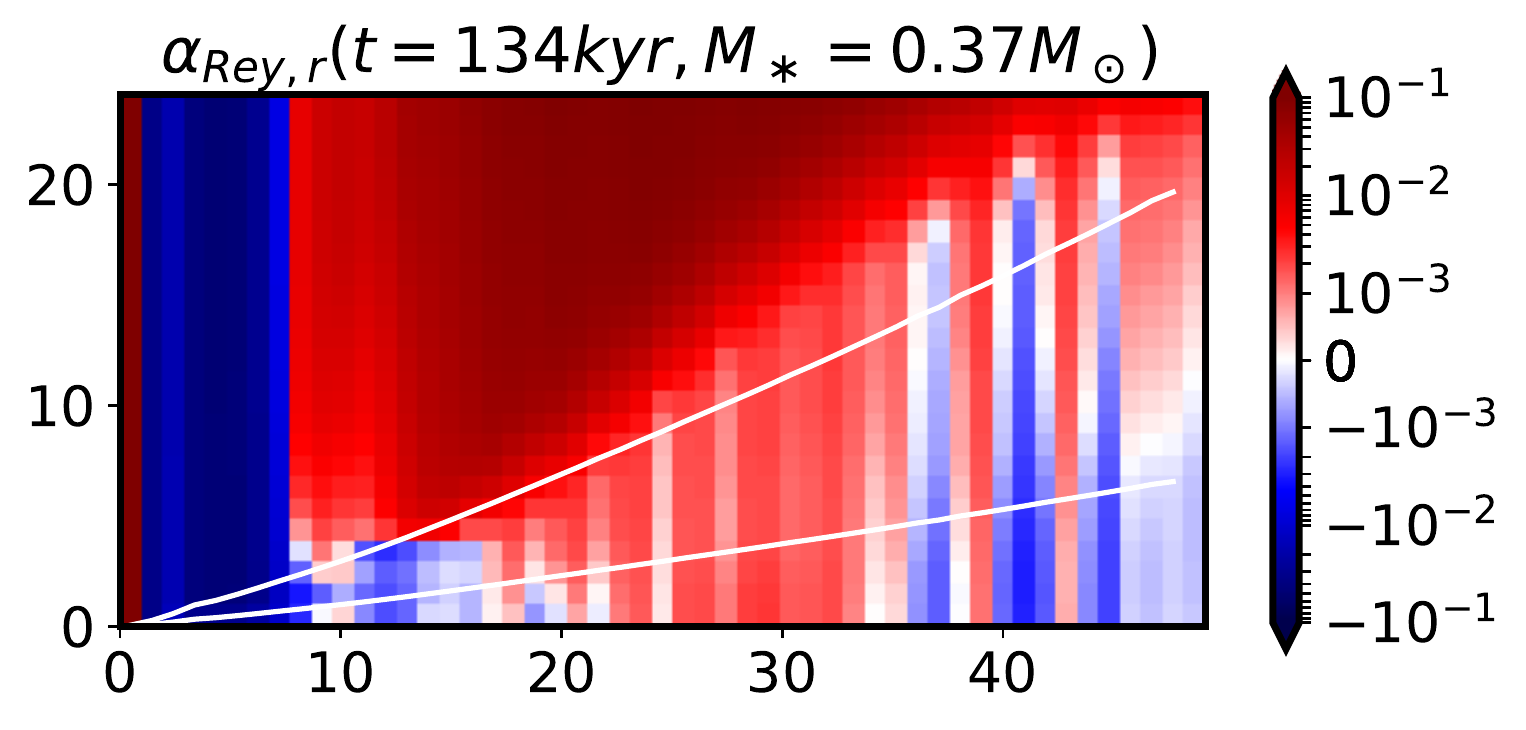} \\

\includegraphics[trim=0 0 0 0,clip,width=0.24\textwidth]{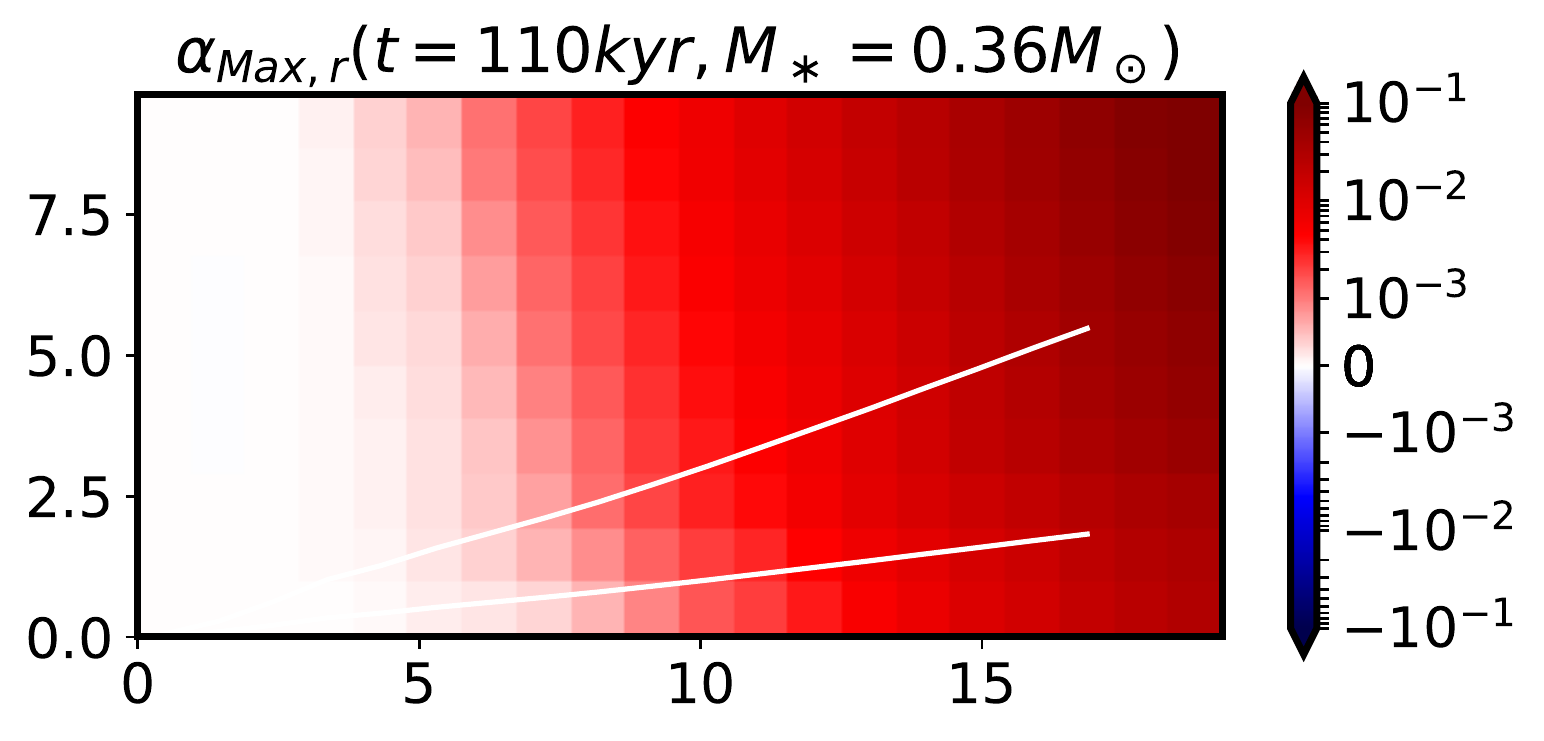}
\includegraphics[trim=0 0 0 0,clip,width=0.24\textwidth]{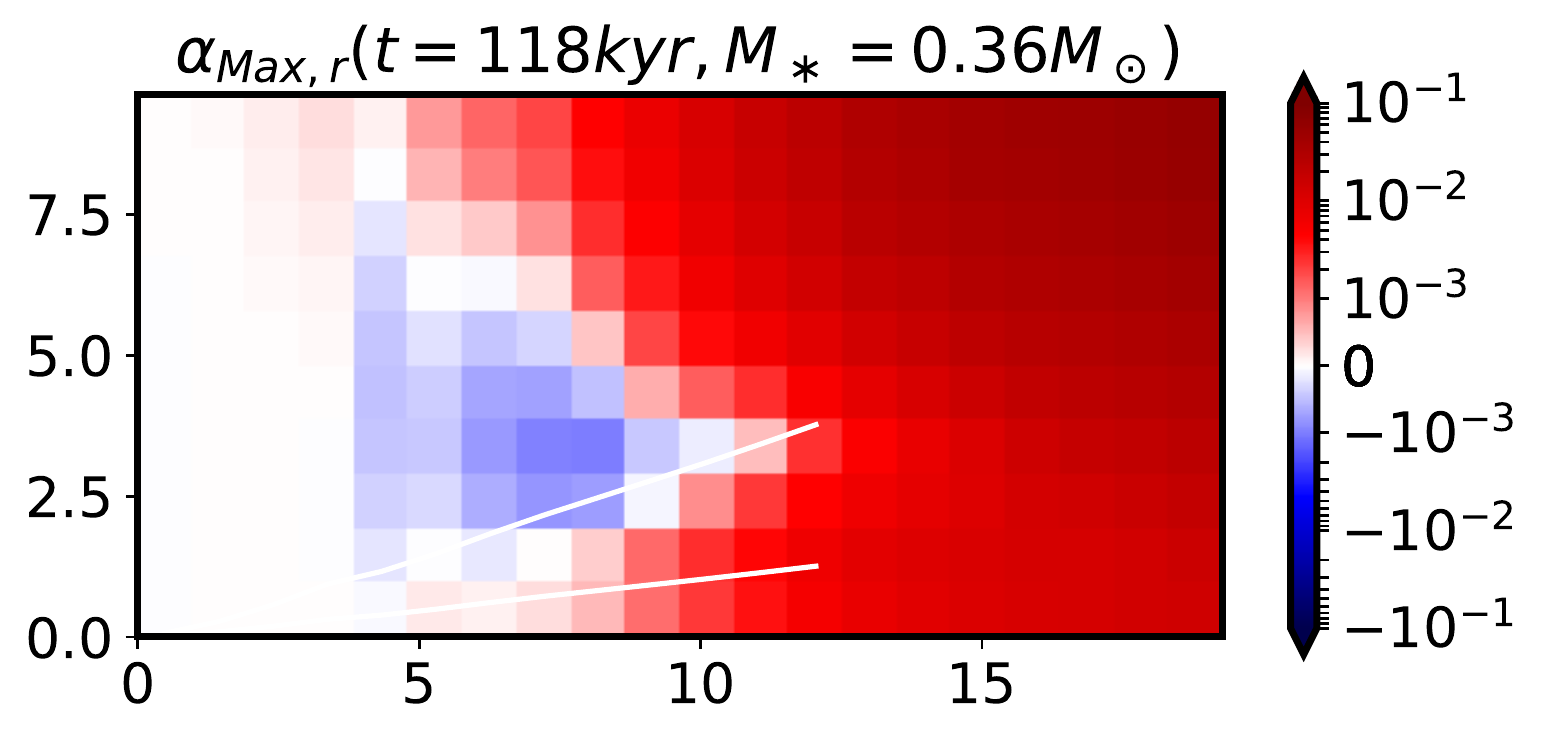}
\includegraphics[trim=0 0 0 0,clip,width=0.24\textwidth]{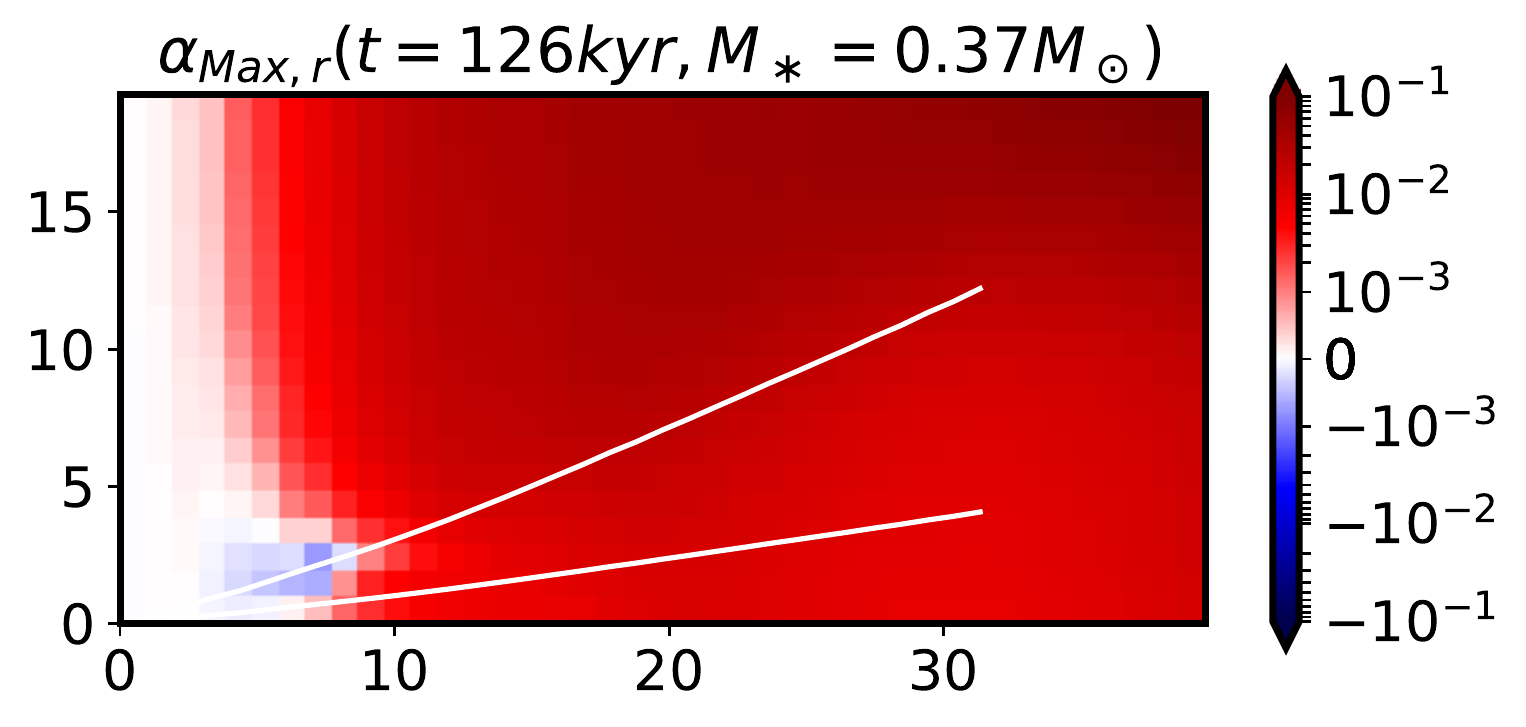}
\includegraphics[trim=0 0 0 0,clip,width=0.24\textwidth]{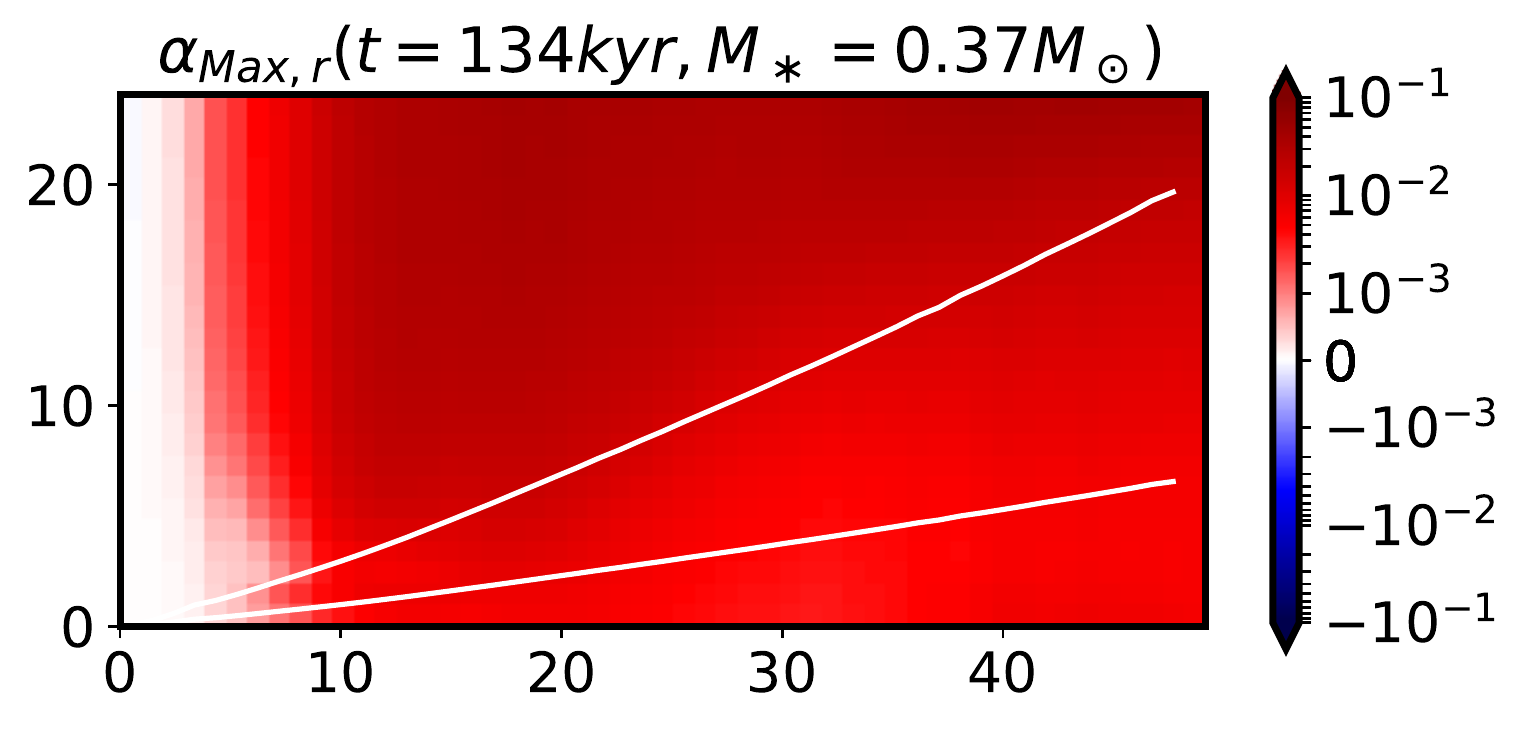}\\

\includegraphics[trim=0 0 0 0,clip,width=0.24\textwidth]{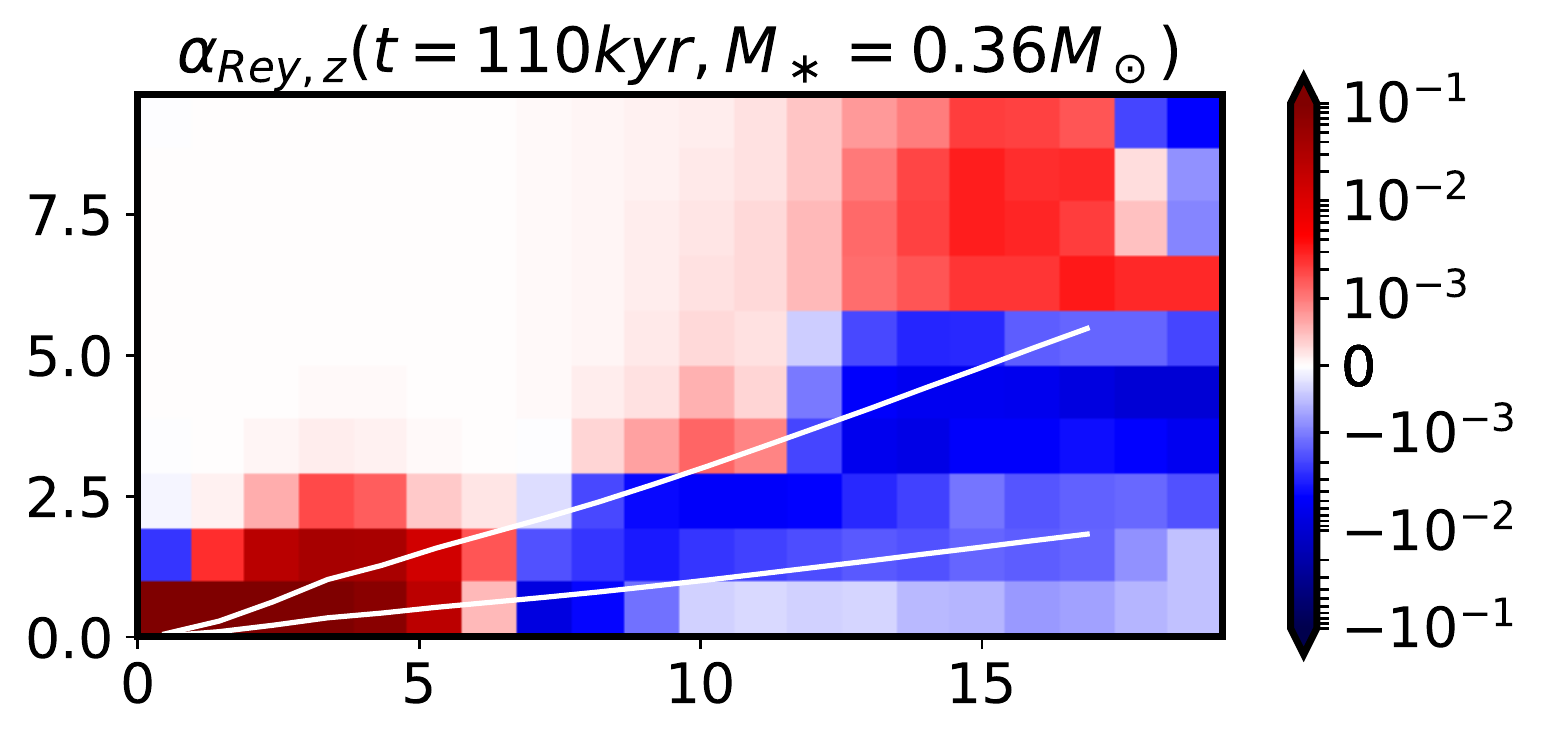}
\includegraphics[trim=0 0 0 0,clip,width=0.24\textwidth]{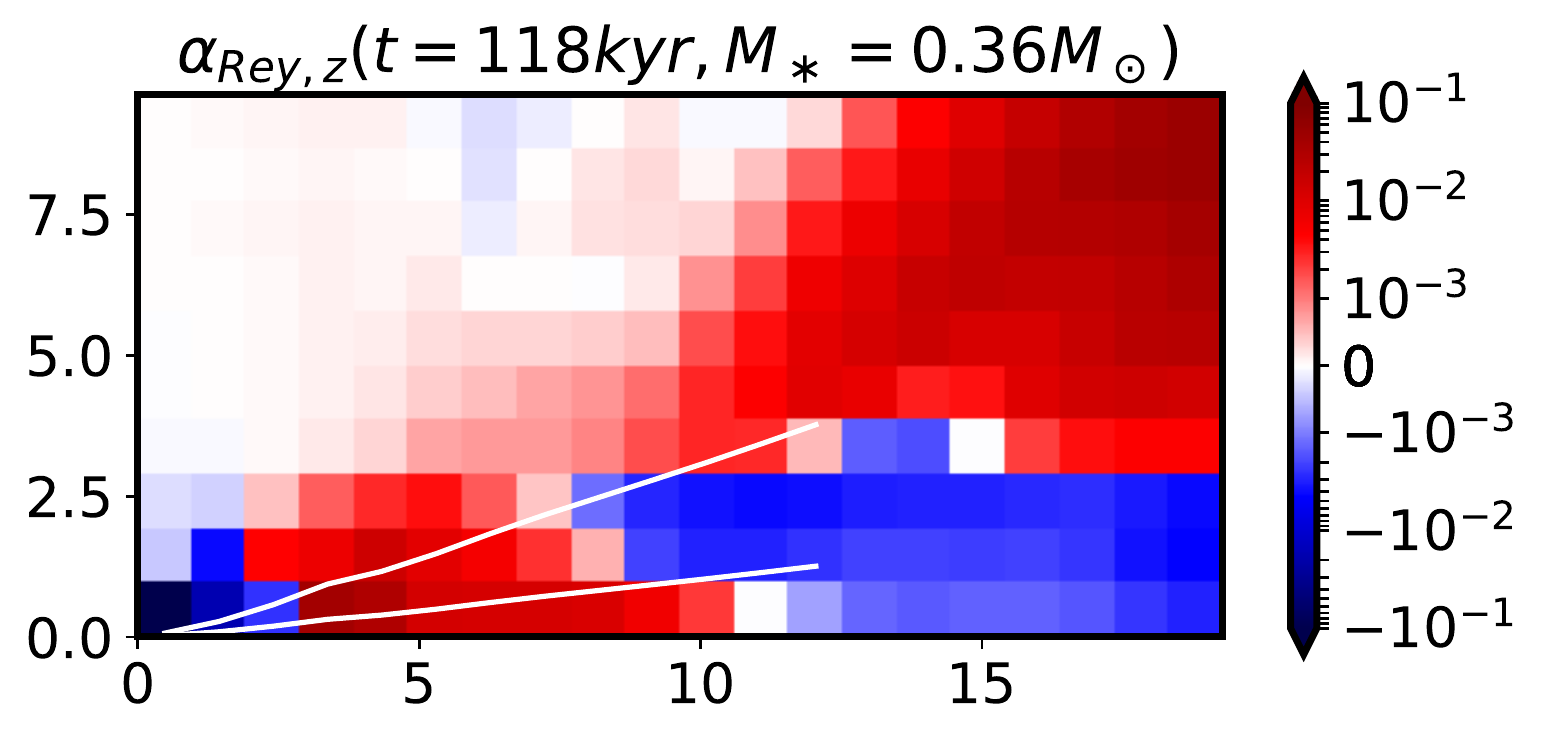}
\includegraphics[trim=0 0 0 0,clip,width=0.24\textwidth]{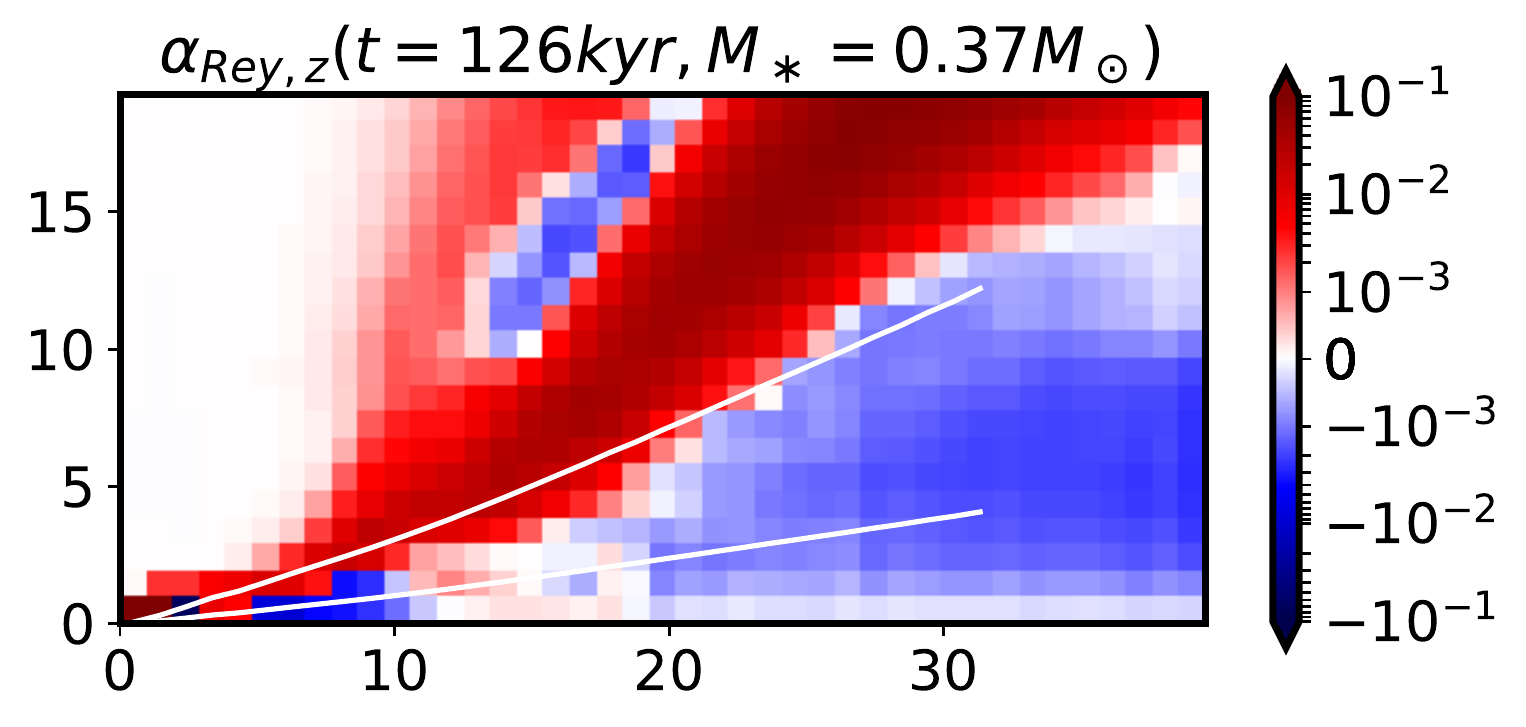}
\includegraphics[trim=0 0 0 0,clip,width=0.24\textwidth]{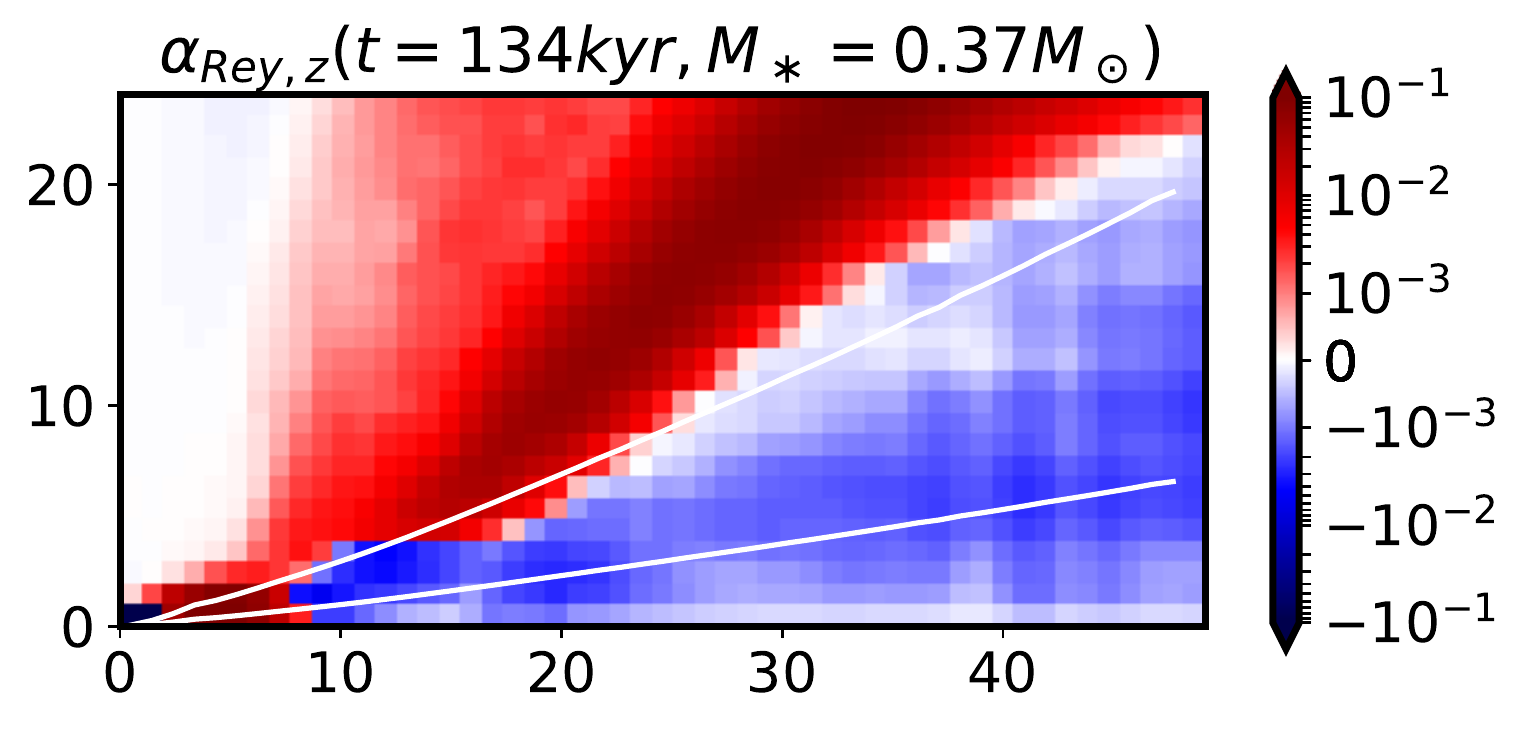}\\

\includegraphics[trim=0 0 0 0,clip,width=0.24\textwidth]{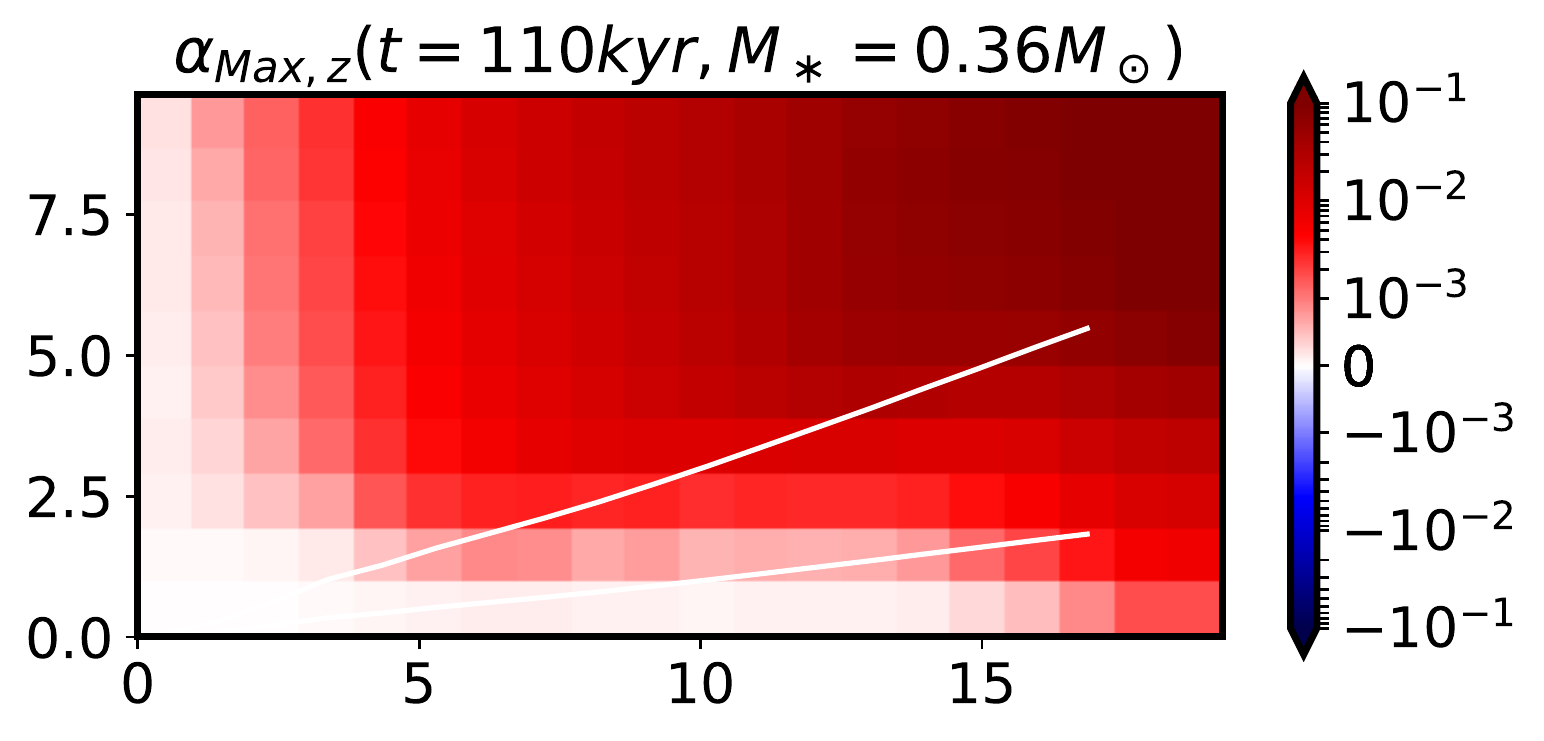}
\includegraphics[trim=0 0 0 0,clip,width=0.24\textwidth]{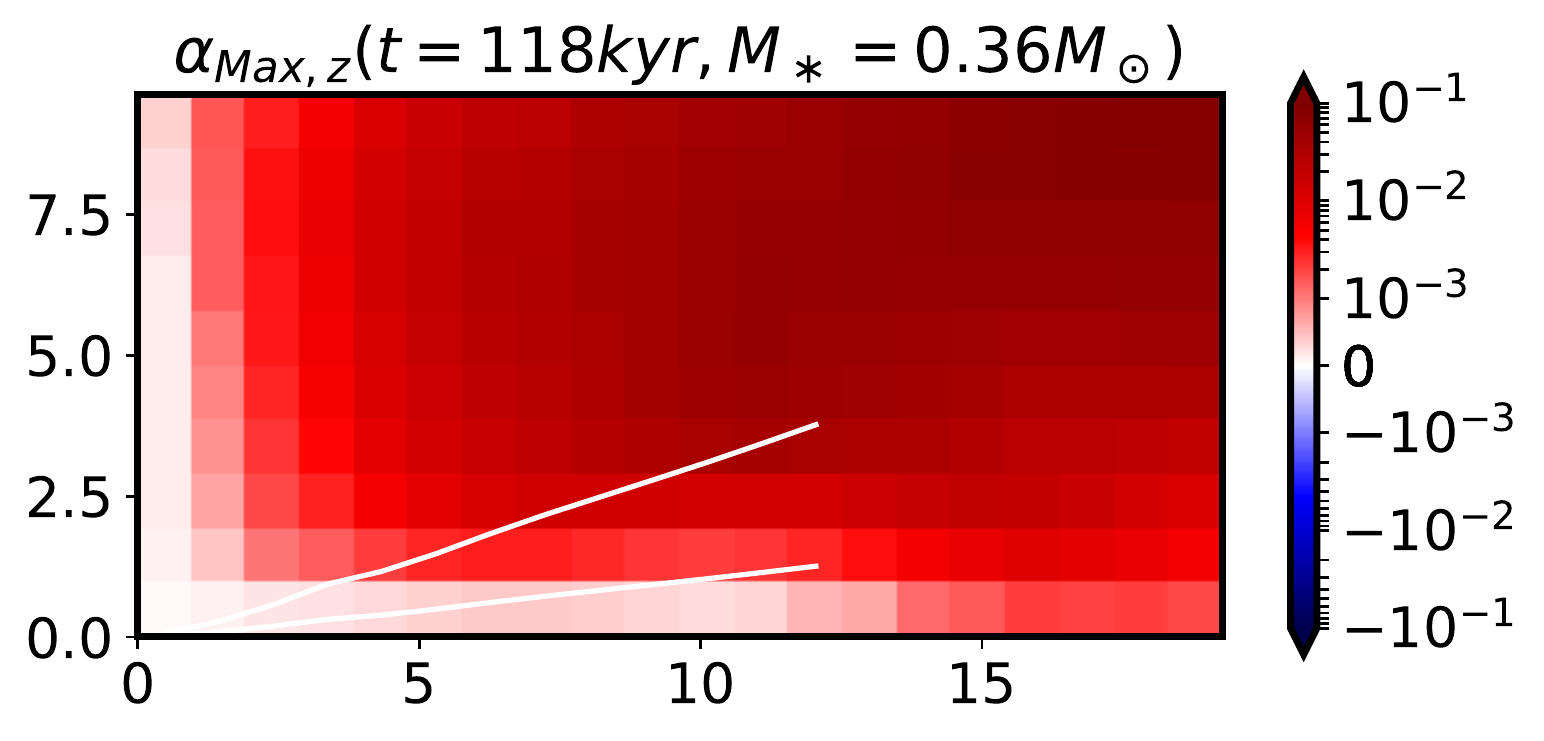}
\includegraphics[trim=0 0 0 0,clip,width=0.24\textwidth]{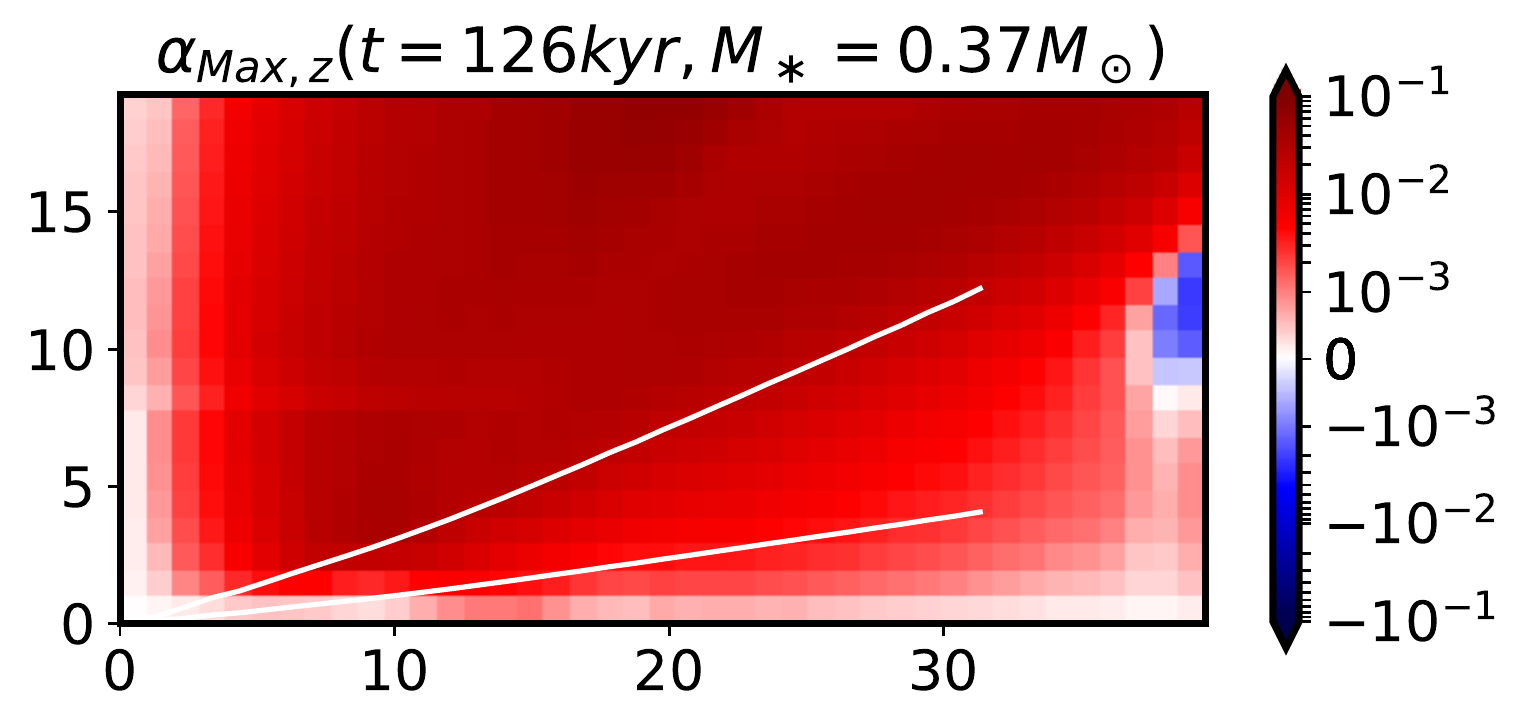}
\includegraphics[trim=0 0 0 0,clip,width=0.24\textwidth]{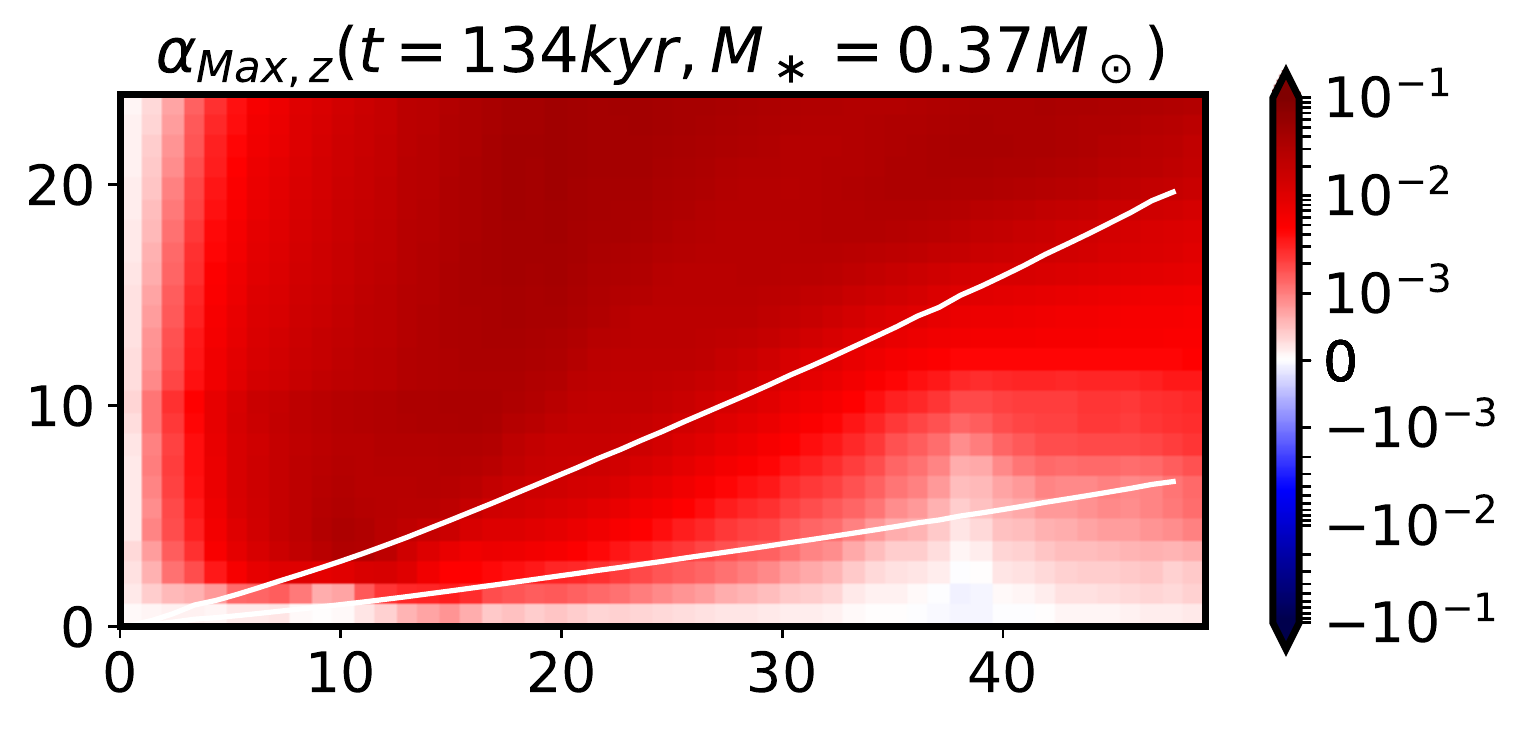}
\caption{$\alpha$ evolution of R\_$\ell$14 ({\it increasing time from left to right}). The four components $\alpha_{{\rm Rey},r}, \alpha_{{\rm Max},r}, \alpha_{{\rm Rey},z}, \alpha_{{\rm Max},z}$  ({\it from top to bottom}). The $r$ components are vertically averaged values between zero and a given altitude, and the $z$ components are local values. The white curves trace the thermal scale height $H$ and $3H$. Positive values ({\it red}) can be roughly interpreted as a radial accretion, and negative values ({\it blue}) as radial expansion. The resolution is 1 AU. As the disk grows in size, the interior of the disk is described with more cells and there is a clearer pattern of $\alpha$. The early disk has relatively strong $\alpha_{\rm Rey}$ compared to its magnetic counter part. It is probably under constant perturbation by the strong infall from the envelope. The first two chosen snapshots are not necessarily representative of this epoch because the behavior of $\alpha_{\rm Rey}$ is very irregular in time, leading to a disk that fluctuates in size around 10 AU. The infall from the envelope weakens later, and the disk becomes more regular, with its inner parts readily accreting. This accretion is mostly caused by vertical  magnetic transport of angular momentum within the envelope. }
\label{fig_alpha_img_evolu}
\end{figure*}

Equation (\ref{eq_Mdot_alpha}) implies that $\alpha_z$ has a stronger effect than $\alpha_r$ by roughly a factor 5 in the case of aspect ratio 0.1 if $\alpha_z \simeq \alpha_r$. 
With the slightly higher value of $\alpha_z$ compared to $\alpha_r$, 
we can infer that most of the radial mass accretion is due to the angular momentum transport along the field lines on the surface of the disk, 
and this accretion occurs mostly in the upper layers (between $H$ and $3H$, or even above) of the disk. 

The mass accretion rate derived from $\alpha$ is shown in Fig. \ref{fig_Mdot_disk} with dotted curves and is compared to the directly measured values later. 
Strong fluctuations are present, while the overall magnitude is comparable with the directly measured values of radial accretion (see Sect. \ref{st_Mdot}).

\begin{figure}[]
\centering
\includegraphics[trim=0 0 0 0,clip,width=0.24\textwidth]{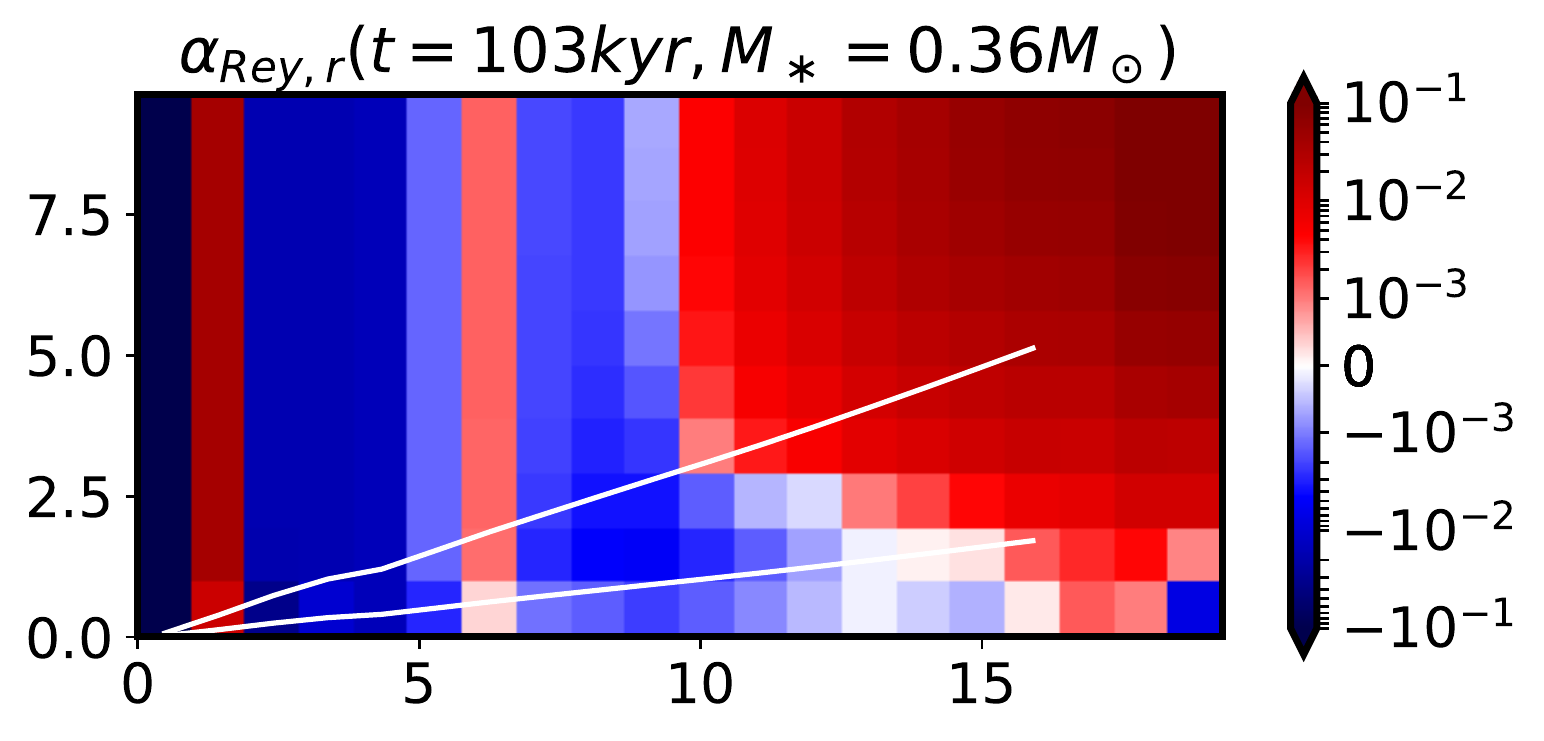}
\includegraphics[trim=0 0 0 0,clip,width=0.24\textwidth]{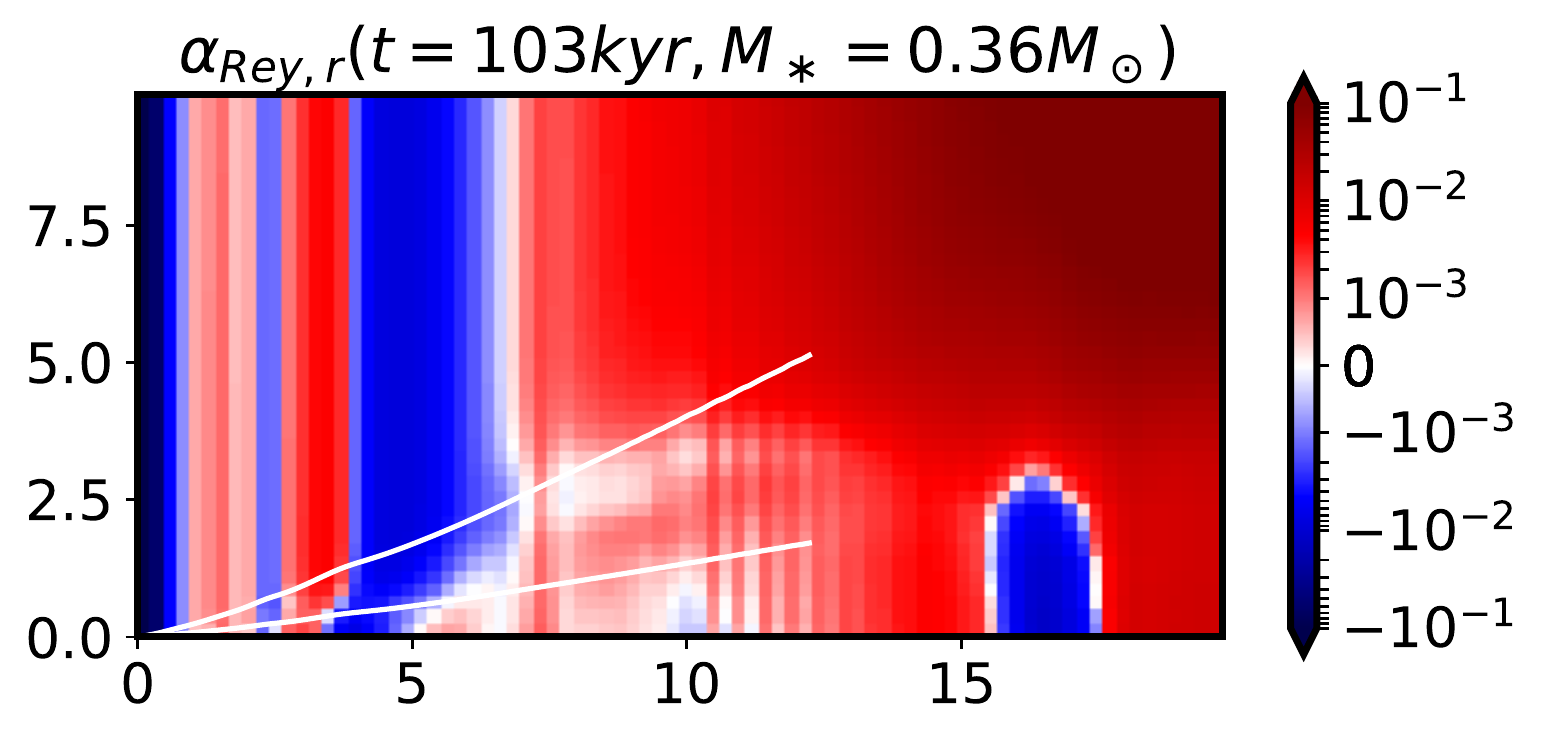} \\

\includegraphics[trim=0 0 0 0,clip,width=0.24\textwidth]{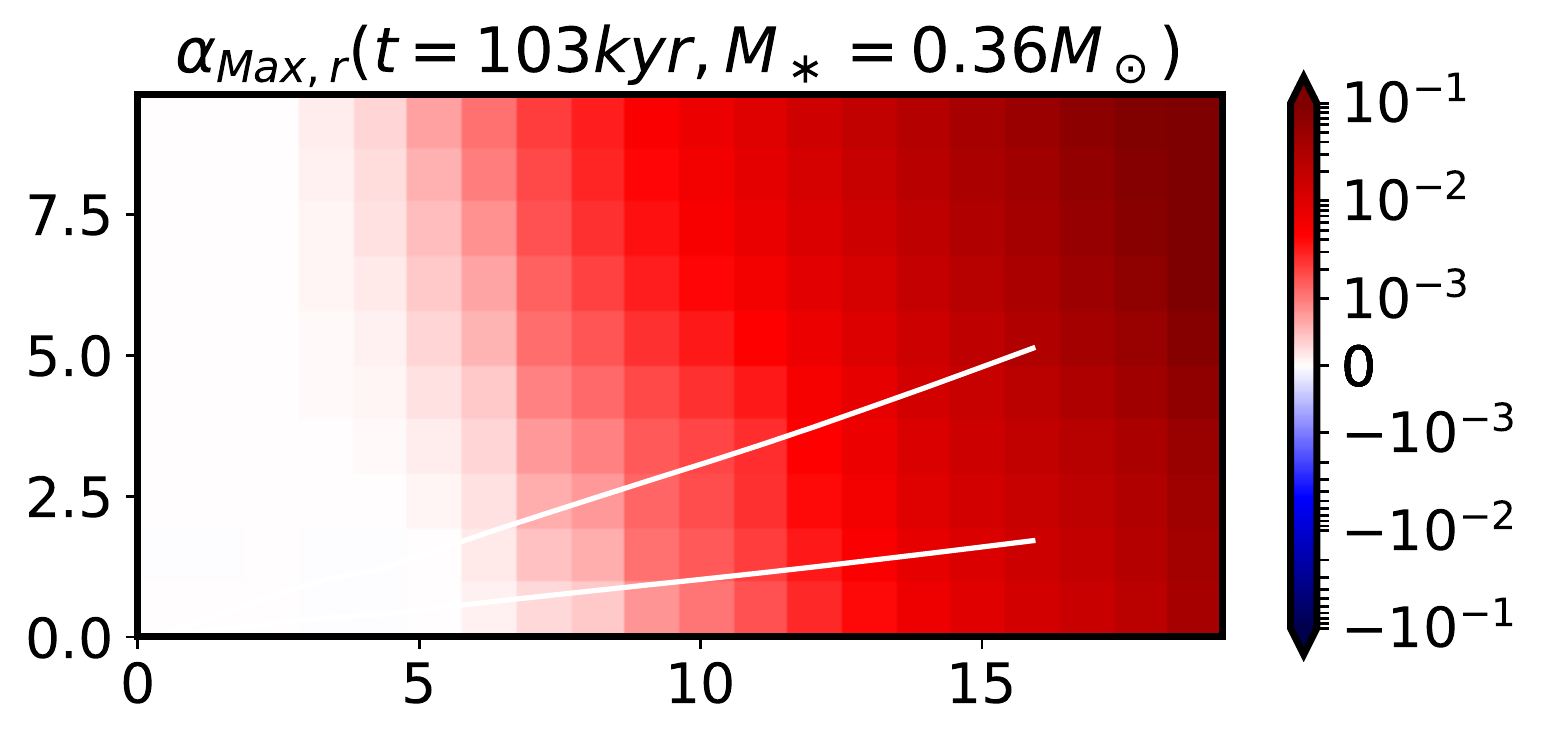}
\includegraphics[trim=0 0 0 0,clip,width=0.24\textwidth]{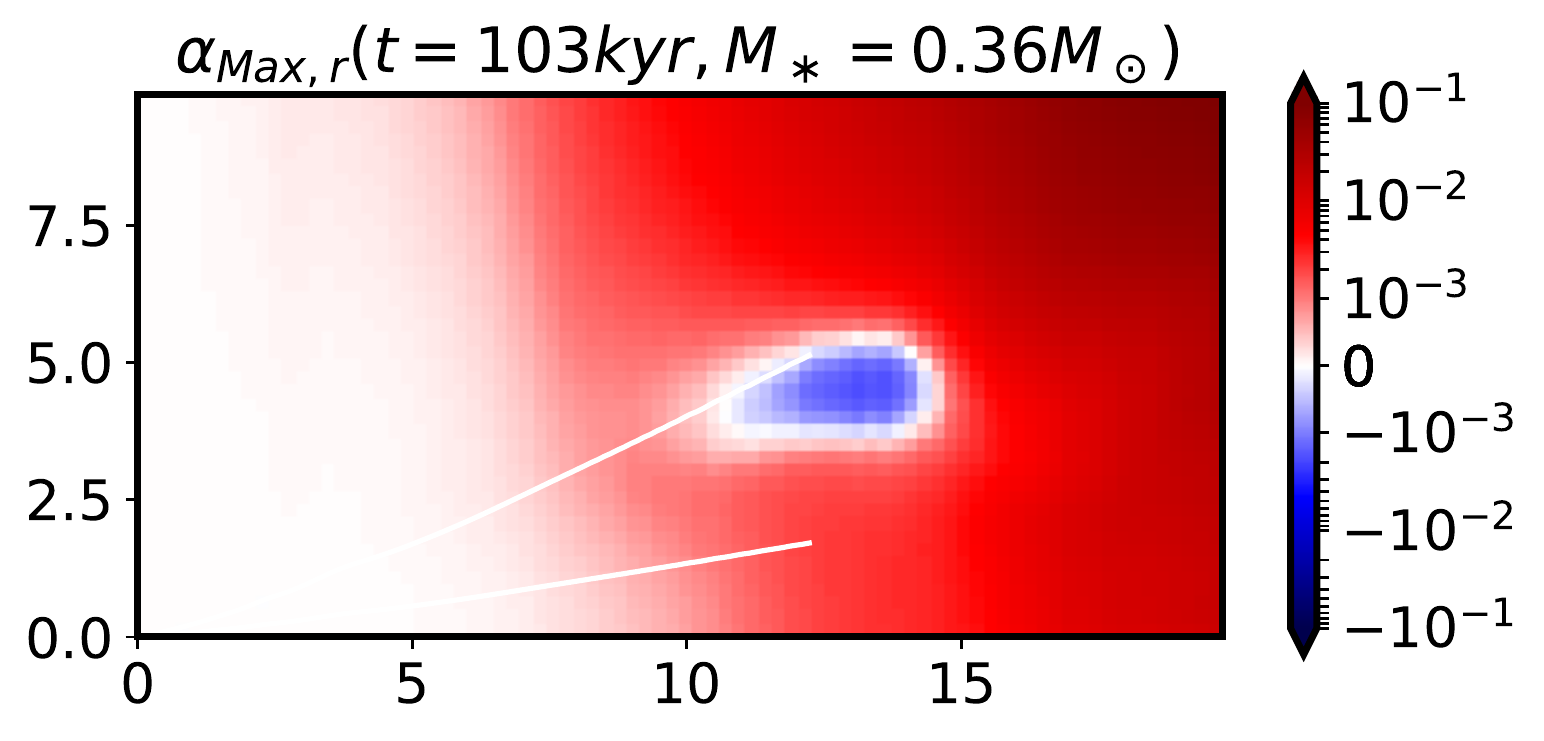}\\

\includegraphics[trim=0 0 0 0,clip,width=0.24\textwidth]{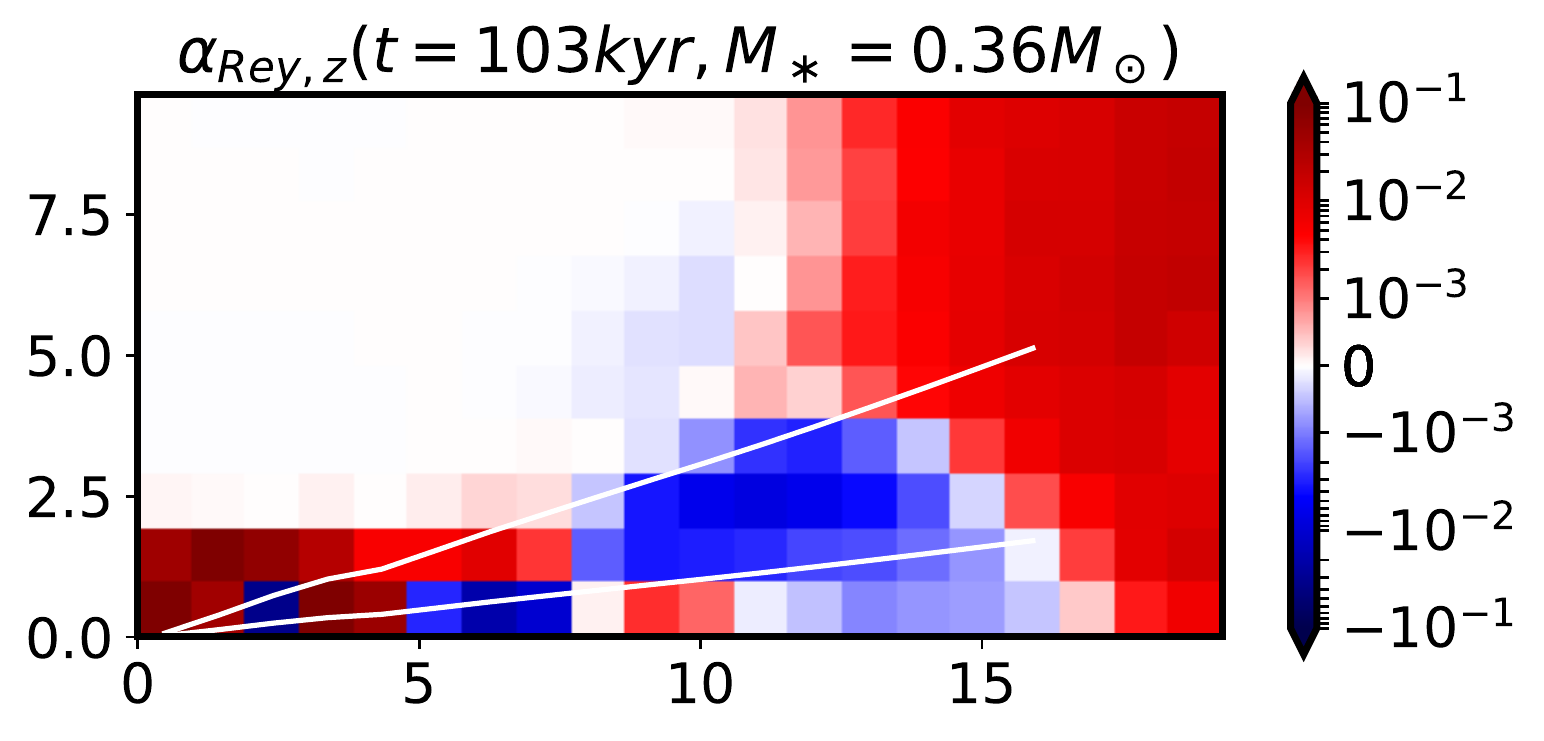}
\includegraphics[trim=0 0 0 0,clip,width=0.24\textwidth]{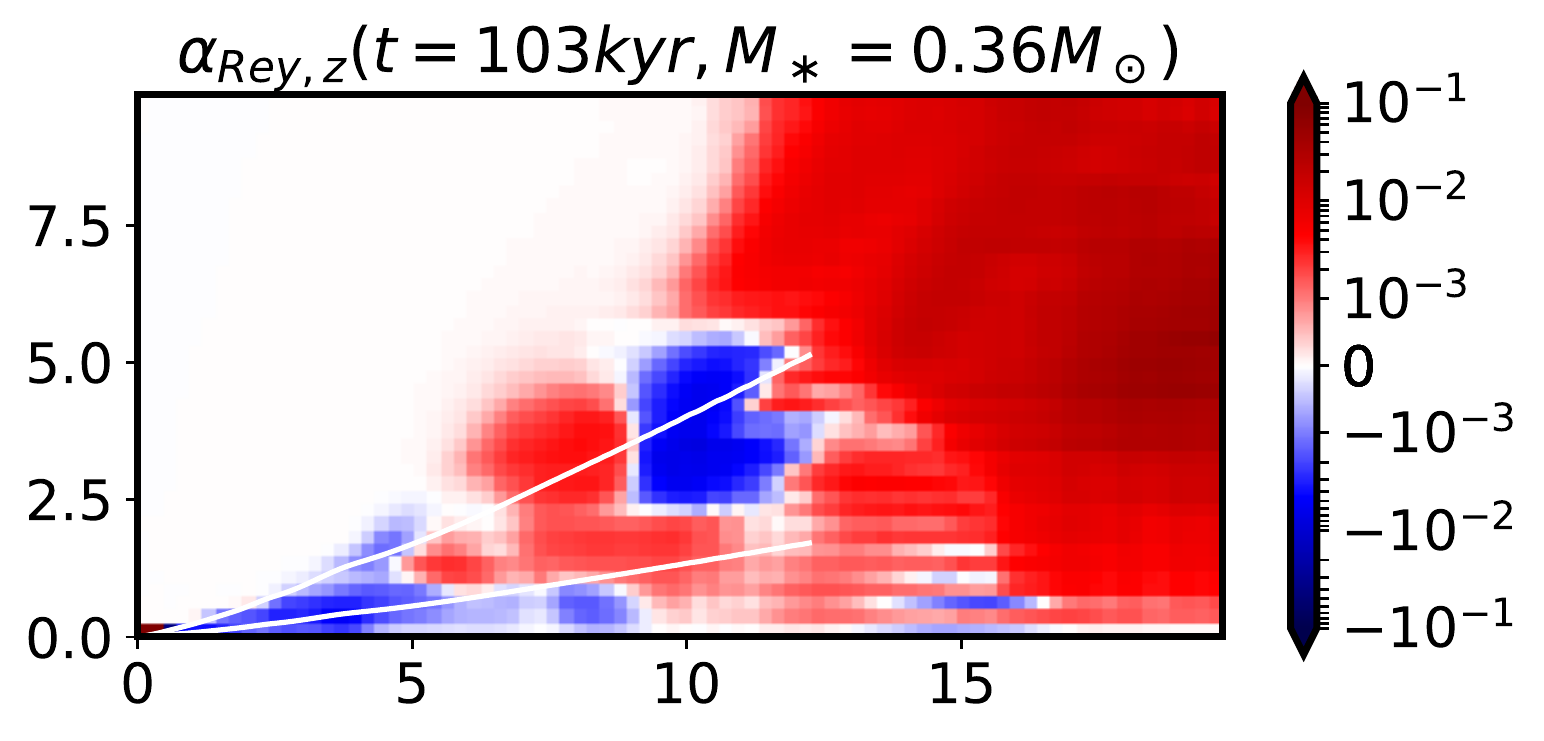}\\

\includegraphics[trim=0 0 0 0,clip,width=0.24\textwidth]{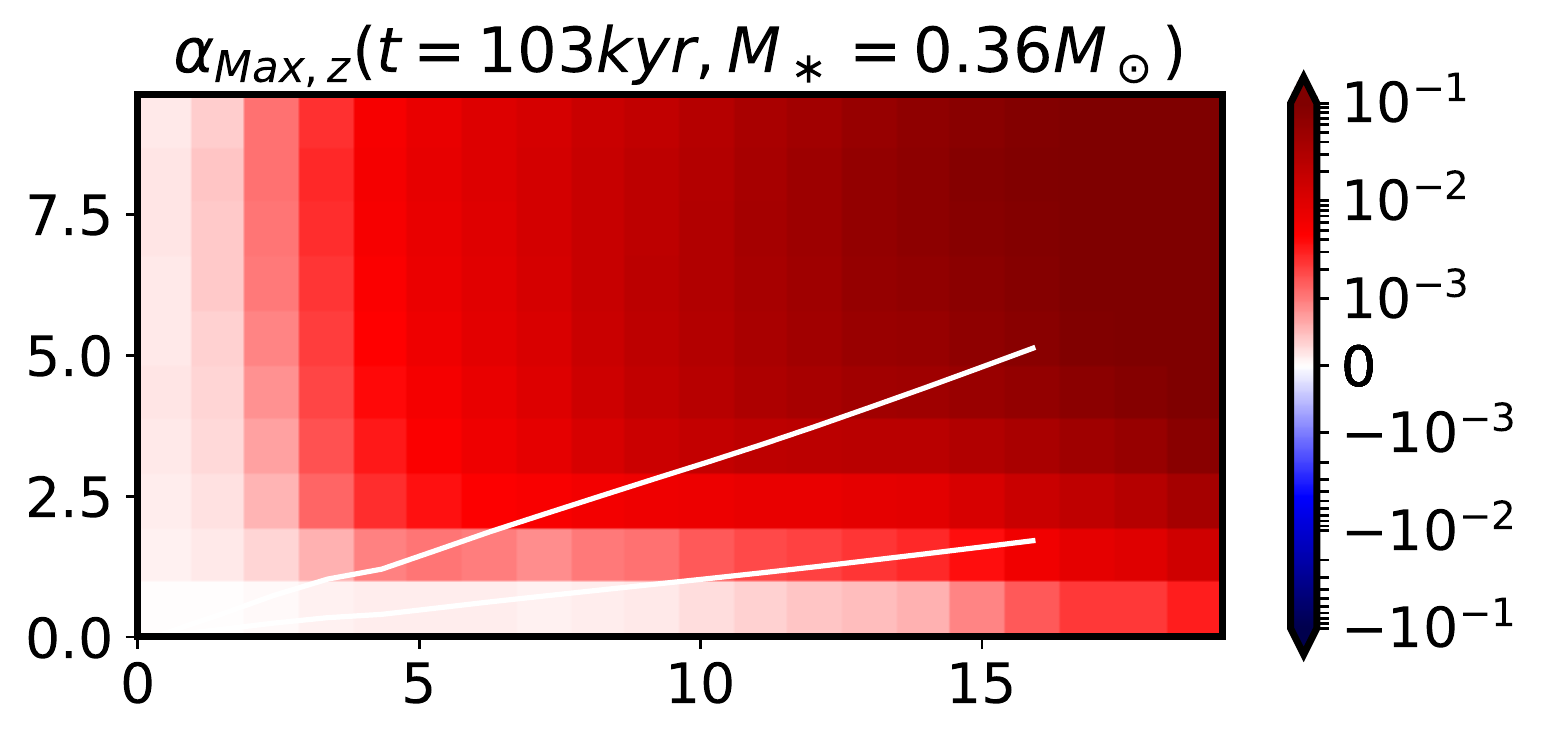}
\includegraphics[trim=0 0 0 0,clip,width=0.24\textwidth]{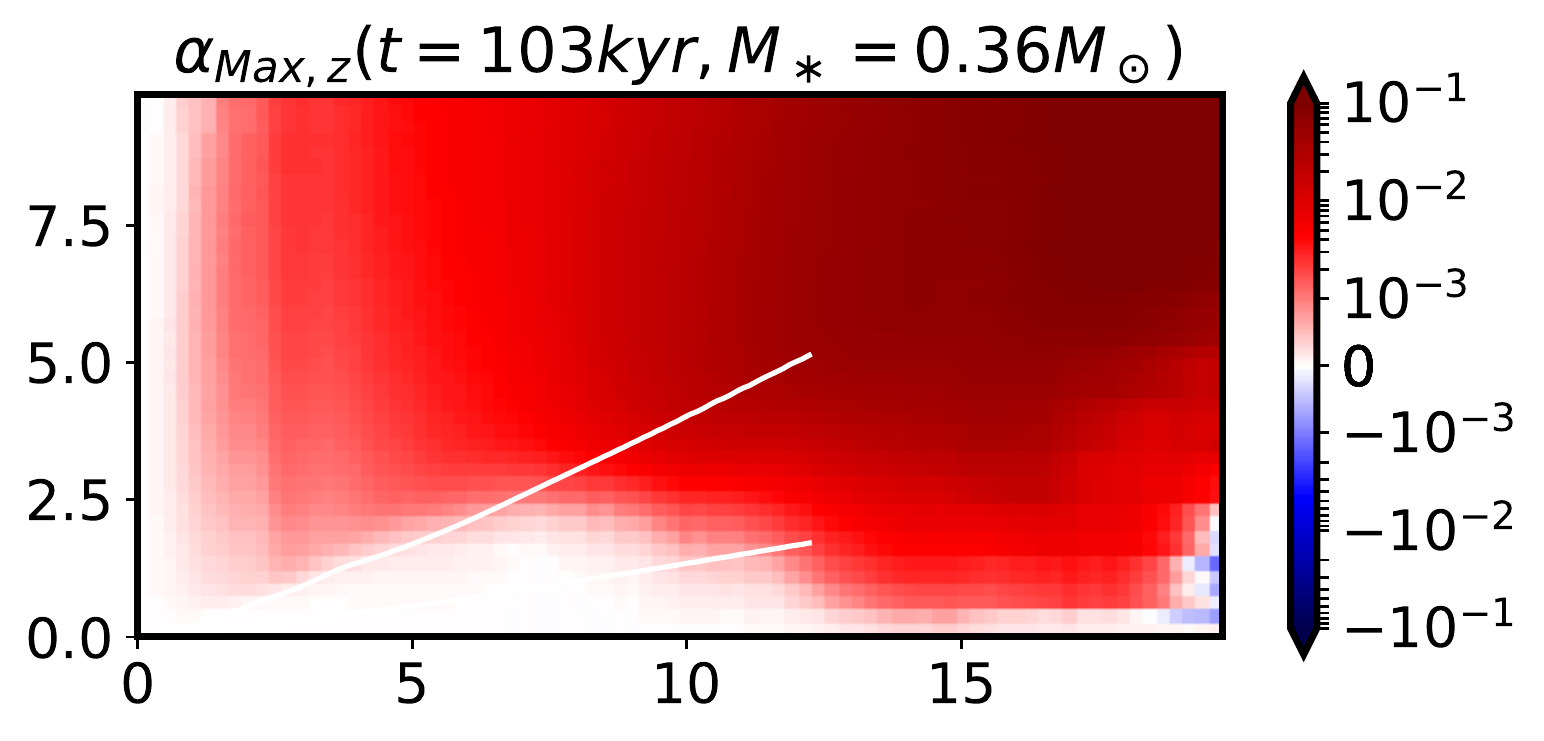}
\caption{Measured $\alpha$ values of R\_$\ell$14 ({\it left}) and R\_40ky\_$\ell$18 ({\it right}) at 103 kyr. The disk radius is about 15 AU. The bins for averaging are chosen to be larger than $dx$ in the high-resolution run in order to remove numerical noise. These figures clearly show that the technique of restart using higher resolutions is needed to study the interior properties of the disk. The $z$ components are more important because they dominate in governing the mass accretion inside the disk. The middle layers of the disk are more irregular with some partially expanding parts, while most of the accretion occurs through the upper layer slightly above $3H$ and reaches a small radius before penetrating the surface of the disk.}
\label{fig_alpha_img_rest2}
\end{figure}

\begin{figure}[]
\centering
\includegraphics[trim=0 0 0 0,clip,width=0.24\textwidth]{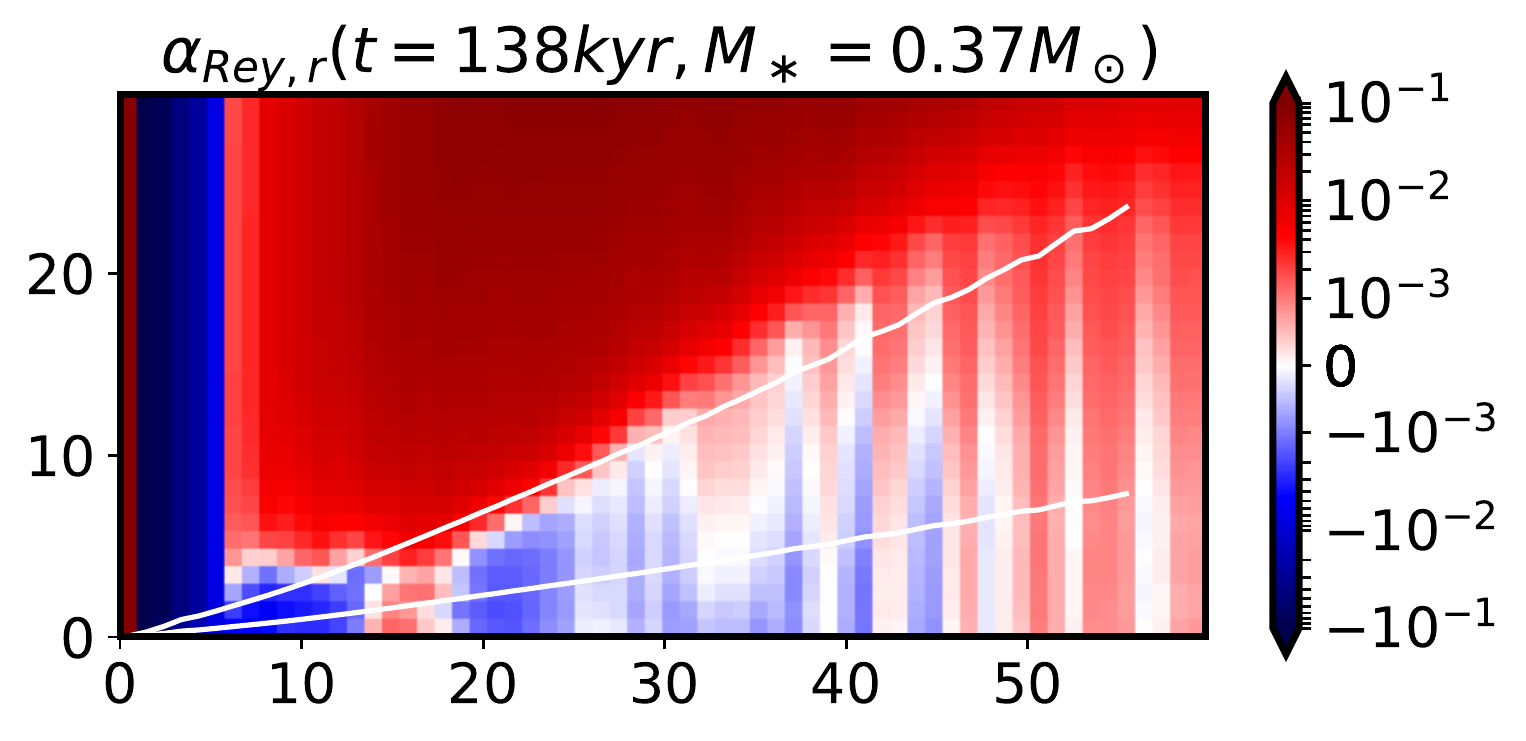}
\includegraphics[trim=0 0 0 0,clip,width=0.24\textwidth]{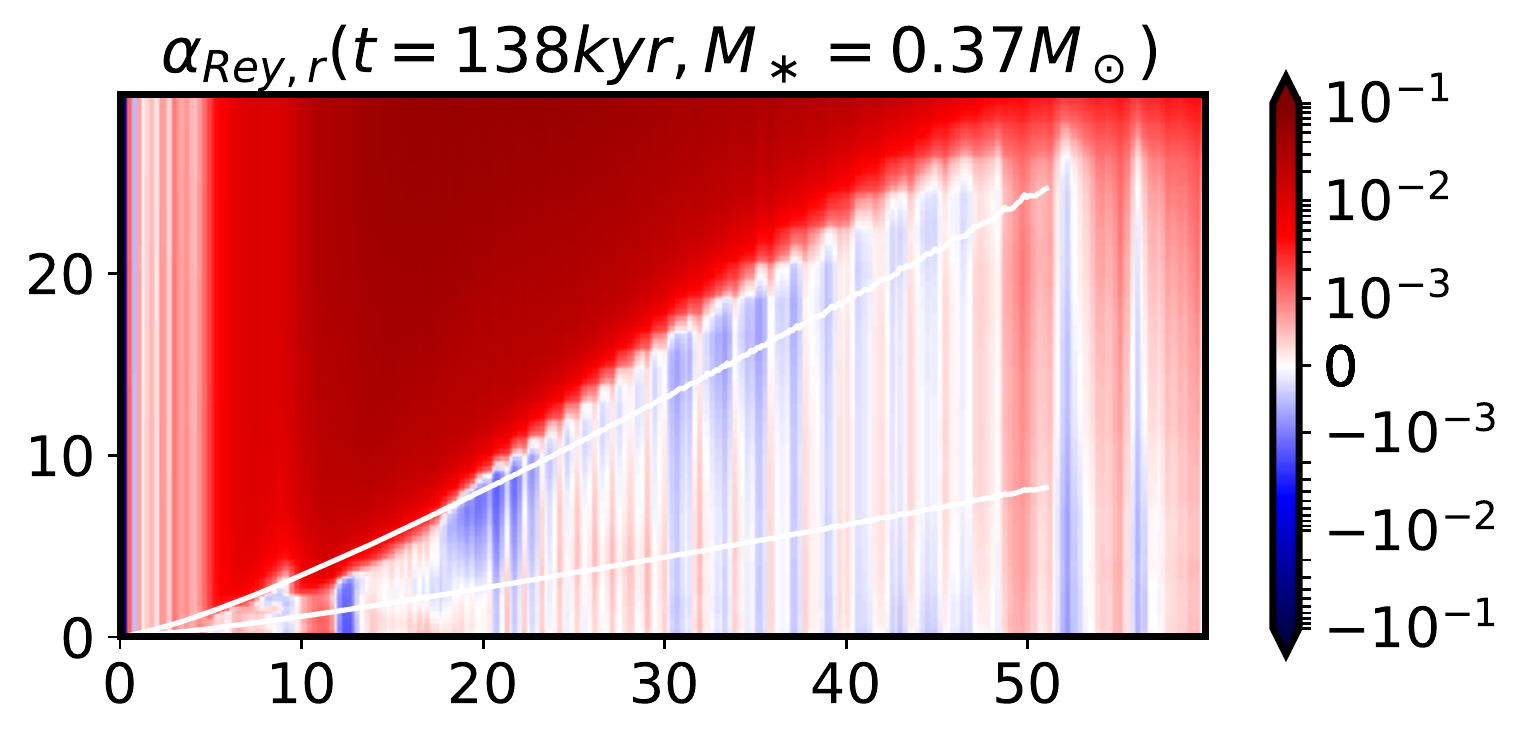}\\

\includegraphics[trim=0 0 0 0,clip,width=0.24\textwidth]{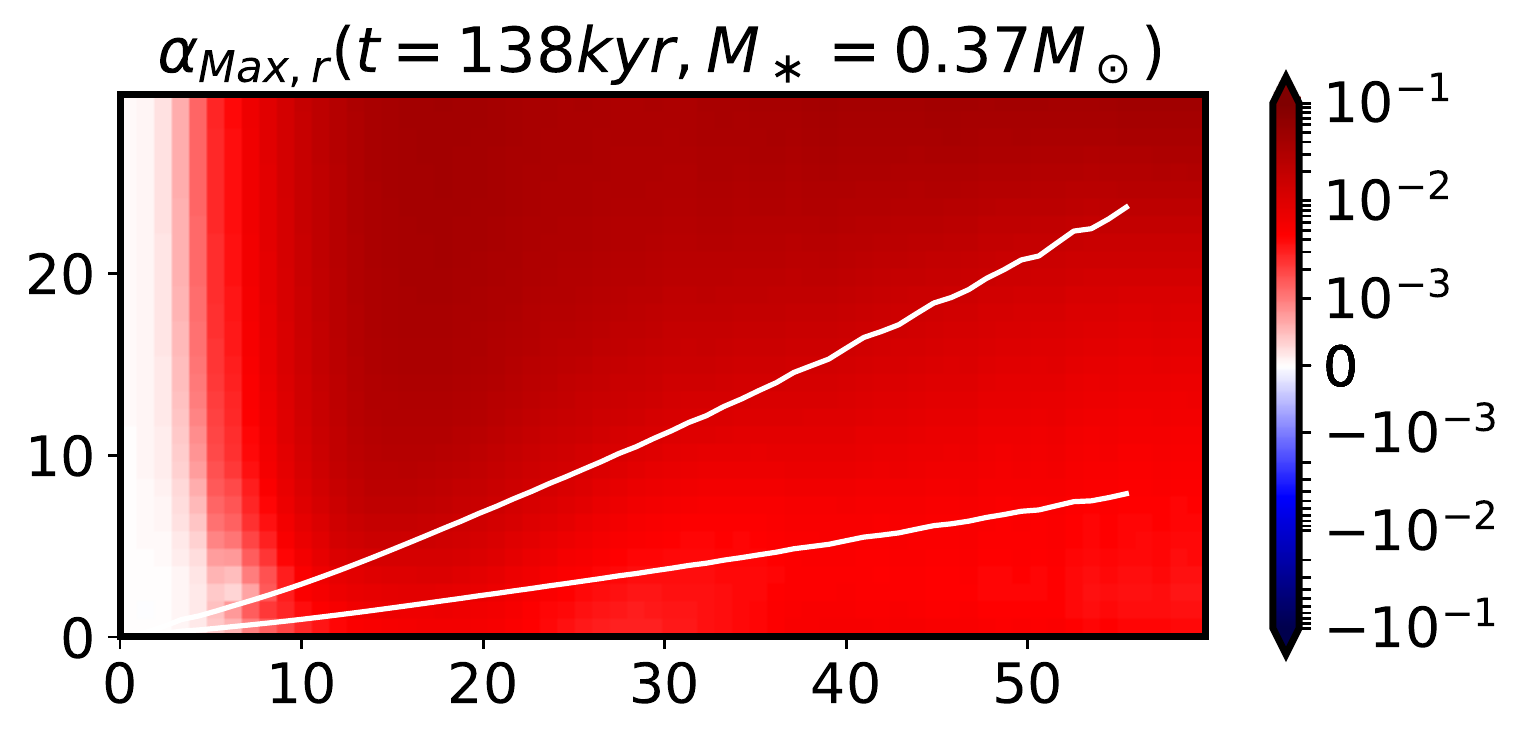}
\includegraphics[trim=0 0 0 0,clip,width=0.24\textwidth]{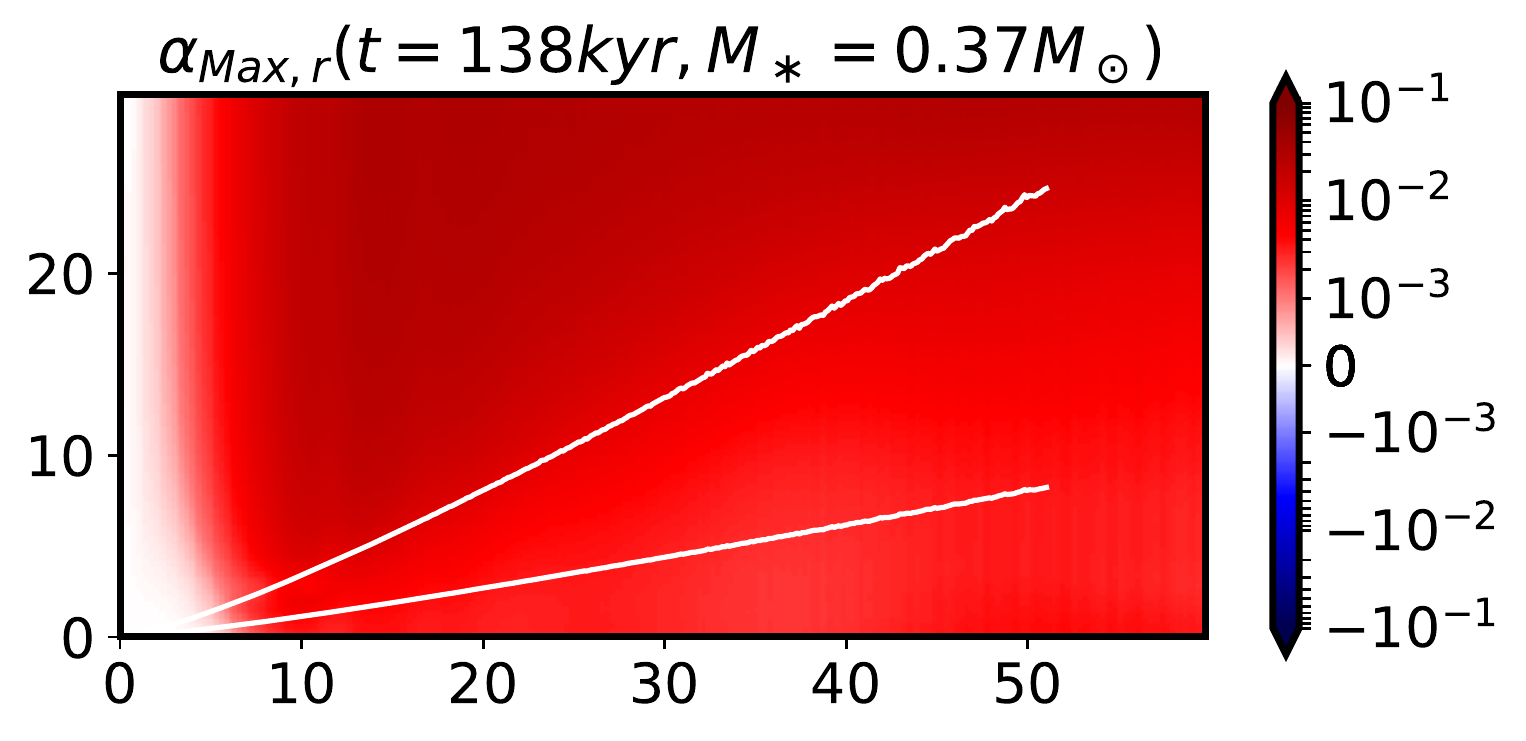}\\

\includegraphics[trim=0 0 0 0,clip,width=0.24\textwidth]{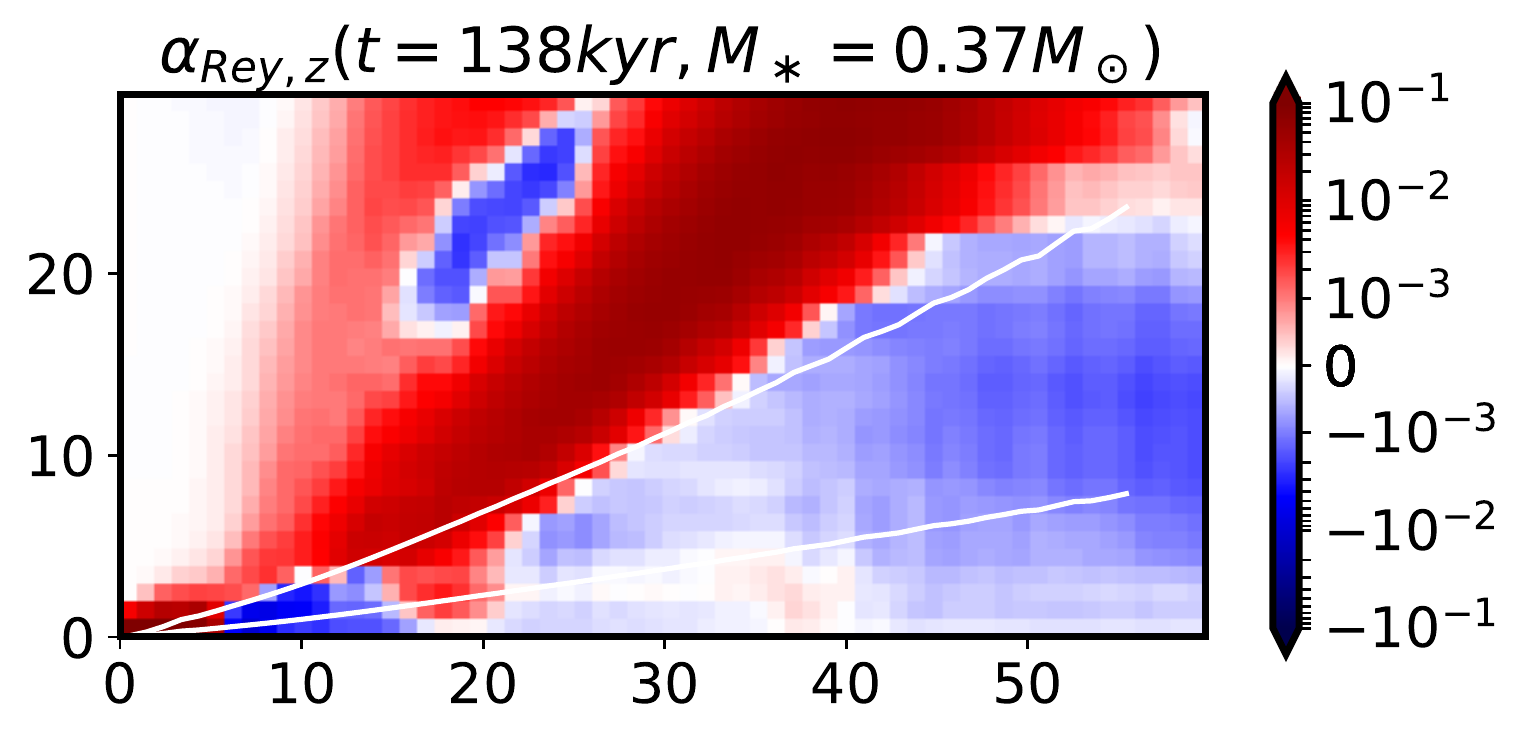}
\includegraphics[trim=0 0 0 0,clip,width=0.24\textwidth]{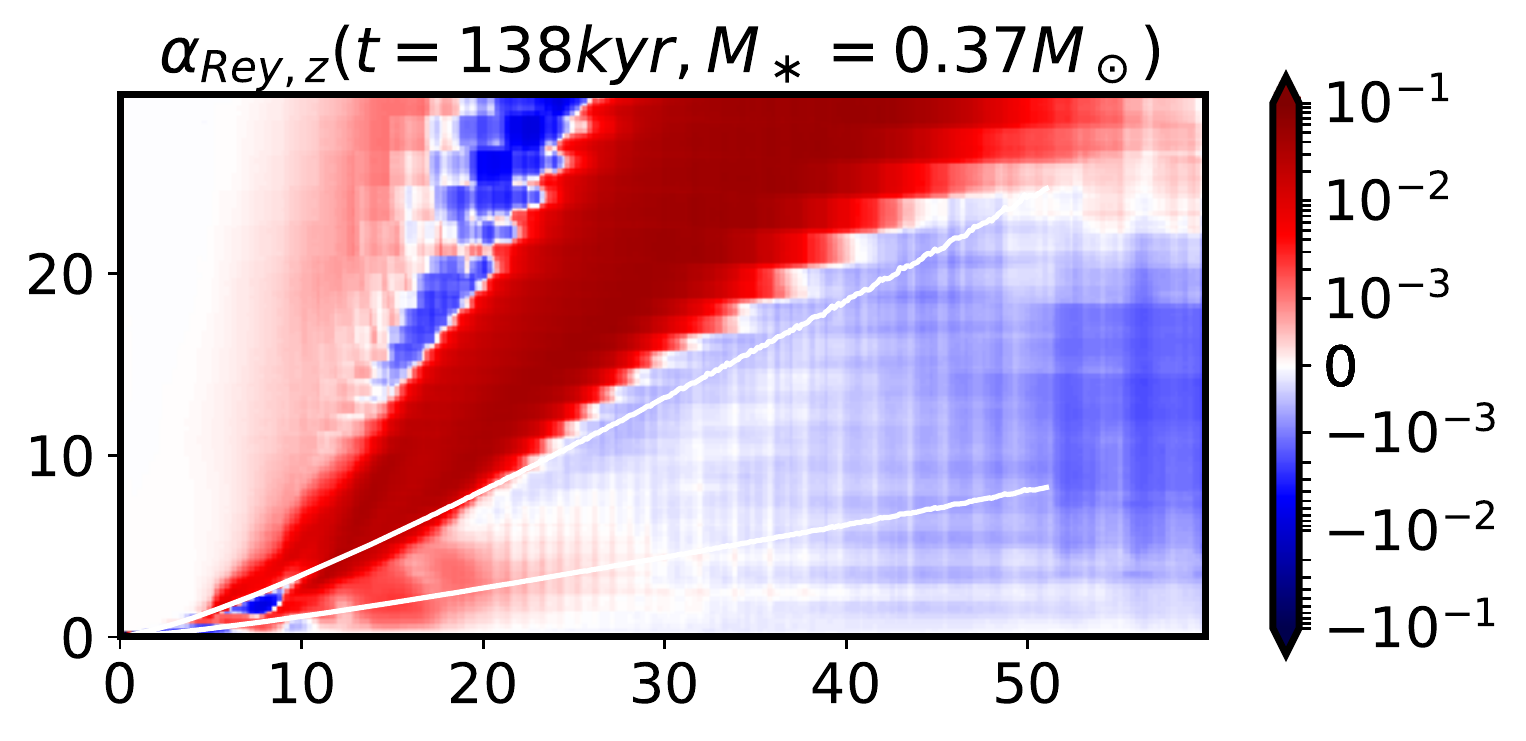}\\

\includegraphics[trim=0 0 0 0,clip,width=0.24\textwidth]{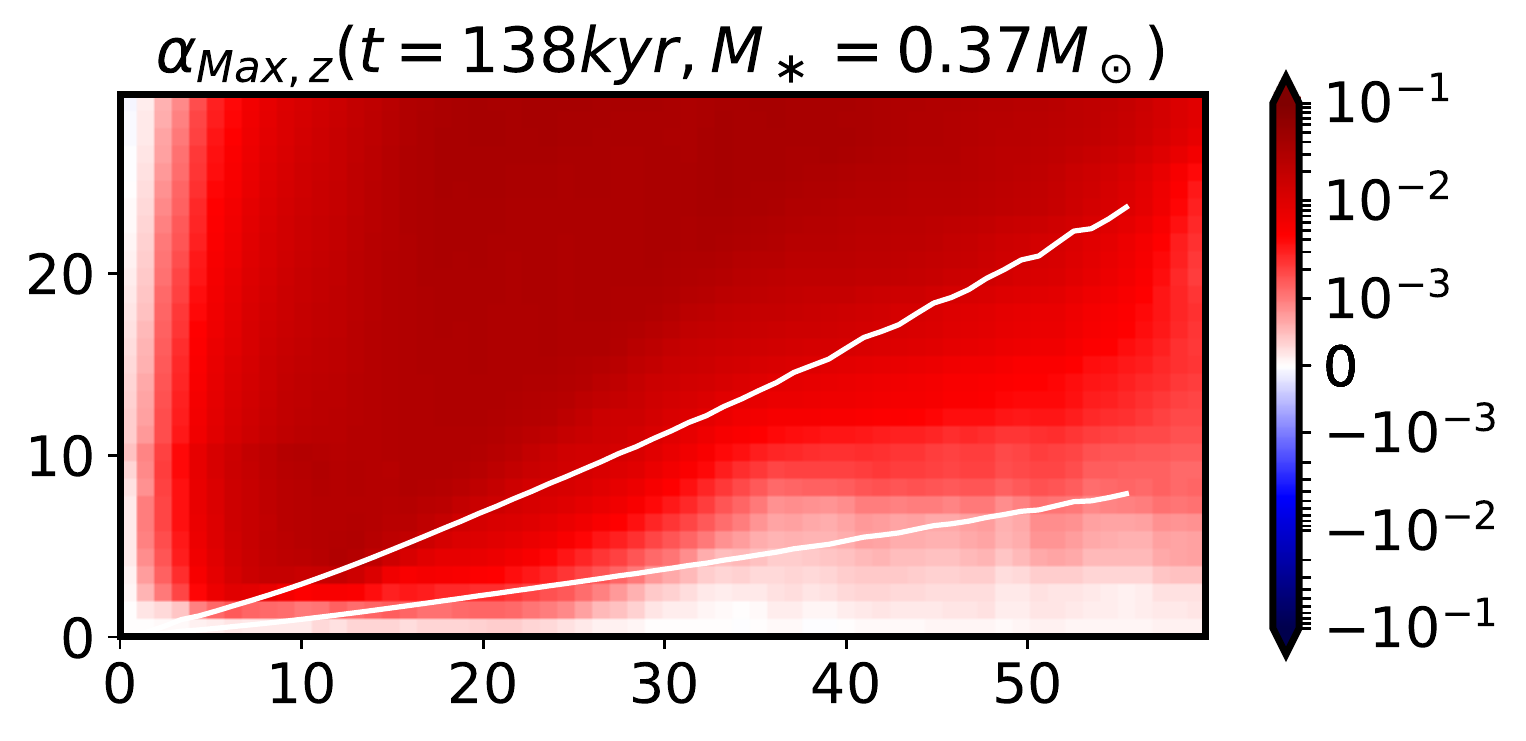}
\includegraphics[trim=0 0 0 0,clip,width=0.24\textwidth]{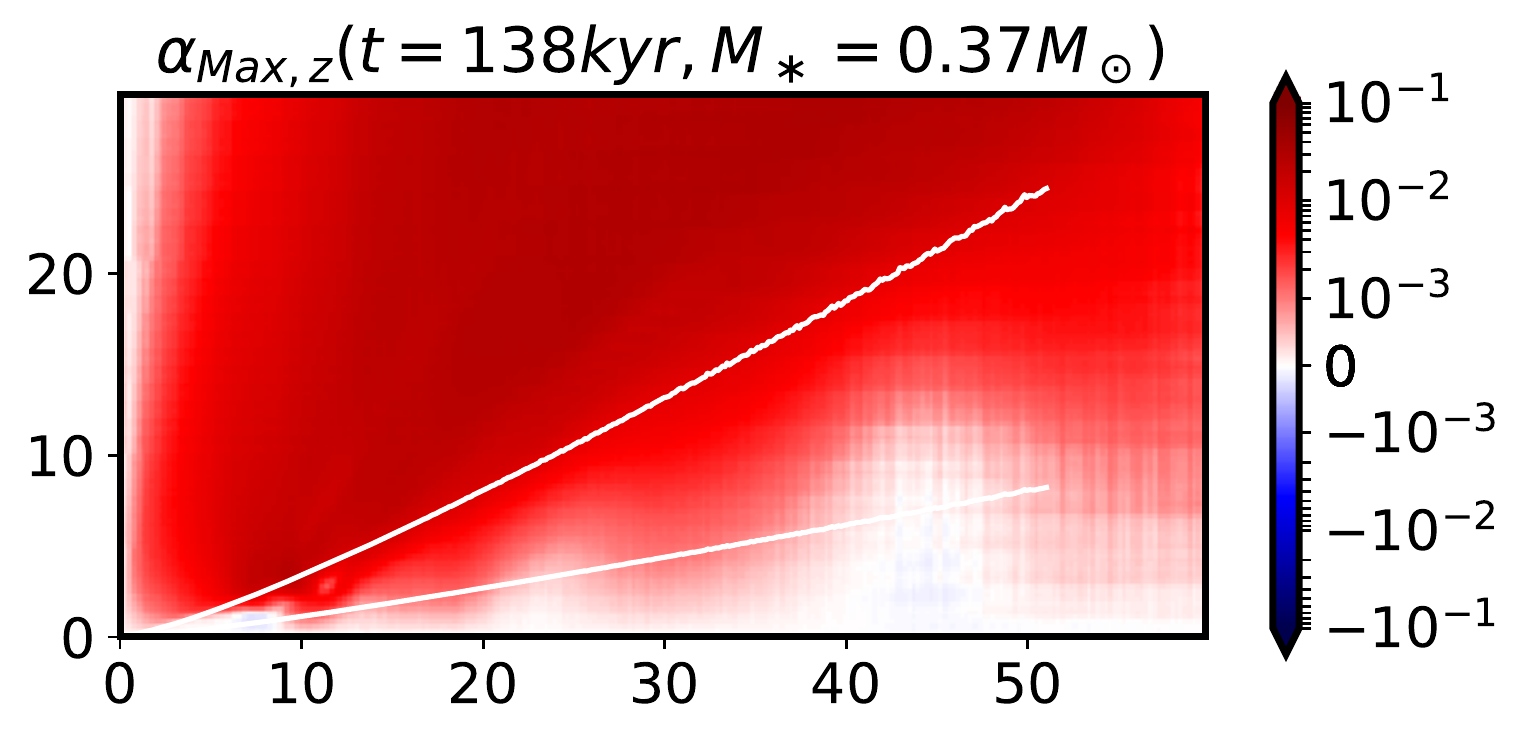}

\caption{Measured $\alpha$ values of R\_$\ell$14 {\it (left)} and R\_80ky\_$\ell$18 {\it (right)} at 138 kyr. Disk radius ($R_{\rm kep}$) grows to 50 AU at this time. The values are evaluated on a grid of 1 AU resolution for both runs. The global patterns are compatible for the two runs with 16 times difference in resolution. The middle layers of the disk are radially structured and partially expending at times, while the accretion occurs through a channel slightly above $3H$ and brings a significant amount of mass from the infalling envelope to a small radius without passing through the disk.  }
\label{fig_alpha_img_rest3}
\end{figure}

The above measurements of $\alpha$ along given surfaces reveal important information: The transport onto and within the disk is highly structured. 
To give a better idea of the transport pattern and the behavior of different stress tensors, 
we show $\alpha$ values on the $r-z$ plane for several snapshots of the canonical run in Fig. \ref{fig_alpha_img_evolu}.
We also examine the effect of high resolution restarts in Figs. \ref{fig_alpha_img_rest2} and \ref{fig_alpha_img_rest3}. 
Positive values are shown in red, corresponding to radial accretion (or angular momentum removal, see Eq. (\ref{eq_Mdot_alpha})), 
and negative values in blue signify radial expansion. 
The total effect receives contributions from all the four terms shown in the figure. 

Figure \ref{fig_alpha_img_evolu} shows that at early times $\alpha_{\rm Rey}$ usually dominates within the disk but does not show a regular pattern (first two snapshots). 
This is probably linked to the strong accretion that constantly perturbs the disk, and we recall that the disk size fluctuates in time under the effect of irregular infall of the envelope. 
Moreover, the disk size is very small at the beginning (10 AU for 1 AU resolution), and a good determination of the interior structure of the disk is in principle not possible. 
At later times, the disk grows to a few dozen AU in size, which results in a more resolved internal structure. 

The middle layers of the disk do not have very well behaved values of $\alpha$ and some parts of the disk midplane can even expand. 
Most of the accretion is channeled through a high-altitude layer of the disk and
moves radially inward before it reaches the surface of the disk. 
This is most clearly shown in the figures of $\alpha_{{\rm Rey},z}$ and $\alpha_{{\rm Max},z}$ as a dark red region slightly above the line of  $3H$.

Figures \ref{fig_alpha_img_rest2} and \ref{fig_alpha_img_rest3} show the comparisons with the canonical run of R\_40ky\_$\ell$18 and R\_80ky\_$\ell$18. 
At 40 kyr, the restart with increased resolution allows us to study the internal structure of the disk without having to evolve the whole simulation at high resolution since the beginning. 
The difference in $\alpha_{{\rm Rey},r}$ is more remarkable but is not the dominant term. 
Accretion occurs mostly at high altitude (see $\alpha_{{\rm Max}, z}$), and the middle layers of the disk are highly structured. 
At 80 kyr, the disk grows much larger, 
and thus even the canonical run allows satisfying evaluation of $\alpha$. 
The similarity of the $\alpha$ patterns in the two runs with a difference of a factor 16 in resolution is a validation of numerical convergence. 
Moreover, the high-resolution restart also allows us to probe much smaller radii and shows that the similar disk and accretion structure continue at smaller radii.

\subsubsection{Toomre $Q$ parameter: Stability and fragmentation}\label{st_Q}

\begin{figure}[]
\centering
\setlength{\unitlength}{0.5\textwidth}
\begin{picture}(1,1.84)
\put(0,1.38){\includegraphics[trim=0 0 0 3,clip,width=0.5\textwidth]{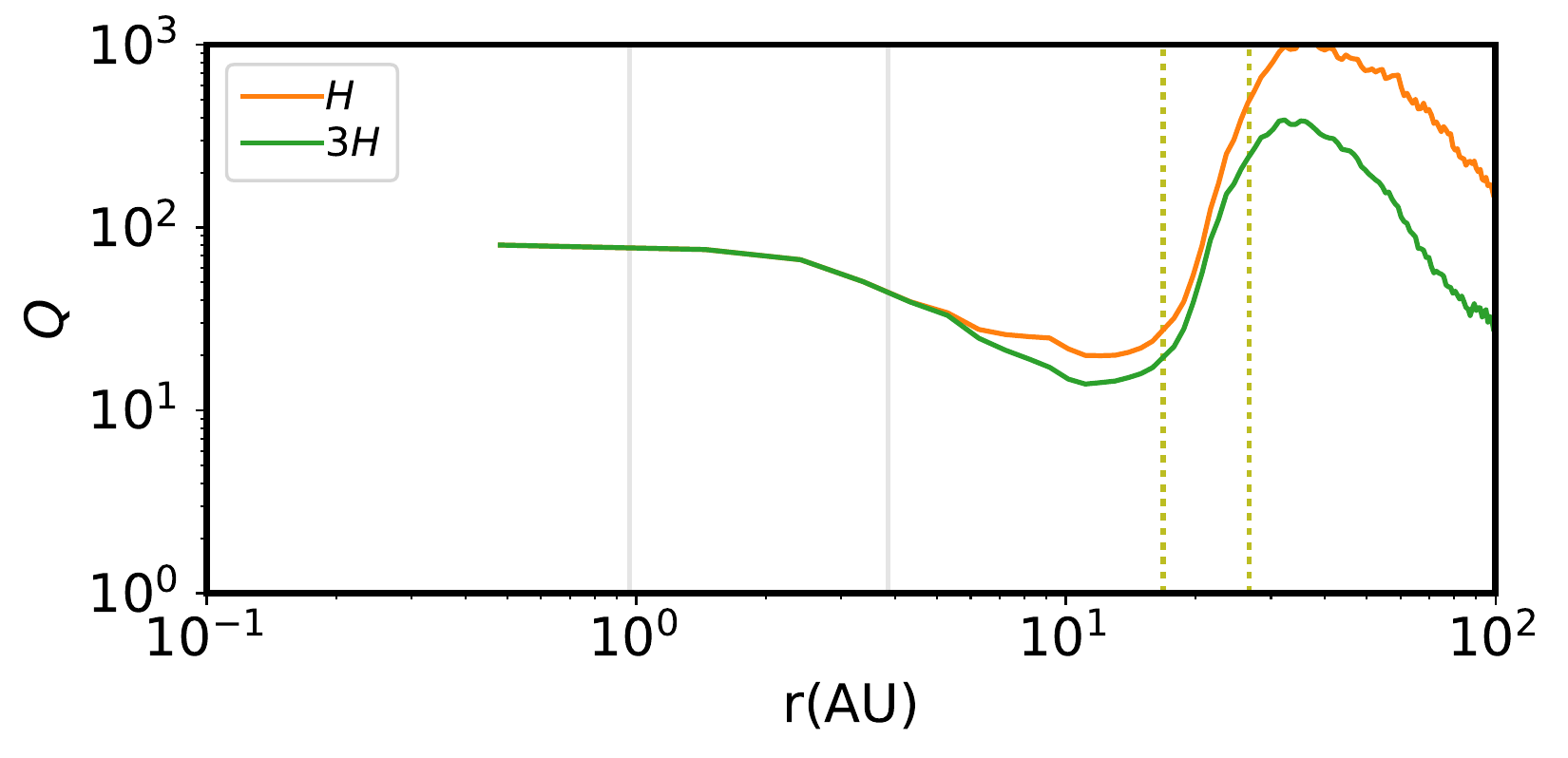}}
\put(0,0.92){\includegraphics[trim=0 0 0 3,clip,width=0.5\textwidth]{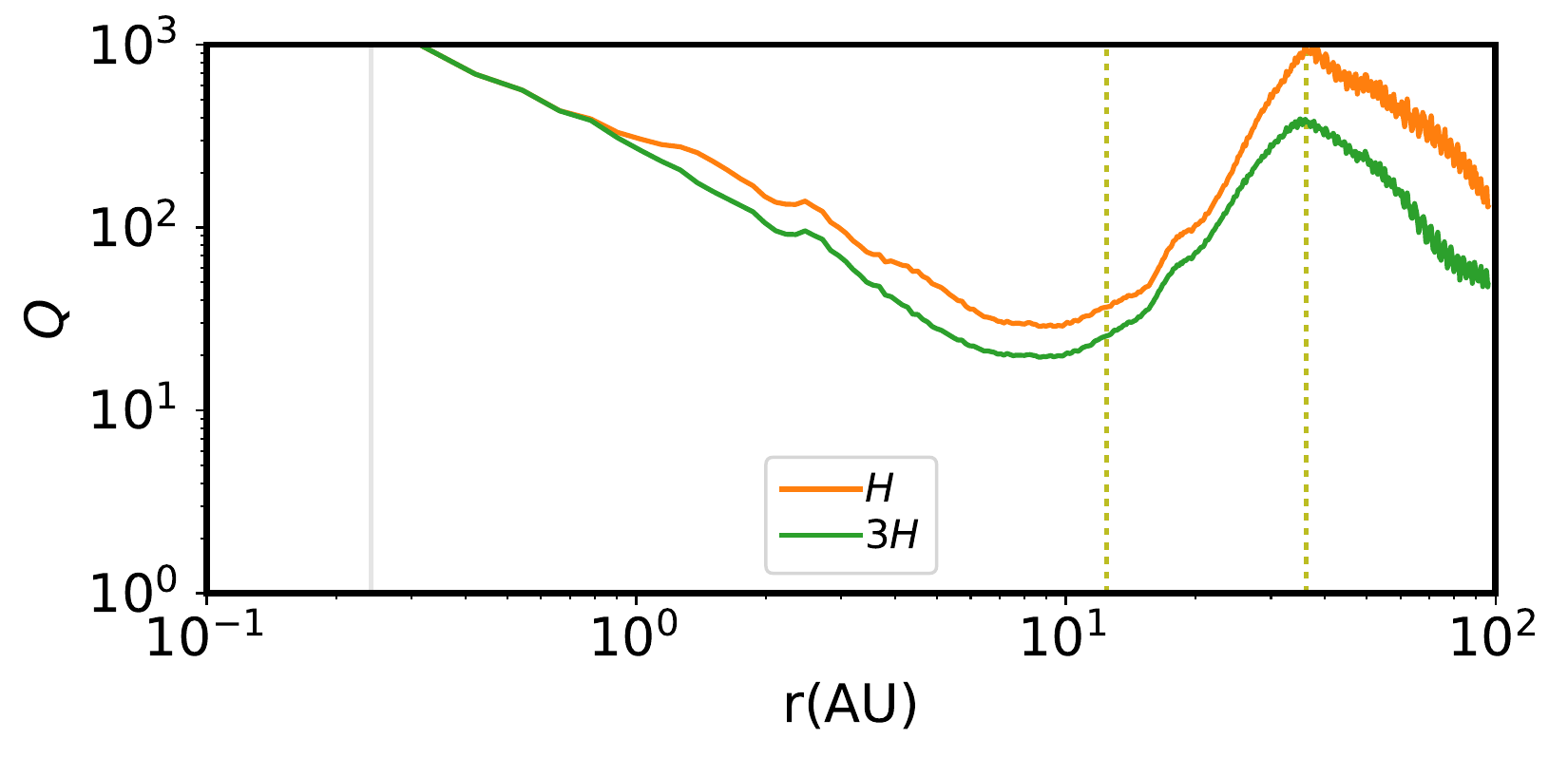}}
\put(0,0.46){\includegraphics[trim=0 0 0 3,clip,width=0.5\textwidth]{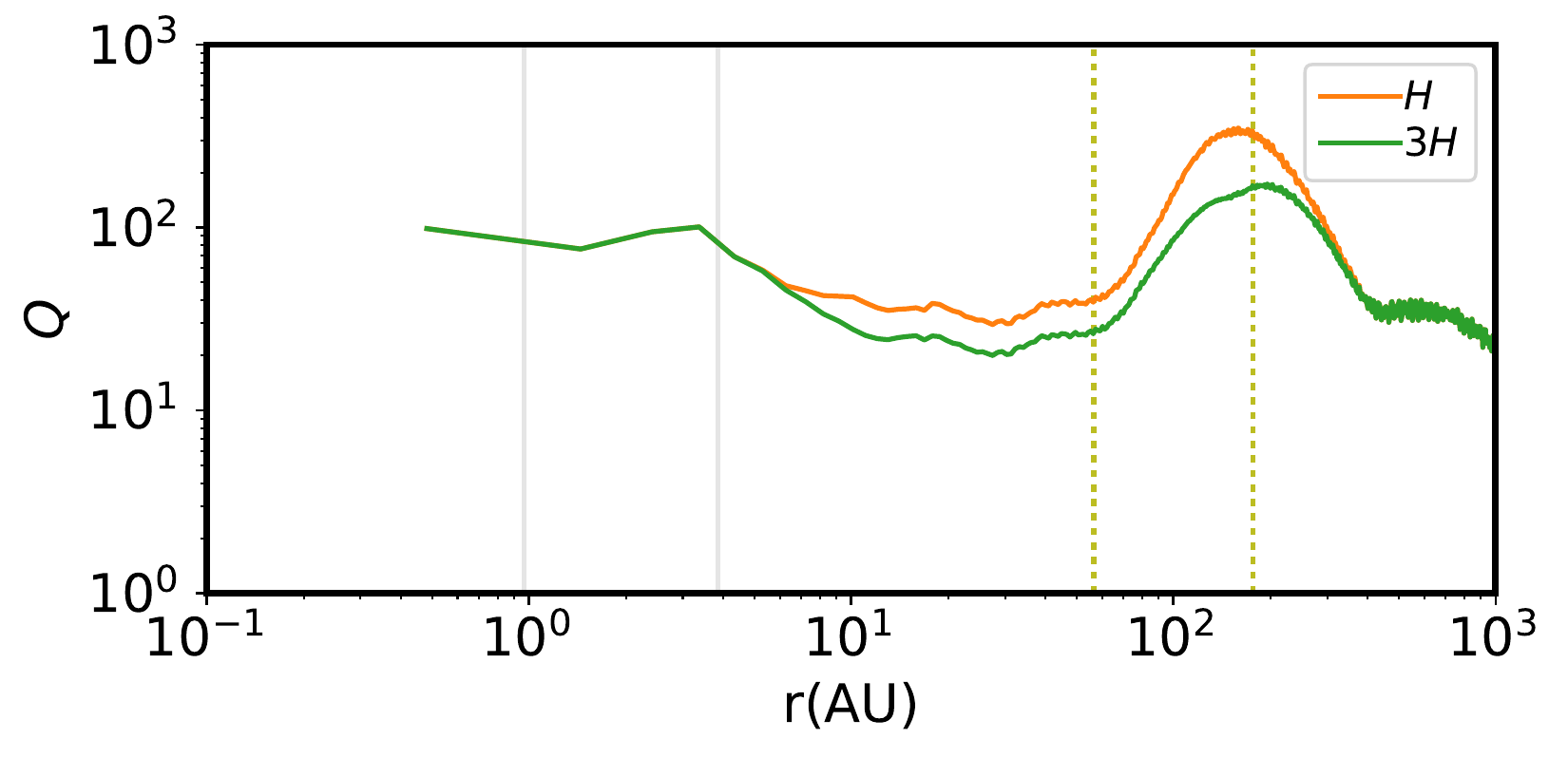}}
\put(0,0){\includegraphics[trim=0 0 0 3,clip,width=0.5\textwidth]{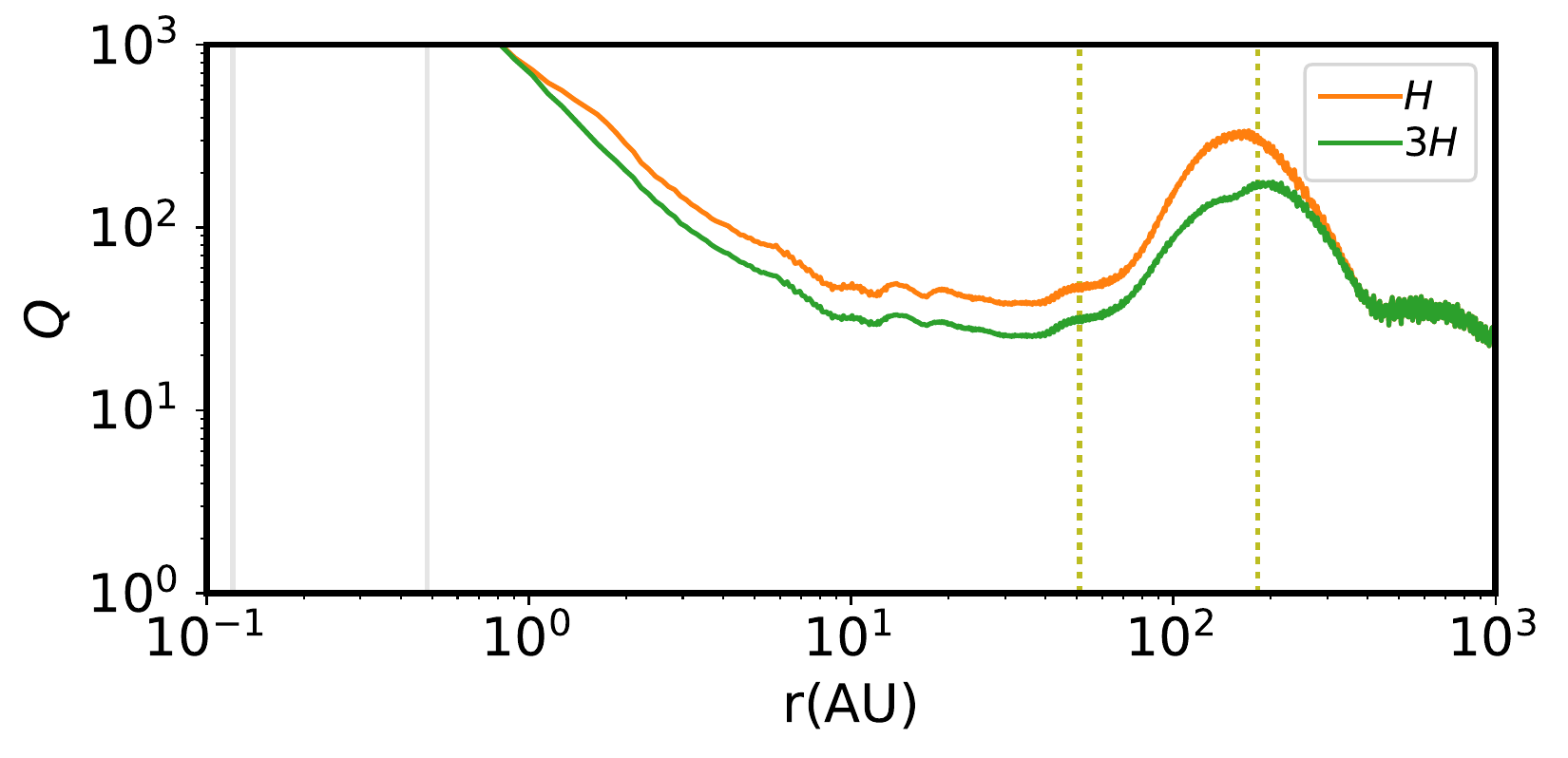}}
\put(0.15,1.50){R\_$\ell$14 (103 kyr)}
\put(0.15,1.04){R\_40ky\_$\ell$18 (103 kyr)}
\put(0.15,0.58){R\_$\ell$14 (138 kyr)}
\put(0.15,0.12){R\_80ky\_$\ell$18 (138 kyr)}
\end{picture}
\caption{Toomre $Q \approx \Omega c_{\rm s} / (\pi G \Sigma)$ parameter evaluated within one (orange) and three (green) times the disk scale heights. The same snapshots are taken as in Fig. \ref{fig_H}. The disk is very stable against self-gravitating fragmentation. The difference between the two curves mainly arises because the total mass is not taken into account for the value within $H$ and $3H$. }
\label{fig_Q}
\end{figure}

The Toomre parameter, $Q$, indicates whether the disk is subject to gravitational fragmentation. 
It is defined as
\begin{align}\label{eq_Q}
Q(r)  = {\kappa(r) c_{\rm s}(r) \over \pi G \Sigma(r)} \approx  {\Omega(r) c_{\rm s}(r) \over \pi G \Sigma(r)},
\end{align}
where 
\begin{align}
\kappa^2 =  {1 \over r^3} {d(r^4\Omega^2) \over dr}={2\Omega \over r} {d(r^2\Omega) \over dr}
\end{align}
is the epicyclic frequency. 
The approximation, $\kappa \approx \Omega$, is valid for a disk in Keplerian rotation, 
therefore valid only within $\sim R_{\rm mag}$. 
The Toomre parameter compares the stabilizing radial shear from differential rotation to its self-gravity that induces local collapse. 
For a value of $Q>1$, the disk is stable against self-gravitating fragmentation. 

Figure \ref{fig_Q} shows the Toomre $Q$ profile for some snapshots. 
The value is very high near the center because of the high temperature, 
and it decreases to reach the lowest point near $R_{\rm kep}$.  
The value of $Q$ increases again in the exterior magnetized part of the disk, 
while the approximation in Eq. (\ref{eq_Q}) starts to be invalid. 
With the high value of $Q$, the whole disk is very stable against self-gravitating fragmentation. 
All values of $c_{\rm s}$, $\Omega$, and $\Sigma$ are evaluated locally within $H$ or $3H$.
The difference in the vertical direction mostly arises because not the entire mass is taken into account when only the region within $H$ is considered.   

The Toomre parameter $Q \sim 10$ near the edge of the Keplerian disk, and it reaches very high values at smaller radii ($10^2-10^3$).
This extreme stability, from high temperature and/or low surface density, should be regarded with caution, and we point out some possible origins.
First, the disk might be overheated with our prescription of protostellar irradiation (see Sect. \ref{st_temperature} and \ref{st_adisk}). 
Second, the disk global profile is likely to be governed by the star-disk boundary condition, or the sink accretion algorithm from a numerical aspect. 
Our simulation has a percent-level disk-star mass ratio, which is somewhat lower than in some other studies \citep[e.g., 30-40\% was found by][]{Tomida2017}.
This result is directly linked to the density threshold for the sink accretion (see Sect. \ref{st_restart}). 
This has a direct consequence on the surface density of the disk, and therefore high $Q$ values could occur if $\Sigma$ is too low.

\subsection{Properties of the flow: Infall from the envelope and accretion within the disk}\label{st_flow}
The primary goal of this study is to follow the building phase of the protoplanetary disk: 
how mass arrives onto the disk and how mass is transported within the disk. 
We therefore start by considering some general properties of the global flow, 
and then continue with a more detailed quantification of the mass flow properties.

\subsubsection{Global properties: Mass transport}\label{st_transport}

\begin{figure}[]
\centering
\setlength{\unitlength}{0.5\textwidth}
\begin{picture}(1,1.24)
\put(0,0.62){\includegraphics[trim=0 5 8 5,clip,width=0.5\textwidth]{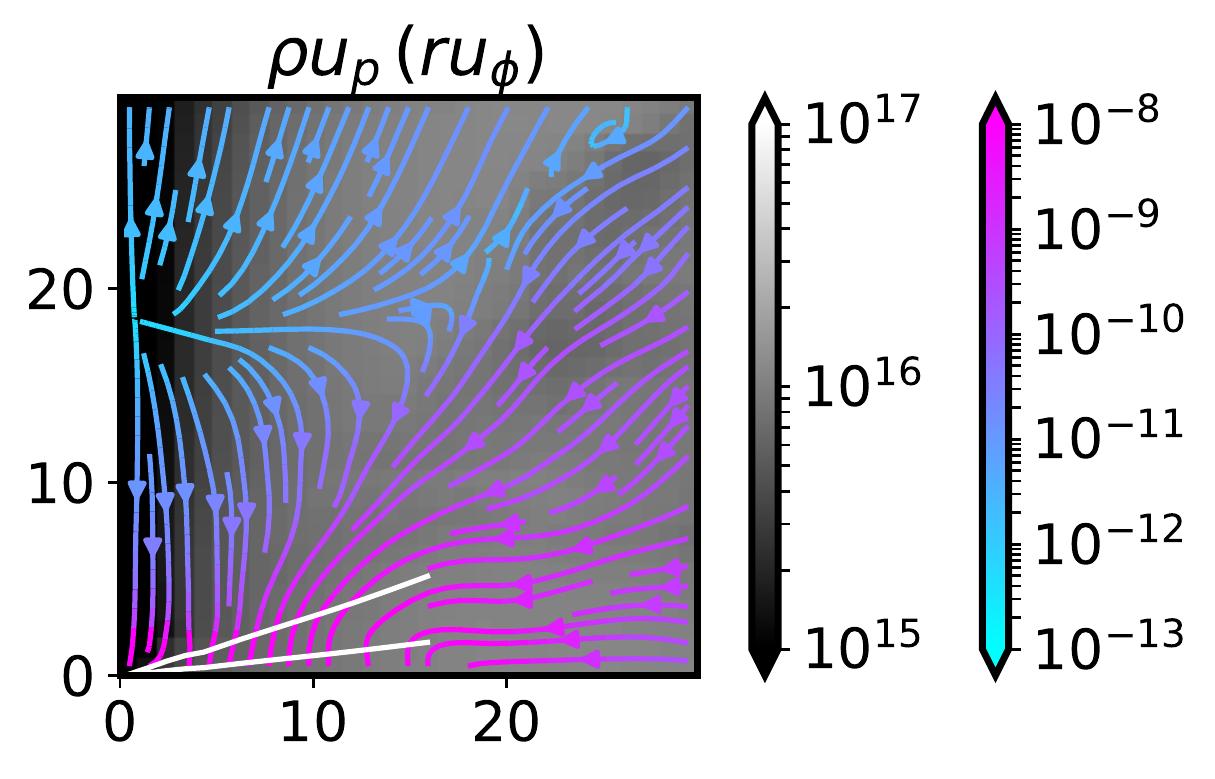}}
\put(0,0){\includegraphics[trim=0 5 8 5,clip,width=0.5\textwidth]{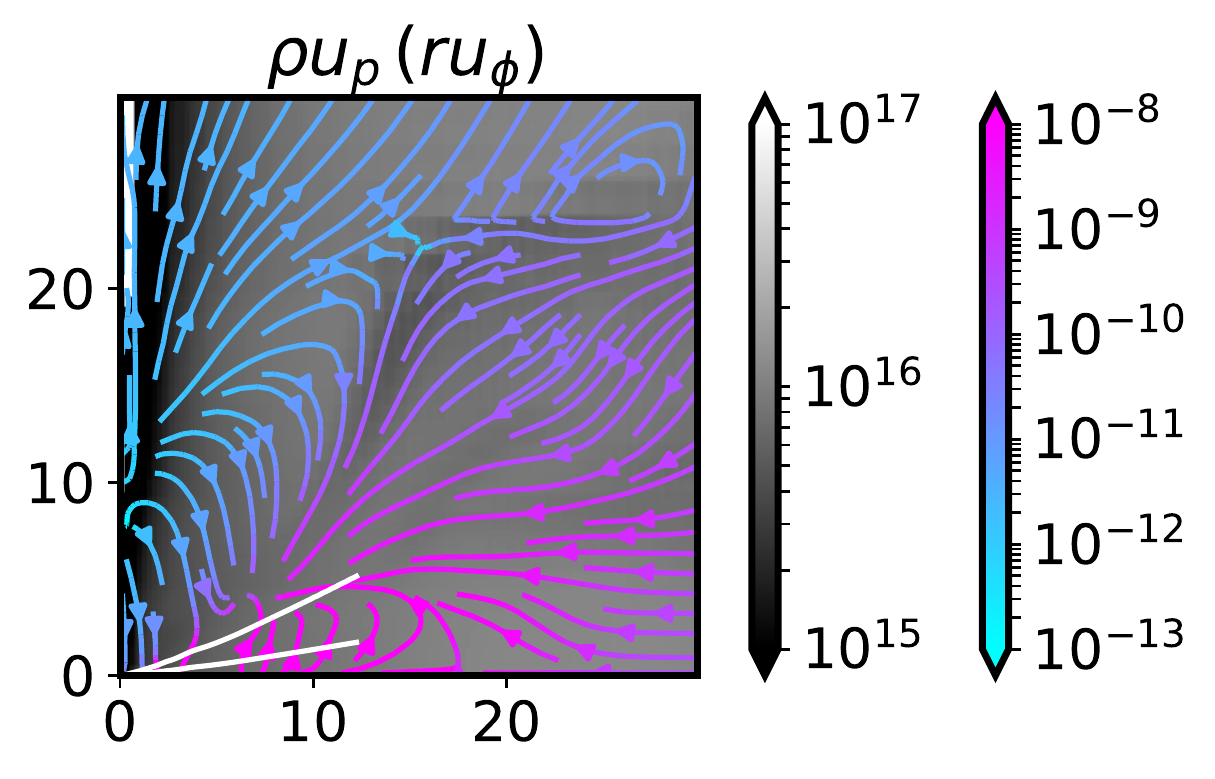}}
\end{picture}
\caption{Poloidal mass flux ($ \langle  \rho \vec{u}_{\vec{p}} \rangle_\phi$) and specific angular momentum ($r \langle u_\phi \rangle_\phi$) around the disk region (color streamlines in the gray image, in SI units), for R\_$\ell$14 ({\it top}) and R\_40ky\_$\ell$18 ({\it bottom}) at simulation time 103 kyr. All quantities are azimuthally averaged and averaged between the upper and lower planes. White curves outline the disk scale height $H$ and $3H$ at $r < R_{\rm kep}$. An outflow is launched at high altitudes at $z\approx 20$ AU in R\_$\ell$14. In the high-resolution run, the flow varies slightly more, but it is qualitatively similar. The outflow is launched closer to the disk ($z\approx 10$ AU) and the disk size is slightly smaller. The outflow has relatively low density and manifests itself as a cavity. Significant infall from the envelope is channeled through a zone of low angular momentum at several scale heights (darker gray region near $z \approx r$) of the disk. The horizontal infall from the envelope reaches the outer edge of the disk (shown in Fig. \ref{fig_40kyr_s} at $r \approx 15$ AU), and the density contrast generates a shock that pushes the gas to higher altitude. This creates a counterclockwise poloidal circulation near the disk edge between 10 and 15 AU in R\_40ky\_$\ell$18 because the outer disk decretes at this instant, while it is not seen in the upper panel. }
\label{fig_mflux_rest2}
\end{figure}

\begin{figure}[]
\centering
\includegraphics[trim=0 0 0 0,clip,width=0.3\textwidth]{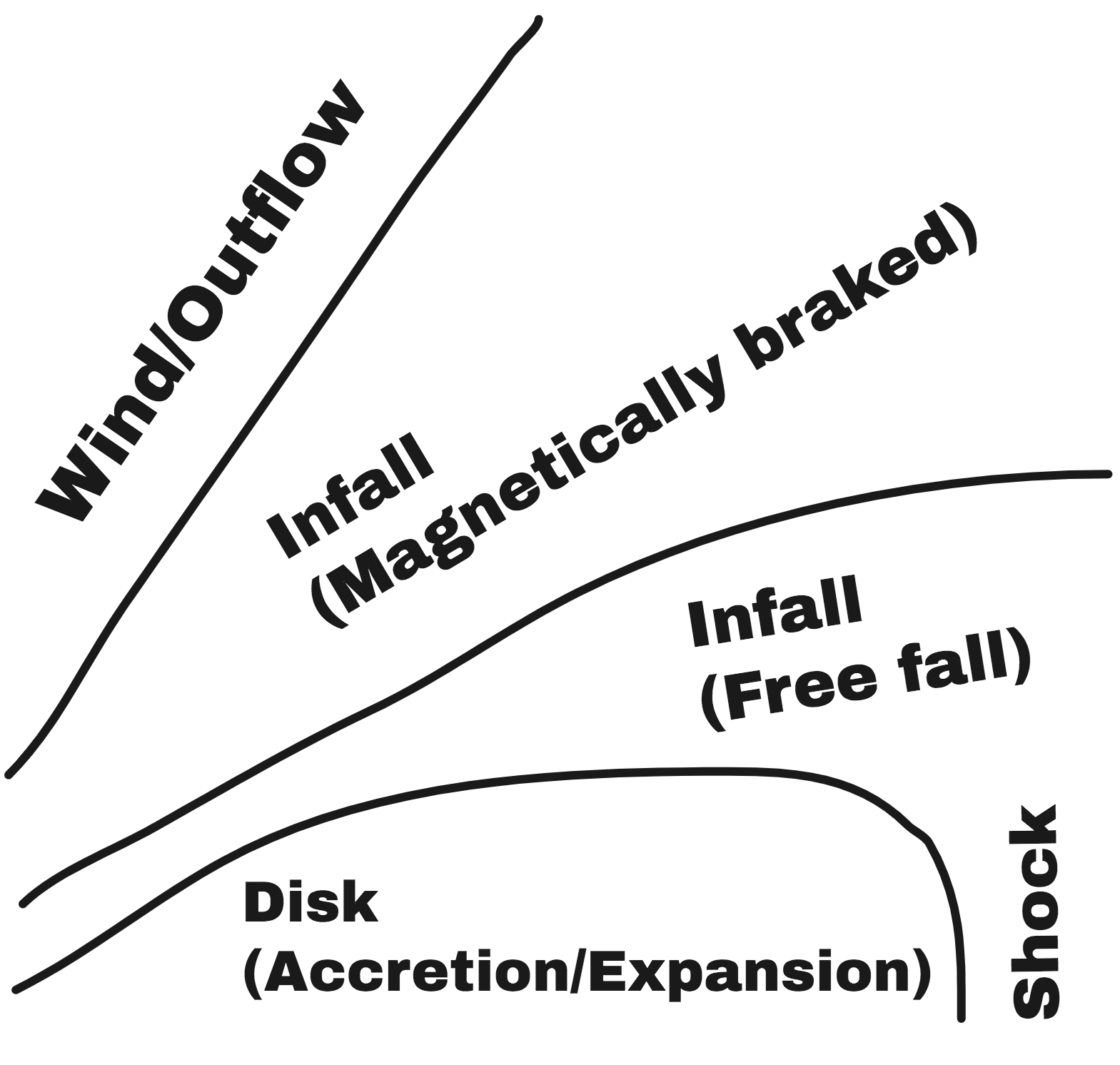}
\caption{Zones showing the main properties of the disk-envelope system in Sect. \ref{st_transport} and \ref{st_jtransport}.}
\label{fig_cartoon}
\end{figure}

Figure \ref{fig_mflux_rest2} shows R\_$\ell$14 (top) and R\_40ky\_$\ell$18 (bottom) at 103 kyr. 
Colored streamlines present the azimuthally averaged poloidal mass flux, $\langle \rho \vec{u}_{\vec{p}} \rangle_\phi$ in addition to the specific angular momentum, $\langle \rho r u_\phi \rangle_\phi/\langle \rho \rangle_\phi$, shown in gray scale. 
White curves mark the disk scale height $H$ and $3H$ up to $R_{\rm kep}$. 
The general behavior is not strongly affected by the resolution.
The flow pattern varies slightly at different instants. 
Nonetheless, the patterns all show some similar characteristics that we present below: 
infall, outflow, shock, and accretion (or sometimes expansion or decretion). 
Figure \ref{fig_cartoon} presents a schematic view of the various zones. 

\begin{itemize}
\item Infall:
The envelope is globally infalling. 
The infall is mostly radial near the equatorial plane. 
Slightly above the disk surface, the vertical mass flux increases. 
Most importantly, the flux intensity varies (change of color) with altitude because of the small radial component that brings the mass flux to smaller radii as it approaches the disk surface from the top. 
Therefore most of the infalling mass reaches small radius of within a few AU before penetrating the disk surface. 
We recall the source function, which is the mass flux crossing the disk surface, that we aimed to measure. 
This type of infalling pattern gives a surface flux that is centrally concentrated and strongly depends on the reference surface (altitude).  
\\
\item Outflow: 
In the axial direction, a low-density outflow cavity is launched at 10-20 AU above the disk midplane, 
and the radial extent increases with the vertical distance from the disk. 
The transition between infall and outflow occurs at high altitudes above the disk, 
roughly near the cone $z \approx r$. 
Immediately below this is a zone with low angular momentum (darker in the gray image), through which significant infall is channeled. 
There might be some circulation and recycling of the wind material across the cavity wall. 
\\
\item Shock: 
We recall from  Fig. \ref{fig_v} that the radial velocity in the equatorial plane reaches a fraction of the free-fall velocity and quickly drops at $\sim R_{\rm mag}$. 
In the post-shock region, the rotation replaces the infall as the dominant kinetic energy. 
A slowly expanding zone appears very often in the magnetized outer region of the disk and meets the radial infall at the shock front. 
This shock is located at $r \sim 15$ AU in both panels of Fig. \ref{fig_mflux_rest2}, where the radial velocity apparently decreases.
In R\_$\ell$14, it is more evident that the gas is pushed upward due to the increased pressure and falls back to the disk surface at smaller radius, while the streamlines in R\_40ky\_$\ell$18 are more regular and there is only very mild upward motion when the gas reaches the shock front.
The radial flow inside the disk is sometimes outward, causing circulation at the edge of the disk: 
matter moves upward at the shock and falls back from the top.  
This behavior has been suggested in observations \citep{Sakai17}, and the shock might be traced with chemical molecules. 
However, in our case, the disk is probably too thin and the shock-generated heat is quickly radiated away. 
\\
\item Accretion: 
Within the disk ($z  \lesssim 3H$), the behavior of the flow is less clear. 
R\_$\ell$14 shows an almost vertical flux downward, while in R\_40ky\_$\ell$18 the flux is upward. 
The disk internal dynamics does not converge numerically, and conclusions cannot be drawn. 
Globally speaking, the innermost parts of the disk (a few AU) are always in radial accretion, which leads to the mass growth of the central star. 
However, 
the radial velocity within the rest of the disk is not strictly negative all the time. 
Outward expansion sometimes occurs in the middle layers of the disk. 
As mentioned earlier, this occurs very often in the magnetized zone of the disk (between $R_{\rm kep}$ and $R_{\rm mag}$). 
However, the disk is confined by the magnetized zone and the infalling envelope 
and thus cannot expand freely. 
\end{itemize}

\begin{figure}[]
\centering
\setlength{\unitlength}{0.5\textwidth}
\begin{picture}(1,1.24)
\put(0,0.62){\includegraphics[trim=0 5 8 5,clip,width=0.5\textwidth]{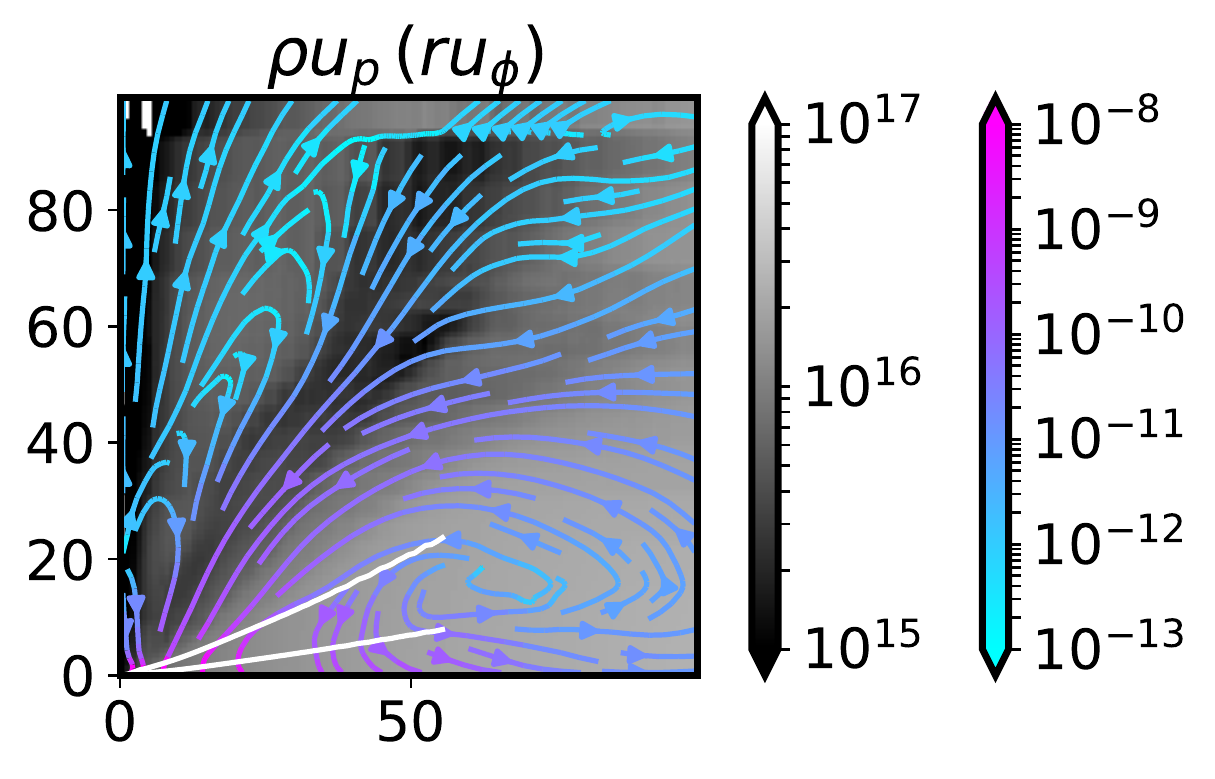}}
\put(0,0){\includegraphics[trim=0 5 8 5,clip,width=0.5\textwidth]{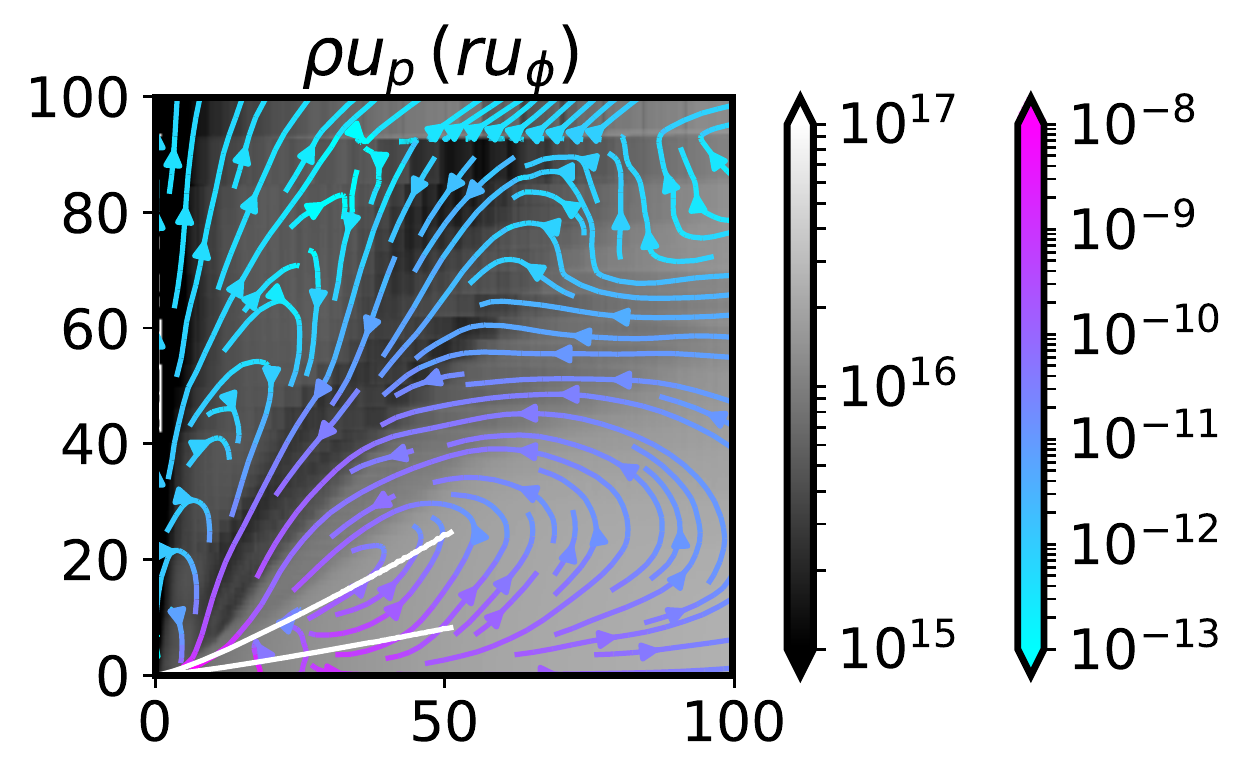}}
\end{picture}
\caption{Poloidal mass flux ($\langle  \rho \vec{u}_{\vec{p}} \rangle_\phi$) and specific angular momentum ($r \langle u_\phi \rangle_\phi$) around the disk region (color streamlines in the gray image, in SI units) for R\_$\ell$14 ({\it top}) and R\_80ky\_$\ell$18 ({\it bottom}) at simulation time 138 kyr.  All quantities are azimuthally averaged and averaged between the upper and lower planes. The white curves outline the disk scale height $H$ and $3H$ at $r < R_{\rm kep}$. An outflow is launched at $z\approx 20$ AU. The disk has grown in size, while the flow pattern stays very similar to that in Fig. \ref{fig_mflux_rest2}. At this moment, the magnetized zone is almost twice the size of the Keplerian disk ($R_{\rm mag} \approx 2R_{\rm kep}$), and the counterclockwise poloidal circulation around $R_{\rm kep}$ is evident in both panels and occupies almost the entire magnetized zone and the outer Keplerian disk.}
\label{fig_mflux_rest3}
\end{figure}

At 103 kyr, the disk radius is roughly 15 AU, while it grows to 50 at 138 kyr (Fig. \ref{fig_mflux_rest3}). 
The general behavior is the same as described above, 
and the region with low angular momentum becomes a clear channel of mass infall. 
The expanding magnetized zone is very large and almost comparable to the disk radius, $R_{\rm kep}$. 
Here again, the disk internal dynamics is not identical in the two runs with different resolutions. 
Nonetheless, the two main remarks remain valid: matter might circulate near the wind-infall wall and at the disk edge.

\subsubsection{Global properties: Angular momentum transport}\label{st_jtransport}

\begin{figure}[]
\centering
\setlength{\unitlength}{0.5\textwidth}
\begin{picture}(1,1.4)
\put(0,1.05){\includegraphics[trim=0 5 8 5,clip,width=0.25\textwidth]{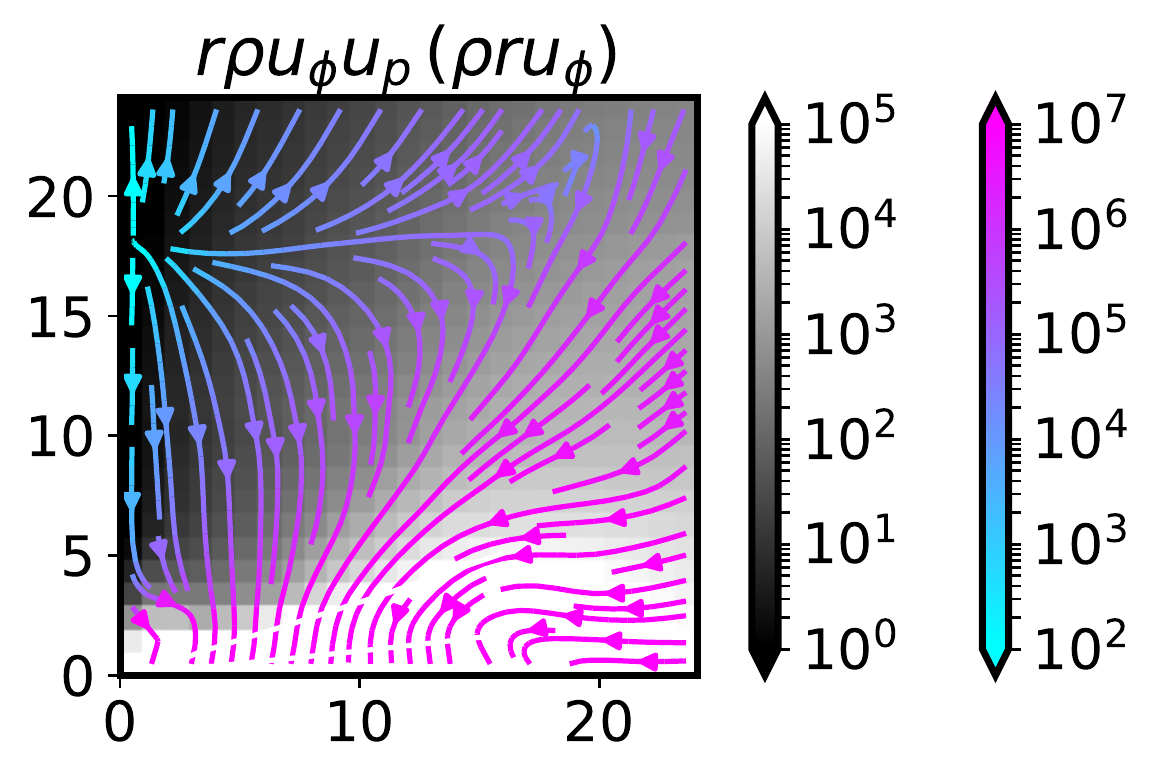}}
\put(0,0.7){\includegraphics[trim=0 5 8 5,clip,width=0.25\textwidth]{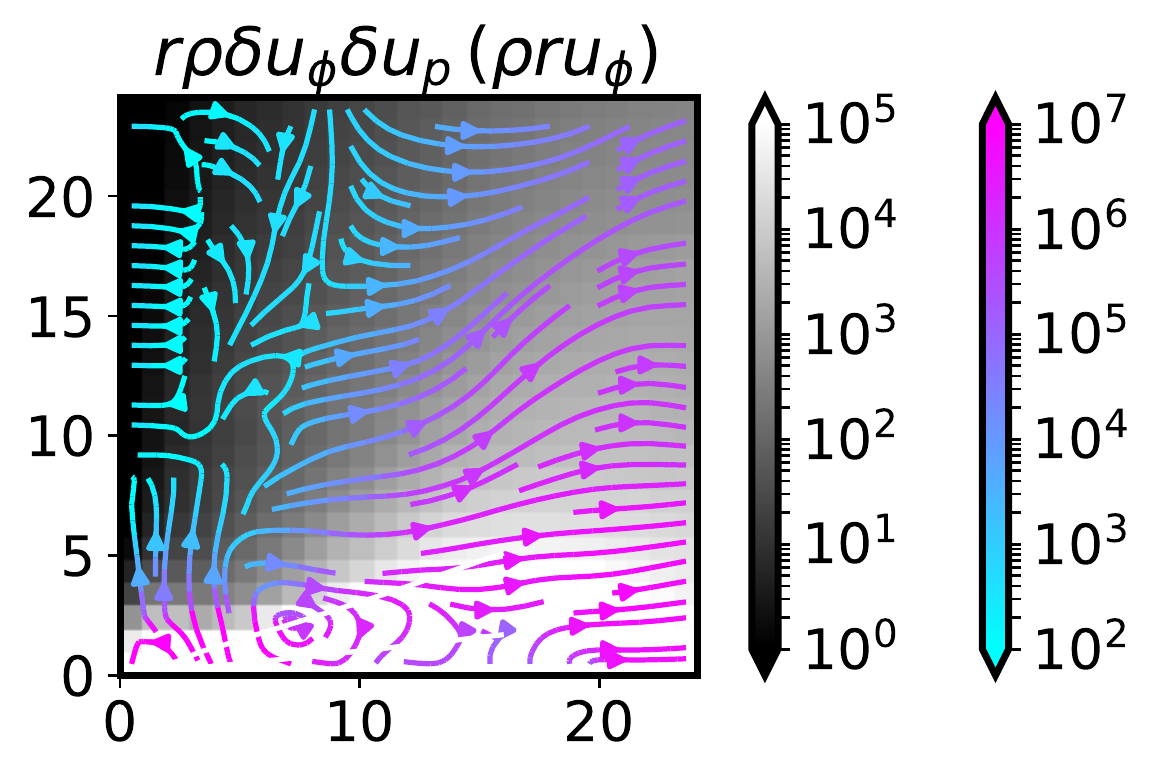}}
\put(0,0.35){\includegraphics[trim=0 5 8 5,clip,width=0.25\textwidth]{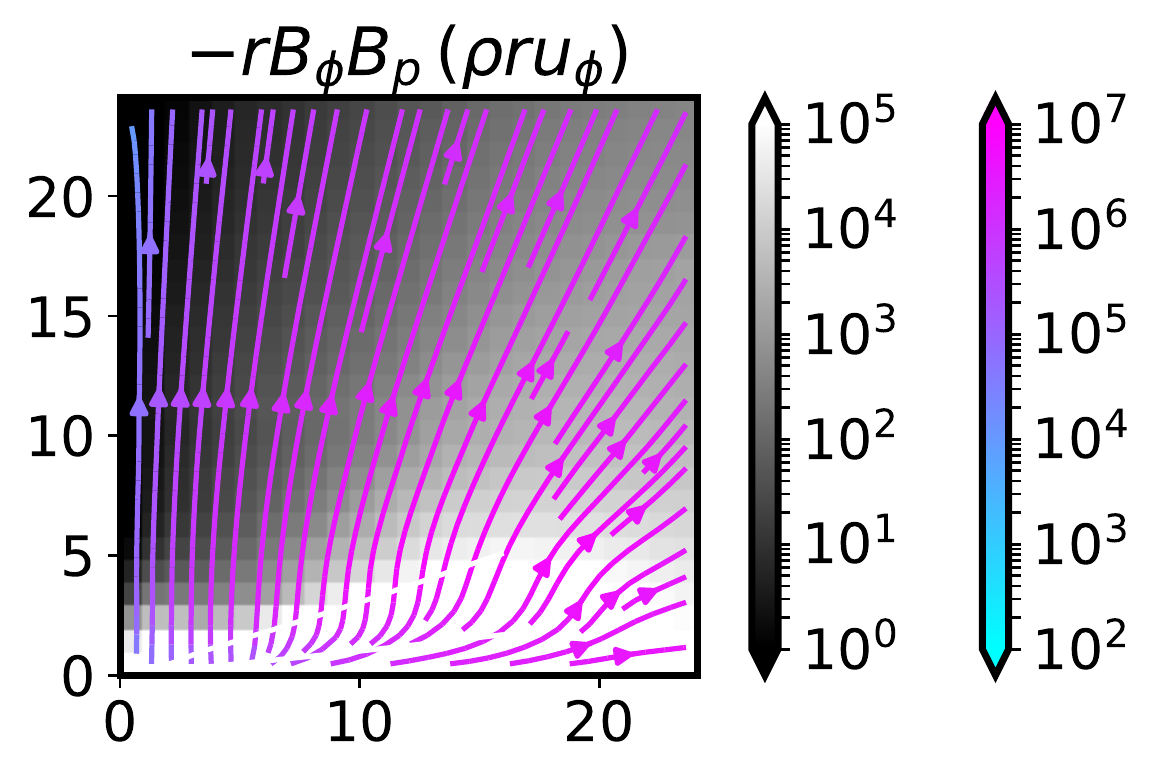}}
\put(-0.08,0.){\includegraphics[trim=0 5 8 5,clip,width=0.26\textwidth]{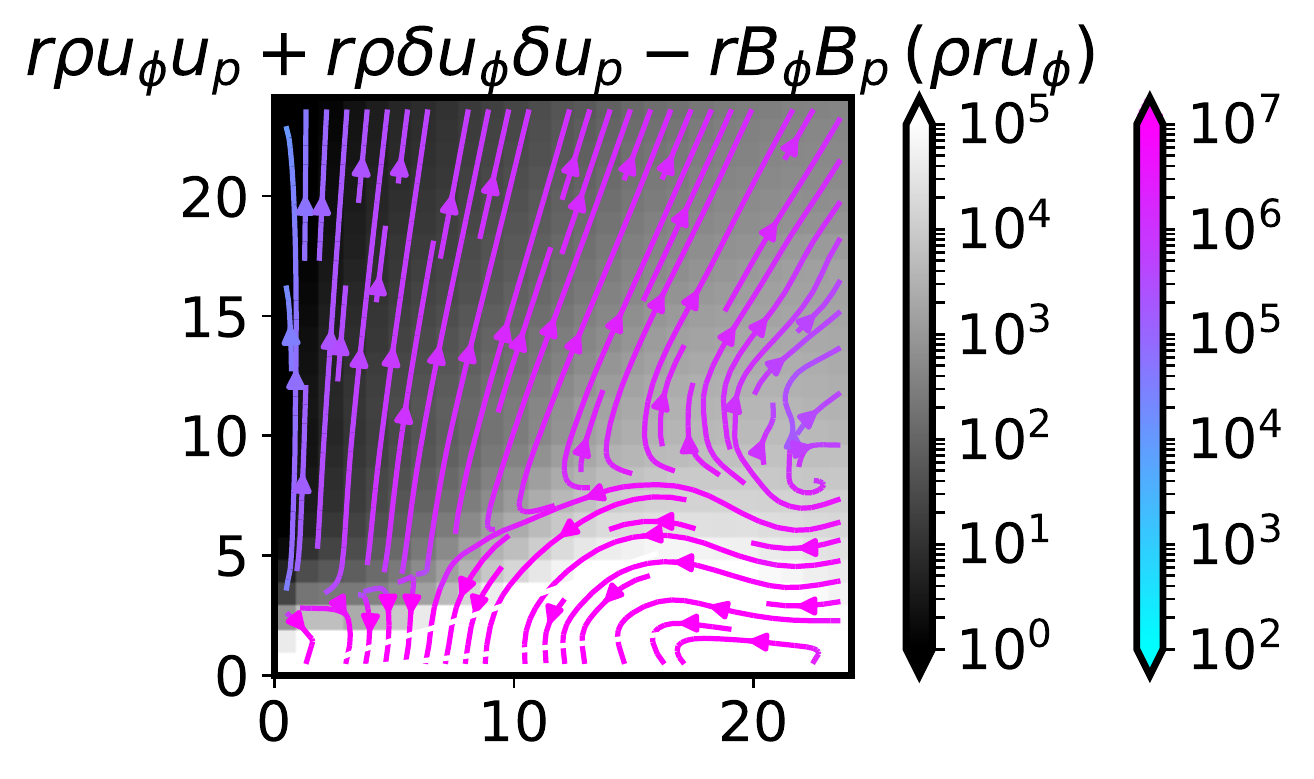}}
\put(0.5,1.05){\includegraphics[trim=0 5 8 5,clip,width=0.25\textwidth]{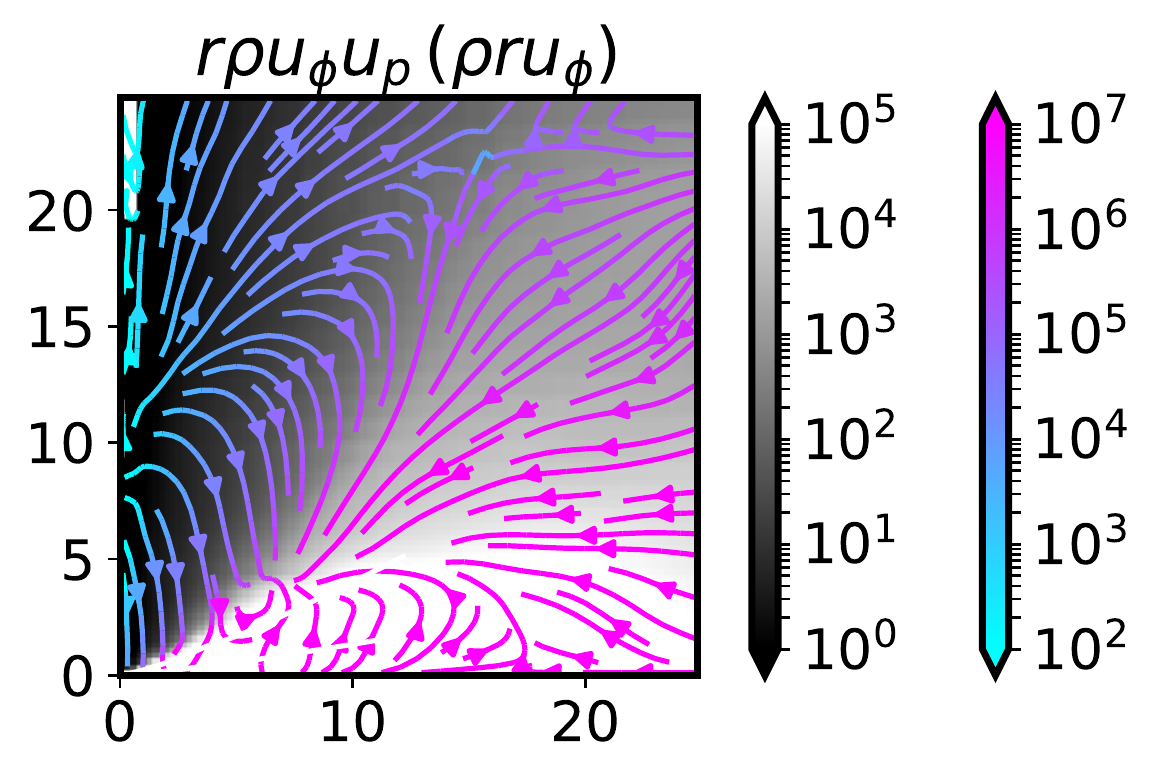}}
\put(0.5,0.7){\includegraphics[trim=0 5 8 5,clip,width=0.25\textwidth]{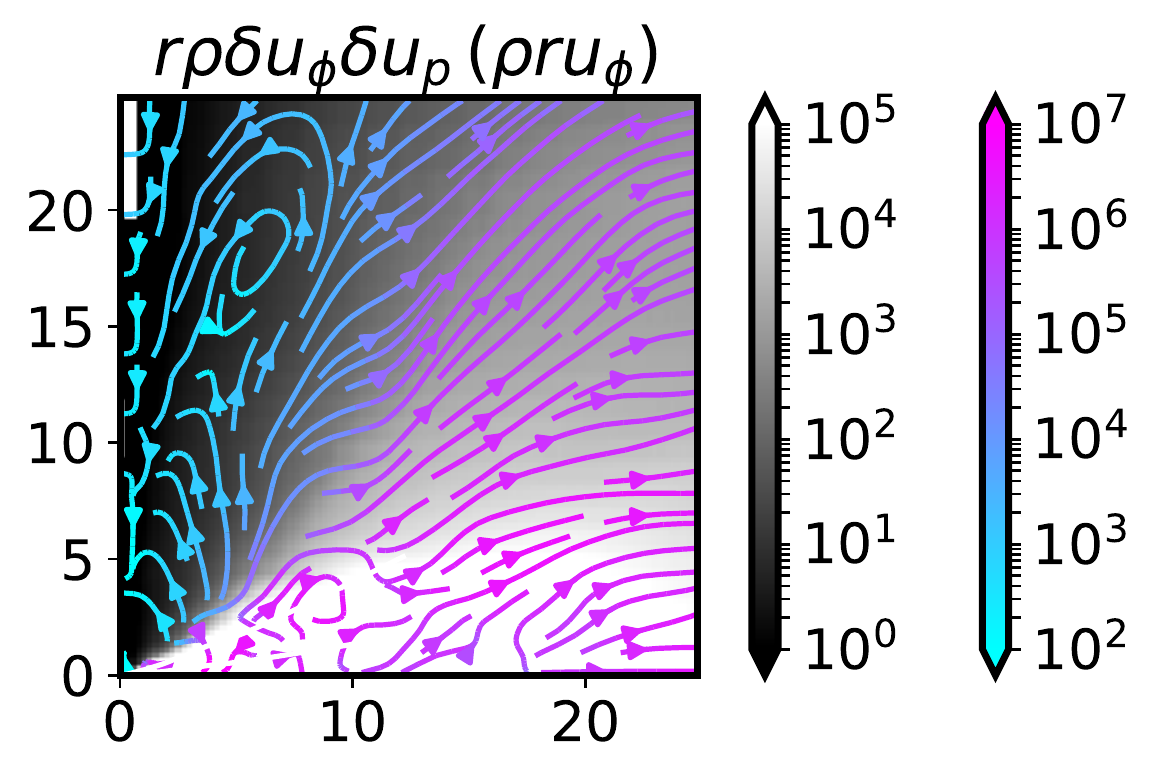}}
\put(0.5,0.35){\includegraphics[trim=0 5 8 5,clip,width=0.25\textwidth]{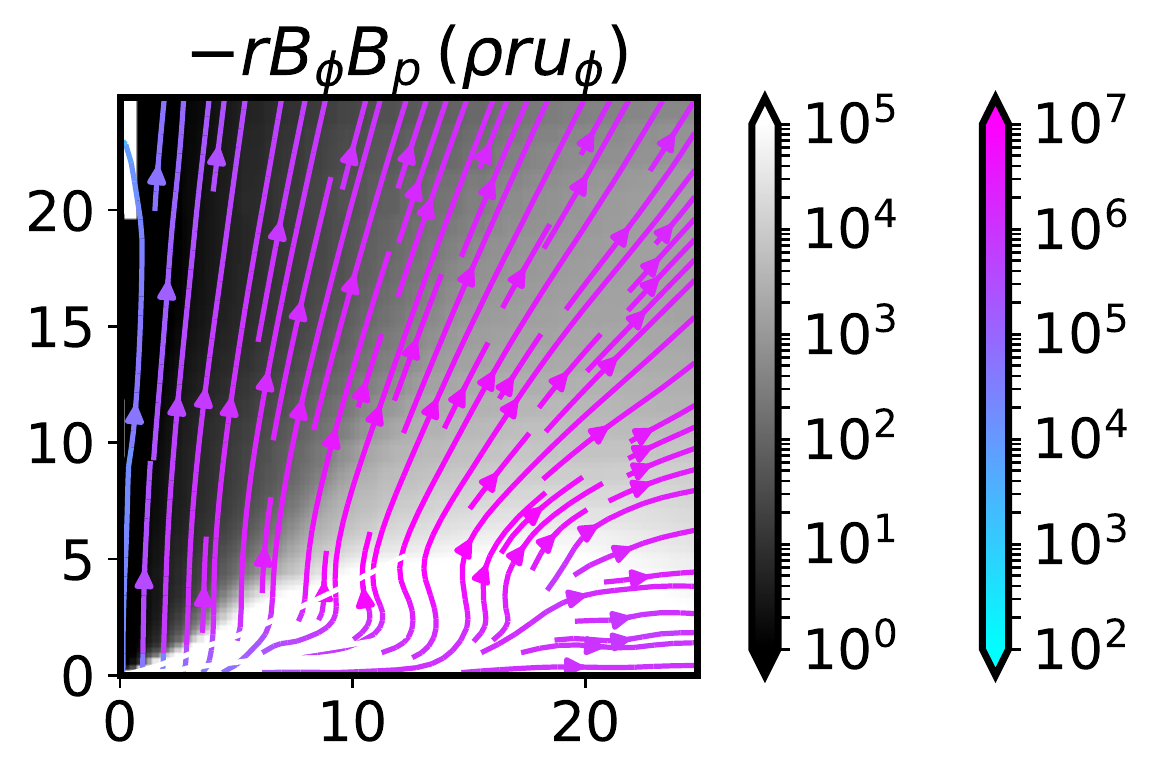}}
\put(0.44,0){\includegraphics[trim=7 5 8 5,clip,width=0.26\textwidth]{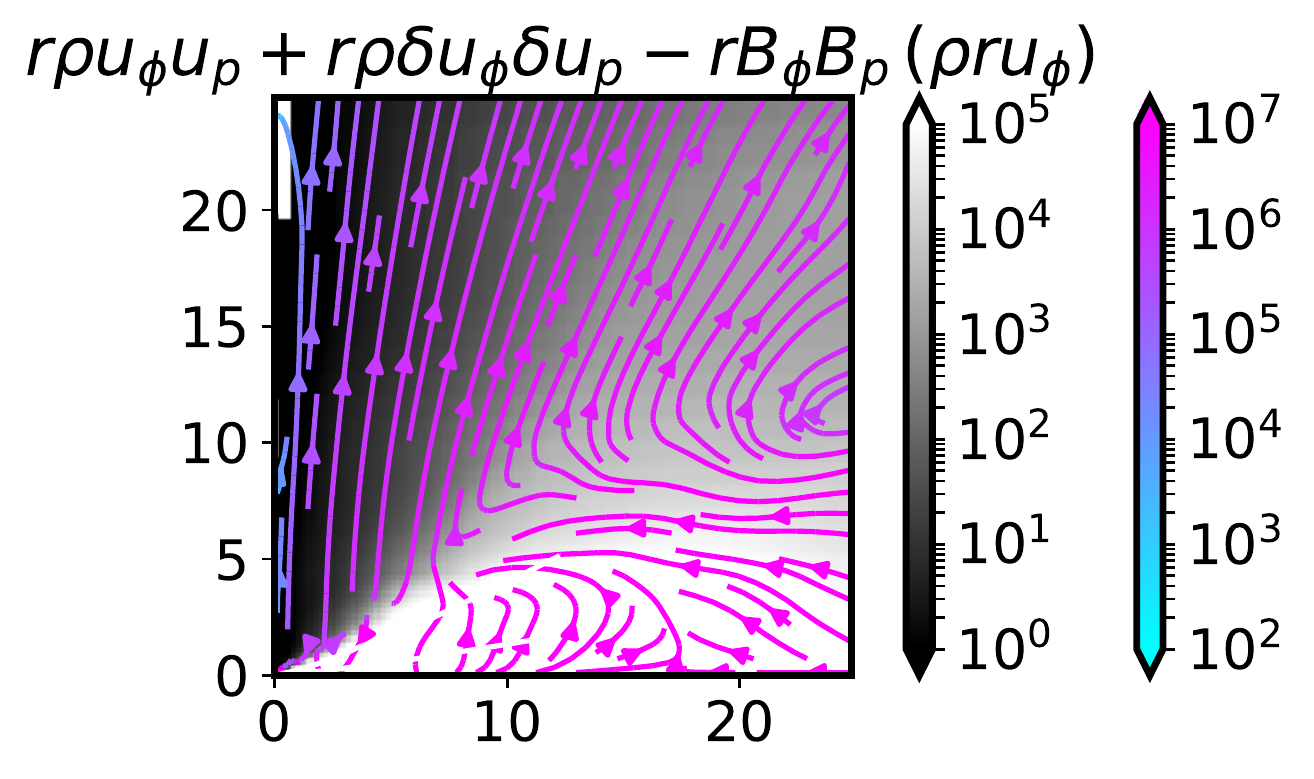}}
\end{picture}
\caption{Angular momentum flux and angular momentum ( $r\langle \rho u_\phi \rangle_\phi$) around the disk region (color streamlines in the gray image, in SI units), for R\_$\ell$14 ({\it left}) and R\_40ky\_$\ell$18 ({\it right}) at 103 kyr. 
All quantities are azimuthally averaged and averaged between the upper and lower planes. 
{\it From top to bottom:} Laminar transport ($r \langle u_\phi \rangle_\phi \langle \rho\vec{u}_{\vec{p}}\rangle_\phi$), turbulent transport ($r\langle\rho \delta u_\phi \delta \vec{u}_{\vec{p}}\rangle_\phi$), magnetic transport ($-r\langle B_\phi \vec{B}_{\vec{p}}\rangle_\phi/(4\pi)$), and their sum.
White curves outline the disk scale height $H$. 
The laminar flow convects mass with positive angular momentum, while the tensor terms (Reynolds and Maxwell) transport the angular momentum outward. In particular, the magnetic tensor dominates in the envelope, and this allows the mass to fall onto the disk through the loss of angular momentum. The general pattern persists with improved resolution. White patches in the gray image are due to locally negative angular momentum. This is an artifact of the averaging because the selected axis is not exactly aligned with the asymmetric outflow.}
\label{fig_jflux_rest2}
\end{figure}

\begin{figure}[]
\centering
\setlength{\unitlength}{0.5\textwidth}
\begin{picture}(1,1.4)
\put(0,1.05){\includegraphics[trim=0 5 8 5,clip,width=0.25\textwidth]{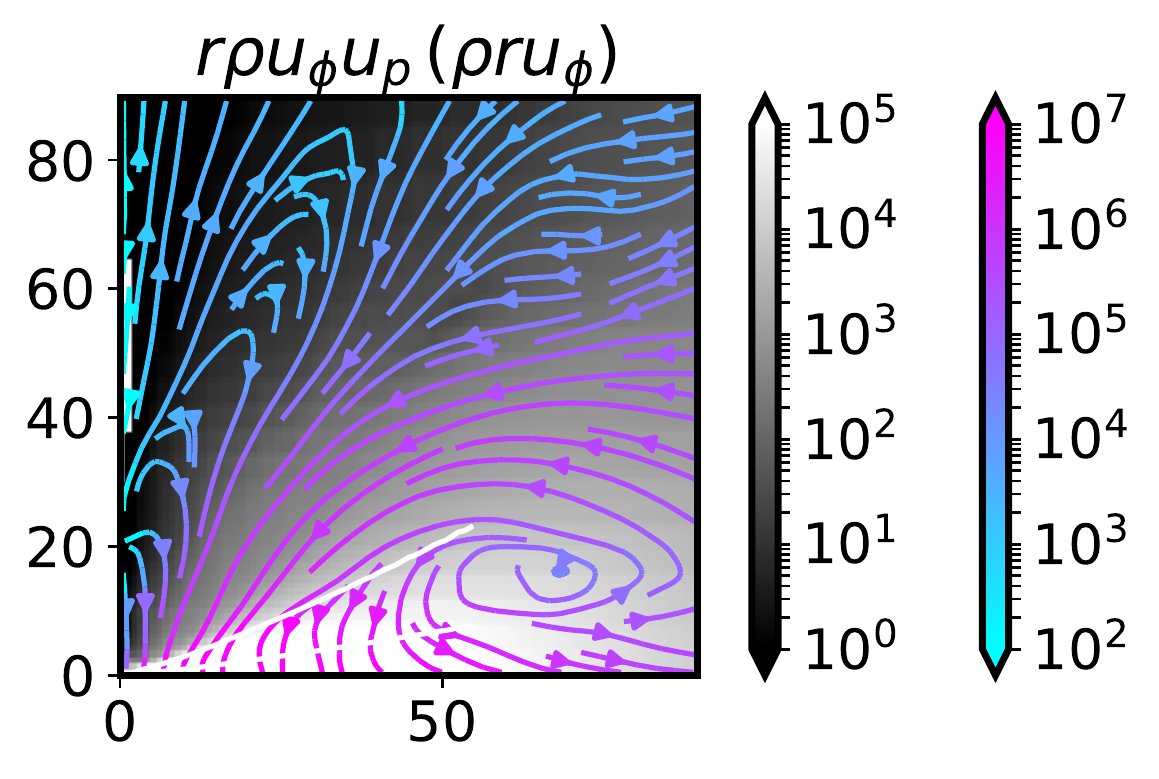}}
\put(0,0.7){\includegraphics[trim=0 5 8 5,clip,width=0.25\textwidth]{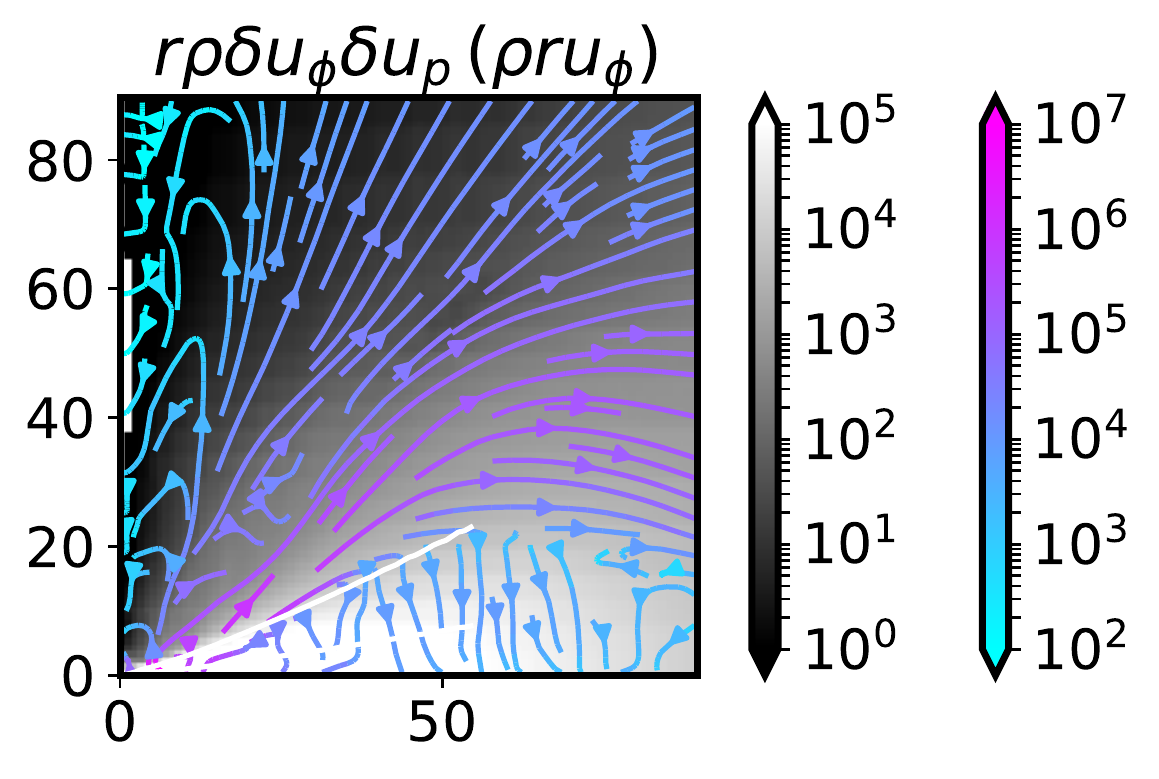}}
\put(0,0.35){\includegraphics[trim=0 5 8 5,clip,width=0.25\textwidth]{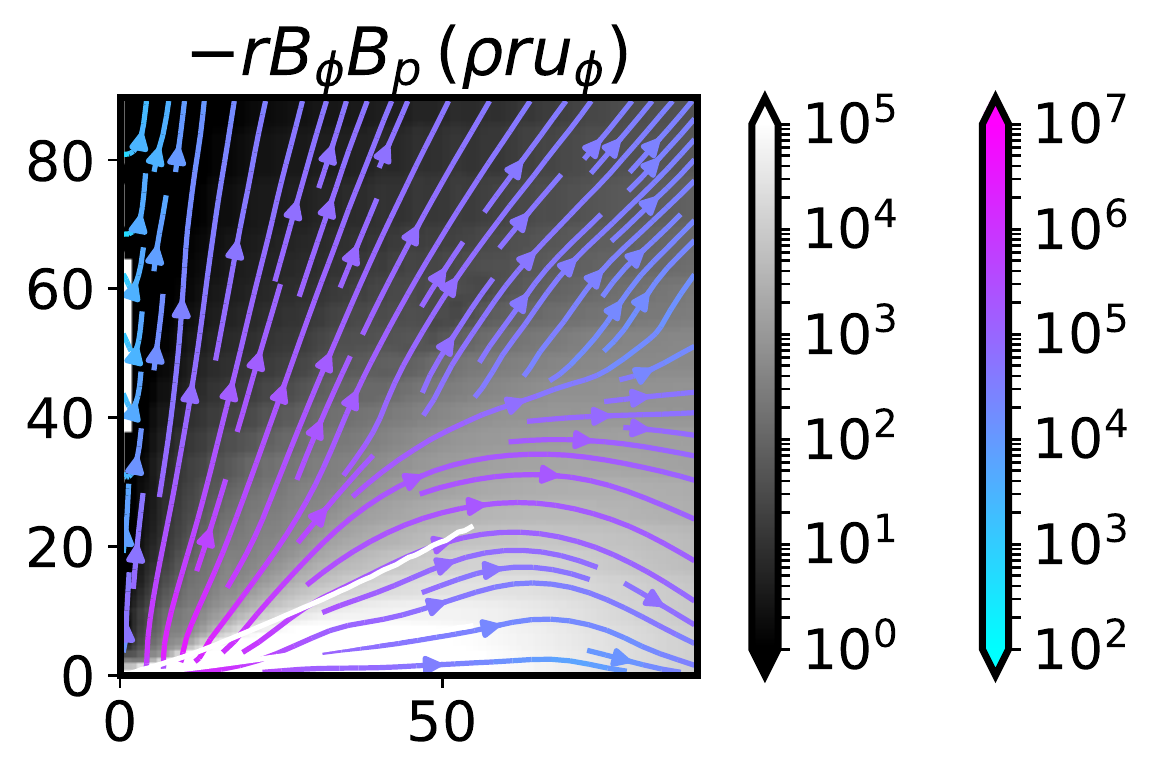}}
\put(-0.05,0.){\includegraphics[trim=0 5 8 5,clip,width=0.26\textwidth]{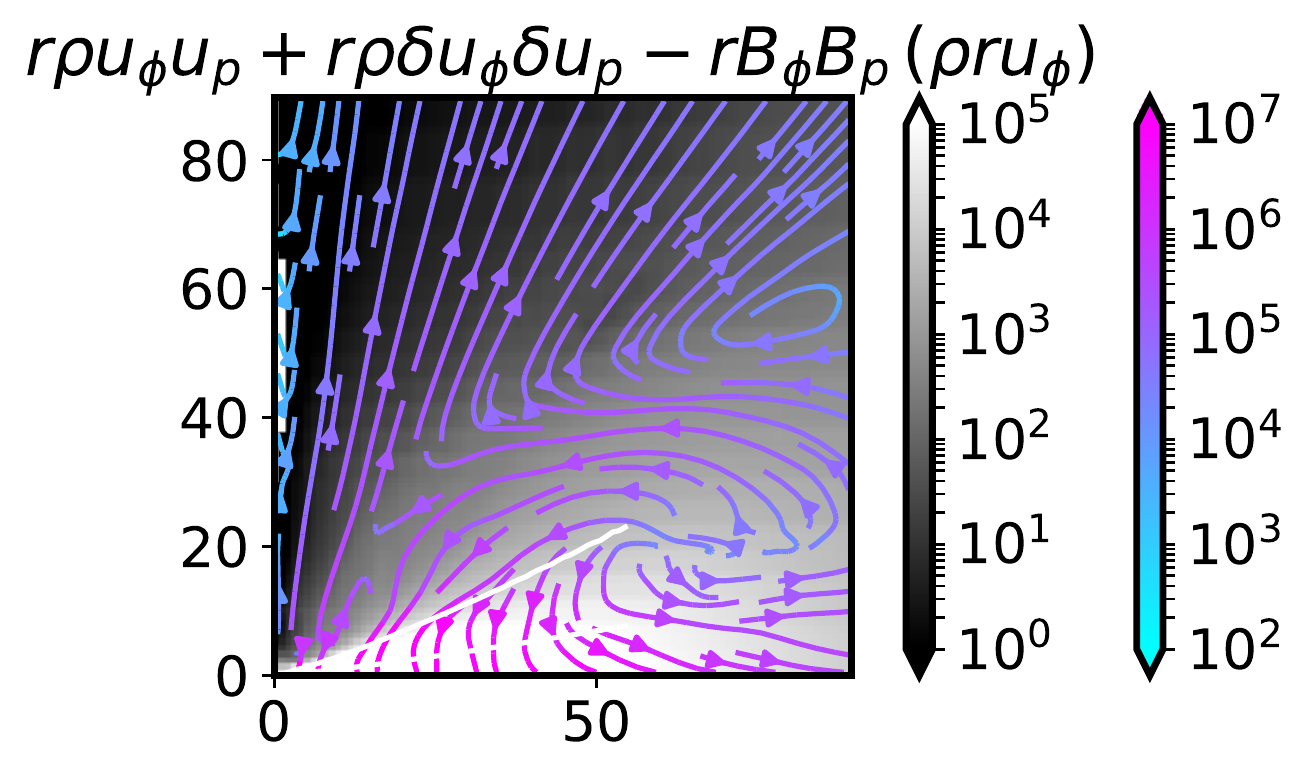}}
\put(0.5,1.05){\includegraphics[trim=0 5 8 5,clip,width=0.25\textwidth]{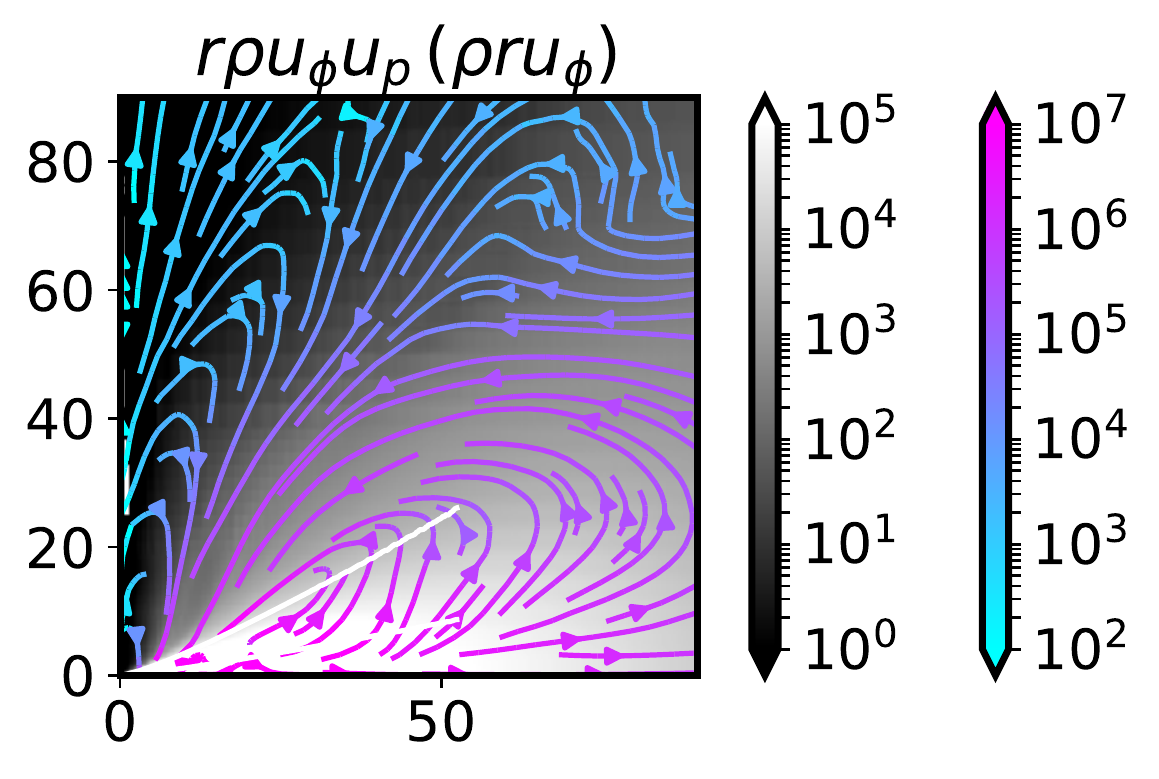}}
\put(0.5,0.7){\includegraphics[trim=0 5 8 5,clip,width=0.25\textwidth]{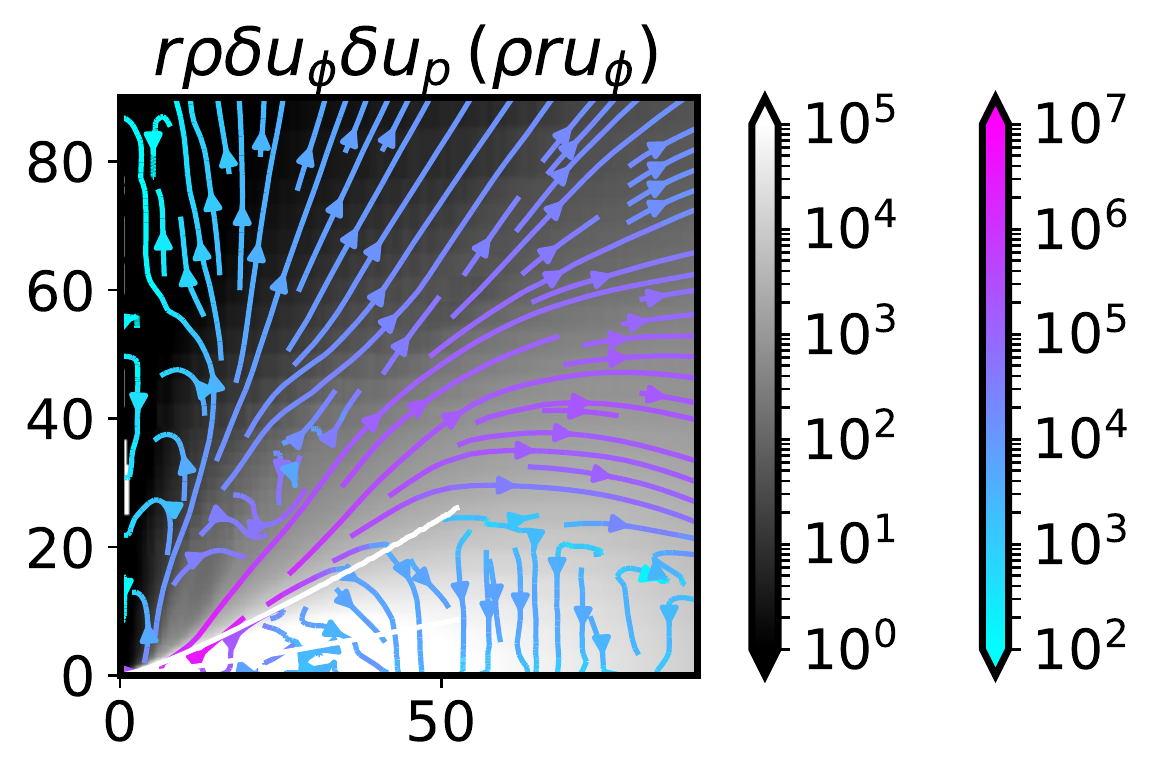}}
\put(0.5,0.35){\includegraphics[trim=0 5 8 5,clip,width=0.25\textwidth]{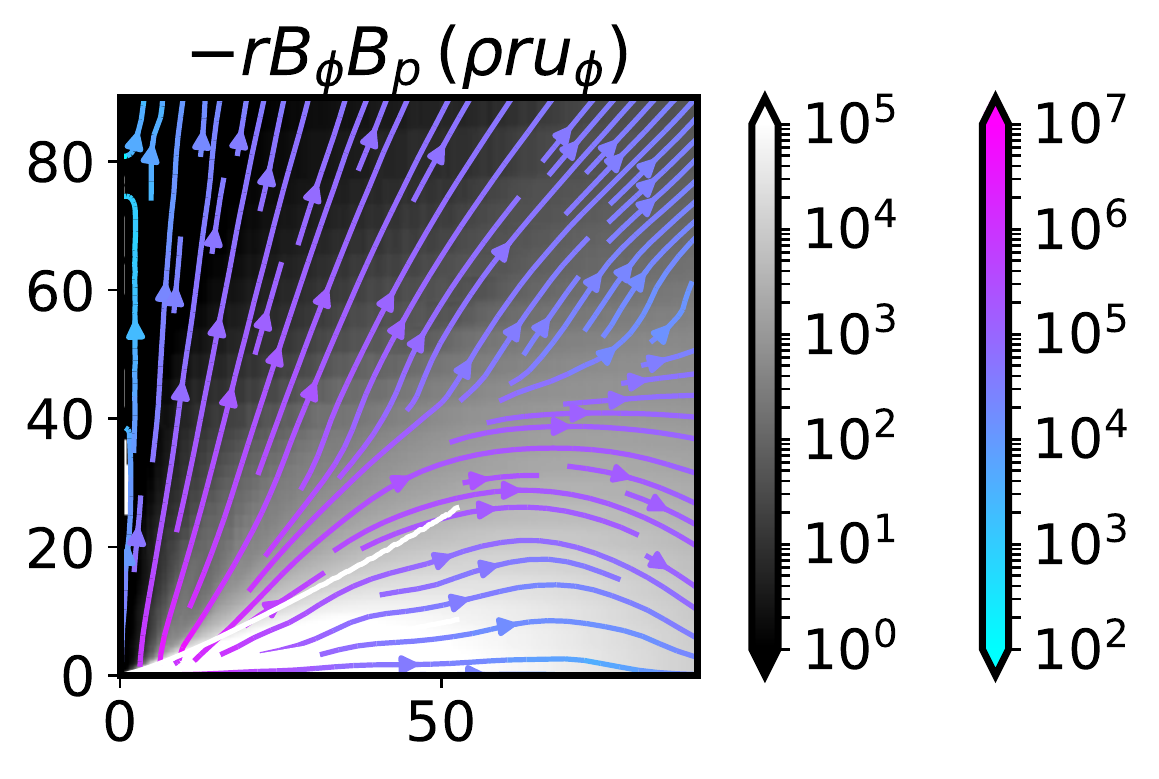}}
\put(0.47,0){\includegraphics[trim=7 5 8 5,clip,width=0.26\textwidth]{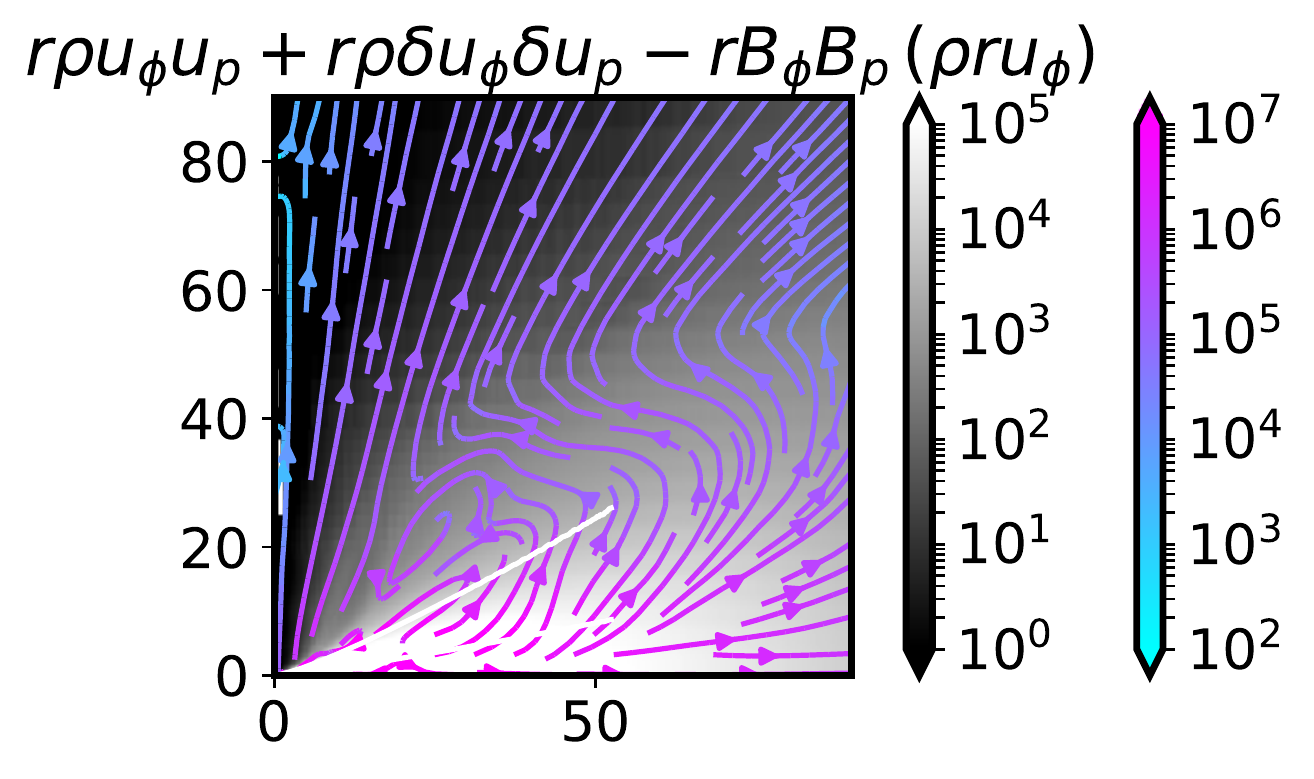}}
\end{picture}
\caption{Angular momentum flux and angular momentum ($r\langle \rho u_\phi \rangle_\phi$) in the disk surrounding (color streamlines in the gray image, in SI units) for R\_$\ell$14 ({\it left}) and R\_80ky\_$\ell$18 ({\it right}) at 138 kyr. The disk scale height $H$ is shown in white. 
All quantities are azimuthally averaged and averaged between the upper and lower planes. 
{\it From top to bottom:} Laminar transport ($r \langle u_\phi \rangle_\phi \langle \rho\vec{u}_{\vec{p}}\rangle_\phi$), turbulent transport ($r\langle\rho \delta u_\phi \delta \vec{u}_{\vec{p}}\rangle_\phi$), magnetic transport ($-r\langle B_\phi \vec{B}_{\vec{p}}\rangle_\phi/(4\pi)$), and their sum. }
\label{fig_jflux_rest3}
\end{figure}

The commonly accepted scenario describes the loss of angular momentum due to various transport mechanisms in the envelope, 
which allows the matter to be accreted within the disk. 
The equation of angular momentum transport is obtained by multiplying the $\phi$ component of the momentum conservation equation by $r$ and averaging azimuthally, which gives 
\begin{align}
{\partial \over \partial t} \langle \rho r u_\phi \rangle_\phi + \vec{\nabla} \cdot \left[ \langle r \rho \vec{u}_{\vec{p}}  u_\phi - r {\vec{B}_{\vec{p}} B_\phi  \over 4\pi} \rangle_\phi \right] = 0.
\end{align}
The second term is the divergence of the poloidal flux of angular momentum. 
By separating the toroidal velocity into the circular and fluctuating components $u_\phi = r\Omega + \delta u_\phi$, 
where $r\Omega = \langle \rho u_\phi \rangle_\phi / \langle \rho \rangle_\phi$, 
we can rewrite the angular momentum flux as
\begin{align}
\mathcal{F}_j &= \langle r \rho \vec{u}_{\vec{p}}  u_\phi- r {\vec{B}_{\vec{p}} B_\phi  \over 4\pi} \rangle_\phi \\
& =  r^2 \Omega \langle  \rho \vec{u}_{\vec{p}} \rangle_\phi + r \langle  \rho \delta\vec{u}_{\vec{p}} \delta u_\phi \rangle_\phi  - r {  \langle \vec{B}_{\vec{p}}B_\phi  \rangle_\phi  \over 4\pi}.  \nonumber 
\end{align} 
The first term is the laminar transport, which describes the the angular momentum flux associated with the mass flux if the momentum is to be frozen within the mass. 
The second term is the turbulent transport term that originated from the cross correlation of the fluctuations. 
This is the main transport mechanism in the $\alpha$-disk model that allows the disk to accrete. 
The last term is the magnetic transports associated with the Maxwell stress tensor, which tends to deviate the movement along the field line direction, 
thus accelerating or braking the rotation. 

In the case of an isolated disk at later phases, the focus lies more on the angular momentum within the disk, and there is no important net mass flux. 
The tensor terms are therefore of main interest. 
In this study, however, the disk actively receives mass from the infalling envelope and the laminar transport must not be neglected. 
Figure \ref{fig_jflux_rest2} shows the laminar, turbulent, magnetic, and total angular momentum fluxes (from top to bottom) at 103 kyr at two resolutions (left and right). 
From the pattern of angular momentum transport, the disk-envelope system can be divided into four zones (see Fig. \ref{fig_cartoon} for an illustration). 

\begin{itemize}
\item The braked infalling layer: This region is dominated by the vertical upward transport of angular momentum by the magnetic field while the velocity is downward. This means that during the infall, the gas is braked by the magnetic field that is not aligned with the flow. This is the dark zone in Fig. \ref{fig_mflux_rest2} where the specific angular momentum is lowest and the mass is strongly infalling.
\item The free infalling layer: The region contained within a few scale heights above the disk surface is dominated by the free infall of mass. The turbulent and magnetic stress tensors transport the angular momentum upward, while this is dominated by the laminar flow that directly advects the angular momentum associated with the gas. In this region, the magnetic field is mostly lost through ambipolar diffusion. This zone is clearly seen by comparing Figs. \ref{fig_mflux_rest2} and \ref{fig_jflux_rest2}, where the mass flux and the total angular momentum flux are almost in the same direction. 
\item The wind cavity: This low-density region is dominated by the vertical upward transport of angular momentum by the magnetic field. Matter moves outward along field lines, and its angular momentum increases. 
\item The disk interior: The disk interior (below a few $H$) is dominated by the laminar transport of the angular momentum. This is linked to the global behavior of the disk and has to be investigated with great care. While radially outward magnetic transport and vertical Reynolds transport exist as well, these two contributions are weak. We refrain from overinterpreting the results because the low- and high-resolution runs are not consistent at this scale. 
\end{itemize}
Throughout the braked layer and the wind cavity, the magnetic angular momentum flux is almost constant. This means that the field lines channel out the angular momentum in an almost stationary way and eventually launch it away through the wind. 

Figure \ref{fig_jflux_rest3} 
shows the same plots at 138 kyr, 
where the disk grows to a larger size.  
The general behavior is the same, while we have better resolved disk internal dynamics, especially in R\_80ky\_$\ell$18. 
The disk midplane at this moment expands. 
It is clear that the angular momentum transport is largely dominated by the laminar transport that simply advects the angular momentum along with the mass. These results show that 
the $\alpha$-disk model is probably not suitable for an embedded disk 
because the disk is not in a stationary state and is very sensitive to irregularities in the infalling envelope. 

\subsubsection{Infall onto the disk: Source function}\label{st_source}

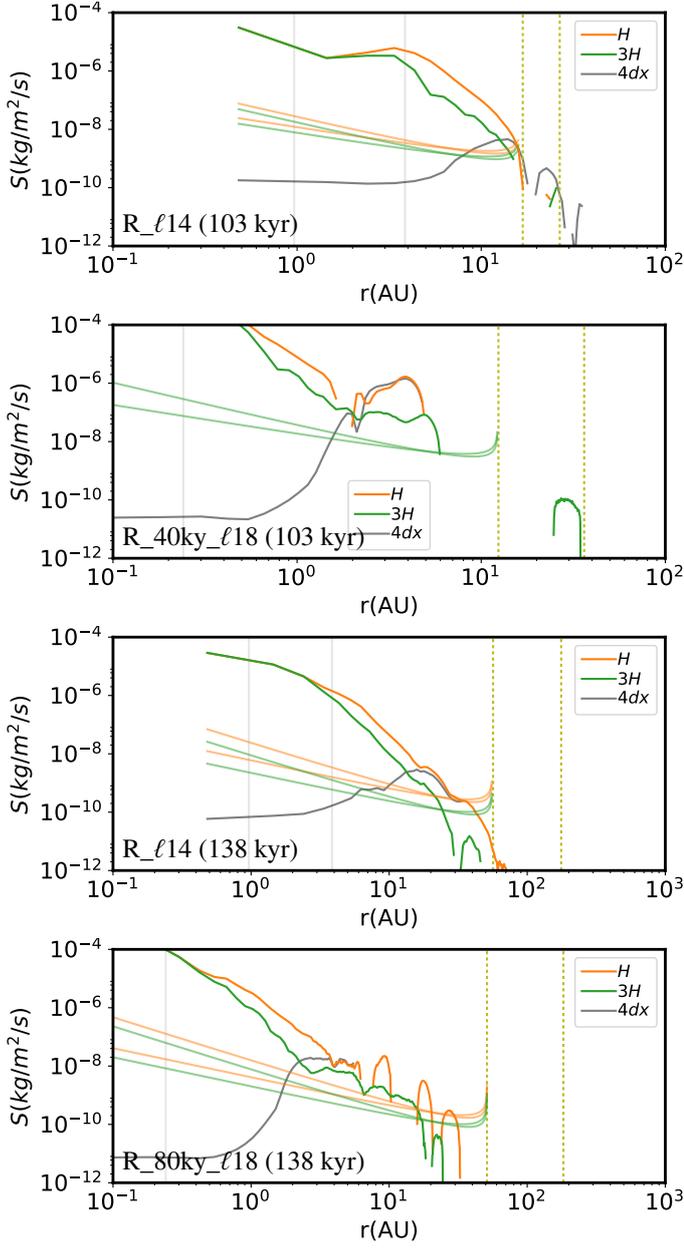
\begin{figure}[]
\centering
\setlength{\unitlength}{0.5\textwidth}
\begin{picture}(1,1.8)
\put(0,1.35){\includegraphics[trim=0 0 0 6,clip,width=0.5\textwidth]{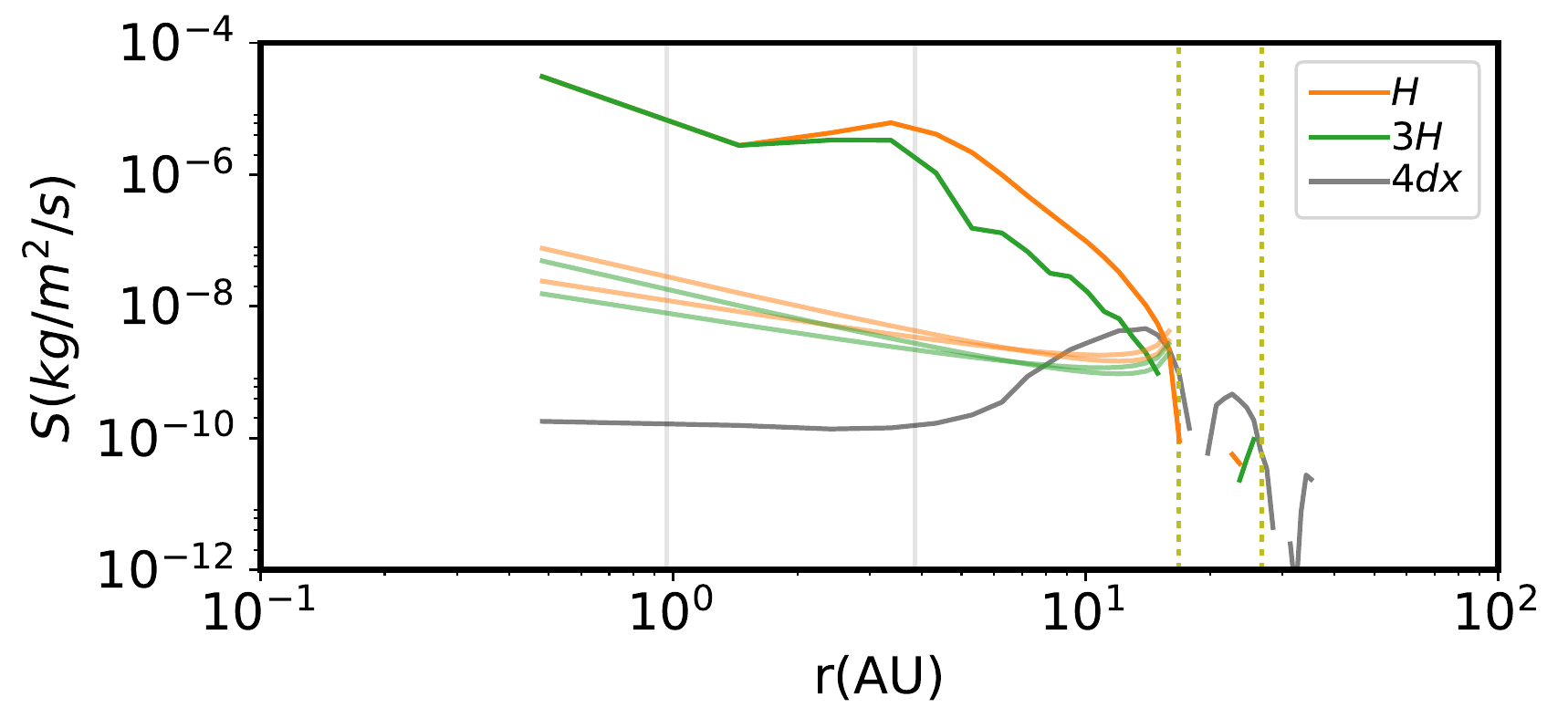}}
\put(0,0.9){\includegraphics[trim=0 0 0 6,clip,width=0.5\textwidth]{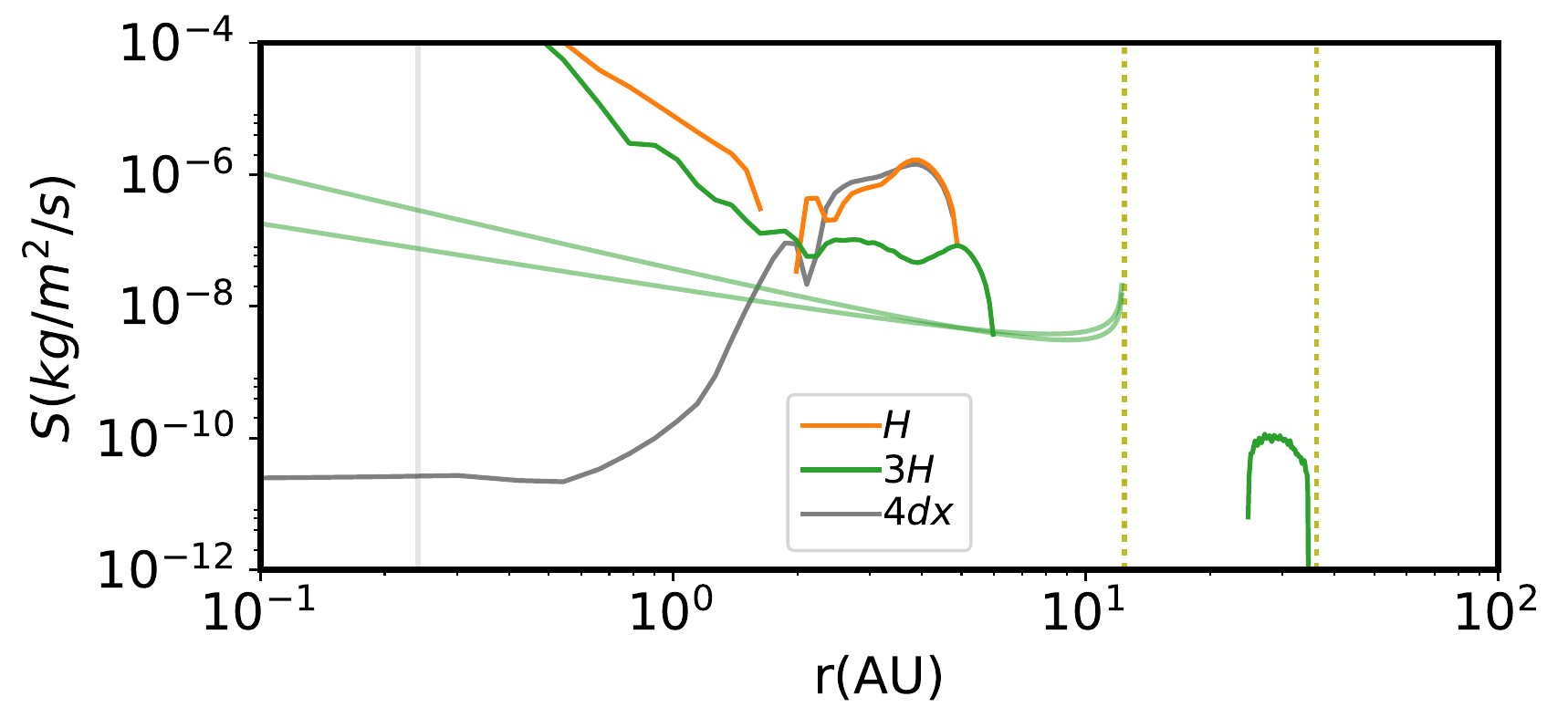}}
\put(0,0.45){\includegraphics[trim=0 0 0 6,clip,width=0.5\textwidth]{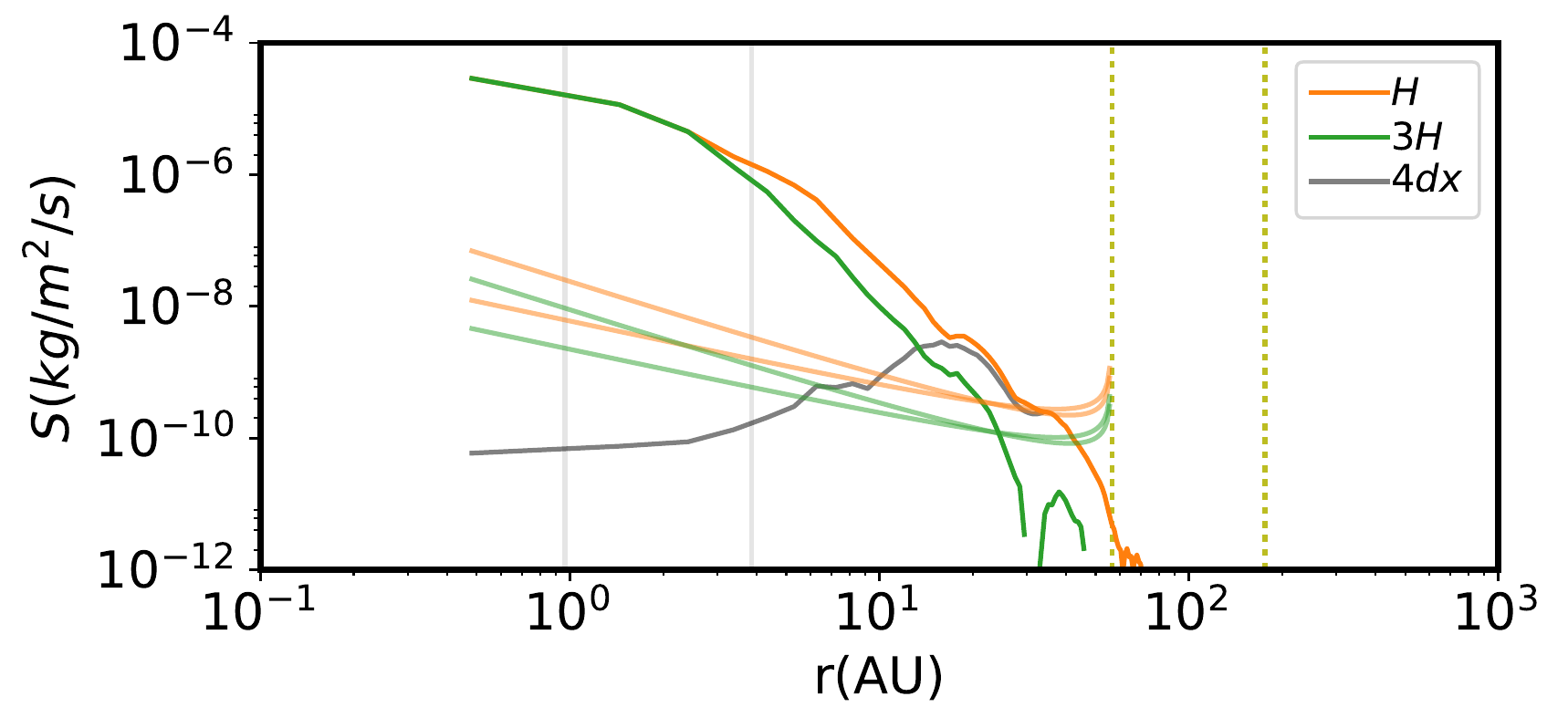}}
\put(0,0){\includegraphics[trim=0 0 0 6,clip,width=0.5\textwidth]{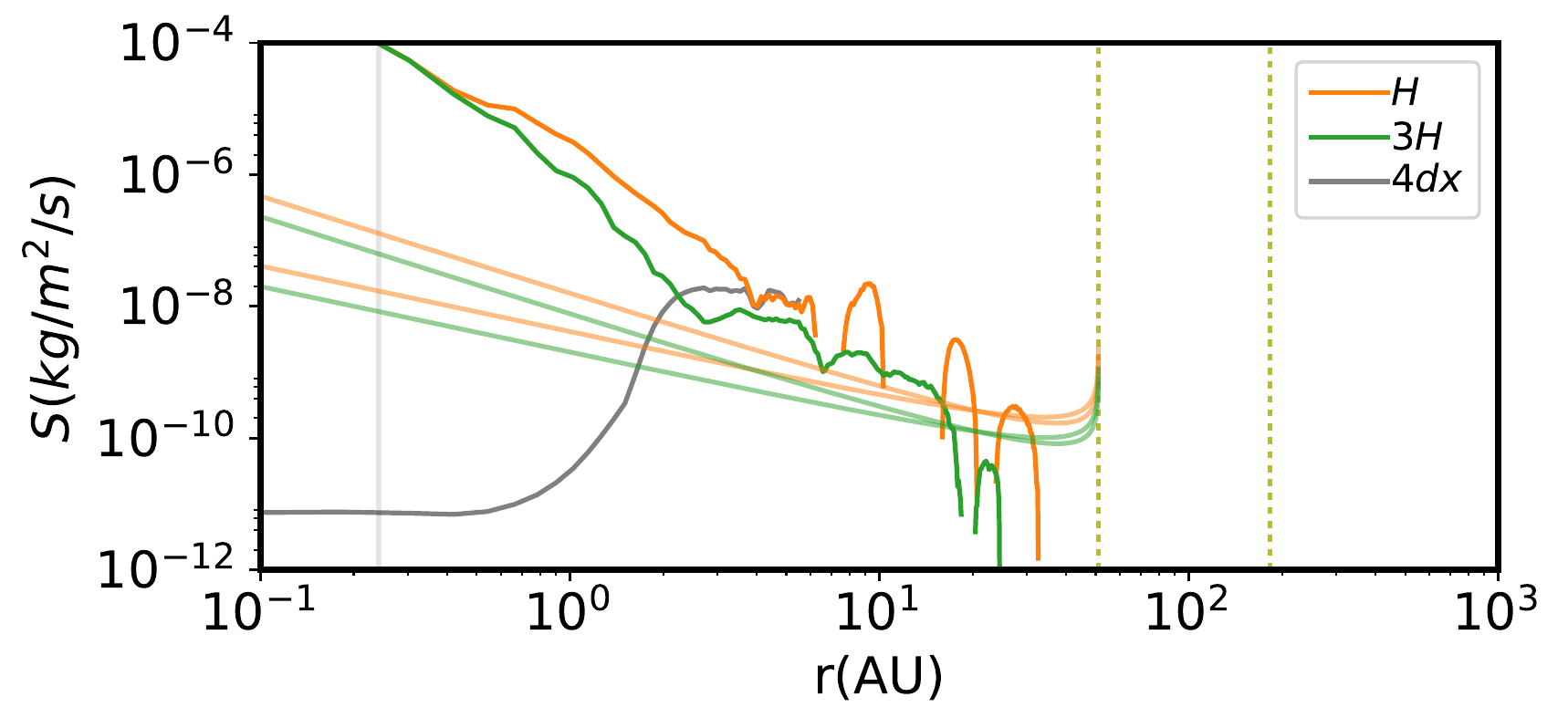}}
\put(0.18,1.47){R\_$\ell$14 (103 kyr)}
\put(0.18,1.02){R\_40ky\_$\ell$18 (103 kyr)}
\put(0.18,0.57){R\_$\ell$14 (138 kyr)}
\put(0.18,0.12){R\_80ky\_$\ell$18 (138 kyr)}
\end{picture}
\caption{Source function $S(r)$ measured at $H$ (orange) and $3H$ (green) in R\_$\ell$14 and R\_40ky\_$\ell$18 at 103 ky, R\_$\ell$14 and R\_80ky\_$\ell$18 at 138 ky ({\it from top to bottom}). The flux measured at the horizontal plane $z=4dx$ is shown in gray. The curves in lighter colors are the analytical functions from pure hydrodynamic predictions that have the same amount of total flux (see Fig. \ref{fig_source_classical}). The source function measured in the simulations is much steeper than that of the classical model, and strong infall occurs in the central part of the disk. At a given radius, the flux is generally greater at lower altitudes, indicating that there is a strong radial flux in the interior of the disk. At small radii (below the sink accretion radius), the measured flux at $H/3H$ is no longer realistic, and we used the measurement at $z=4dx$ to calculate the total flux. }
\label{fig_Source}
\end{figure}

The source function, $S(r)$, in units of mass per unit time per unit surface, 
describes the rate and the spatial distribution of mass arrival from the envelope onto the disk surface.  
The surface we refer to when we discuss the mass flux is always the projection of the disk surface onto the equatorial plane, 
which is equal to the actual (inclined) disk surface multiplied by the $\cos(\theta)$ of the local inclination angle. 
This choice is made such that the description is consistent with the classical model that regards the disk as infinitely thin \citep[e.g., ][]{Hueso05}. 

The disk scale height $h_{\rm p}$ (see Sect. \ref{st_height}) was used to define the surface at which the source function was evaluated. 
To avoid numerical errors due to the small-scale fluctuations of $h_{\rm p}$, 
we first fit the scale height to a power-law expression,
\begin{align}
H = \eta \left( {r \over 1~ {\rm AU}} \right)^p, 
\end{align}  
such that it became a monotonically increasing smooth function. 
The fitted parameters are shown in Fig. \ref{fig_H}. 
The disk aspect ratio is almost constant at 10 \%.
We chose to measure the source function at $h_{\rm s} = H$ and  $h_{\rm s} = 3H$. 
The source function is therefore measured as 
\begin{align}
S(r) &= - \rho (r,z) \left[u_z (r,z) - u_r  (r,z) \tan(\theta(r))\right] \\
&= - \rho(r, h_{\rm s}(r)) \left[ u_z (r, h_{\rm s}(r)) - u_r  (r, h_{\rm s}(r)) {dh_{\rm s} \over dr}(r) \right]. \nonumber
\end{align}

Figure \ref{fig_Source} shows snapshots of the measured $S(r)$ at $H$ (orange) and $3H$ (green) at two resolutions. 
Closer to the equatorial plane, the source function is more centrally concentrated, as we discussed earlier; this is due to the converging flow pattern (see Sect. \ref{st_alpha} and \ref{st_jtransport}). 
When we compare this with the source functions from \citet{Ulrich76} and \citet{Hueso05}, 
the measured $S(r)$ is much steeper, 
meaning that the infall onto the disk is more centrally concentrated for our simulated disk. 
Particularly, this means that most of the disk at $r \gtrsim 3-4$ AU does not receive much material and a significant amount of mass falls straight onto the central star. 
We note that the choice of reference plane for the mass flux measurement is not a simple matter. 
The lower panel of Fig. \ref{fig_Source} shows that the mass flow crosses the disk surface and moves upward, giving negative flux. 
A reference surface therefore has to be sufficiently far away from the midplane to avoid being affected by the disk internal motions.  

Moreover, because the size of the sink particle accretion radius is finite, the flux within the disk scale height at small radius is not physical. 
We therefore measured the flux at the horizontal surface $z=4dx$ (shown in gray). 
It is evident that the extreme high flux values at small radii are not physical. 
Very close to the disk center, the measurement of the vertical and horizontal flux becomes sensitive to the exact selection of the axis, and there might be some confusion between the two. 
When we calculated the total mass flux, we therefore used $\min(4dx,H)$ for the disk surface. 
This introduces some uncertainties in the measurement, but it is the best we can do at this given resolution with the constraints from the algorithm. 

\subsubsection{Accretion inside the disk}\label{st_Mdot}

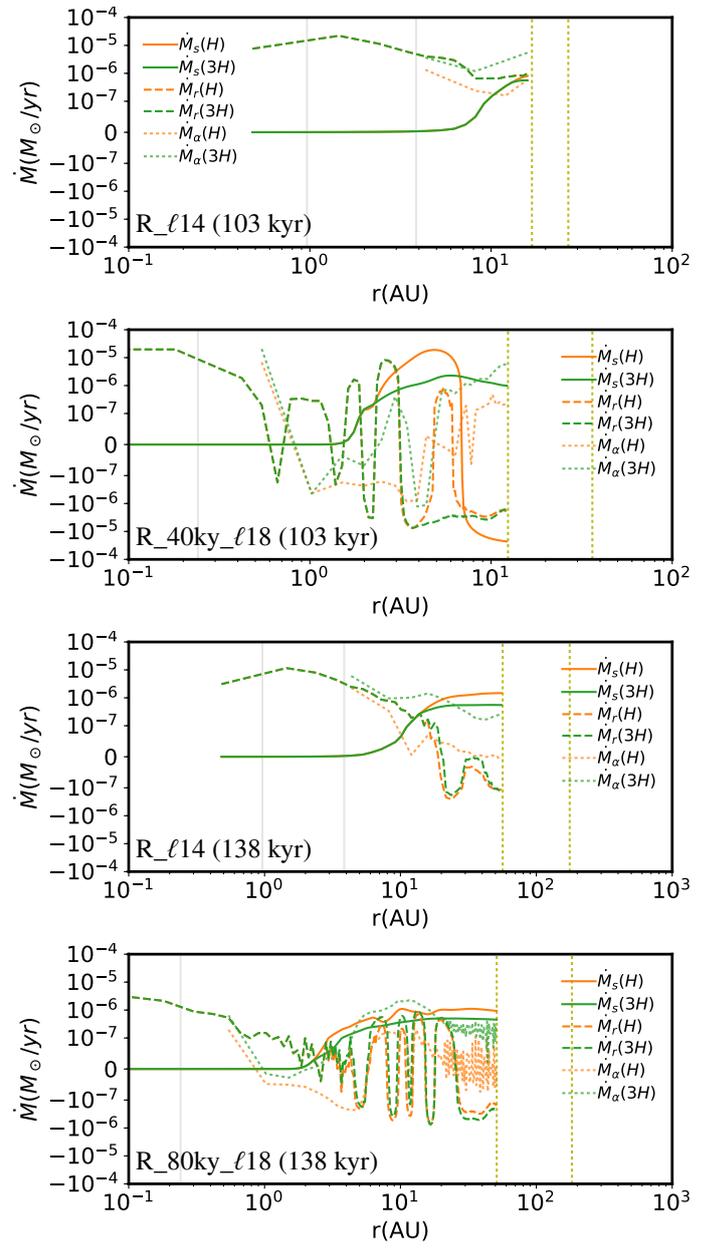
\begin{figure}[]
\centering
\setlength{\unitlength}{0.5\textwidth}
\begin{picture}(1,1.8)
\put(0,1.35){\includegraphics[trim=0 0 0 6,clip,width=0.5\textwidth]{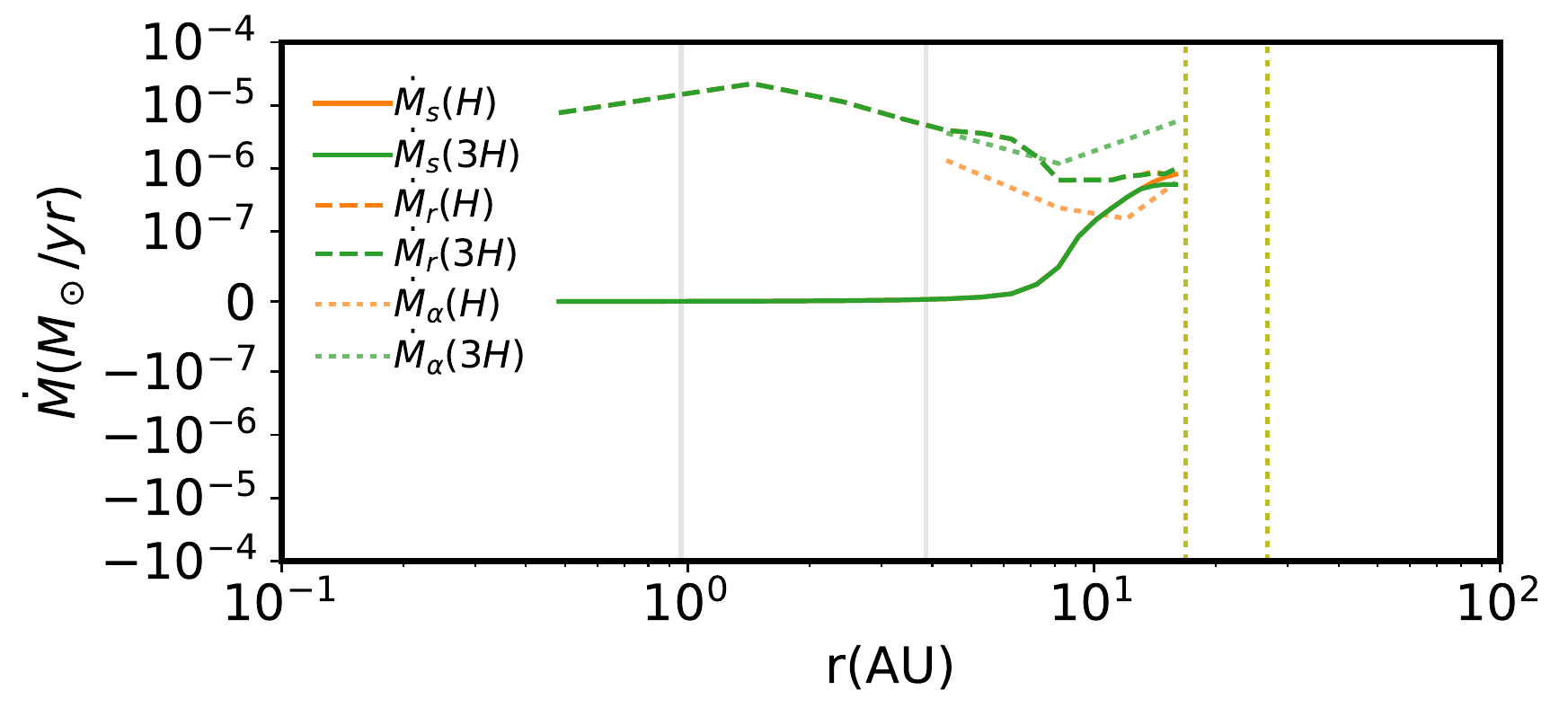}}
\put(0,0.9){\includegraphics[trim=0 0 0 6,clip,width=0.5\textwidth]{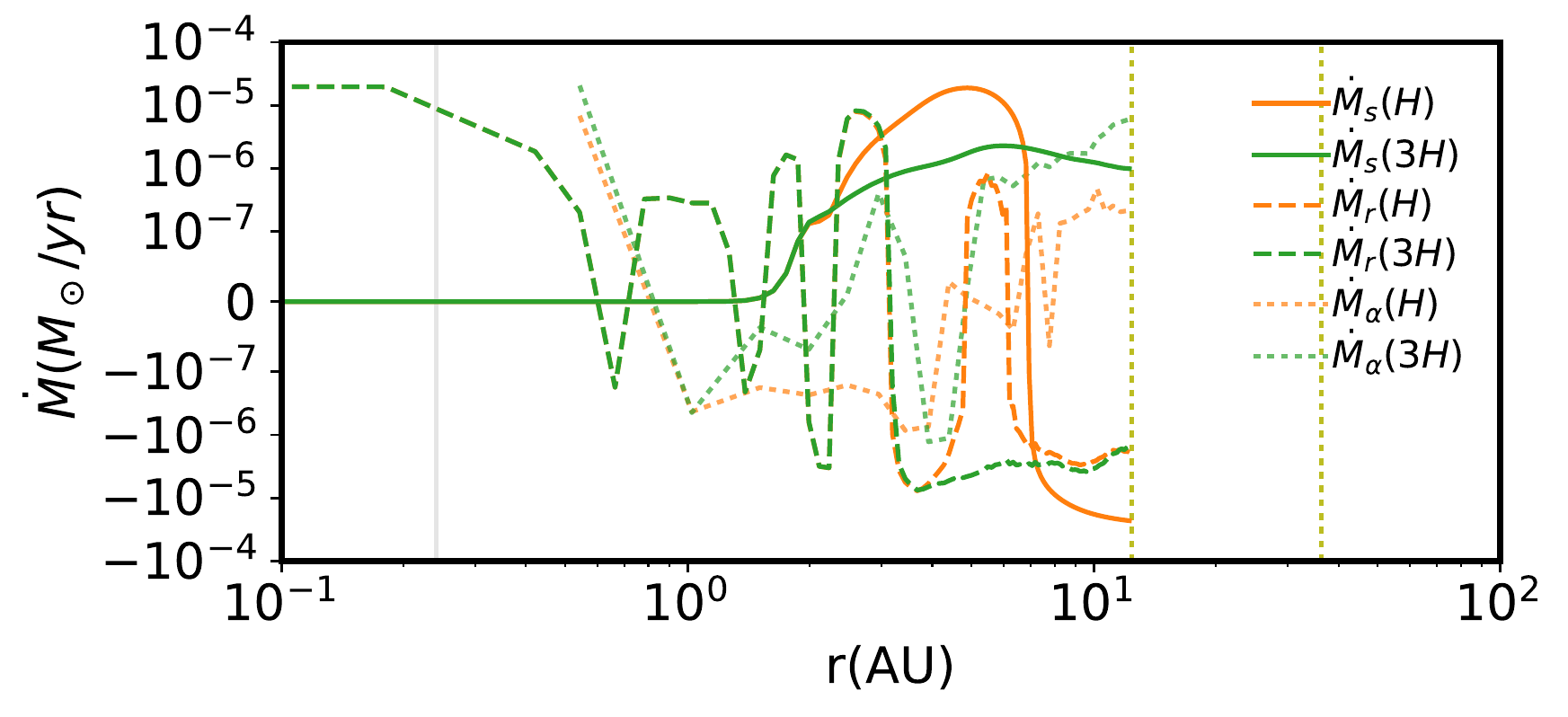}}
\put(0,0.45){\includegraphics[trim=0 0 0 6,clip,width=0.5\textwidth]{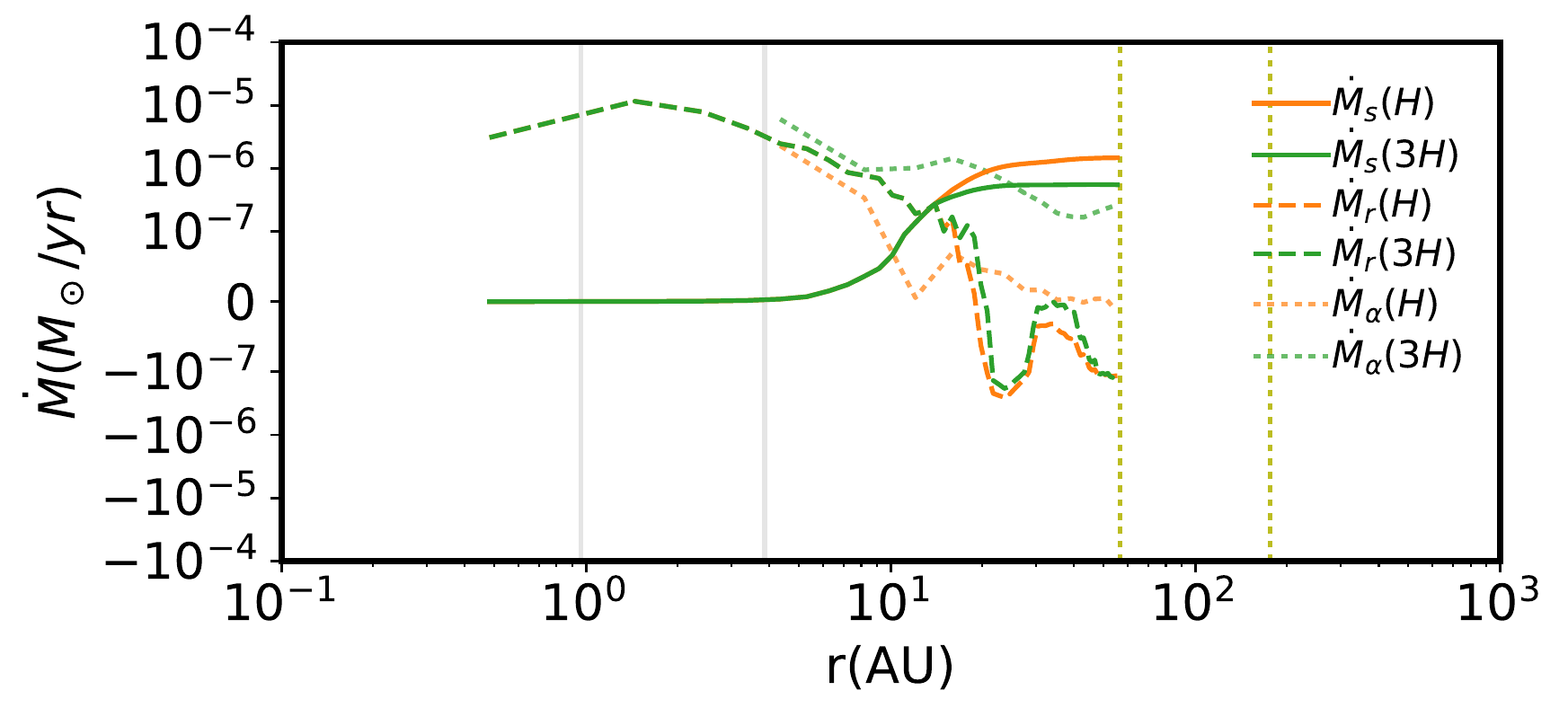}}
\put(0,0){\includegraphics[trim=0 0 0 6,clip,width=0.5\textwidth]{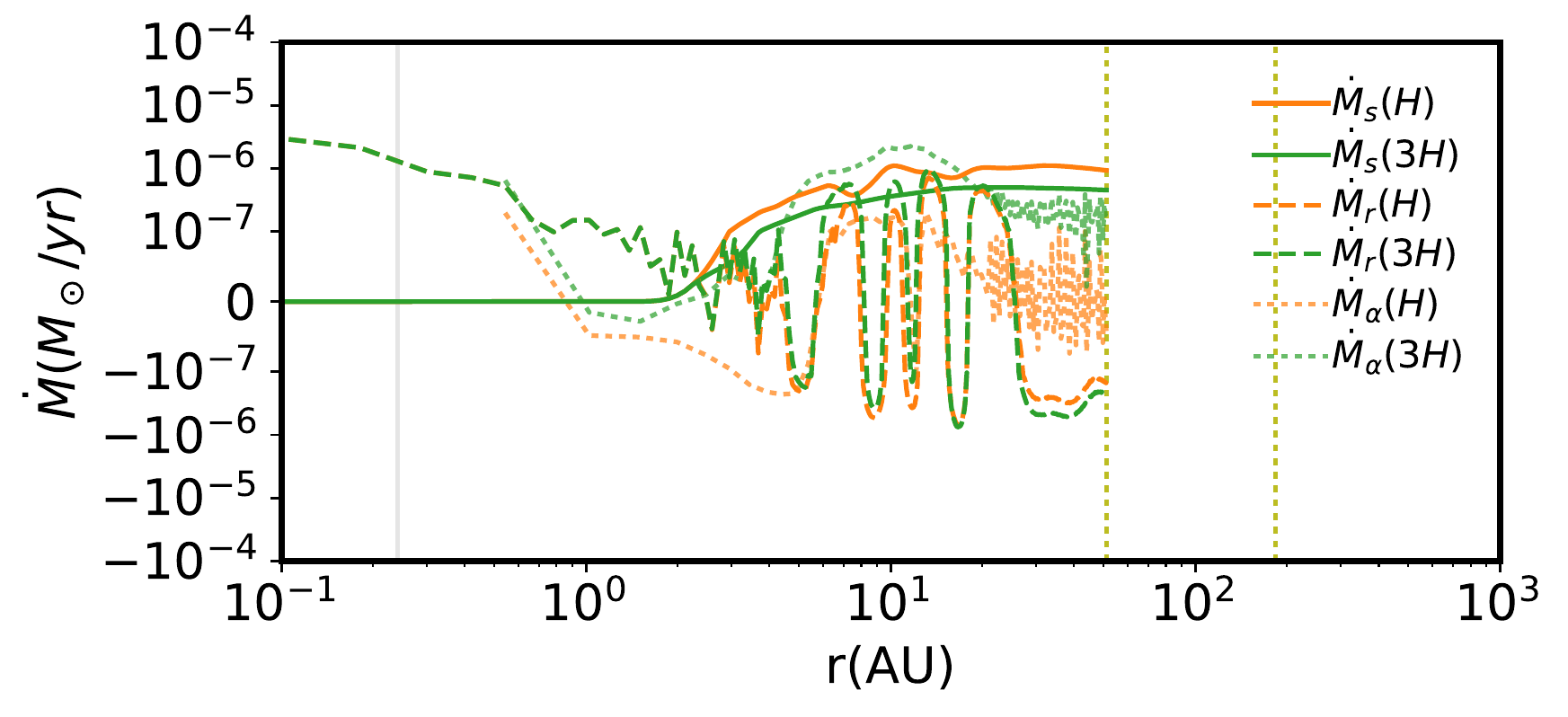}}
\put(0.19,1.47){R\_$\ell$14 (103 kyr)}
\put(0.19,1.02){R\_40ky\_$\ell$18 (103 kyr)}
\put(0.19,0.57){R\_$\ell$14 (138 kyr)}
\put(0.19,0.12){R\_80ky\_$\ell$18 (138 kyr)}
\end{picture}
\caption{Mass accretion rate measured across the disk surface ($\dot{M}_s (<r) $), measured radially inside the disk ($\dot{M}_r(r)$), and inferred from the measured $\alpha$  ($\dot{M}_\alpha(r)$) for R\_$\ell$14 and R\_40ky\_$\ell$18 at 103 ky, R\_$\ell$14 and R\_80ky\_$\ell$18 at 138 ky ({\it from top to bottom}). Both positive and negative values with $|\dot{M}| > 10^{-7}$ are plotted in logarithmic scale, and this is connected with a linear range in between. Most of the surface flux arrives where the slope of $\dot{M}_s (<r) $ is the steepest. Below this radius, the disk readily accretes in the radial direction only at a very small radius (<1AU), and the outer part can have complex motions. The radial mass flux derived from $\alpha$ is not exactly the same as $\dot{M}_r(r)$, although the amplitudes are comparable. }
\label{fig_Mdot_disk}
\end{figure}

The straightforward measure of the mass accretion rate, $\dot{M}_r$, is the mass that crosses radius $r$ inward per unit time. 
This is calculated as
\begin{align}
\dot{M}_r(r) = - 2\pi r \sum_{-h_{\rm s}}^{h_{\rm s}} \langle \rho u_r(r,z) \rangle_\phi dz,
\end{align}
where $\rho u_r$ is the local radial mass flux. 
This radial accretion rate is shown in Fig. \ref{fig_Mdot_disk} with dashed curves for the integrated values within $H$ (orange) and $3H$ (green).
At small radii, the radial flux within $\pm 4dx$, which is larger than the scale height, is shown, and therefore the green curves are superimposed on the orange curves.
As discussed earlier, the disk can sometimes be in partial expansion (negative $\dot{M}_r(r)$), 
while in general, the inner part of the disk (below $\sim 5$ AU) is normally accreting, with small local fluctuations. 
The discrepancy between the $H$ and $3H$ curves shows that the radial accretion within the disk mostly occurs in the upper layers.
This is shown in the two lower panels of Fig \ref{fig_Mdot_disk}. 
Beyond $\sim 10$ AU, the green ($3H$) curve is often above the orange ($H$) curve, meaning more accretion (positive) or less decretion (negative). 
In the region slightly beyond the sink accretion radius (second vertical gray line), the value of $\dot{M}_r(r)$ increases toward the center of the disk in all four panels, 
meaning that either the disk is being emptied or that there is a surface source.

To examine this, we also calculated the surface mass accretion rate, that is, 
the source function integrated over the disk surface. 
We chose to perform this integration between $0$ and $r$, giving
\begin{align}
\dot{M}_s(r) = \sum^{r}_{0} S(r) 2\pi r dr.
\end{align}
This is shown in the same figure with solid curves for source functions measured at $H$ (orange) and $3H$ (green).
Because of the numerical limits described above, we used the flux at the surface $z=4dx$ when $H$ was smaller than this value. 
For R\_$\ell$14 at 103 kyr, the disk is so small that $H<4dx$ for almost the entire disk. 

Most of the mass penetrates the disk surface where the slope of $\dot{M}_s(r)$ is the steepest. 
The total mass that falls onto the disk, with significant contribution from this zone at a few AU, is broadly comparable with the mass that falls radially inward onto the star ($\dot{M}_r(r)$ at small radii). 
As a general remark, the infalling material from the envelope can reach quite a small radius of a few AU before it penetrates the disk surface. 
Inside the disk, the inward accretion is well established only for the innermost of the disk, 
while some of the mass that falls from the surface onto the disk might move outward. 
This expansion reaches a shock generated by the radial infall form the envelope near the equatorial plane and brings the material to high altitudes, 
possibly generating a circulation in the outer part of the disk. 

Radial accretion behavior is present in all panels in the marginally resolved region, that is, a few times the sink accretion radius. Nonetheless, we would like to caution that this zone is primarily dominated by the numerical viscosity, and this is probably why the accretion is readily established. 
This should not have strong consequences on the long-term stellar mass growth because the infall from large scales is the dominant factor. However, this numerical effect does determine whether the evolution of the star-disk system can be described on short timescales, which remains a computational challenge \citep[e.g.,][]{Kuffmeier18}. 
In the outer part of the high-resolution runs, both restarts show oscillating radial motions. 
This again supports our proposition that the disk is highly sensitive to the environment, while the role of the disk as a mass buffer between the envelope and the star is less evident. 

Figure \ref{fig_Mdot_disk} also shows $\dot{M}_\alpha$, the mass accretion rate estimated from the $\alpha$ parameter (see Sect. \ref{st_alpha}, Eq. (\ref{eq_Mdot_alpha})).
The values of $\dot{M}_r$ and $\dot{M}_\alpha$ do not match perfectly. 
However, they are similar in magnitude. 

\section{Comparison to standard alpha-disk models}\label{st_adisk}
The present paper shows that the disk formed during the infall of the dense core has a number of similarities to but also differences with classical isolated disks assumed in  standard alpha-disk models. We report the most notable similarities below.

\begin{enumerate}
\item{The Keplerian disk structure is established rapidly in less than 61 kyr, a timescale shorter than the envelope infall timescale.}

\item{The disk mass is about $1\%$ of the stellar mass after about 80 kyr (Fig.  \ref{fig_Mdisk}). This is a typical disk-to-star mass ratio \citep[like for the MMSN, see, e.g.,][]{Hayashi81}. However, we note that the stellar mass is only $0.2~\Ms$  at the end of the simulation so that we cannot confirm at the moment that this results is valid up to one solar mass. It is therefore premature to state that there is enough mass to build the S. S. Moreover, we caution again that the disk mass is (unfortunately) sensitive to the disk inner boundary condition ($n_{\rm acc}$). }

\item{The midplane of the disk is almost devoid of MHD turbulence with $\alpha_r < 10^{-4}$. This is in agreement with previous studies arguing that the midplane of a protoplanetary disk should be dead because of lack of ionization \citep[see][for a review]{Turner14}. Several authors showed that low values of $\alpha$ in the disk midplane can lead to efficient planetesimal formation \citep{Drazkowska14, Drazkowska18, Charnoz19}. The resulting disk might be the birthplace of a massive planetesimal population. However, we caution that the Toomre  $Q$ parameter is high in our simulation. Planetesimal formation can be allowed if $\Sigma$ is higher in reality (see Sect. \ref{st_Q}), or if the metallicity is enhanced toward the disk midplane with dust enrichment as high as a factor $\simeq 2$, as indeed suggested recently by \citet{lebreuilly2019,lebreuilly2020}.}
\end{enumerate}

We report the notable differences with classical alpha-disk models below.
\begin{enumerate}
\item{
The disk in our simulation is truncated at about 20 AU through magnetic braking, consistent with \citet{Hennebelle16}. 
This may imply that the classical picture of a viscously spreading disk up to several dozen or 100 AU
is maybe not the general rule, whereas a disk like this does exist \citep[see, e.g., ][]{Isella10}. 
However, recent ALMA observations showed that the majority of class 0 disk are likely small \citep{Andrews18,Maury19}. 
The absence of viscous spreading may preclude the outward transport of CAIs because of outward viscuous relaxation of the disk \citep{Yang12,Pignatale18}. 
However, alternative solutions could exist because transient outward-directed flows are detected in the disk (see Sect. \ref{st_flow}) and might drive CAIs outward from the protostar. 
Alternatively, the disk might expand in a later phase when the envelope is almost exhausted. 

Interestingly, the truncation of the disk at 20 AU agrees with some versions of the Nice model, 
as well as the more generalized giant planet instability model that appeared later. 
In the recent view of S. S. formation, the giant planets formed in
a much more compact configuration than today, within about 15-20 AU of
the Sun \citep[see, e.g.,][]{Gomes05,Tsiganis2005, MorbidelliNesvorny2020}. The
main argument for this is the necessity of forming a low-mass Kuiper Belt
with a correct orbital architecture in which an accumulation of bodies
is observed in mean-motion resonance with Neptune. To reach the
observed state, it is necessary to invoke an outward migration of
Uranus and Neptune starting from 15-20 AU. When the gas
disperses, the four giant planets are still in a compact configuration,
with Jupiter and Saturn close to their 3:2 mean motion resonance. This
scenario agrees with disk and planet evolutionary models \citep{Bitsch15}, at least for Jupiter and Saturn.

The Nice model originally aimed to explain the lunar bombardment \citep{Gomes05}. The key idea is that the orbits of giant planets were unstable during the S.S.  evolution. If the giant planets were formed in a more compact configuration than today \citep[as suggested in the Nice model, or as a result of evolution within the protoplanetary disk, see][]{KleyCrida08,Walsh11}, then they should at some moment have experienced a global dynamical instability, which led to a very short and violent reorganization of their orbits \citep{Deienno17,Clement18}. During this rearrangement, small-body belts were destabilized, resulting in massive bombardment of nearby planets. For example, the outward migration of Neptune may have pushed the original Kuiper belt outward, which might initially have been contained within about 20 AU and not at the current position of about 50 AU. It is therefore possible that the young S.S. was more compact than today, and reached its current dimensions after such an instability \citep{Levison08}. 

In the original version of the Nice model, the giant-planets instability occurred at 800 Myr after CAIs. 
More recent models of giant planet instability, using the configuration of the four giant planets at the
end of the solar nebula obtained from hydrodynamical simulations of
planet migration, also converge to a formation of Neptune within 25 AU
\citep{NesvornyMorbidelli2012}.
\citet{Ribeiro20} suggested that this event could have occurred as early as in the first million years. 
Forming Uranus and Neptune at 30 AU and beyond would have disrupted
the Kuiper belt and in particular the population of dynamically cold
objects (low orbital eccentricity and inclination). This suggests that the disk itself was
initially in a compact configuration and presumably did not extend
to 100 AU.

Nonetheless, the disk in our simulation also experienced a rapid grow in size after $\sim 120$ kyr. This suggest that the disk size is sensitive to the properties of the infalling material, which can be highly variable in time \citep[see also][]{Vorobyov15a, Kuffmeier17}. Therefore more constraints are needed to reconstruct the history of the S. S.
}
 
\item{The formed disk is significantly hotter than classical isolated disks. 
The snow line starts at about 10 AU and shifts inward down to 5 AU, 
where it then remains constant (until the end of the simulation) while the accretion drops to about $10^{-7} \Ms/{\rm yr}$ .
This temperature profile is notably higher than in passive radiatively heated disks \citep[see, e.g.,][]{Chiang97}, 
where the snow line falls around 2 AU (independently of the accretion rate, and assuming a $2.5 ~R_\odot$ star with $T=4000$K),  
or isolated disk with both viscous and radiative heating: for example, \citet{Bitsch15} reported that the snow line lay between 2 and 5 AU for an accretion rate lower than or equal to $10^{-7}\Ms/{\rm yr} $, consistent with the detailed model of \citet{DAlessio98}, who reported a snow line around 2 AU for an accretion rate of $\sim 10^{-7}\Ms/{\rm yr}$. 
A snow line located at 5 AU would be the ideal position to trigger fast planetesimal formation of the Jupiter core, as it has been demonstrated that streaming instability and planetesimals formation is strongly enhanced at the snow line \citep{Drazkowska17, Charnoz19}.

Nonetheless, the temperature of the disk should be interpreted with caution because it is subject to some model uncertainties.  
First, the prescription by \citet{Kuiper13} was used to describe the protostellar irradiation on the Hosowaka track. This heating might be excessive and overheats the surrounding of the protostar. Fig. \ref{fig_snowline} shows that the snow line is shifted to 2-3 AU in the absence of this heating term. 
Second, the the radiative transfer was treated with a flux-limited diffusion scheme, which is mostly valid for infrared radiation, while optical photons may escape more easily. 
In reality, the absorption of optical radiation by dust  and the re-emission in infrared wavelengths is not complete and needs to be further investigated with care.
}

\item{The disk surface density does not follow the classical -1.5 exponent of the MMSN \citep[][]{Hayashi81} but is closer to -0.5, with a sharp drop beyond $R_{\rm kep}$. This recalls observed protoplanetary disks, see, for instance,  \citet{Isella10}, who suggested that the surface densities have an exponent >  -1 and that the outer edge of the disks is between 23 and 27 AU.  }

\end{enumerate}

\section{Comparison with similar numerical works}\label{st_comp_simu}

We can draw essential conclusions from previous numerical studies and our current work. 
Primarily, the disk sensitively responds to its environment, that is, the collapsing prestellar core. Therefore class 0/I disks should not be studied in the same way as isolated disks. 
Second, whether the disk has significant effects on the evolution of the star remains debatable. 
Bursty mass accretion may result from fragmentation within the disk, while the numerical origin of these gravitational instabilities is not yet completely ruled out. 
An alternative way of regarding the disk as an accessory of the protostar and an inevitable existence due to residual angular momentum is emerging. 
Evolution of the star is determined mostly from the core scale and is partly mediated by the disk. This deviates from the widely accepted view that mass first falls onto the disk and then the star accretes from the disk.

\citet{Vorobyov10, Vorobyov15b} and \citet{Vorobyov15a} employed a 2D thin-disk approximation. These models cannot describe the mass accretion that reaches the star through upper layers and everything seems to transit through the disk. In the embedded phase, the stability conditions in the disk are mostly reflected in the fluctuations (bursts) of the mass accretion rate, while the accretion rates from the envelope onto the disk and that from the disk onto the star are broadly consistent  (for class 0/I, where envelope infall is significant). 

\citet{Kuffmeier18} resimulated small regions around six sink particles selected from a giant molecular cloud simulation at increased resolution of 2 AU for 200 kyr. One of these runs was further evolved for 1000 yr at 0.06 AU resolution. These parameters are similar to ours, and we share some common findings. First, the overall mass accretion history is likely determined by large-scale flow properties \citep[see also][]{Kuffmeier17} instead of the viscous accretion process in the disk. Second, the sink mass accretion mediated through the disk is bursty, although it might be of numerical origin. Finally, significant mass infall arrives onto the star without transiting radially through the disk.

\section{Conclusions}\label{st_conc}

We have simulated the collapse of a magnetized prestellar core, 
considering the nonideal MHD effects of the ambipolar diffusion that removes the magnetic field from the gas. 
The temperature was calculated with a flux-limited diffusion scheme for the radiative transfer. 
We followed the evolution during the first 100 kyr after protoplanetary disk formation, 
and we performed high-resolution restarts in parallel to study the numerical convergence as well as the detailed interior structure of the disk. 
This is one of the first works to try to self-consistently form a protoplanetary disk and resolve the interior structure at the same time. 
Nonetheless, we recall that the numerical convergence is not strictly demonstrated in this work and challenges remain.

The collapse problem is known to be computationally challenging because the dynamical ranges are wide. 
For this reason, the formation of the protoplanetary disk has rarely been followed for a long duration. 
On the other hand, the detailed structure of the disk interior is usually studied with a prescribed disk 
or a shearing box that contains a small region of the disk. 
In this study, we have followed the evolution of the disk during a relatively long period of time at 1 AU resolution. 
The disk is formed self-consistently from a prestellar core that is a few thousand AU in radius. 
This compromise in resolution allowed us to reach more evolved stages within a reasonable computation time. 
In order to study the detailed interior dynamics of the disk, 
we selected snapshots from the canonical simulation and reran with increased resolution down to 0.06 AU. 

From this work, we were able to follow the time evolution of disk global properties such as its mass and size. 
The disk is established roughly 60 kyr after the beginning of the simulation and grows slowly in size. 
The mass accretion rate is a decreasing function of time, and the disk gradually approaches an isolated state. 
The global appearance (density structure) of the disk is quasi-stationary and has mass $1.5~\%$  of the stellar mass, 
which is massive enough with respect to the star to form planets. 
However, the dynamics is likely to be predominantly governed by infall from the envelope, which inherits sometimes irregularities from the prestellar core. 
As a result, the transport of matter and angular momentum is mostly laminar and regulated by the magnetized infall. 
The $\alpha-$disk model that describes the turbulent transport, linked to the local surface density and temperature, is probably not applicable for a disk in the embedded phase. 

Because of the small size of the disk (<25 AU) and the strong MHD turbulence in the outer regions, 
all planets must form within $\sim 30$ AU ($R_{\rm kep}$). 
This is qualitatively consistent with the modern view of S. S. formation,  
where all planets are suggested to have formed well inside 20 AU \citep[e.g.,][]{Raymond17, Morbidelli18}
and most planetesimals may have accreted up to  30 AU \citep{Gomes04, Nesvorny13, Nesvorny18}. 
The temperature profile evolution can also be extracted and allows determining the migration of the snow line. 
The radial temperature profile is consistent with a power-law exponent between -1 and -0.5. This is somewhere between a purely irradiated disk and a viscously heated disk. However, the physical origin of this profile is likely the radiative diffusion in the disk radial direction instead of the aforementioned mechanisms.
The snow line starts at about 10 AU and ends at about 5 AU after about 100 kyr evolution. This is much farther away than commonly assumed for isolated and passively irradiated disks. 
This result should be regarded with care because it might have some numerical origins. First of all, the flux-limited diffusion scheme method treats only infrared radiation and thus could trap more energy than it should in reality. Second, the temperature is highly sensitive to the density (optical depth) of the disk, which in turn depends on the inner boundary conditions of the disk. 
Moreover, we recall that the stellar mass has only reached $\sim 0.3~\Ms$ in this study. Initial conditions with higher mass should be explored further.

The disk has an inner part with a shallow power-law surface density profile ($\gtrsim -1$ exponent) and is almost in Keplerian rotation. 
While the inner part is warm and almost demagnetized, an outer magnetized zone links the disk to the infalling envelope. 
The size of this zone varies significantly in time, 
from a few AU to almost comparable to the radius of the disk. 
The pattern of mass infall from the envelope is much more complicated than that described by the classical models that conserve angular momentum. 
The infall arrives mostly from a channel at altitude $\gtrsim 3H$, 
reaching the interior of the disk without penetrating the disk surface at large radius, which was also suggested by \citet{Kuffmeier18}. 
The midplane of the disk, on the other hand, can sometimes expand. 
The expanding material from the disk midplane sometimes exceeds the Keplerian radius, $R_{\rm kep}$, of the disk and meets the infalling material in the magnetized zone between $R_{\rm kep}$ and $R_{\rm mag}$. 
Some material is pushed up vertically and falls down again through the disk surface at a smaller radius. 
This might generate some circulation of material, 
which remains to be verified using tracer particles or passive scalar fields. 
In particular, the source function evaluated from our simulations is much more centrally peaked than in classical models, 
and most of the mass that finally rests inside the star does not transit through the whole disk.

\begin{acknowledgements}
We thank the referee, Enrique Vazquez-Semadeni, for thorough comments that have significantly improved the manuscript.
This work was granted access to HPC
   resources of CINES under the allocation x2014047023 made by GENCI (Grand
   Equipement National de Calcul Intensif). 
Y.-N. Lee acknowledges the financial support of the UnivEarthS Labex program at Sorbonne Paris Cit\'e (ANR-10-LABX-0023 and ANR-11-IDEX-0005-02), MOST (108-2636-M-003-001, 109-2636-M-003-001), and the MOE Yushan Young Scholar program. 
\end{acknowledgements}

\bibliographystyle{aa}
\bibliography{disc}

\appendix
\section{Identification of the disk}\label{ap_id}
Because we are interested in the interior properties of the disk as well as in its accretion onto the central star, 
we chose the sink particle as the center of the star-disk system. 
Despite some non-axisymmetric oscillation of the disk, 
the gravitational effects from the central star cancel out when calculating azimuthally averaged quantities, 
$\alpha$ in particular. 
This allows a discussion of the disk properties in coherence with the axisymmetric disk models (which is the case for almost all existing models).
The disk is identified in two steps. 
The first step is based on pure density thresholding, 
and the second step is more refined and takes the radial and vertical mass distributions of the disk into account (see Sect. \ref{st_result}). 

The disk is significantly marked by a density jump, which is used as a simple first identification. 
The rotating dense disk is surrounded by a collapsing diffuse envelope. 
The procedure is as follows:
\begin{itemize}
\item[1.]Calculate the inverse cumulative density function (CDF) of mass against density, that is, 
the amount of mass that is at a density above a certain value, 
\begin{align}
\mathcal{M}(\rho) = \int_\rho^\infty dm.
\end{align}
This results in a decreasing function of $\rho$ that has a plateau, which corresponds to the density jump between the disk and the envelope. 

\item[2.]Define the disk threshold $\rho_{\rm th}$ density as the point where the derivative of the CDF is highest, that is, where the CDF is flattest. This writes 
\begin{align}
\rho_{\rm th} = \arg \max d\log\mathcal{M}(\rho) / d\log \rho.
\end{align}

\item[3.]Select all the cells that have density $\rho \ge  \rho_{\rm th}$.
\item[4.]Calculate the angular momentum of these cells with respect to the sink particle and define this as the disk axis.
\end{itemize}

A simple density thresholding is robust to the turbulent fluctuations for most of the time, but not always. 
As the high-density regions dominate the mass inside the disk, 
the rotation axis we defined is robust to slightly different definitions of the disk.
The rotation in the initial condition is aligned to the $-z$ axis, and the axis of the disk is very close to this direction. 

\section{Disk vertical temperature distribution}\label{ap_Tz}
\begin{figure}[]
\centering
\includegraphics[trim=0 0 0 0,clip,width=0.5\textwidth]{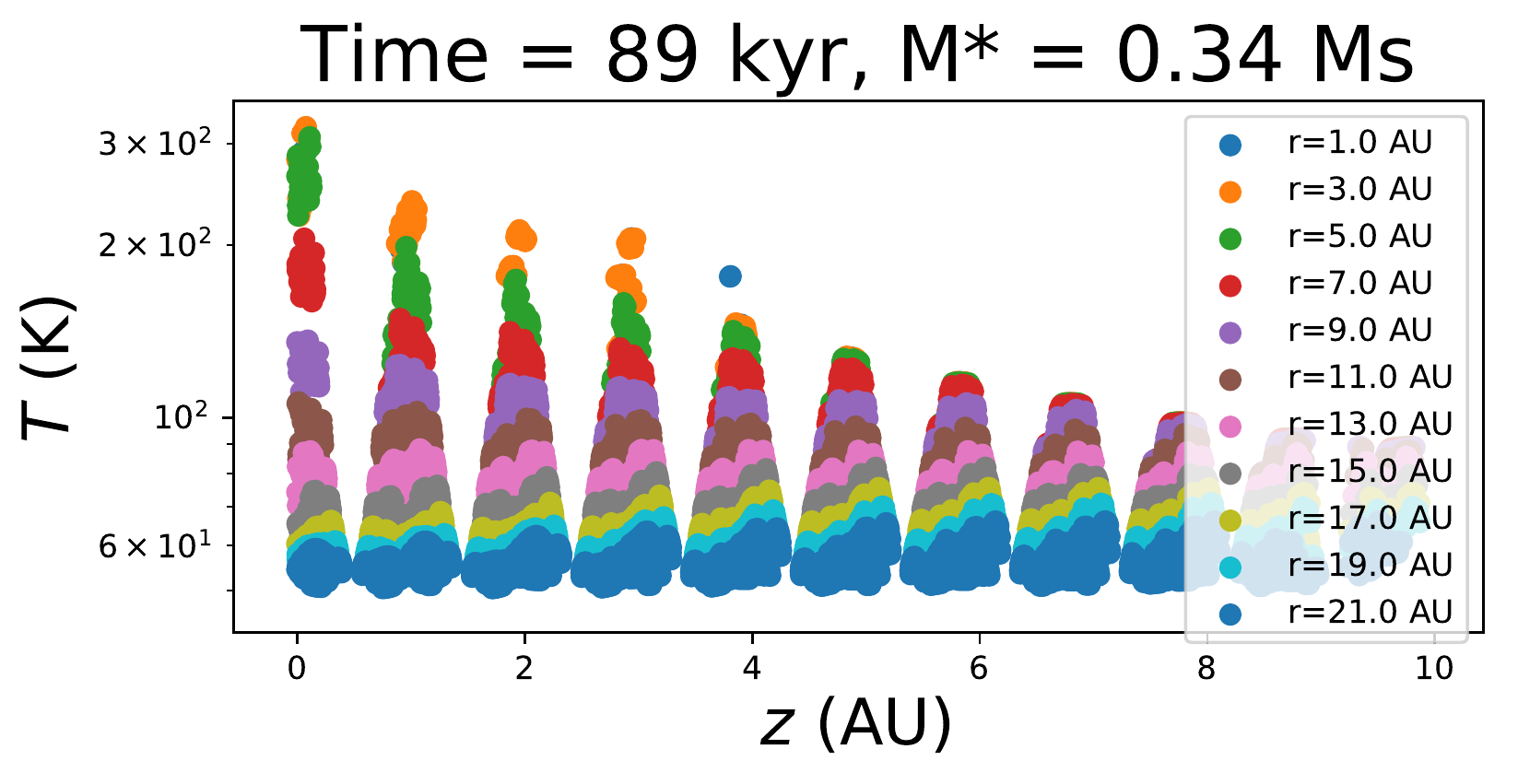}
\caption{Temperature plotted against altitude, $z$, at various distance to the central star, $r$, from R\_$\ell$14 (the disk aspect ratio is $\sim 0.1$ (Sect. \ref{st_height}). Every dot corresponds to one cell in the simulation, without any averaging. The discrete distribution comes from the slight difference between the grid direction and the selected disk axis.}
\label{fig_T_z}
\end{figure}

Figure \ref{fig_T_z} shows and example of disk vertical temperature distribution, where the temperature from each cell is plotted directly. Within a few times the scale height, the disk is basically isothermal, and decrease in temperature is only significant far away from the disk midplane. Vertical temperature variation is more clearly seen at small radii, that is, within 7 AU, where the vertical structure of the disk is not well resolved, however. The temperature dispersion among cells at same radial distance is sometimes more prominent than the vertical variation. 

\section{Stress tensors linked to the transport}\label{ap_transport}
Here we self-coherently define the stress tensors used in this work to avoid confusion. 
Following \citet{Balbus98}, the angular momentum conservation averaged in azimuthal direction becomes
\begin{align}\label{eq_angmom}
{\partial \over \partial t} \langle \rho r^2 \Omega \rangle_\phi = \nabla \cdot r [ \langle \rho r \Omega \vec{u}_{\vec{p}} \rangle_\phi + \vec{T} ] = 0,
\end{align}
where $\vec{T}$ is the poloidal stress tensor, 
\begin{align}
\vec{T} = \langle \rho u_\phi \vec{u}_{\vec{p}} - B_\phi \vec{B}_{\vec{p}} /4\pi \rangle_\phi,
\end{align}
and $\vec{W}$ = $\vec{T} /   \langle \rho \rangle_\phi$. 
Instead of integrating over the whole vertical domain, we integrated over the disk scale height $[-H, H]$, and defined a mass-weighted mean
\begin{align}
\langle X \rangle_\rho = {1 \over 2\pi \Sigma_H \Delta r} \int_{-H}^H \int_{r-\Delta r/2}^{r+\Delta r/2} \int_0^{2\pi} \rho X d\phi dr dz,
\end{align}
where $\Delta r$ is a radial averaging scale chosen such that the profile is not significantly affected by fluctuations. 
The choice of $\Delta r = dx$ is reasonable for most variables, while for the stress tensors, higher values are used, and this is discussed in Appendix \ref{ap_alpha_dx}. 
It is easily shown from this definition that the disk surface density within a the thickness $\pm H$, $\Sigma_H$, 
can be found by calculating $\langle 1 \rangle_\rho=1$. 

Based on this averaging, the equation of mass conservation becomes
\begin{align}\label{eq_mean_mass}
{\partial \Sigma_H \over \partial t} + {1\over r} {\partial (r\Sigma_H \langle u_r \rangle_\rho ) \over \partial r}  
+ \langle \rho u_z \rangle_\phi^H -  \langle \rho u_z \rangle_\phi^{-H}= 0,
\end{align}
where the superscript $\pm H$ denotes the values at $\pm H$. 
The second term is the radial mass flux, 
and the last two terms are the flux across the disk surfaces.

Likewise, Eq. (\ref{eq_angmom}) becomes
\begin{align}\label{eq_mean_angmom}
&{\partial  \over \partial t} \left(\Sigma_H r^2 \langle \Omega \rangle_\rho \right) 
+ {1\over r} {\partial \over \partial r}  \left(r^3\Sigma_H \langle \Omega u_r \rangle_\rho + r^2 \Sigma_H \langle W_{r\phi} \rangle_H \right) \\
& + r^2   \left(\Omega^H \langle \rho u_z \rangle_\phi^H - \Omega^{-H} \langle \rho  u_z \rangle_\phi^{-H}\right) 
+ r \left(\langle \rho W_{z\phi} \rangle_\phi^H -  \langle \rho W_{z\phi} \rangle_\phi^{-H}\right)=0, 
\nonumber
\end{align}
where we also define a vertical density-weighted average as
\begin{align}
\langle X \rangle_H = {1 \over \Sigma_H \Delta r} \int_{-H}^H \int_{r-\Delta r/2}^{r+\Delta r/2}  \langle\rho\rangle_\phi X dr dz.
\end{align}
The second term in the first line contains the the radial transport due to the laminar flow and the stress tensors, 
while the second line corresponds to the same contributions across the disk surface. 

Assuming a stationary disk, we ignored the time derivatives. 
Combining Eqs. (\ref{eq_mean_mass}) and (\ref{eq_mean_angmom}) and assuming that the disk is symmetric with respect to the midplane ($\Omega^H=\Omega^{-H}$, we recall that $\Omega=\langle u_\phi \rangle_\phi / r$ by definition), 
we can derive
\begin{align}\label{eq_disk_stat}
&\Sigma_H \langle u_r \rangle_\rho {\partial \over \partial r}  \left( r^2 \Omega^H \right) 
+  {1 \over r} {\partial \over \partial r}  \left( r^2 \Sigma_H \langle W_{r\phi} \rangle_H \right)  \\
&+  r \left(\langle \rho W_{z\phi} \rangle_\phi^H -  \langle \rho W_{z\phi} \rangle_\phi^{-H}\right) 
- {1 \over r} {\partial \over \partial r} \left( r^3 \Sigma_H \left\langle {\partial \Omega \over \partial z} \int \rho u_r dz \right\rangle_\rho \right) = 0. \nonumber 
\end{align}
The last two terms disappear when the integration is performed over the whole vertical extent with vanishing boundary conditions and the rotation profile is assumed vertically invariant. 

The radial mass accretion rate within the disk
\begin{align}
\dot{M}(r) = - 2\pi R \Sigma_H \langle u_r \rangle_\rho
\end{align}
no longer has a simple expression. 
We limited our discussion to the vertical extent where the disk is practically Keplerian, and thus the last term in Eq. (\ref{eq_disk_stat}) can be neglected. 
We can then infer Eq. (\ref{eq_Mdot_alpha}).

There are two contributions to the mass accretion due to angular momentum removal. 
The first is in the radial direction and the contribution is integrated across the disk thickness, 
for which we can define a mean $\alpha$ value such that 
\begin{align}
\alpha_r =  {\langle W_{r\phi} \rangle_H \over  \langle c_{\rm s}^2 \rangle_\rho} = {\hat{T}_{r\phi}^H \over \rho_H \langle c_{\rm s}^2 \rangle_\rho},
\end{align} 
where $\rho_H = \Sigma_H / (2H)$ is the disk vertical mean density.
The second contribution comes from the angular momentum that is lost due to vertical transport across both surfaces of the disk, 
and we define the corresponding $\alpha$ as 
\begin{align}
\alpha_z = {(T_{z\phi}^H -  T_{z\phi}^{-H})  \over  \rho_H \langle c_{\rm s}^2 \rangle_\rho }  = 
{(\langle \rho \rangle_\phi^H W_{z\phi}^H -  \langle \rho \rangle_\phi^{-H}W_{z\phi}^{-H})  \over  \rho_H \langle c_{\rm s}^2 \rangle_\rho }. 
\end{align}
The normalization is always performed with respect to the bulk-averaged pressure. 

\section{Numerical discretizing effects on the evaluation of the stress tensor}\label{ap_alpha_dx}
The azimuthally averaged stress tensor of the velocity is defined locally in a narrow range of radius $\Delta r$ around position $r$, 
such that rapid radial fluctuations are smoothed out, 
\begin{align}
W_{\vec{p}\phi,\rm loc} ={1\over \epsilon} \int_{r-\epsilon/2}^{r+\epsilon/2} \langle \delta u_{\vec{p}} \delta u_\phi \rangle_\phi dr, 
\end{align}
where $\epsilon$ is a minimum value such that the function is smooth. 
When numerically evaluating the stress tensor, this average is performed within a practically larger region. 
We discuss here the effects of discretized calculation on the stress tensor. 
When calculating the stress tensor in a larger finite region, there is an additional velocity difference with respect to the mean velocity in the whole region. 
We have thus
\begin{align}
W_{\vec{p} \phi, \rm fin} &= {1\over \Delta r}  \int_{r-\Delta r/2}^{r+\Delta r/2}\langle (\Delta u_{\vec{p}} +\delta u_{\vec{p}}) (\Delta u_\phi + \delta u_\phi) \rangle_\phi dr\\
&= \overline{W}_{\vec{p}\phi, \rm loc} +{1\over \Delta r}  \int_{r-\Delta r/2}^{r+\Delta r/2} \langle \Delta u_{\vec{p}} \Delta u_\phi  \rangle_\phi dr \nonumber\\
&= \overline{W}_{\vec{p}\phi, \rm loc} + W_{\vec{p}\phi, \rm cor}, \nonumber
\end{align}
where $\Delta$ specifies the difference between the mean value at $r$ and the mean value averaged in $[-\Delta r/2, \Delta r/2]$, and $\overline{\cdot}$ stands for the radially averaged value. 
Below we perform a calculation with some simplifying assumptions to estimate the effect of averaging in a finite region on the radial component of the tensor. 

First of all, we assumed Keplerian rotation such that the azimuthal average gives
\begin{align}
\langle u_\phi \rangle_\phi (r)  = r\Omega (r) = \sqrt{GM_\ast \over r}.
\end{align}
In the radial direction, the accretion velocity is given by
\begin{align}
\langle u_r \rangle_\phi (r)  =- { (r^2\Sigma W_{r\phi, \rm loc})^\prime  \over  r\Sigma(r^2\Omega)^\prime}
\sim - \eta_1 {W_{r\phi, \rm loc} \over r\Omega} = - \eta_1 W_{r\phi, \rm loc}  \sqrt{r\over GM_\ast},
\end{align}
where $\eta_1$ is a factor on the order of unity from the unknown profiles of $\Sigma$ and $W_{r\phi, \rm loc}$.
By assuming the average inside $\Delta r$ is the value at $r$,  
we can calculate
\begin{align}
\Delta u_r \Delta u_\phi  
&= -\eta_1 W_{r\phi, \rm loc} \left(\sqrt{r+\Delta r/2}-\sqrt{r}\right) \left( {1\over\sqrt{r+\Delta r/2}} - {1\over \sqrt{r}}\right) \\
&= -\eta_1 W_{r\phi, \rm loc} \left(2 - {2r+\Delta r/2 \over \sqrt{(r+\Delta r/2)r}}\right). \nonumber
\end{align}
To calculate the average in the whole region,
we assumed that  $W_{r\phi, \rm loc} (r)$ and $\Sigma (r)$ vary slowly and can be locally approximated as a constant.
We also neglected the disk geometry and performed  a linear integration along $r$. 
The correction term for averaged stress tensor calculated in a non-negligible finite region of $[-\Delta r/2, \Delta r/2]$ is thus
\begin{align}
W_{r\phi, \rm cor} &= {1\over \Delta r}  \int_{r-\Delta r/2}^{r+\Delta r/2}\langle \Delta u_r \Delta u_\phi \rangle_\phi dr\\
& = {\eta_2  \over \Delta r} \int\limits_{-\Delta r/2}^{\Delta r/2} \langle \Delta^\prime u_r \Delta^\prime u_\phi \rangle d\Delta^\prime r \nonumber \\
&= \eta_1\eta_2 W_{r\phi, \rm loc}  \left[ {1\over 48} \left({\Delta r\over r}\right)^2 -  {1\over 128} \left({\Delta r\over r}\right)^3  +  \mathcal{O}\left({\Delta r\over r}\right)^4\right], \nonumber 
\end{align}
where $\eta_2$ is also a factor on the order of unity from the simplifications. 
The stress tensor that we measure from the simulations is therefore
\begin{align}
W_{r\phi, \rm fin}  = \overline{W}_{r\phi, \rm loc} + W_{r\phi, \rm cor} 
\approx \overline{W}_{r\phi, \rm loc} \left[1+{\eta_1\eta_2\over 48}\left({\Delta r\over r}\right)^2 \right].
\end{align}
As long as $\Delta r/ r$ is small and $W_{r\phi, \rm loc}$ varies slowly, the correction term remains negligible and does not affect the estimation of the stress tensor. 
Nonetheless, the choice of $\Delta r/ r$ still has to be reasonable such that the simplifying assumptions stay valid.  

\begin{figure}[]
\centering
\includegraphics[trim=0 0 0 0,clip,width=0.5\textwidth]{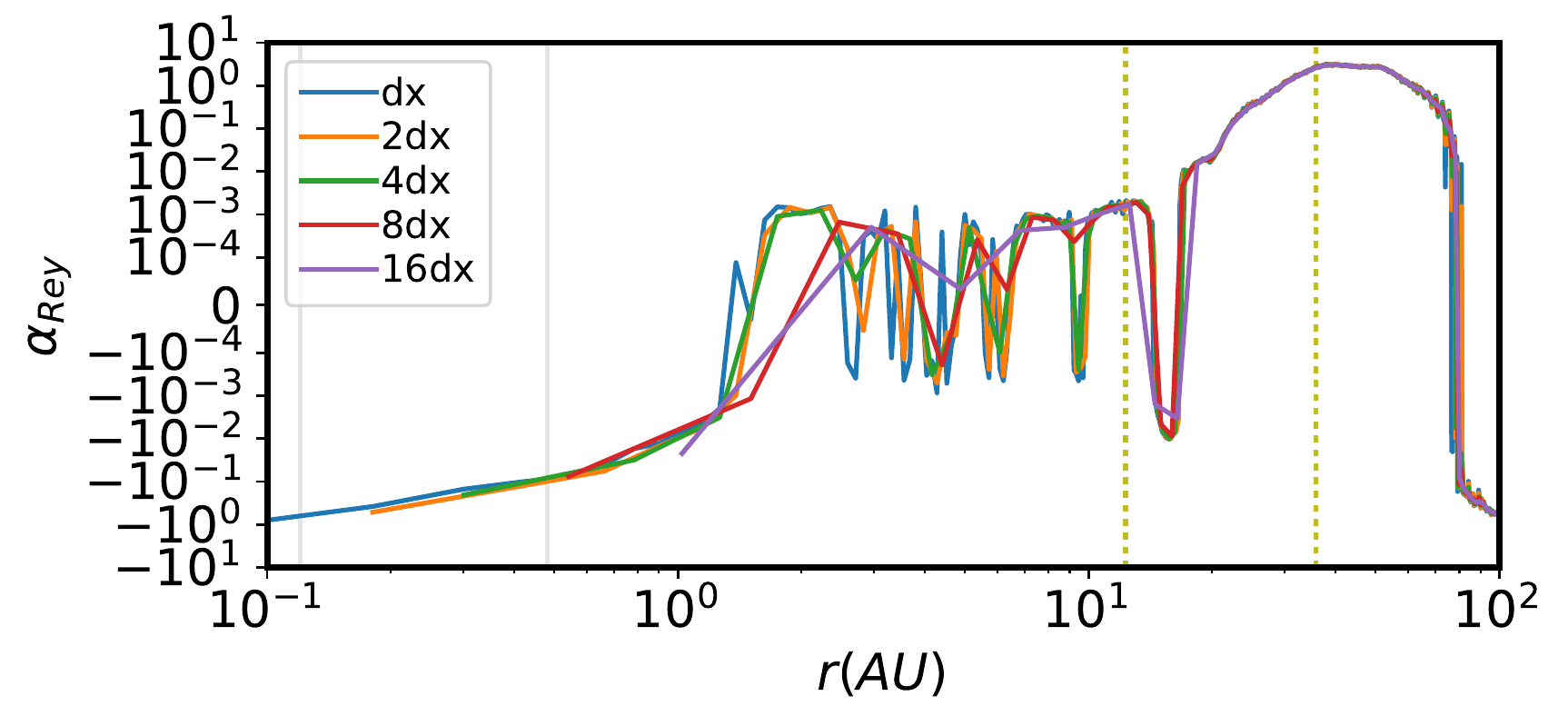}
\includegraphics[trim=0 0 0 0,clip,width=0.5\textwidth]{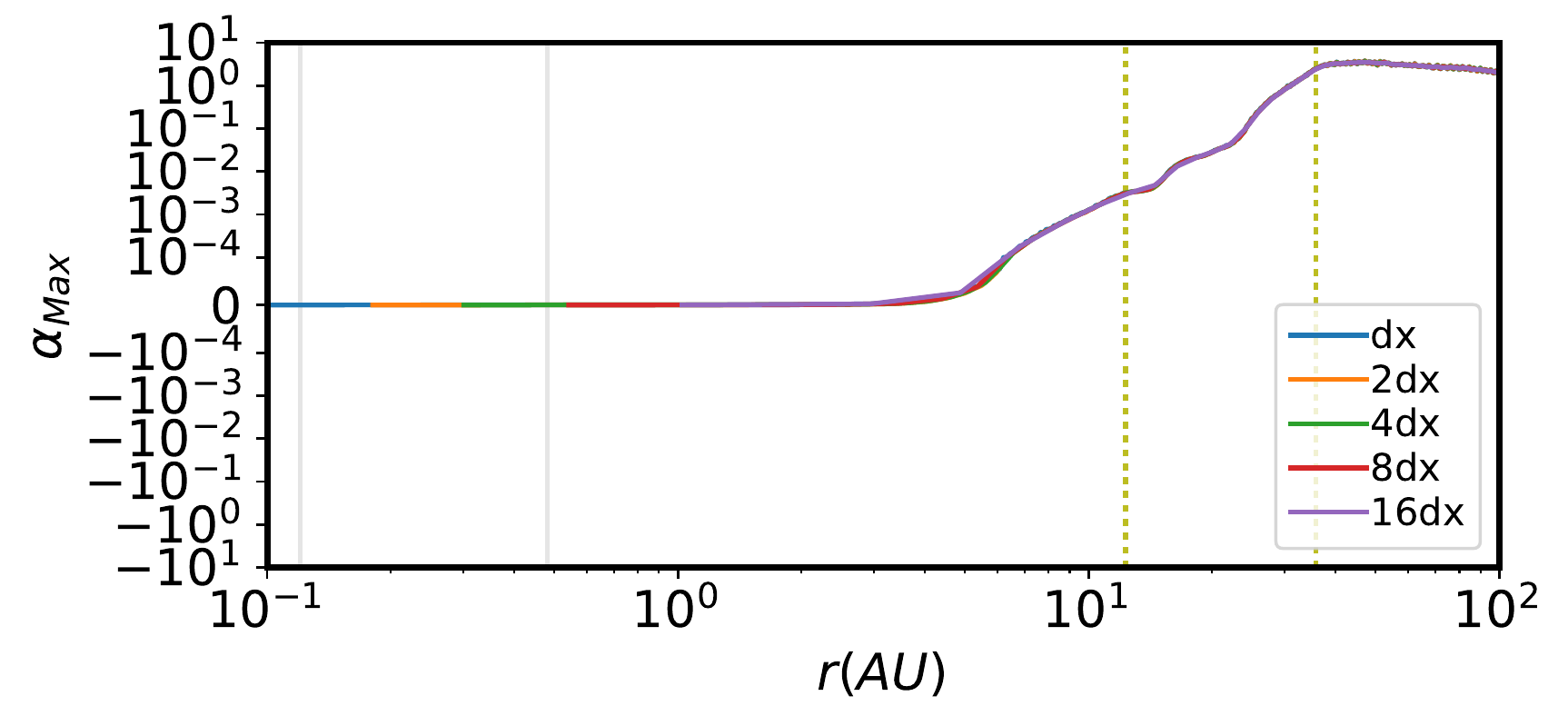}
\includegraphics[trim=0 0 0 0,clip,width=0.5\textwidth]{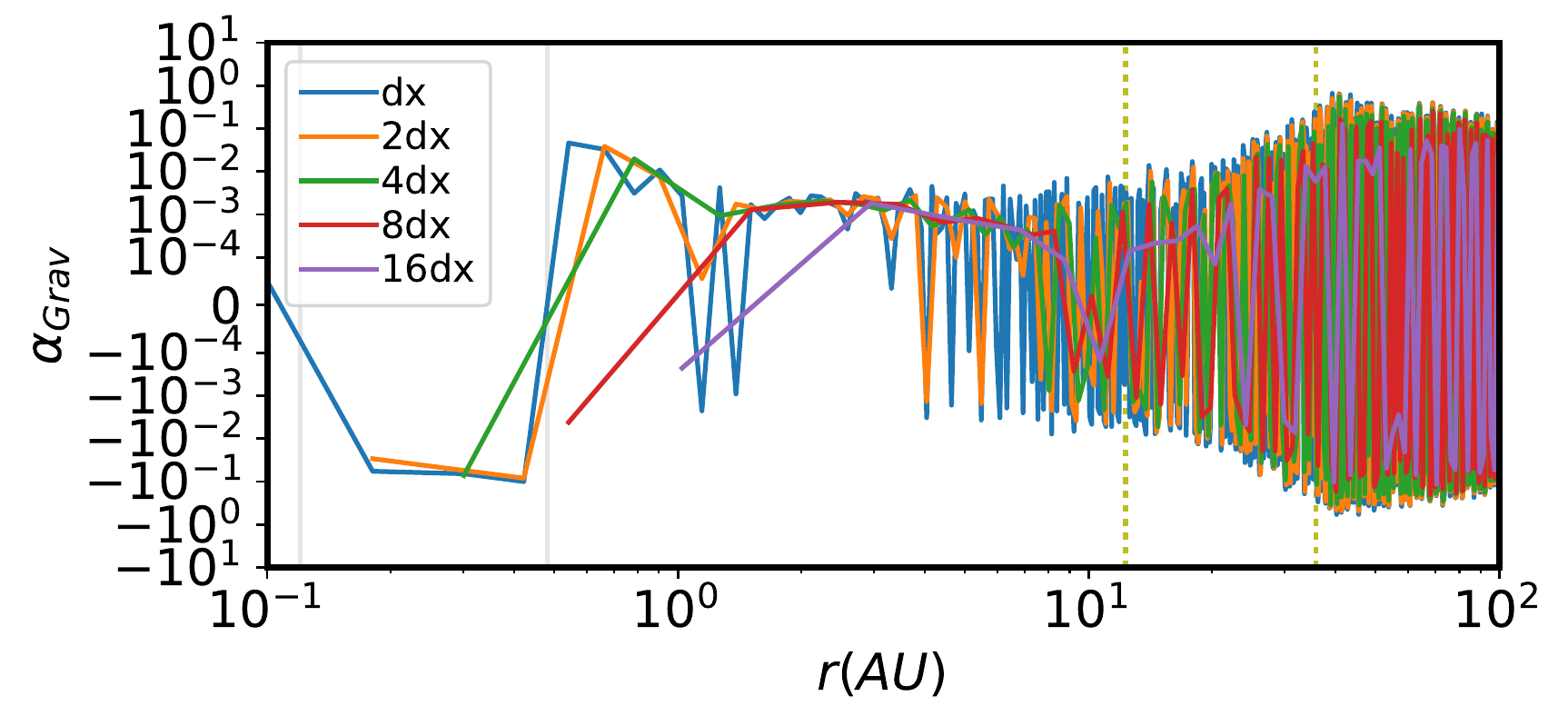}
\caption{{\it Top:} Reynods $\alpha$ calculated at various bin sizes in $r$. The function fluctuates strongly at small bin sizes and becomes smooth for bin sizes $\ge 4dx$. {\it Bottom:} Gravitational $\alpha$ calculated at various bin sizes in $r$. The function fluctuates strongly between positive and negative values regardless of the increased bin size. It is impossible to measure the gravitational $\alpha$ in the simulations in this study, probably because of Cartesian discretization. The resolution $dx = 0.12$ AU.}
\label{fig_alpha_bins}
\end{figure}

When the $\alpha$ values are measured from our simulations, the size of the binning has to be carefully chosen. 
As shown in Fig. \ref{fig_alpha_bins} (top panel), 
the value of the Reynolds $\alpha$ fluctuates strongly when the bin size in $r$ is taken to be the smallest cell size. 
Nonetheless, by increasing the bin size, the measured $\alpha$ converges toward a smooth function. 
This function does not seem to be dependent of the bin size as long as the bin size $\ge 4dx$, confirming the above discussions. 

On the other hand, this does not seem to be true for the gravitational $\alpha$. 
The measurement remains noisy regardless of the increased bin size. 
This might be a result of noise introduced by the Cartesian discretization for the disk system.

\end{document}